\documentclass[final,5p,times]{elsarticleMod}
\usepackage{amsmath,array}
\usepackage{amsfonts}
\usepackage{amssymb}
\usepackage{blindtext}
\usepackage{eqexpl}
\usepackage{float}
\usepackage{lineno}
\usepackage{graphicx}
\usepackage{caption}
\usepackage{physics}
\usepackage{subfig}
\usepackage{mathtools,minitoc}
\usepackage{multirow,multicol}
\usepackage{chngcntr}
\usepackage[dvipsnames]{xcolor}
\usepackage[colorlinks=true, linkcolor = blue]{hyperref} 
\hypersetup{
pdftitle={Vector Boson Scattering Processes: Status and Prospects -- Reviews in Physics 8 (2022) 100071 -- [arXiv:2106.01393]}, pdfauthor={Buarque Franzosi, Gallinaro, Ruiz, et al}, 
colorlinks=true, linkcolor=ForestGreen, citecolor=ForestGreen,
filecolor=ForestGreen, urlcolor=ForestGreen}

\graphicspath{{Plots/}}

\newcommand{\bea}{\begin{eqnarray}}
\newcommand{\eea}{\end{eqnarray}}

\counterwithin{figure}{section}
\counterwithin{equation}{section}
\counterwithin{table}{section}
\newcommand{\ie}{{\em i.e.}}
\newcommand{\eg}{{\em e.g.}}

\eqexplSetIntro{where:} 
\eqexplSetDelim{=} 

\newcommand{\invfb}{~\text{fb}^{-1}}

\newcommand{\ZZ}{\ensuremath{ZZ}}
\newcommand{\WZ}{\ensuremath{WZ}}
\newcommand{\WW}{\ensuremath{WW}}
\newcommand{\SSWW}{\ensuremath{W^{\pm}W^{\pm}}}
\newcommand{\Wg}{\ensuremath{W\gamma}}
\newcommand{\Zg}{\ensuremath{Z\gamma}}
\newcommand{\calO}{\mathcal{O}} 
\newcommand{\mm}{\mu^+\mu^-}
\newcommand{\LambdaEW}{\Lambda_{\textrm{EW}}}
\newcommand{\LambdaQCD}{\Lambda_{\textrm{QCD}}}
\newcommand{\muEW}{\mu_{\textrm{EW}}}
\newcommand{\muQCD}{\mu_{\textrm{QCD}}}
\newcommand{\lsea}{\ell_{\textrm{sea}}}
\newcommand{\lval}{\ell_{\textrm{val}}}
\newcommand{\mpm}{\mu^\pm}
\newcommand{\mpmm}{\mu^+\mu^-}
\newcommand{\sss}{\scriptscriptstyle}
\newcommand{\OO}{\ensuremath{\mathcal{O}}}
\newcommand{\Op}[1]{\OO_{\sss #1}}
\newcommand{\pdp}{\ensuremath{\varphi^\dagger\varphi}}

\begin{document}
\begin{frontmatter}
\title{Vector Boson Scattering Processes: Status and Prospects}

\author[gothenburg,chalmers]{Diogo Buarque Franzosi (ed.)}
\ead{diogo.buarque.franzosi@fysik.su.se}
\author[lip]{Michele Gallinaro (ed.)}
\ead{michgall@cern.ch}
\author[ifj]{Richard Ruiz (ed.)}
\ead{rruiz@ifj.edu.pl}

\author[cern]{Thea K. Aarrestad}
\author[pav]{Mauro Chiesa}
\author[milano]{Flavia Cetorelli}
\author[uclouvain]{Antonio Costantini}
\author[wue]{Ansgar Denner}
\author[fr]{Stefan Dittmaier}
\author[wue]{Robert Franken}
\author[milano]{Pietro Govoni}
\author[pit]{Tao Han}
\author[durham]{Ashutosh~V.~Kotwal}
\author[sichuan]{Jinmian Li}
\author[shef]{Kristin Lohwasser}
\author[cern]{Kenneth Long}
\author[pit]{Yang Ma}
\author[uclouvain]{Luca Mantani}
\author[eth]{Matteo Marchegiani}
\author[fr]{Mathieu Pellen}
\author[wue]{Giovanni Pelliccioli}
\author[oxford]{Karolos Potamianos}
\author[desy]{J\"urgen Reuter}
\author[wue]{Timo Schmidt}
\author[mil]{Christopher Schwan}
\author[warsaw]{Micha{\l} Szleper}
\author[ucl]{Rob Verheyen}
\author[pit]{Keping Xie}
\author[sichuan]{Rao Zhang}

\address[durham]{Department of Physics, Duke University, Durham, NC 27708, USA }
\address[desy]{Deutsches Elektronen-Synchrotron (DESY) Theory Group, Notkestr. 85, D-22607 Hamburg, Germany}
\address[cern]{European Organization for Nuclear Research (CERN) 
  CH-1211 Geneva 23, Switzerland}
\address[chalmers]{Department of Physics, Chalmers University of Technology,
Fysikg{\aa}rden 1, 41296 G\"oteborg, Sweden}
\address[eth]{Swiss Federal Institute of Technology (ETH) Z{\"u}rich, Otto-Stern-Weg 5, 8093 Z{\"u}rich, Switzerland}
\address[fr]{Universit\"at Freiburg,        Physikalisches Institut,
        Hermann-Herder-Stra\ss{}e 3,        79104 Freiburg,        Germany}
\address[gothenburg]{Physics Department, University of Gothenburg, 41296 G\"oteborg, Sweden}
\address[lip]{Laborat\'orio de Instrumenta\c{c}\~{a}o e F\'isica Experimental de Part\'iculas (LIP), Lisbon, Av. Prof. Gama Pinto, 2 - 1649-003, Lisboa, Portugal}
\address[ifj]{Institute of Nuclear Physics, Polish Academy of Sciences, ul. Radzikowskiego, Cracow 31-342, Poland}
\address[ucl]{University College London, Gower St, Bloomsbury, London WC1E 6BT, United Kingdom}
\address[uclouvain]{Centre for Cosmology, Particle Physics and Phenomenology {\rm (CP3)},\\
Universit\'e Catholique de Louvain, Chemin du Cyclotron, B-1348 Louvain la Neuve, Belgium}
\address[milano]{Milano - Bicocca University and INFN, Piazza della Scienza 3, Milano, Italy}  
\address[mil]{Tif Lab, Dipartimento di Fisica, Universit\`a di Milano and INFN,
    Sezione di Milano,
    Via Celoria 16,
    20133 Milano,
    Italy}
\address[oxford]{Department of Physics, University of Oxford, Clarendon Laboratory, Parks Road, Oxford OX1 3PU, UK}    
\address[pav]{Dipartimento di Fisica, Universit\`a di Pavia and INFN, Sezione di Pavia, Via A. Bassi 6, 27100 Pavia, Italy}
\address[pit]{Pittsburgh Particle Physics, Astrophysics, and Cosmology Center, Department of Physics and Astronomy,\\
	 University of Pittsburgh, Pittsburgh, PA 15260, USA
}
\address[shef]{Department of Physics, Sheffield University, UK}
\address[sichuan]{College of Physics, Sichuan University, Chengdu 610065, China}
\address[warsaw]{National Center for Nuclear Research, ul.~Pasteura 7, 02-093 Warszawa, Poland}
\address[wue]{Universit\"at W\"urzburg, Institut f\"ur Theoretische Physik und Astrophysik,
  Emil-Hilb-Weg 22, 97074 W\"urzburg, Germany}

\journal{CP3-21-14, DESY-21-064, IFJPAN-IV-2021-8, PITT-PACC-2106, VBSCAN-PUB-04-21} 

\begin{abstract}
Insight into the electroweak (EW) and Higgs sectors can be achieved through measurements of vector boson scattering (VBS) processes. The scattering of EW bosons are rare processes that are precisely predicted in the Standard Model (SM) and are closely related to the Higgs mechanism. Modifications to VBS processes are also predicted in models of physics beyond the SM (BSM), for example through changes to the Higgs boson couplings to gauge bosons and the resonant production of new particles. In this review, experimental results and theoretical developments of VBS at the Large Hadron Collider, its high luminosity upgrade, and future colliders are presented.
\end{abstract}

\begin{keyword}
Standard Model, Beyond Standard Model, Vector Boson Scattering, Vector Boson Fusion, Colliders
\\
Journal: \href{https://doi.org/10.1016/j.revip.2022.100071}{Reviews in Physics 8 (2022) 100071}
\\
ArXiv: \href{https://arxiv.org/abs/2106.01393}{2106.01393}
\end{keyword}

\end{frontmatter}

\tableofcontents
\newpage
\part[Introduction]{Introduction}

The importance of the vector boson scattering (VBS) and vector boson fusion (VBF) processes in understanding the electroweak (EW) sector is well-established in high energy physics. At scattering energies far above the EW scale, the longitudinal modes of the weak bosons are manifestations of the Nambu-Goldstone bosons originating from the spontaneous breaking of EW symmetry. Probing their interactions therefore helps unveil the dynamics behind the Higgs mechanism~\cite{Lee:1977yc,Lee:1977eg,Chanowitz:1985hj,Chanowitz:1984ne}. Moreover, VBS is relevant to testing the gauge structure of EW interactions due to the contribution of quartic interactions and their interplay with trilinear couplings, which leads to potentially large gauge cancellations. These are unique features of VBS processes. It is thus extremely important to achieve precise predictions in the Standard Model (SM) of particle physics as well extensively investigate beyond-the-Standard Model (BSM) scenarios, both in the context of specific models and effective field theory (EFT) frameworks. The importance of VBS physics has also been well-documented in the literature as reviewed by Refs.~\cite{Rauch:2016pai,Green:2016trm,Alessandro:2018khj,Bellan:2019xpr,Baglio:2020bnc,Covarelli:2021gyz}.

Measurement of VBS processes are at last possible at the Large Hadron Collider (LHC), where the unprecedented energies and integrated luminosities of data sets allowed for the first observations of such rare interactions~\cite{Sirunyan:2017ret,Sirunyan:2017fvv,Aaboud:2018ddq,Sirunyan:2019ksz,Aaboud:2019nmv}. The Run~3 of the LHC's data taking is scheduled to start in 2022. And with the High-Luminosity extension (HL-LHC), the available data will enable the community to move beyond first observations and into the domain of precision measurements, aiming at the ambitious goal of isolating the transverse and longitudinal components of VBS. Future colliders can likewise improve this picture in several aspects, including direct exploration of when EW symmetry is approximately restored.

In light of the experimental, phenomenological, and theoretical advancements surrounding VBS, as well as the mandates of the 2020 European Strategy Update~\cite{Strategy:2019vxc,EuropeanStrategyGroup:2020pow}, it is timely to produce a picture of the current state-of-the art in the field with a focus on the instruments and concepts that will be needed for future developments. We start in Part \ref{part:LHC}, where current measurements and analysis techniques of VBS are summarized. In Part \ref{part:projections}, projections for the HL-LHC are reviewed for both the SM and BSM scenarios. In Part \ref{part:future}, the nature of VBS processes itself and prospects for measurements at future colliders are discussed. Finally, we conclude with an outlook in  Part \ref{part:final_conclusion}.

\part[VBS at the LHC]{VBS at the LHC}\label{part:LHC}

\section[Current results on vector boson scattering]{\large Current results on VBS from the LHC}\label{sec:lhc_results}

At the LHC, the ATLAS and CMS Collaborations first studied diboson production via VBS in the $\SSWW$, $W\gamma$, and $Z\gamma$ final states at $\sqrt{s}=8$ TeV in the LHC Run~1~\cite{Khachatryan:2014sta,Aad:2014zda,Aaboud:2016uuk,Khachatryan:2016vif,Aaboud:2016ffv,Khachatryan:2017jub,Aaboud:2017pds} using the leptonic decay modes of weak bosons. These states were further investigated during the first full year of data taking at 13 TeV, when the $\SSWW$, $WZ$, and $\ZZ$  channels were first observed~\cite{Sirunyan:2017ret,Sirunyan:2017fvv,Aaboud:2018ddq,Sirunyan:2019ksz,Aaboud:2019nmv}, also using the leptonic decay modes. The first study of VBS using semi-leptonic decays was also made with this data set, targeting $\WZ/\WW$ events where one of the boson decays to a boosted hadronic jet. A summary of these results can be found in Ref.~\cite{Gallinaro:2020cte} and references therein.

With the nearly $\mathcal{L}=140$ fb$^{-1}$ of data collected by each of the ATLAS and CMS collaborations during the LHC's second run of data-taking (Run~2), the experimental outlook for searches and measurements of VBS processes is very bright. The initial searches have rapidly been extended, with several results reaching high-statistical significance for observation and becoming measurements in their own right. In particular, the $WZ$ and $\SSWW$ processes have been observed and measured differentially~\cite{Sirunyan:2017ret,Aaboud:2019nmv}. The search for the VBS $\ZZ$ production with leptonic decays was also updated with $\mathcal{L}=137\invfb$ by CMS~\cite{Sirunyan:2019der,Sirunyan:2020alo}, exceeding the 3$\sigma$ threshold for evidence but not yet reaching a 5$\sigma$ observation. A 5$\sigma$ observation of $ZZ$ in the leptonic channels was achieved with $\mathcal{L}=139\invfb$ by ATLAS~\cite{Aad:2020zbq}. In addition, the first experimental search for polarized VBS production was performed in the {\SSWW} channel~\cite{Sirunyan:2020gvn}, a measurement widely considered a major priority for the HL-LHC. Additional results with $\mathcal{L}=137\invfb$ have been released~\cite{ATLAS:2018iui,CMS:2021slu} and others are expected soon.

The following summary briefly discusses the current state of the experimental studies of VBS processes. While some focus is placed on results obtained by the CMS collaboration, comparable findings have been achieved by ATLAS. It is worth stressing that having results from both experiments is of utmost importance for addressing the physics goals of the LHC program and the high-energy community as a whole. Further details can be found in the relevant references.

\subsection{Results on scattering of massive vector bosons}
\label{sec:vbsresults}

\begin{figure}[!t]
    \includegraphics[width=\columnwidth]{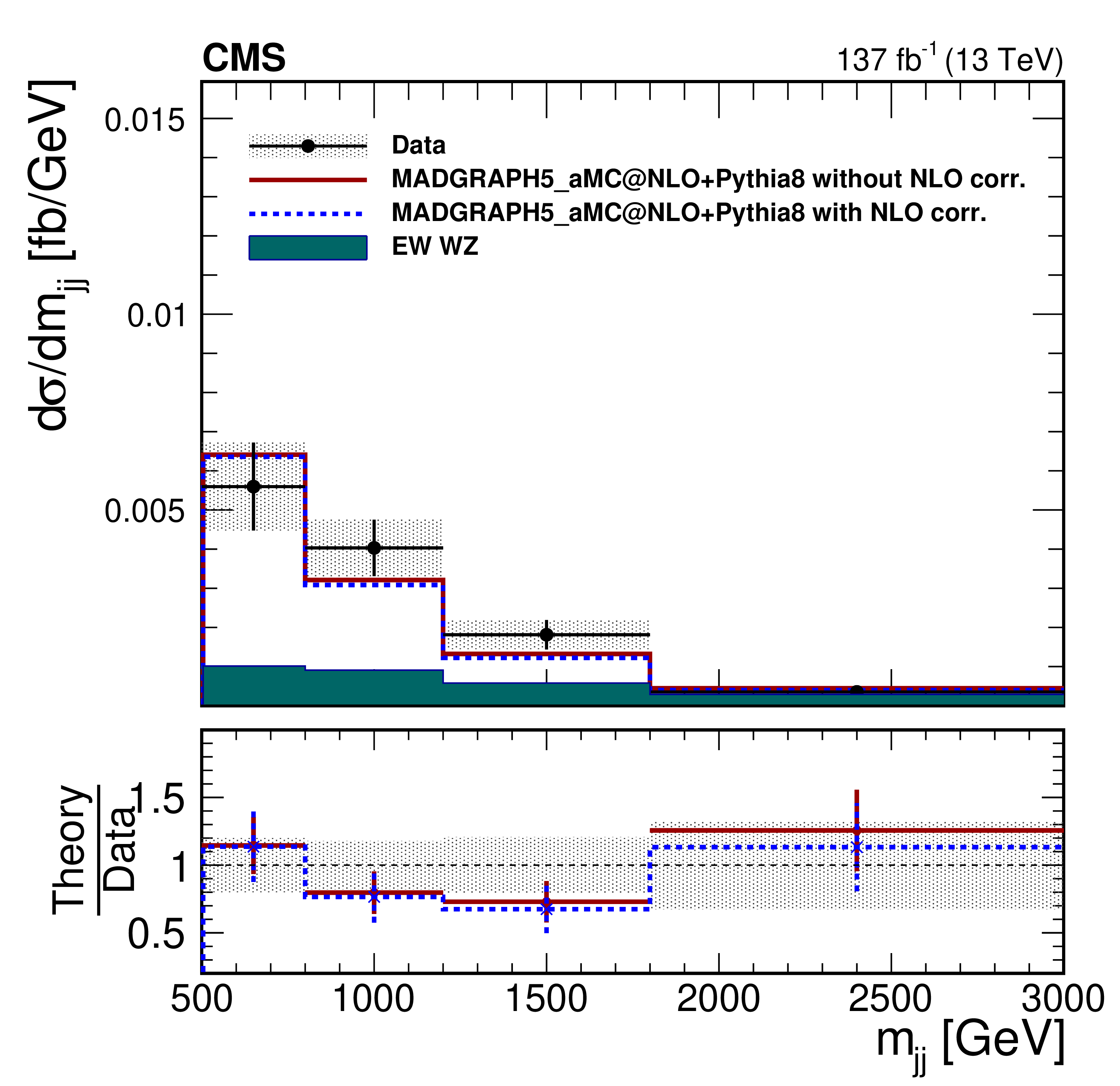}
    \caption{The unfolded results for the {\WZ{jj}} process from Ref.~\cite{Sirunyan:2020gyx}. The contribution of the EW production and the impact of NLO EW corrections are shown.}
    \label{fig:wwwz}
\end{figure}

Due to its distinctive signature and low background, the same-sign $\SSWW\to\SSWW\to\ell^\pm_i\ell^\pm_j\nu\nu$ process is the ``golden'' channel of VBS processes. Unlike other VBS modes, the quantum chromodynamics (QCD)-induced production of $\SSWW$ events is significantly subdominant with respect to the EW contribution. Therefore, the dominant backgrounds are VBS-like events from $WZ$ production (EW or QCD-induced) or events with non-prompt leptons imitating true same-sign lepton events. Consequently, VBS $WZ$ production is a significant background to the $\SSWW$ process. Because estimating the $WZ$ process as background is analogous to measuring the VBS $WZ$ process as signal, the CMS Collaboration has performed these measurements simultaneously, fully exploiting the nearly 140 fb$^{-1}$ as a tool to probe the SM~\cite{Sirunyan:2020gyx}. Similarly, ATLAS reports the observation of $WZ$ scattering in the $3\ell\nu$ channel with a statistical significance exceeding $5\sigma$ using 36 fb$^{-1}$ of 13 TeV data~\cite{ATLAS:2018mxa}; sensitivity to the semi-leptonic channel is presently at the $2.5\sigma$ level with the same data set~\cite{ATLAS:2019thr}.

\begin{figure*}[!t]
    \includegraphics[width=\columnwidth]{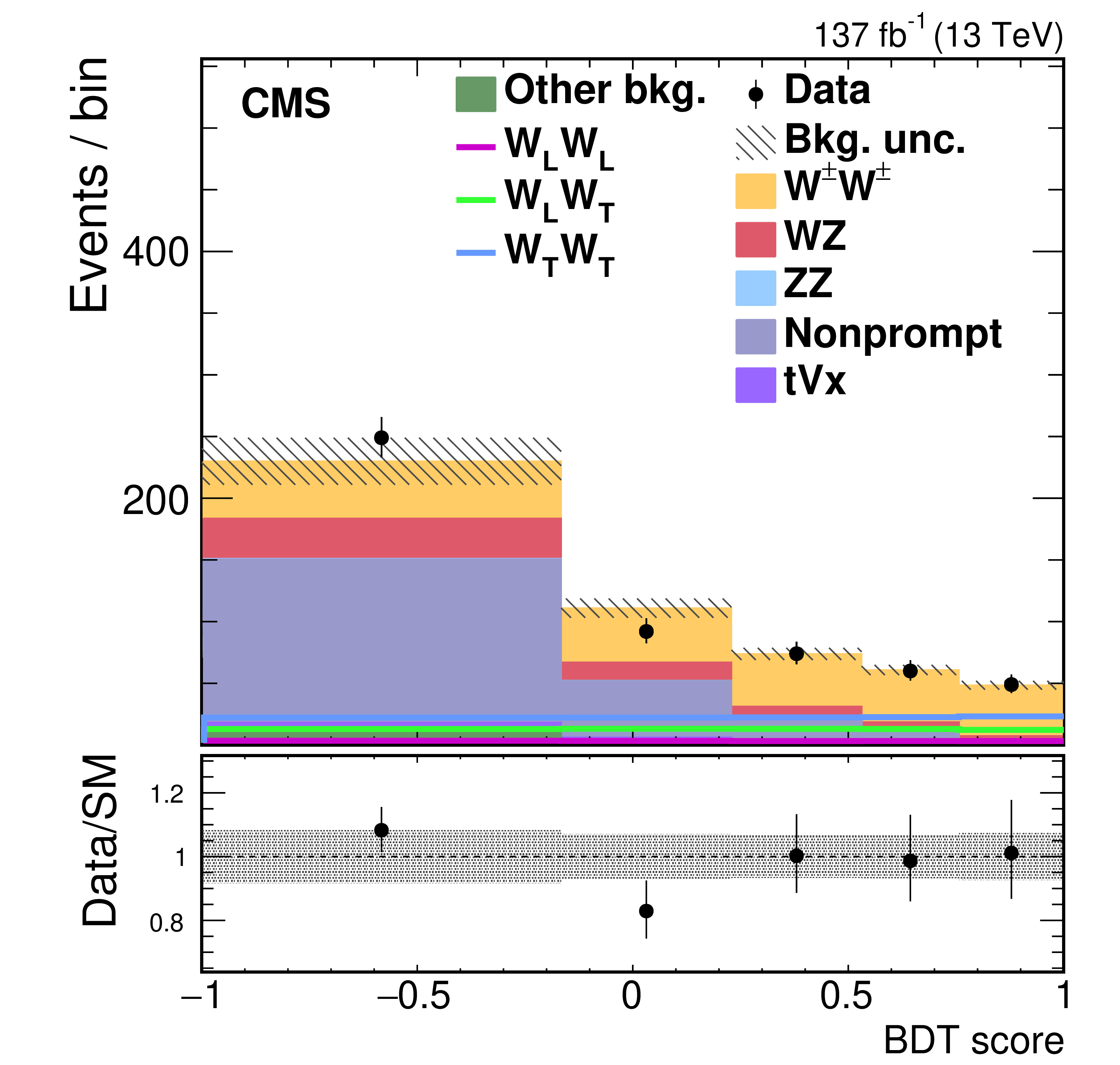}
    \includegraphics[width=\columnwidth]{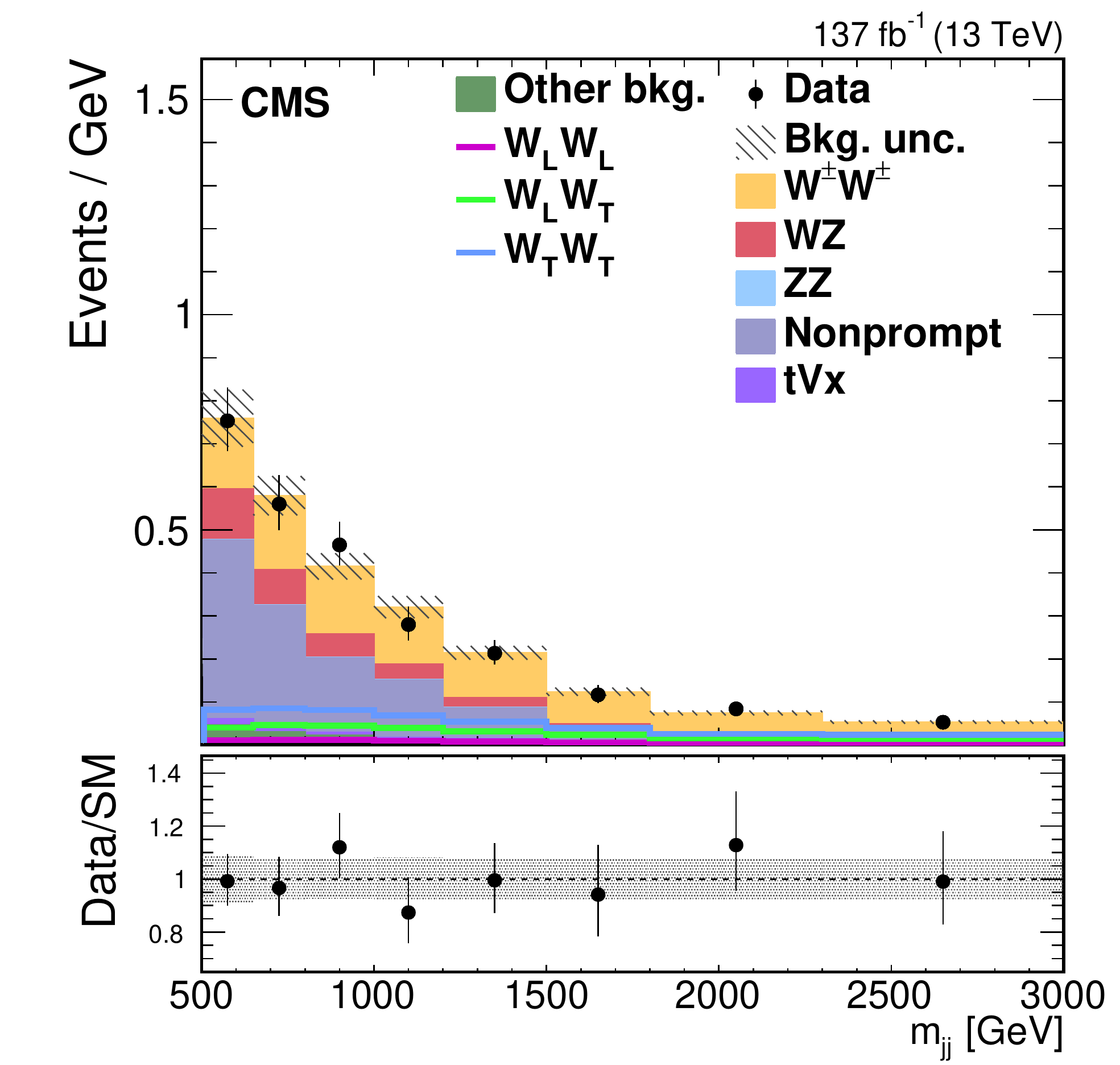}
    \caption{\emph{Left:} The distribution of the BDT discriminant used to extract the signal strength for the polarized components of the {\WW} cross  section. 
    \emph{Right:} Dijet invariant mass distribution. Images from Ref.~\cite{Sirunyan:2020gvn}.}
    \label{fig:ssww_pol}
\end{figure*}

In the CMS analysis, the $\SSWW$ and $WZ$ processes are treated as independent signals in a single maximum likelihood fit. The likelihood is built from binned distributions of the invariant mass of the two VBS-tagged jets $m_{jj}$ and the dilepton mass $m_{\ell\ell}$ in the $\SSWW$ signal selection region, as well as from the output of a boosted decision tree (BDT) discriminant in the $WZ$ signal region. 
Several observables are used as the inputs to the BDTs, including the jet and lepton transverse momenta, dilepton mass and transverse momentum, transverse $WW$ mass, difference in azimuthal angle between leptons and between jets, and others~\cite{Sirunyan:2020gyx}.
Control regions are also used to constrain the modeling of top quark and non-prompt lepton backgrounds. The resulting measurements have a statistical significance of well over 5$\sigma$ for each process, and unfolded differential distributions are produced for comparison with theoretical predictions. In particular, the impact of the next-to-leading (NLO) EW corrections, which are known to be larger than pure QCD corrections for VBS processes~\cite{Biedermann:2016yds,Frederix:2018nkq,Chiesa:2019ulk}, are studied, as shown in Fig.~\ref{fig:wwwz}.
(For related details on theoretical predictions, see Section~\ref{sec:thpred}.)

This analysis was recently extended to perform the first study of a VBS process separated by the polarization of the vector bosons~\cite{Sirunyan:2020gvn}. Because the polarization is not a Lorentz-invariant quantity, its definition is reference frame dependent. Therefore, the analysis independently measures the cross sections of polarized states using the polarization defined in the parton-parton center-of-mass frame and in the W boson pair center-of-mass frame. The analysis exploits BDTs trained to distinguish between the polarization states. Figure \ref{fig:ssww_pol} shows the BDT score (left) and the dijet invariant mass distribution (right). The BDT distribution summarizes the discrimination of signal and background. Due to the lack of statistical power, the combined cross section for at least one longitudinally polarized boson is reported, with a statistical significance of 2.3$\sigma$ (3.1$\sigma$) observed (expected) in the $WW$ center-of-mass frame.

Searches have also been performed for VBS production of {\ZZ}  pairs~\cite{Sirunyan:2020alo,Aad:2020zbq} using the full Run~2 data set. While the production rate for {\ZZ} events is lower than for {\WZ} and {\SSWW},  its four-lepton signature is very clean, providing an ideal experimental handle in an otherwise complicated hadronic environment. The QCD-induced {\ZZ} process is the overwhelming background in this channel, and includes a contribution from the loop-induced $gg\to ZZ$ process. This is a small component of the inclusive {\ZZ} production rate but can be relevant in VBS-like regions of phase space. This contribution is estimated using state-of-the-art merged simulations for  accurate modeling of the ${\ZZ}jj$ state~\cite{Caola:2015psa,Caola:2016trd,Li:2020nmi}. The analysis is driven by a matrix-element likelihood approach. The resulting measurement provides observed (expected) evidence of VBS {\ZZ} production at 4.0 (3.5) standard deviations.

\subsection{Results on vector boson scattering with photons}

\begin{figure}[!t]
    \includegraphics[width=\columnwidth]{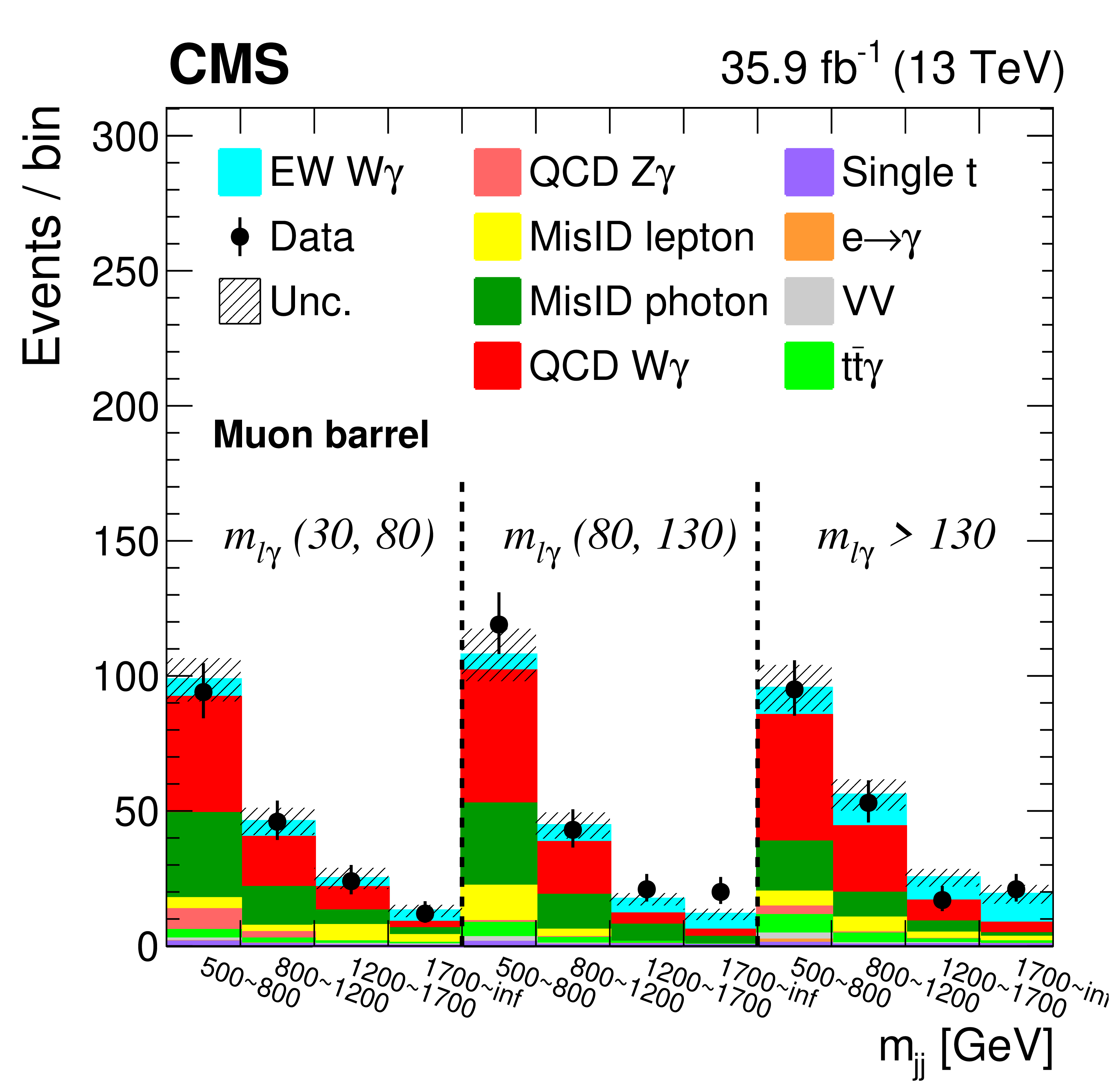}
    \caption{The two-dimensional distribution of $m_{jj}$ and $m_{\ell\gamma}$
    used to extract the EW {\Wg} cross section, from Ref.~\cite{Sirunyan:2020azs}}
    \label{fig:ssww_pol2}
\end{figure}

The ATLAS and CMS Collaborations have also performed measurements of the VBS production of a massive vector boson and a photon in the {\Wg}~\cite{Sirunyan:2020azs} and {\Zg}~\cite{ATLAS:2019qhm,Sirunyan:2020tlu,ATLAS:2021pdg} channels. These measurements were performed using approximately 36 fb$^{-1}$ or the full Run 2 data set, and exploit similar analysis strategies. Considerable backgrounds due to non-prompt photons are estimated using data-driven techniques. The QCD-induced production is the leading background in both cases. The background is estimated using Monte Carlo simulation with the normalization constrained with control regions in data. Results are extracted from a binned maximum likelihood fits, where the likelihood is built from a two-dimensional distribution of selected events (Fig.~\ref{fig:ssww_pol2}). The {\Zg} analysis from CMS exploits the $m_{jj}$ and the rapidity separation of the tagged jets $\Delta\eta_{jj}$, whereas the $m_{jj}$ and the photon and lepton invariant mass $m_{\ell\gamma}$ are used in the {\Wg} analysis.  Results are combined with the corresponding measurements made at 8 TeV~\cite{Khachatryan:2016vif,Khachatryan:2017jub}, under the assumption of the SM production rate, to reach high statistical significance of around 5$\sigma$. Separately, ATLAS has observed the $Z\gamma\to\nu\overline{\nu}\gamma$ final state with more then $5\sigma$ statistical sensitivity~\cite{ATLAS:2021pdg}.

\subsection{Constraints on anomalous couplings}\label{sec:lhc_results_smeft}

\begin{figure*}[!t]
    \centering
    \includegraphics[width=\textwidth]{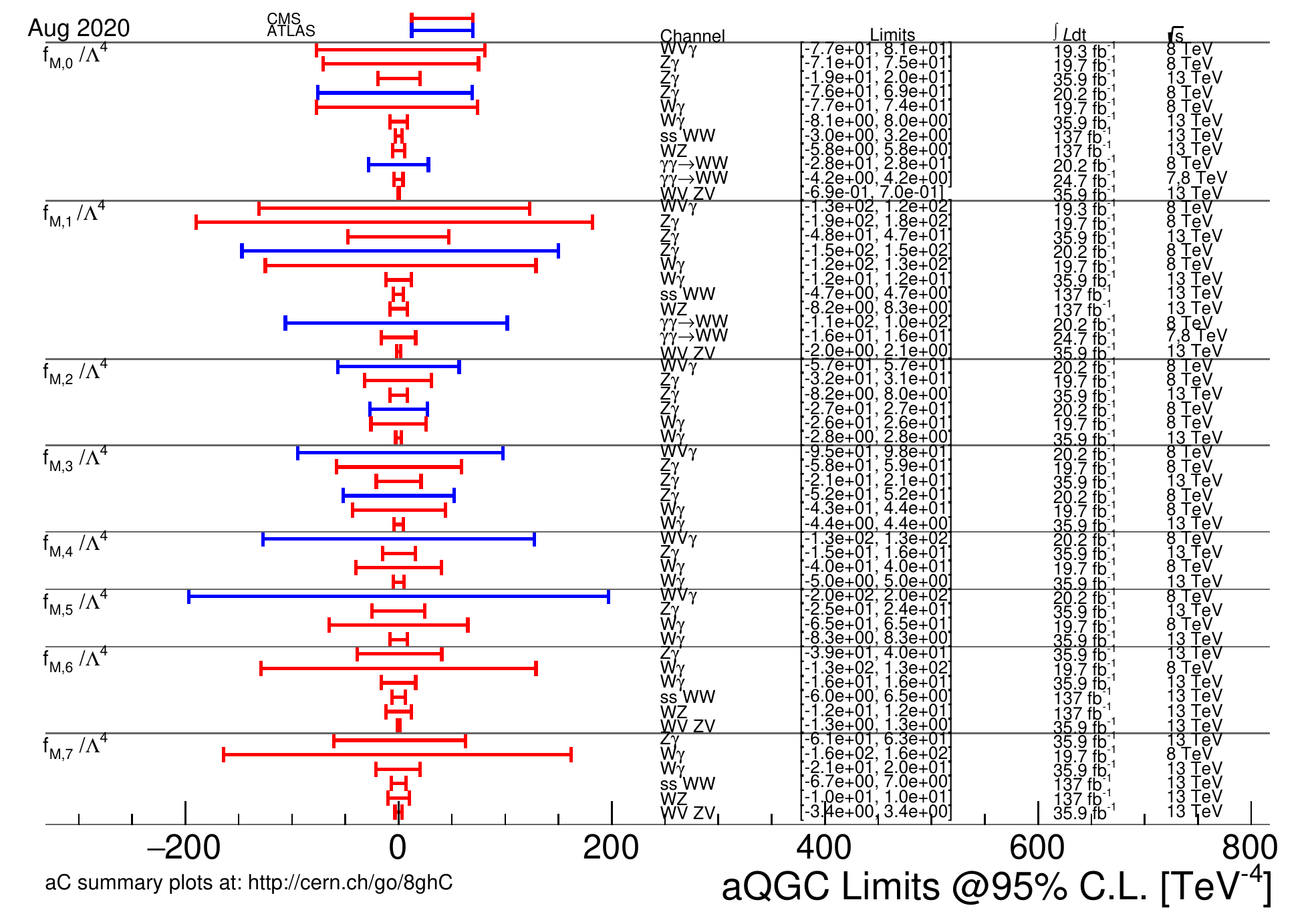}
    \caption{Summary of constraints on the couplings to the EFT operator $F_{M,7}$
        from experimental measurements. For details, see the URL at
        \url{http://cern.ch/go/8ghC}.}
    \label{fig:cmseftlimits}
\end{figure*}

All channels are used to place constraints on anomalous dimension-8 EFT interactions. In general, the operator formulation proposed in Ref.~\cite{Eboli:2006wa} is used. Variables sensitive to the total energy of the interaction, for example, the diboson mass, are exploited to search for signs of higher-energy modifications to the vector boson interactions. A summary of the current constraints at 95\% Confidence Level (C.L.) by CMS for the mixed longitudinal and transverse operator $F_{M,7}$ is shown in Fig.~\ref{fig:cmseftlimits}. Though they are not yet sensitive to the SM VBS production rates, the production of diboson events with one vector boson decaying hadronically is a particularly powerful channel for these searches due to the higher cross section from hadronic vector boson decays. Results using these final states are presented in Ref.~\cite{Sirunyan:2019der}. Related constraints on anomalous couplings using VBS signatures have been reported by ATLAS using the full Run 2 data set~\cite{ATLAS:2020nzk}.

\subsection{Summary}

In summary, the ATLAS and CMS Collaborations have recently produced many new results studying diboson production via VBS processes. These results include the first observations of several new processes, measurements of helicity-polarized cross sections, and constraints on anomalous couplings. Most of these measurements have exceeded the traditional metric for observations, and several have sufficient accuracy to enable unfolded results. Following the LHC Run 2, these measurements are quickly leaving the era of discovery and are approaching a new era of precision measurements. VBS measurements will soon be sensitive to the polarization fractions of the production mechanism, providing an important probe of the EW sector of the SM. For prospects at the HL-LHC, see Part~\ref{sec:vbs_proj}.


\section[Polarization and $\tau$ lepton studies in VBS]{\large Polarization and $\tau$ lepton studies in VBS}\label{sec:sec:vbs_tau_pol}

The study of the interactions between vector bosons offers an important test of the SM for its sensitivity to gauge boson self-couplings. It provides a direct probe of the triple and quartic gauge boson couplings. Precise measurements can shed light on the EW symmetry breaking  and probe new physics processes at multi-TeV energy scales. The scattering amplitude of vector bosons, such as $V^+V^-\rightarrow V^+V^- (V=W,Z)$, is expected to increase with center-of-mass energy. Unless there is a cancellation, it will eventually diverge and will lead to a violation of unitarity~\cite{Lee:1977yc,Lee:1977eg}. The divergent behavior at high energies is more acute for the scattering of longitudinally polarized vector bosons. However, this divergence is canceled by the contribution of the Higgs boson exchanges in the $s$ and $t$ channels, provided that the Higgs boson couplings to vector bosons are those expected in the SM. Any modification of these couplings would interfere in this delicate cancellation. If the Higgs boson's couplings to weak bosons deviate from SM predictions, cancellations may not be as effective and the diboson scattering amplitude may increase with energy. 

In addition, there are many BSM scenarios that predict a cross section increase of VBS processes, through extended Higgs sectors or new resonances. Hence, VBS processes provide both a window to new physics processes as well as a constraint on fundamental Higgs boson properties and anomalous Higgs couplings. However, measuring VBS processes is experimentally challenging due to the small cross sections.
VBS processes are rare and have been studied in different final states. Final states with light leptons, \ie, electrons and muons, are those where the signal-over-background ratio is larger and have been studied first. However, $\tau$ leptons are more complicated to identify due to the larger backgrounds (misidentification rates for hadronic $\tau$ decays are typically one or two orders of magnitude larger than for electrons or muons) and have not been included in these studies yet.

In addition to inclusive and differential cross sections, polarization measurements of $W$ and $Z$ bosons in diboson production provide stringent tests of the SM and its couplings. Polarization studies provide an additional handle and a further discrimination between signal and background. In particular, the study of longitudinally polarized VBS processes, that is to say when VBS is mediated by initial- and final-state weak bosons that are in their longitudinal helicity state, is an important probe of particle interactions at the highest energies, both at the LHC and at future colliders, and it will help further understanding fundamental interactions and the SM gauge structure.

\subsection{Polarization}
One of the most promising ways to measure VBS processes uses events containing two leptonically-decaying, same-sign $W^\pm$ bosons produced in association with two jets, $pp\to W^\pm W^\pm j j$. However,  the  angular  distributions of the leptons in the $W$ boson rest frame, which are commonly used to fit polarization fractions, are not readily available in this process due to the presence of two neutrinos in the final state. Advanced analysis methods that make use of Machine Learning and other similar techniques can be used to reconstruct the angular distributions from measurable event kinematics and study the polarization fractions. It is also possible to define reference frames that are more accommodating to experimental limitations~\cite{BuarqueFranzosi:2019boy,Ballestrero:2020qgv}.

Polarization measurements are not unique and depend on the reference frame in which they are defined. Polarization fractions and kinematic distributions that define polarization vectors in different reference frames have been employed. The laboratory frame is a natural choice. A definition of the polarization observables in the diboson center-of-mass frame has the advantage that the line-of-flight of the two bosons also defines the longitudinal polarization vectors, and the decay products are directly related to the scattering process. The latter may be better suited to search for deviations from the SM.

Precise calculations at the NLO in QCD of SM polarization observables are available for various multi-boson processes~\cite{Melia:2011tj,Ballestrero:2018anz,Baglio:2018rcu,Baglio:2019nmc,Denner:2020bcz,Denner:2020eck}. Predictions with NLO in EW~\cite{Baglio:2018rcu,Baglio:2019nmc} and at next-to-next-to-leading order (NNLO)  in QCD~\cite{Poncelet:2021jmj} are also available, as is the automation at LO for an arbitrary scattering process~\cite{BuarqueFranzosi:2019boy}. Separate polarization measurements for each of the $W^\pm/Z$ bosons might be helpful in investigating CP violation in the interaction between gauge bosons. In the longer term, measuring the scattering of longitudinally polarized vector bosons will provide a fundamental test of the EW breaking. 

Measurements of polarized cross sections for VBS processes will be an exciting component of the Run 3 and HL-LHC programs. 
These analyses will complement on-going LHC measurements of $W$ polarization fractions in the $W$+jets process \cite{Chatrchyan:2011ig,ATLAS:2012au} and in $t\bar{t}$ events ~\cite{Aad:2020jvx}, as well as measurements of angular coefficients of lepton pairs in $Z$ boson production~\cite{Aad:2016izn, Khachatryan:2015paa}. 
Generically, distributions of angular observables are extremely sensitive to the final state considered and discrepancies from the SM predicted behavior can be used to look for the presence of new interactions. The polarization of a gauge boson can be determined from the angular distribution of its decay products. At the Born level, the expected angular distribution for massless fermions in the rest frame of the parent $W$ boson is given in terms of its helicity fractions~\cite{Ellis:1991qj,Bern:2011ie,Stirling:2012zt}. These are commonly represented  by the longitudinal helicity fraction $f_0$, and the  left-handed and right-handed transverse helicity fractions $f_L$ and $f_R$, respectively. 

In inclusive $Z$ events, the  angular  distributions  of  charged  lepton  pairs  produced  via the Drell-Yan neutral current process provide a portal to precisely determine the production process through spin-correlation effects between the initial-state partons and the final-state leptons~\cite{Collins:1977iv,Kajantie:1978yp,Collins:1978yt}. In this context, we briefly comment that reports by Refs.~\cite{Aad:2016izn, Khachatryan:2015paa} of apparent discrepancies between data and theory expectations for the so-called Lam-Tung relation~\cite{Lam:1980uc}, which relates angular coefficients of lepton pairs in the Drell-Yan process, can be alleviated by recent improvements to the predictions' formal accuracy~\cite{Gauld:2017tww}.

Polarization measurements of the three helicity fractions of the $W$ and $Z$ bosons were also performed in $WZ$ production at $\sqrt{s}=13$~TeV~\cite{Aaboud:2019gxl}. 
Importantly, this is a stepping stone to measuring helicity fractions in triboson processes, which are related to VBS by crossing symmetry. An analysis of angular distributions of leptons from decays of $W$ and $Z$ bosons was performed in $WZ$ events, and integrated helicity fractions in the detector fiducial region were measured for the $W$ and $Z$ bosons separately. Of particular interest, the longitudinal helicity fraction of pair-produced vector bosons was also measured. The measurements are dominated by statistical uncertainties. Nonetheless, good agreement of the measured helicity fractions of both the $W$ and $Z$ bosons with  predictions is observed.

These studies can be extended to VBS processes and, in particular, to VBS processes that include $\tau$ leptons in the final state. This is something that has not been done yet, but it would certainly be an important study to perform.

\subsection{The $\tau$ lepton and polarization studies}

Recent checks of lepton flavor universality violation have sparked renewed interest in measurements involving $\tau$ leptons~\cite{Aaij:2019wad,Lees:2013uzd,Sato:2016svk,Aaij:2021vac}. The $\tau$ lepton is the most massive lepton and a third-generation particle. It therefore plays an important role as a probe in BSM searches. Due to their large mass, $\tau$ leptons may be particularly sensitive to BSM interactions. For example: the existence of a heavy charged Higgs boson $H^\pm$ may give rise to anomalous $\tau$ lepton production
in $WZ$ scattering~\cite{Djouadi:2005gj,Branco:2011iw,Degrande:2015xnm}. Similarly, the violation of lepton number symmetry can lead to the anomalous production of same-sign $\tau^\pm\tau^\pm$ pairs in same-sign $W^\pm W^\pm$ scattering~\cite{Fuks:2020att,Fuks:2020zbm,Cirigliano:2021peb}. 
Furthermore, due to its short lifetime, spin information is preserved in the decay products of $\tau$ leptons.
VBS polarization studies involving $\tau$s therefore provide additional information on the underlying production processes as well as additional discriminating power against background processes.

With larger data samples in the Run 3 and HL-LHC eras it is expected that $\tau$ leptons will also be included and observed in the studies of VBS processes. The $\tau$ lepton is identified through its visible decay products, either hadrons or leptons. The hadronic decay modes of  $\tau$ leptons represent approximately 65\% of all $\tau$ decays, mostly in one-prong ($\approx $50\%) and three-prong ($\approx $15\%) decays, and are characterized by a narrow jet signature. Their identification and reconstruction are difficult and subject to larger backgrounds from incorrectly identified jets. The remaining 35\% are leptonic decays, \ie, decaying to electrons or muons with lower $p_T$ and the corresponding neutrinos, and are difficult to distinguish from prompt electron and muon production.

Polarization studies incorporating VBS and $\tau$ leptons build on ongoing measurements at the LHC. For example: the $\tau$ polarization was measured by ATLAS in $W\rightarrow \tau\nu$ decays as the asymmetry of cross sections of left-handed and right-handed $\tau$ production, from the energies of the decay products in hadronic $\tau$ decays with a single final state charged particle (one prong)~\cite{Aad:2012cia}. The resulting  measurement is in agreement with SM predictions. The $\tau$ decay mode exhibiting the highest sensitivity to the $\tau$ polarization is $\tau^\pm\rightarrow h^\pm\nu$, where $h^\pm=\pi^\pm, K^\pm$, with a branching ratio of about 11.5\%. In the $\tau$ rest frame, the neutrino (always left-handed) is preferentially emitted opposite to the $\tau$ spin orientation. The $\tau$ polarization value $P$ provides insight into the Lorentz structure of the $\tau$ production mechanism, and gives an indication of the parity violation in the interaction. The angle between the $\tau$ direction of flight and the hadronic decay products in the $\tau$ rest frame is used as the primary observable sensitive to $\tau$ polarization. The $\tau$ polarization is inferred from the relative energy fraction carried by charged and neutral hadrons in the hadronic $\tau$ decays, and is measured by fitting the observed charged asymmetry distribution.

\begin{figure}[t!]
\centering\includegraphics[width=\columnwidth]{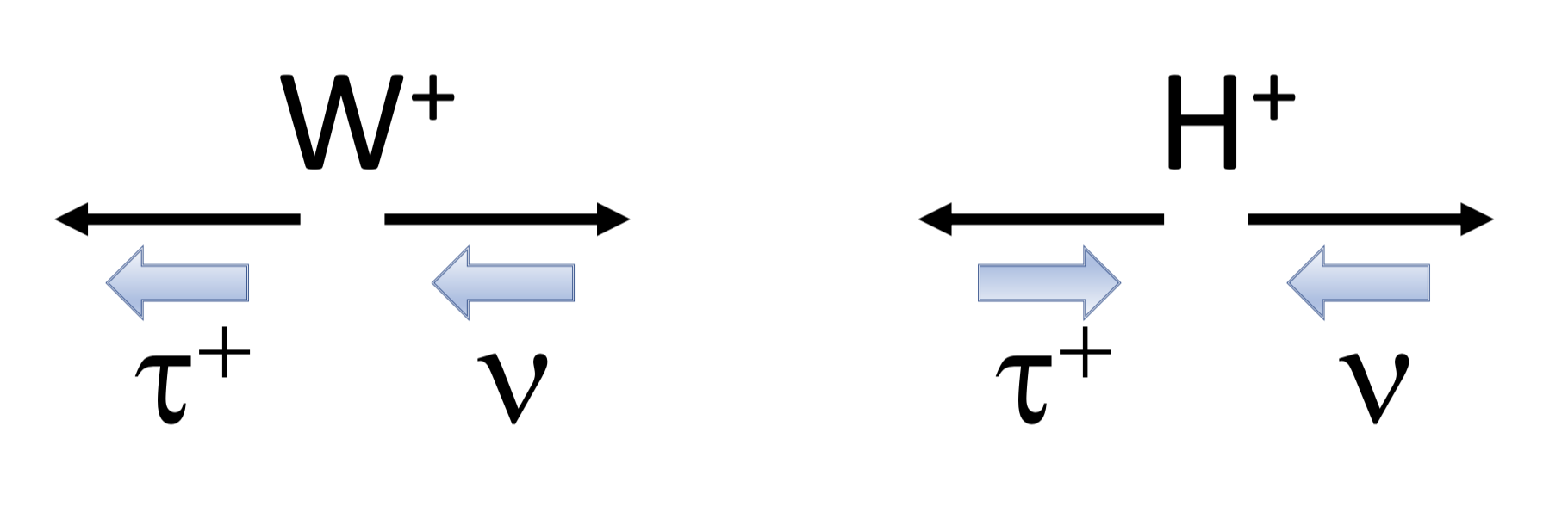}\vspace{-.25cm}
\caption{A cartoon depicting a spin-1 W boson (left) and a spin-0 charged Higgs boson (right) decaying in their rest frames to a $\tau$ lepton and a neutrino. The resulting helicity (larger light-colored shaded arrow) of the $\tau$ lepton depends on the originating boson: the $\tau^+$ polarization is $P=+1$ ($P=-1$) if from a $W^+$ ($H^+$).}
\label{fig:tau_helicity}
\end{figure}

In $W\rightarrow\tau\nu$ decays, the $W^+$ ($W^-$) is expected to couple exclusively to a right(left)-handed $\tau^+$ ($\tau^-$) corresponding to a $\tau$ polarization of $P=+1$, up to corrections of order $\mathcal{O}(m_\tau^2/m_W^2)$. On the other hand, since neutrinos are left-handed, a Minimal Supersymmetric SM (MSSM) charged Higgs boson would couple to left(right)-handed $\tau^+$ ($\tau^-$) $\tau$ leptons leading to a prediction of $P=-1$. This is exemplified in Fig.~\ref{fig:tau_helicity}. The method used for extracting the $\tau$ polarization is independent of the production mode and can be applied to other processes.

The polarization of $\tau$ leptons produced in $Z/\gamma^*\rightarrow\tau\tau$ decays was also measured~\cite{Aaboud:2017lhv}. Results are in agreement with SM predictions. The final state includes a hadronically decaying $\tau$ lepton with a single charged particle, accompanied by another $\tau$ decaying leptonically. The leptonic decay is used to trigger, select, and identify the candidate events, while the hadronic decay serves as a spin analyzer.

A good understanding of $\tau$ polarization is a powerful discriminating tool in these and other processes containing a $\tau$ lepton in the final state, such as same-sign $W^\pm W^\pm$ VBS and charged Higgs production.  Similar studies may be performed for future measurements in the decays of the Higgs boson or other final states with high invariant masses.  In particular, it may help to distinguish decays of heavy particles where the same final states involving $\tau$ leptons are predicted but with different helicity configurations,  such as for separating $Z$ and $H$ or new bosons, or for distinguishing $W$ and charged Higgs bosons~\cite{CMS:mxa}. Larger statistics to be able to properly select the desired decay modes and a good control of systematic uncertainties are necessary ingredients.

In order to enhance the sensitivity to anomalous quartic gauge couplings (aQGCs), studies of EW boson production through VBS processes include leptons and jets of large transverse momenta in the large diboson mass region $m_{VV}$. Specific, boosted-object reconstruction algorithms have been developed to retain high efficiency for high-$p_T$ objects. These techniques include reconstruction of hadronically decaying $\tau$s as well as electrons, muons, jets, and bosons~\cite{Sirunyan:2019jbg,Aad:2020ldt}. Efforts have been deployed to study proton-proton collisions at the LHC.  At future, higher energy colliders, further understanding and improvements of these reconstruction tools will become of pivotal importance to be able to fully explore the data at the highest energies.

\subsection{Summary}
Polarization studies of vector boson final states may provide further understanding of fundamental rules governing particle interactions. 
Some polarization studies of VBS processes are discussed in Section~\ref{sec:ML}.
Polarization properties carry information of the interaction process and can be studied in the final-state decay products, provided an appropriate understanding of the event kinematics. Detailed studies have been performed in VBS processes and their sensitivity is limited by the size of the data samples. Large data samples are needed to properly model and disentangle the longitudinal and transverse polarization distributions, and kinematic characteristics.  The $\tau$ lepton, the heaviest and perhaps the most intriguing of the leptons, is so far missing from these studies. Its inclusion will certainly offer an additional handle to probe the SM. With the larger data samples imminently expected at the LHC Run~3 and further in the future at the HL-LHC and at future colliders, the study of $\tau$ leptons in VBS process final states may provide additional sensitivity in the search for new physics processes.

\section[Precise theoretical predictions for VBS]{\large Precise theoretical predictions for VBS}\label{sec:thpred}

\begin{table*}
\resizebox{\textwidth}{!}{
\begin{tabular}
{lccccccc}
\hline\hline
Process & $\sigma_{\mathrm{LO}}^{{\cal O}(\alpha^6)}$~[fb] &
$\sigma_{\mathrm{NLO,EW}}^{{\cal O}(\alpha^7)}$~[fb] 
& $\delta_{\mathrm{EW}}~[\%]$ 
& $\delta_{\mathrm{EW}}^{\mathrm{log,int}}~[\%]$
& $\delta_{\mathrm{EW}}^{\mathrm{log,diff}}~[\%]$ &
$\langle M_{4\ell}\rangle$~[GeV]
& Ref.
\\
\hline
 $pp \to \mu^+ \nu_\mu e^+ \nu_{e} jj$ ($W^+W^+$) &
$\phantom{1}1.5348(2)$ & $\phantom{1}1.2895(6)$  & {$-16.0$}
& $-16.1$   &  $-15.0$ &  $390$  
&\cite{Biedermann:2016yds}
\\[.5ex]
 $p p \to \mu^+ \mu^- e^+ \nu_{e} jj$ ($ZW^+$) &
$0.25511(1)$ & $0.2142(2)$ & {$-16.0$}
 &  $-17.5$&  $-16.4$& $413$ 
& \cite{Denner:2019tmn}
\\[.5ex]
$p p \to \mu^+ \mu^- e^+ e^- jj$ ($ZZ$) &
$0.097681(2)$ & $0.08214(5)$ & {$-15.9$} 
 &  $-15.8$&  $-14.8$& $385$
& \cite{Denner:2020zit}
\\
\hline\hline
\end{tabular}
}
\caption{EW corrections to fiducial cross sections of different VBS processes at $\sqrt{s}=13$ TeV.}
\label{tab:ewcofid}
\end{table*}

In this section we describe recent developments and challenges in making precise
theoretical predictions for VBS at the LHC. We make some remarks on possible future developments.

\subsection{Electroweak corrections to VBS}
The increasing experimental precision of VBS measurements requires adequate theoretical predictions. NLO QCD corrections to VBS for leptonic final states and the corresponding irreducible background processes have existed for many years including their matching to parton showers. For a recent review, see Ref.~\cite{Rauch:2016pai}. They are available in \textsc{VBFNLO}~\cite{Baglio:2014uba} as well as the general-purpose generators \textsc{MadGraph5\_aMC@NLO}~\cite{Stelzer:1994ta,Alwall:2014hca}, \textsc{Sherpa}~\cite{Bothmann:2019yzt} and \textsc{POWHEG}~\cite{Nason:2004rx,Frixione:2007vw,Alioli:2010xd}.

EW corrections to VBS processes, on the other hand, have only been calculated recently. The EW corrections to same-sign $WW$ scattering have been published in Ref.~\cite{Biedermann:2016yds}, those for $WZ$ scattering in Ref.~\cite{Denner:2019tmn}, and the ones for VBS into $ZZ$ bosons in Ref.~\cite{Denner:2020zit}. The calculation of EW corrections for the scattering of opposite-sign $W$ bosons is ongoing work. For same-sign $WW$ scattering an event generator has been made available based on Powheg~\cite{Nason:2004rx,Frixione:2007vw} and Recola~\cite{Actis:2012qn,Actis:2016mpe} with Collier~\cite{Denner:2016kdg} for the processes $pp\to\ell^\pm\nu_\ell \ell^{\prime\pm}\nu_\ell' jj$, with $\ell, \ell' = e, \mu$ including EW corrections and matched to a QED parton shower and interfaced to a QCD
parton shower \cite{Chiesa:2019ulk}.

The relative EW corrections to fiducial cross sections of VBS processes in the SM turn out to be around $-16\%$, independent of the specific process and the details of event selection. While the EW corrections $\delta_{\mathrm{EW}}$ for the three processes in Table \ref{tab:ewcofid} are very close to each other, this is accidental and the expected spread is at the level of a few per cent. The large universal EW corrections for VBS processes can be explained by a Sudakov approximation applied to the $VV\to VV$ sub-processes. 

Following Refs.~\cite{Denner:2000jv,Accomando:2006hq}, the leading logarithmic corrections to the scattering of transverse vector bosons, which is the dominant contribution, can be cast into a simple universal correction factor
\begin{equation}
 \delta_{\mathrm{LL}} = \frac{\alpha}{4\pi} \left\{
- 4 C_W^{\mathrm{EW}} \log^2 \left(\frac{Q^2}{M_W^2}\right)
+  2b_W^{\mathrm{EW}} \log \left(\frac{Q^2}{M_W^2}\right) \right\}.
\label{eq:LLcorr}
\end{equation}
It includes all logarithmically enhanced EW corrections apart from the angular-dependent sub-leading soft-collinear logarithms and applies to all VBS processes that are not mass suppressed, owing to the fact that these scattering processes result from the same $\mathrm{SU}(2)_\mathrm{w}$ coupling. The constants are given by $C_W^{\mathrm{EW}} = 2/s_\mathrm{w}^2$ and $ b_W^{\mathrm{EW}} = 19/(6 s_\mathrm{w}^2)$, where $s_\mathrm{w}$ represents the sine of the weak mixing angle.  Further, $Q$ is a representative scale of the $VV \to VV$ scattering process, which is conveniently chosen as the four-lepton invariant mass $M_{4\ell}$.  Using $Q=M_{4\ell}$ event-by-event results in the numbers for $\delta^{\mathrm{log,diff}}_{\mathrm{EW}}$ shown in the 6th column of Table \ref{tab:ewcofid}, which agree within $2\%$ with the full NLO results. Applying eq.~(\ref{eq:LLcorr}) directly to the fiducial cross section with the average values for $M_{4\ell}$ obtained from a leading order (LO) calculation (see 7th column of Table \ref{tab:ewcofid}) yields the numbers for $\delta^{\mathrm{log,int}}_{\mathrm{EW}}$ shown in the 5th column of Table \ref{tab:ewcofid}.

Since the leading logarithmic corrections are universal and only depend on the gauge structure of the theory and the external particles, one expects similar corrections in extensions of the SM that do not modify the gauge sector. For the scattering of longitudinal gauge bosons, smaller corrections are expected since the related coupling factors are smaller. EW corrections to distributions reach $-30\%$ to $-40\%$ in high energy tails in the TeV range.

\subsection{Complete NLO corrections to VBS}

Besides VBS diagrams, other diagrams contribute to the same physical final states in scattering processes of the form $pp\to l_1\bar{l}_2l_3\bar{l}_4jj$. At leading order the cross sections for processes of the type $pp\to l_1\bar{l}_2l_3\bar{l}_4jj$ receive pure EW contributions of orders $\alpha^6$, QCD-induced contributions of order $\alpha^4\alpha_s^2$, and interference contributions of order $\alpha^5\alpha_s$. At NLO there are corresponding contributions of orders $\alpha^7$, $\alpha^6\alpha_s$, $\alpha^5\alpha_s^2$, and $\alpha^4\alpha^3_s$. While the ${\cal  O}(\alpha^7)$ contributions can be viewed as EW corrections to the LO EW process and the ${\cal O}(\alpha^4\alpha^3_s)$ contributions as QCD corrections to the LO QCD-induced process, the other contributions cannot be classified simply as EW or QCD corrections in general. For instance: the contribution at ${\cal O}(\alpha^6\alpha_s)$ contains QCD corrections to the LO EW process but also EW corrections to the LO interference. A unique assignment can be made within the so-called VBS approximation, where interference between different kinematic channels are neglected.

For same-sign $WW$ scattering, the full NLO corrections of all four orders have been calculated \cite{Biedermann:2017bss}. Since this process is dominated by the EW diagrams at LO, the magnitude of the corrections at orders  ${\cal O}(\alpha^3\alpha^2_s)$ and  ${\cal O}(\alpha^4\alpha^3_s)$ is small, \ie\ below a per cent for the fiducial cross section and at the level of one per cent for distributions. This will be different for other VBS processes where the QCD-induced contribution is larger than the EW contribution. The calculation of these types of corrections is ongoing.

\subsection{Quality of the VBS approximation}

In the past, many calculations for VBS have used the ``VBS approximation.'' In this approximation only the squares of $t$- and $u$-channel contributions are taken into account, while interference between these channels as well as $s$-channel contributions are omitted ($|t|+|u|$ approximation). In some calculations the $s$-channel contributions are included ($|s|+|t|+|u|$ approximation), while interference is still neglected. The quality of these approximations has been investigated in Ref.~\cite{Ballestrero:2018anz} for same-sign $WW$ scattering by comparing the results of different calculations using different approximations. While this study confirmed that the accuracy of the VBS approximation is better than about 1\% at LO, it revealed larger differences at NLO QCD. 
In a more inclusive region, \eg, $M_{jj}>200\,\mathrm{GeV}$, $\Delta y_{jj}>2$, the differences at NLO in QCD amount to $-6\%$ for the $|t|+|u|$ approximation and $+2.6\%$ for the $|s|+|t|+|u|$ approximation, and grow up to $\mp20\%$ (same direction) for some distributions.
In a typical fiducial region for VBS, \ie, where $M_{jj}>500\,\mathrm{GeV}$, $\Delta y_{jj}>2.5$, the $|t|+|u|$ and $|s|+|t|+|u|$ approximations lead to cross sections that differ from the exact NLO calculation by about $2\%$ and $1\%$, respectively. These differences are comparable to residual scale uncertainty at NLO.
For some differential distributions, these different approximations can lead to differences as large as $\pm10\%$. Details are discussed in Ref.~\cite{Ballestrero:2018anz}.

These larger differences at NLO can be attributed to contributions of triple vector-boson production with an additional gluon. While $VVV$ contributions are effectively suppressed by the VBS cuts on the two tagging jets, the additional gluon jet reduces the efficiency of these cuts. This has been confirmed in \cite{Denner:2020zit}, where $24\%$ NLO QCD corrections have been found for VBS into $ZZ$ for a loose VBS cut of $M_{jj}>100\,\mathrm{GeV}$. Similar conclusions have also been reported in Ref.~\cite{Fuks:2019clu} for same-sign $WW$ scattering and the impact of Drell-Yan-like topologies. Thus, one should either take into account triple vector-boson production in the theoretical prediction or use tight VBS cuts.

\subsection{Polarized VBS}

A primary goal in VBS experiments is to measure the scattering of longitudinal vector bosons, which is very sensitive to the  EW symmetry breaking and to physics beyond the SM owing to strong unitarity cancellations within the SM. However, the definition of polarized cross sections for massive vector bosons is not unique, since the vector bosons are unstable, and moreover any definition of their polarization is linked to a certain frame.

In the literature, different definitions of vector-boson polarization have been discussed. A short overview can be found, for instance, in Ref.~\cite{Denner:2020bcz}. A polarization definition based on projections on the LO decay-angle distributions has been put forward in Ref.~\cite{Baglio:2018rcu,Baglio:2019nmc}. This method is tailored to inclusive LO distributions due to the resonant vector-boson diagrams. It is only applicable to one polarized vector boson and fails for tight cuts, sizable background, or large NLO corrections. The calculation of polarized cross sections has been automated for LO calculations in the \textsc{MadGraph5\_aMC@NLO} Monte Carlo event generator~\cite{BuarqueFranzosi:2019boy}, employing decay chains in the narrow-width approximation and including spin correlations via the \textsc{Madspin} package~\cite{Artoisenet:2012st}.  A definition of polarized cross sections based on the pole approximation has been proposed and applied to VBS at LO in
\cite{Ballestrero:2017bxn,Ballestrero:2019qoy,Ballestrero:2020qgv}.

Based on Refs.~\cite{Ballestrero:2017bxn,Ballestrero:2019qoy}, a proposal was made in Ref.~\cite{Denner:2020bcz} that splits the vector boson productions process from the irreducible background and defines polarized matrix elements for  vector boson production. It is applicable to arbitrary processes, multiple resonances, and NLO calculations. It allows for defining polarizations in arbitrary frames and for separating the irreducible background and interference between polarized matrix elements. The method has already been applied to vector boson pair production at NLO QCD~\cite{Denner:2020bcz,Denner:2020eck} and will be used to calculate NLO corrections to polarized VBS in the future. 

\subsection{Summary}
In this short  section, we summarized the status of precision predictions for VBS in the SM with a focus on recent computational developments. In this dynamic field more progress can be expected in the near future. This concerns in particular polarized VBS, semileptonic final states, matching to EW parton showers, and precision predictions for VBS in extended models.


\part[VBS prospects for the HL-LHC]{VBS prospects for the HL-LHC}\label{part:projections}

\section[Experimental projections for the HL-LHC]{\large Experimental projections for the HL-LHC}\label{sec:vbs_proj}

\begin{figure}[!t]
    \centering
    \includegraphics[width=\columnwidth]{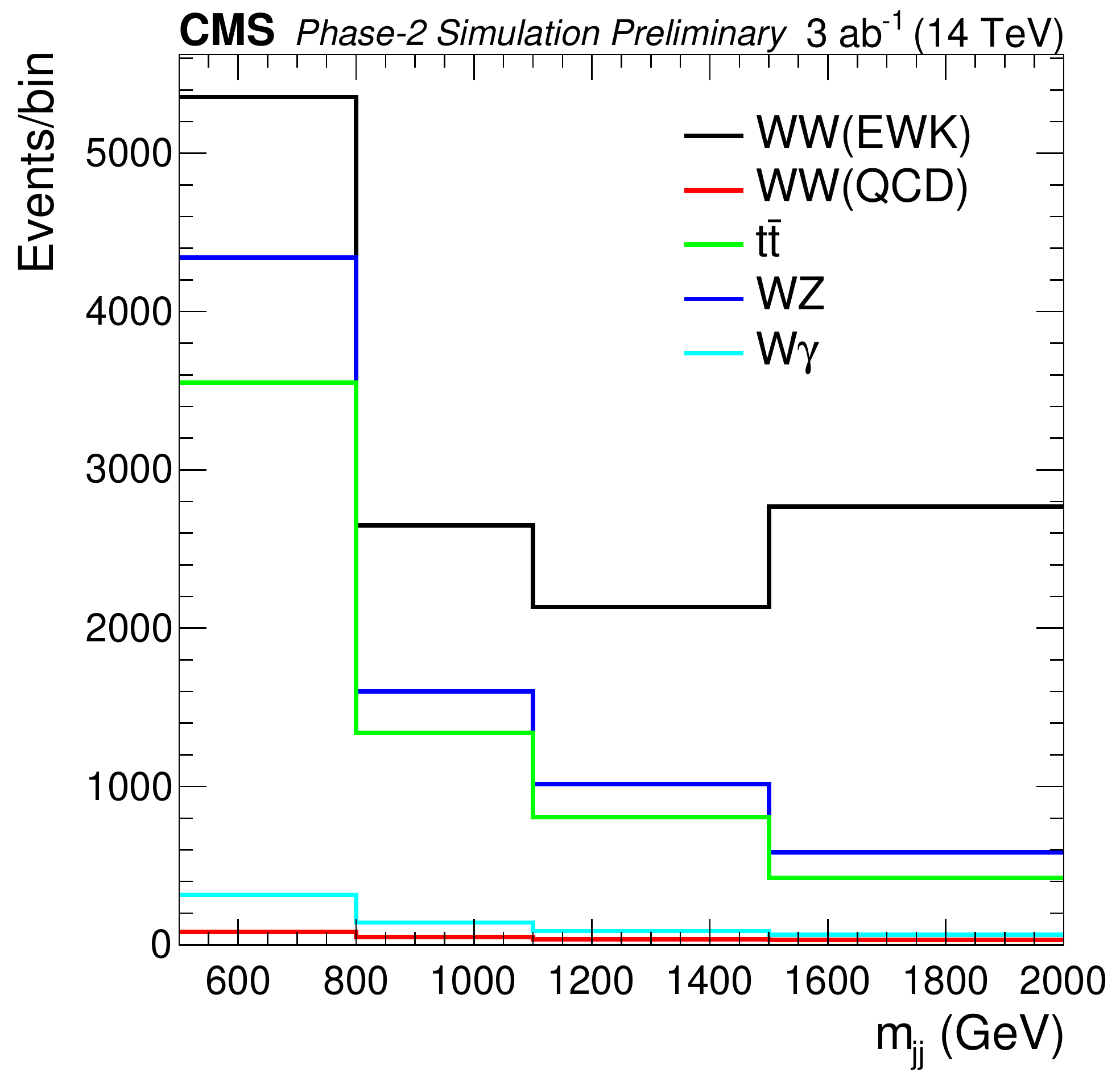}
    \caption{Invariant mass of the two leading jets of the $W^\pm W^\pm jj$ process  at $\sqrt{s}$=14~TeV and normalized to an integrated luminosity of 3000 fb$^{-1}$, from Ref.~\cite{CMS:2018zxa}.}
    \label{fig:WW_mjj}
\end{figure}


The HL-LHC will collide protons at $\sqrt{s}=$14 TeV, with a projected peak instantaneous luminosity of about $5\cdot 10^{34}$ cm$^{-2}$s$^{-1}$, and which will be increased to about $7.5\cdot 10^{34}$ cm$^{-2}$s$^{-1}$~\cite{Apollinari:2017lan}. These rates are about three-to-four times higher than the peak luminosity of Run~2, and will accumulate to a  total integrated luminosity of about 3 ab$^{-1}$. The ATLAS and CMS detectors will be upgraded to cope with the new HL-LHC operating conditions. These upgrades include extended geometric coverage and finer detector resolution, and will improve sensitivity to VBS processes.
See Section~\ref{sec:cms_upgrade} for more details on the detector upgrades.

VBS analyses, which at the moment have uncertainties that are statistically dominated, will benefit from the larger amount of data collected and the higher center-of-mass energy. Moreover, detector upgrades, such as the extension of tracker coverage and the addition of timing detectors, will help in the rejection of  additional leptons and of pileup jets, reducing background contamination. Finally, an increase of statistics will allow better calibration, hence a reduction of experimental uncertainties is also expected. In the following we summarize the anticipated HL-LHC performance with
the CMS (see Section~\ref{sec:sec:vbs_cms_proj}) and
ATLAS (see Section~\ref{sec:sec:vbs_atlas_proj}) detectors.

\begin{figure}[!t]
    \centering
    \includegraphics[width=\columnwidth]{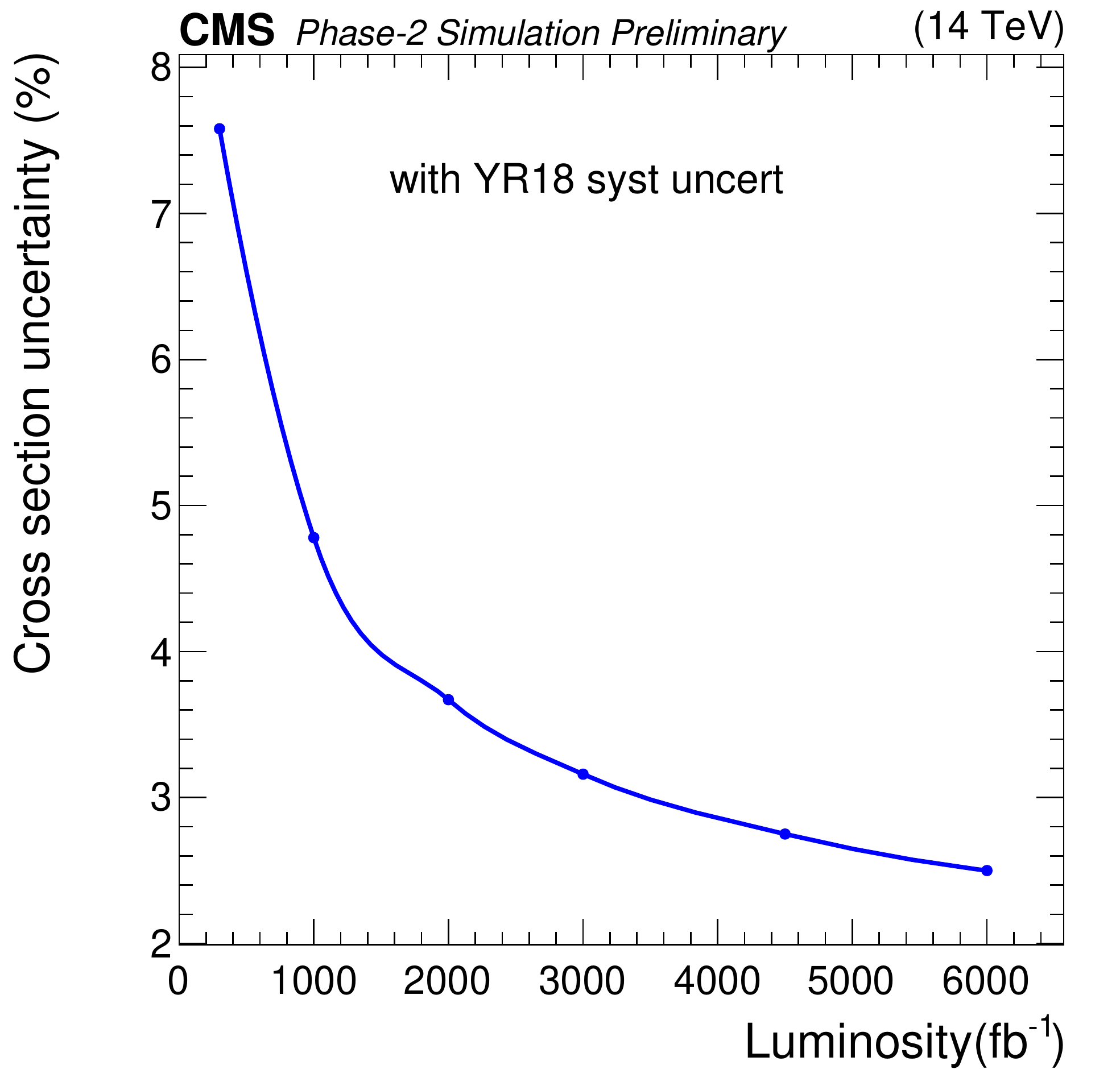}
    \caption{Total cross section uncertainty of the $W^\pm W^\pm jj$ process  as a function of the integrated luminosity, considering systematic uncertainties as in the YR18  scenario~\cite{Dainese:2019rgk}.
    From Ref.~\cite{CMS:2018zxa}.
    }
    \label{tab:WW_proj}
\end{figure}

\subsection{\textbf{\large Projections with the CMS detector}} 
\label{sec:sec:vbs_cms_proj}
Here, we summarize the HL-LHC projections at the CMS experiment for three VBS processes, $W^{\pm}$W$^{\pm}jj$ (Section~\ref{sec:ssWW}),  $WZjj$   (Section~\ref{sec:WZjj}), and $ZZjj$ (Section~\ref{sec:ZZ}), and compare them with the latest published results at 13~TeV with the full Run~2 data sets ($\mathcal{L}\approx$137~fb$^{-1}$). The aim is to understand where these analyses stand now and what would be their perspective for the future. 

\subsubsection{$W^\pm W^\pm jj$}\label{sec:ssWW}

The first VBS process observed with the CMS experiment was the production of two same-sign $W$ bosons  $W^\pm W^\pm jj$ \cite{Sirunyan:2017ret} in the fully leptonic final state (for further details see Section~\ref{sec:vbsresults}). The prospects of measuring VBS production of $W^\pm W^\pm jj$ at the HL-LHC ($\sqrt{s}$ = 14 TeV) has been studied by the CMS Collaboration \cite{CMS:2018zxa}. Results are obtained in the fully leptonic final state, using full simulation of the upgraded Phase-2 CMS detector and accounting for an average number of 200 proton-proton interactions per bunch crossing.

In the HL-LHC scenario, the  $W^\pm W^\pm jj$  would strongly benefit from the increased integrated luminosity, due to its very low cross section $\mathcal{O}$(1~fb). Moreover, the upgrade of the CMS detector, such as the extension of the tracker coverage (improvement of lepton identification) and the new forward High Granularity Calorimeter HGCAL (improving identification of jets), will help with rejecting backgrounds more efficiently, allowing a better discrimination of the signal. The main uncertainties affecting the analysis are treated following the Yellow Report 18 (YR18)~\cite{Dainese:2019rgk} recipe: theoretical uncertainties are halved with respect to the current values, while experimental uncertainties are scaled by 1/$\sqrt{L}$ (where $L$ is the integrated luminosity), until they reach the limit of accuracy estimated with the Phase-2 detector.

\begin{figure}[!t]
    \centering
    \includegraphics[width=\columnwidth]{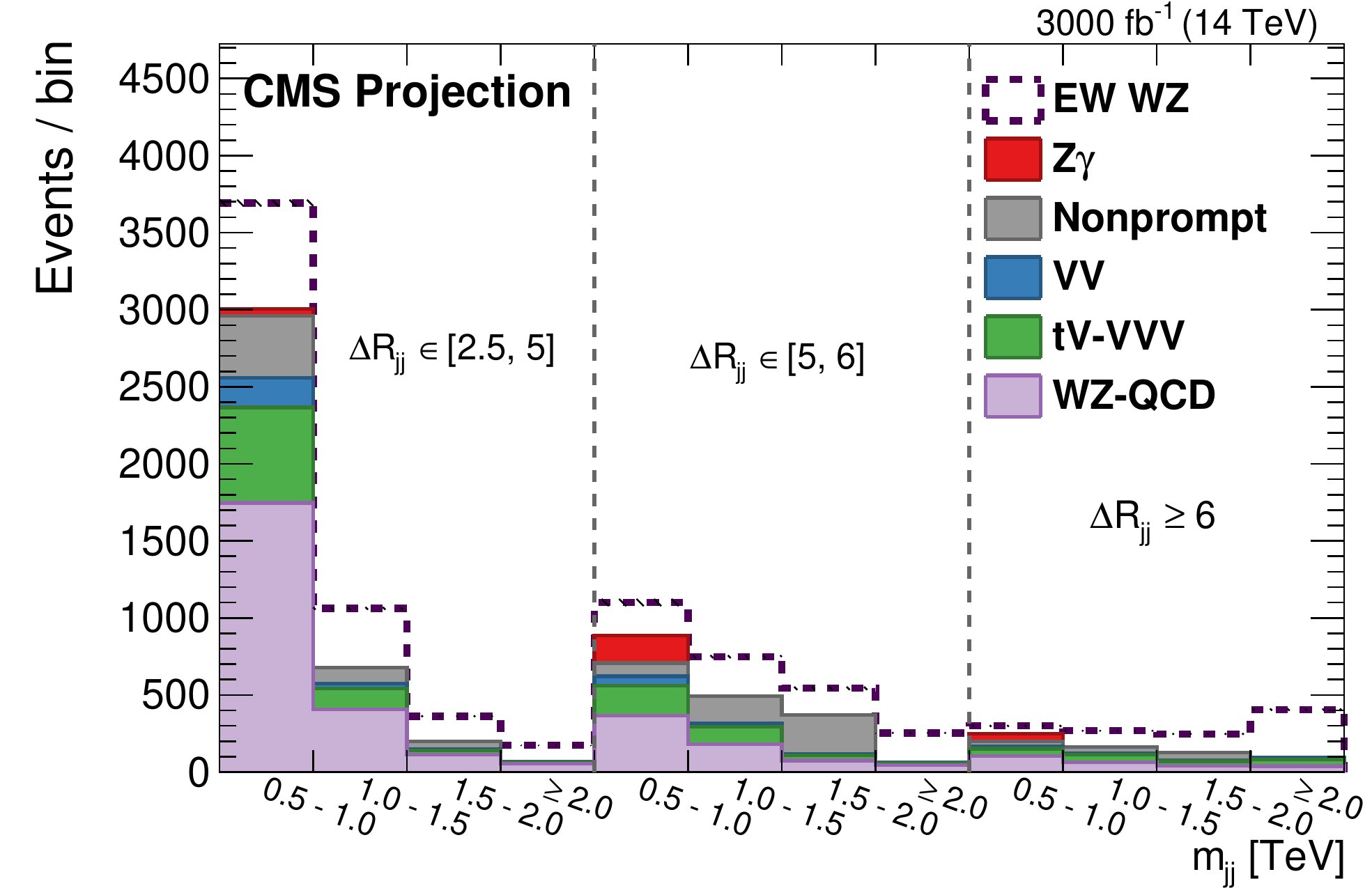}
    \caption{Distribution of m$_{jj}$ in bins of $\Delta R_{\rm jj}$, \ie the distance (in the $\eta,\phi$ space) between jets. From Ref.~\cite{CMS:2018ylh}. }
    \label{fig:WZ_var}
\end{figure}

The final state contains events with two leptons ($e / \mu$) with the same charge, significant missing transverse energy (from $W$ boson decays), and two high-energy jets with a high invariant mass and large separation in pseudorapidity. More details on the analysis are documented in Ref.~\cite{CMS:2018zxa}. 

The invariant mass distribution of the two leading jets (Fig.~\ref{fig:WW_mjj}) is fitted using a binned maximum likelihood approach in which all the three lepton flavor categories ($ee$, $\mu \mu$, $e\mu$) are treated as independent channels. All the systematic uncertainties are inserted as log-normal distributions in the form of nuisances, and  their correlations are taken into account. Despite the requirement of same-sign leptons, top quark pair production is the dominant background of this analysis. The wrong identification of jets containing a $b$ quark is one of the most relevant systematic uncertainties. For the same reason, the wrong identification of jets as leptons, which enhances the so-called ``fake" background, also plays an important role.

Figure~\ref{tab:WW_proj} shows the total cross section uncertainty decreasing as a function of the integrated luminosity, and reaching a value of 3\% for 3000 fb$^{-1}$. A recent CMS result in this channel at $\sqrt{s}$=13~TeV  with the full Run~2 data set (137~fb$^{-1}$)  reports an uncertainty of 11\% on the total cross section~\cite{Sirunyan:2020gyx}. When these results are re-scaled according to the YR18 prescriptions, a good agreement is found with the earlier predictions~\cite{CMS:2018zxa}. This confirms the expectations of reducing the total uncertainty in this channel by almost a factor of 4,  namely down to 3\% for 3000 fb$^{-1}$.

Polarization studies' projections are also performed in Ref.~\cite{CMS:2018zxa}. The longitudinal component of same-sign $WW$ scattering is only 6-7$\%$ of the total cross section. Thus, even at the HL-LHC isolating the longitudinal component in this analysis will be strongly limited by statistics. The expected significance for an integrated luminosity of 3000~fb$^{-1}$ is expected to reach 2.7 standard deviations, and exceed $3\sigma$ when combining CMS and ATLAS results.

\subsubsection{$WZjj$}\label{sec:WZjj}
The first evidence of EW $WZjj$ production in the fully leptonic channel with three leptons at 13 TeV was published by the CMS Collaboration with 2016 data (36~fb$^{-1}$) in Ref.~\cite{Sirunyan:2019ksz} and by the ATLAS Collaboration in Ref.~\cite{Aaboud:2018ddq}. The observation with a statistical significance of 6.8 standard deviations was extracted by a simultaneous fit with the $W^{\pm}W^{\pm}$jj channel using the full Run~2 data set \cite{Sirunyan:2020gyx}. Extrapolations of this analysis to the HL-LHC have shown the potential gain thanks to the larger data samples, since the analysis is still statistically limited~\cite{CMS:2018ylh}.

\begin{figure}[!t]
    \centering
    \includegraphics[width=\columnwidth]{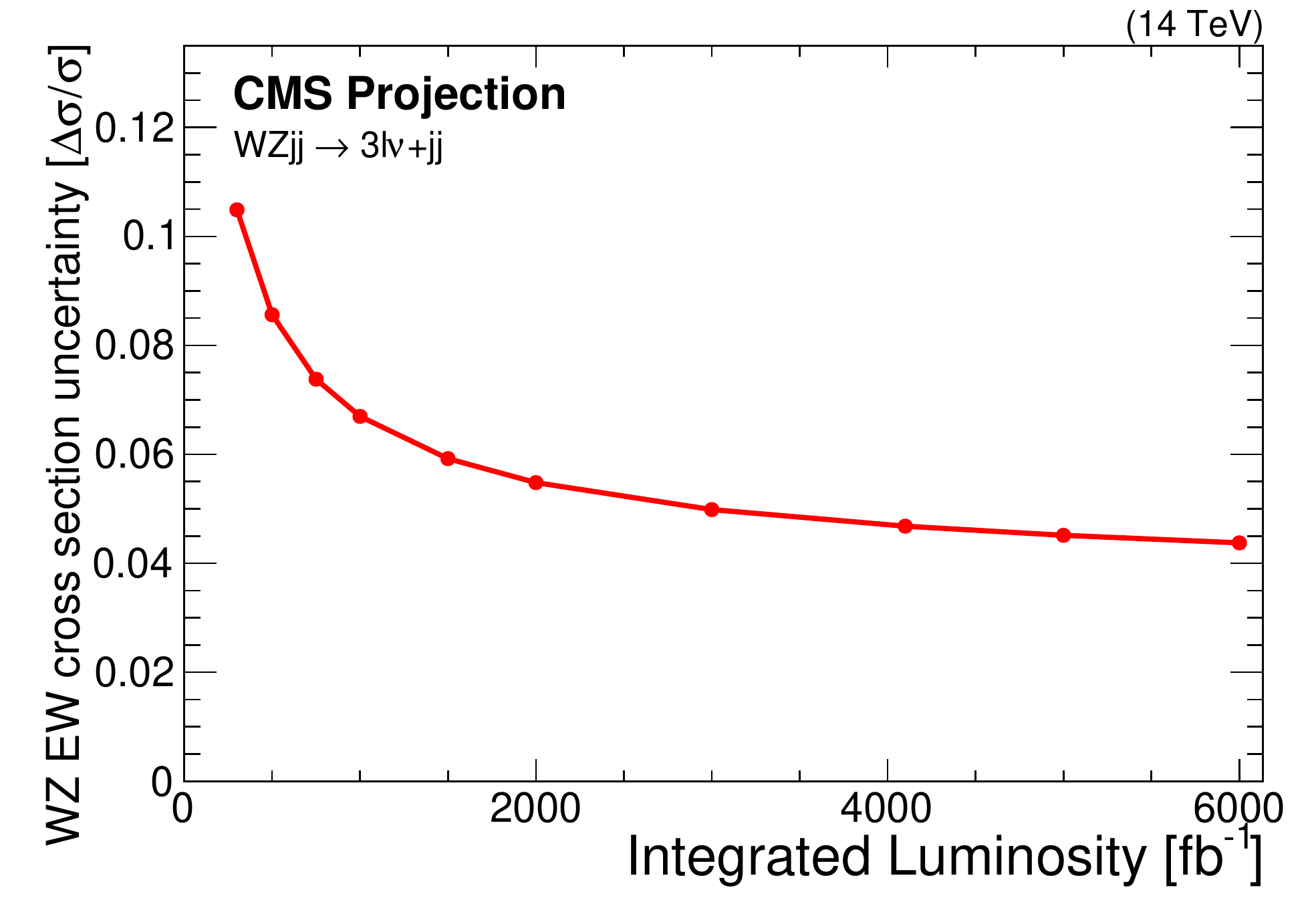}
    \caption{Total cross section uncertainty as a function of integrated luminosity for the WZjj production process. From Ref.~\cite{CMS:2018ylh}.}
    \label{fig:WZ_proj}
\end{figure}

Going from 13 TeV to 14 TeV, VBS cross sections increase by 8-20\%. In particular, the EW component increases by 16\%, while the QCD component (the irreducible background) increases only by 8\%. Moreover, the pseudorapidity coverage extension of the tracker detector would lead to an increase of the event yields, depending on the precise final state, by about 5-8\%. Projections have been obtained using simulated signal and background samples at 13 TeV with a full reconstruction of the Phase-1 detector used in Run~2, scaled by the corrections due to cross section, acceptance, and luminosity increase. The analysis selects events with three charged leptons, missing transverse energy, and two high-energy and well-separated jets. As to charged leptons, one pair of same-flavor and opposite-sign  leptons is expected from the $Z$ boson decay, while the third lepton is from the $W$ boson decay. Events are split into four categories with respect to the flavors of final state leptons: $eee$, $ee\mu$, $\mu\mu e$, $\mu\mu\mu$. Since the fraction of the EW component in inclusive $WZjj$ increases with the invariant mass (m$_{jj}$) and the angular separation $\Delta$R ($\eta$, $\phi$) of the jets, the binned maximum likelihood fit is performed on the 2D distributions of these two variables (Fig.~\ref{fig:WZ_var}), simultaneously for the four independent lepton flavor categories. All the systematic uncertainties are taken into account as nuisances in the fit, as well as their correlations. Input values for most of these uncertainties, both  experimental and theoretical, are expected to be reduced to 1\% at the HL-LHC. Among the theoretical uncertainties, renormalization and factorization scale choices, \ie, QCD scale choices, play the biggest role. For the experimental uncertainties, the jet energy scale and resolution are the largest.  More details on the analysis strategy, the fit and nuisances treatment are discussed in Ref.~\cite{CMS:2018ylh}. Finally, results are summarized in Fig.~\ref{fig:WZ_proj}, which shows the total cross section uncertainty as a function of the integrated luminosity. At 3000~fb$^{-1}$, the total uncertainty would approach 5\%, about a fifth of the current Run~2 precision of $\approx$23\%~\cite{Sirunyan:2020gyx}.

\begin{figure}[!t]
    \centering
    \includegraphics[width=\columnwidth]{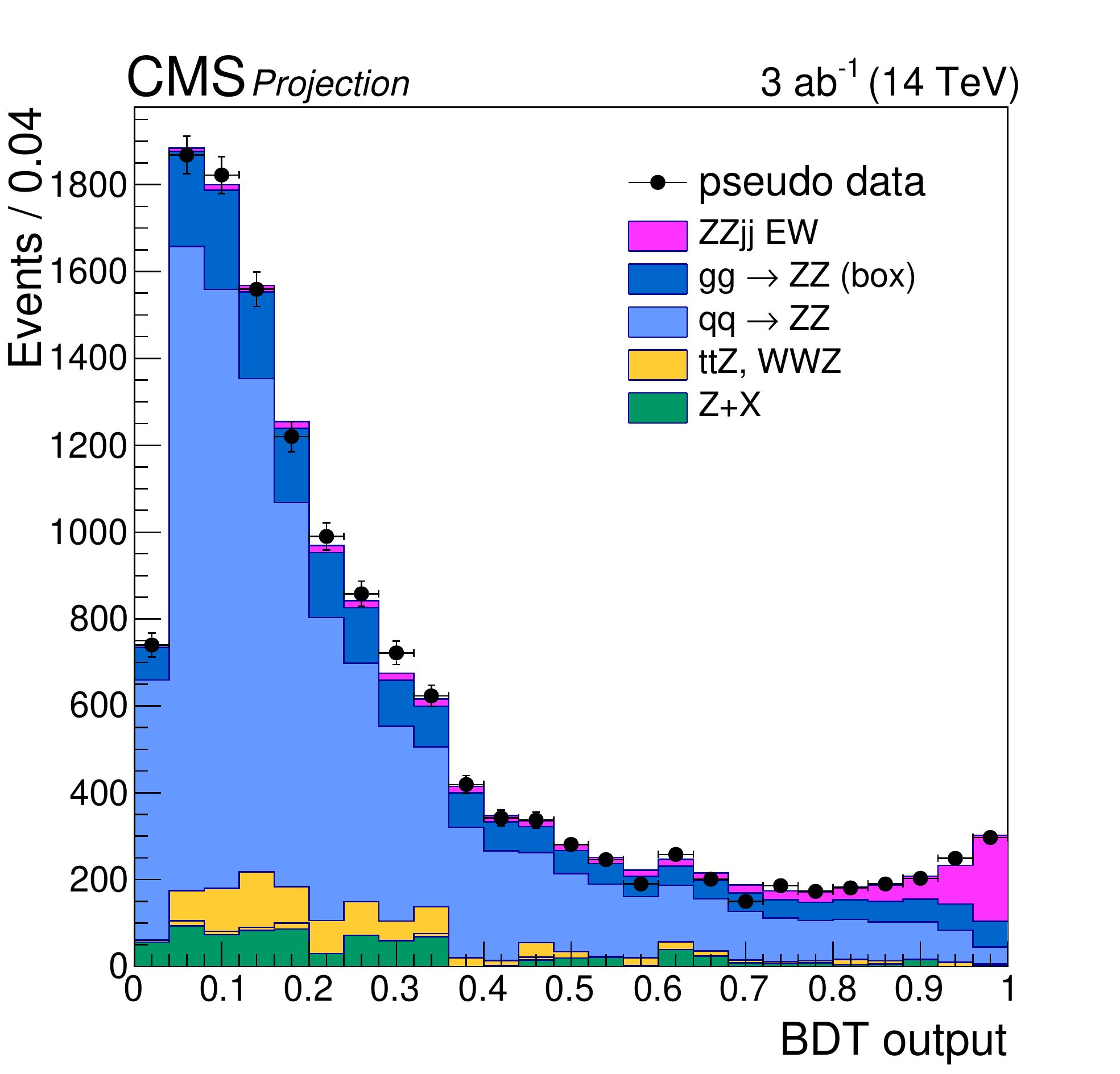}
    \caption{BDT discriminator output for a luminosity of 3000~fb$^{-1}$ at $\sqrt{s}=14$~TeV, for signal and for the backgrounds (from Ref.~\cite{CMS:2018mbt}).
    The purple filled histogram represents the EW signal (ZZjj inclusive process), the other histograms represent the different backgrounds.}
    \label{fig:ZZ_bdt}
\end{figure}

\subsubsection{$ZZjj$}\label{sec:ZZ}
The VBS production of a pair of $Z$ bosons in association with two jets ($ZZjj$) was studied  in the fully leptonic final state by the CMS Collaboration in Ref.~\cite{Sirunyan:2017fvv} with the 2016 data set, and a claim for evidence has been recently published by the CMS Collaboration with the full Run~2 data set~\cite{Sirunyan:2020alo}. Related to this, the ATLAS Collaboration has reported an observation of the same channel with the full Run 2 data set in Ref.~\cite{Aad:2020zbq}. This channel has a small cross section but results in a clean final state with a small contamination of reducible background. The fully leptonic final state allows also the complete reconstruction of the particles' kinematics, making $ZZjj$ a suitable candidate for polarization studies.

\begin{figure}[!t]
    \centering
    \includegraphics[width=\columnwidth]{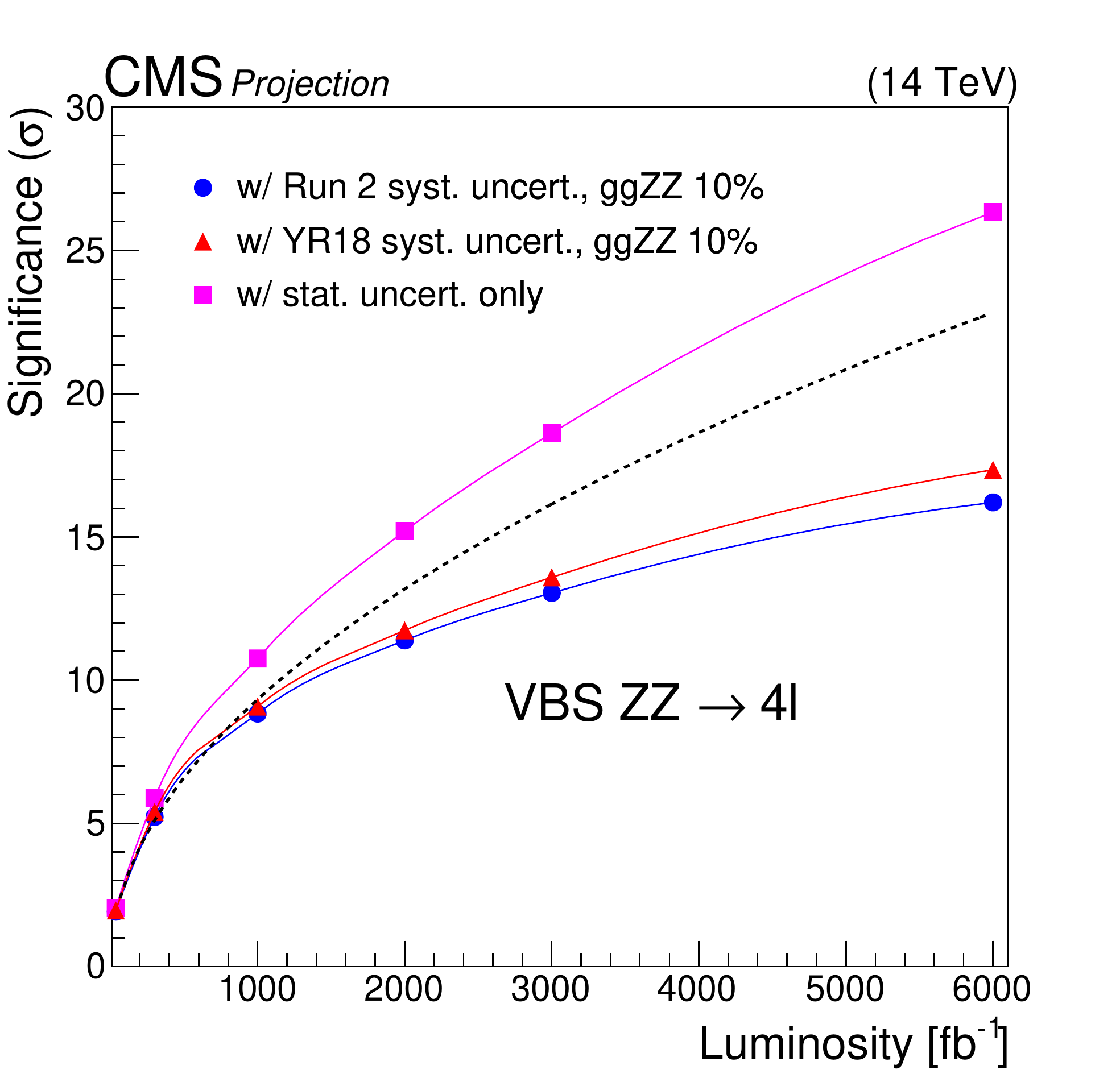}
    \caption{Significance as a function of the integrated luminosity in the Run~2 (blue line, circles) and YR18 (red line, triangles) scenarios. The QCD ggZZ background yield uncertainty is set to 10~\%. Results considering only the statistical uncertainties are also included (magenta line and squares), as well as those obtained by scaling the statistical uncertainties of the 2016 results by the luminosity ratio (dashed black line). From Ref.~\cite{CMS:2018mbt}.
    }
    \label{fig:ZZ_proj}
\end{figure}

The prospects  for measuring both inclusive and longitudinal polarized cross sections of the $ZZjj$ channel at the HL-LHC with the CMS experiment are described in Ref.~\cite{CMS:2018mbt}. The study starts from the analysis of the 2016 data set (36~fb$^{-1}$)~\cite{Sirunyan:2017fvv} and focuses on a final state with two lepton pairs (same-flavor and opposite charge sign from the $Z$ boson decay candidate) and two jets. The Run~2 results are extrapolated to HL-LHC by taking into account the larger integrated luminosity, the increase of center-of-mass energy to 14 TeV, and the extension of the tracker acceptance up to $\eta$=3. The signal's cross section increase is estimated to be about 15\%, while it is of the order of 17\% (13\%) for the QCD $qqZZ$ ($ggZZ$) process. Moreover, the signal yield is expected to increase by up to 20\% thanks to the larger acceptance.  Unfortunately the QCD-induced production of $qqZZ$ is expected to have an increase of 10\% higher than the signal, while the $ggZZ$ background is less relevant in the projected results. A BDT is used to disentangle the EW and QCD components of $ZZjj$. The BDT distribution in Fig.~\ref{fig:ZZ_bdt} is then fitted with a maximum likelihood approach, in which systematic uncertainties are considered as nuisance parameters and profiled.

Regarding systematic uncertainties, two different scenarios have been considered.
In the first scenario (``Run~2''), the systematic uncertainties remain the same as the ones quoted in Ref.~\cite{Sirunyan:2017fvv}. In the second scenario (``YR18")~\cite{Dainese:2019rgk}, the theoretical uncertainties are halved and the experimental ones are scaled down by a factor 1/$\sqrt{L}$. In both scenarios, the theory uncertainty on the loop-induced production of $ZZjj$ has the largest impact; its input value is fixed at 10\%. Results that include the two scenarios are shown in Fig.~\ref{fig:ZZ_proj}. A new prediction, obtained by scaling the YR18 projection with the most recent results (a 4$\sigma$ significance) with the full Run~2 data set in the same final state~\cite{Sirunyan:2020alo}, shows a better performance with respect to the old one. The better performance is due to the fact that, instead of a BDT, a matrix element discriminant is used in the Run~2 analysis~\cite{Sirunyan:2020alo} to separate the EW and QCD components of the $ZZjj$ process. New projections show that the 5$\sigma$ threshold can be exceeded in the early stage of HL-LHC.

Analogously, the VBS Z$_L$Z$_L$ signal component is separated from the VBS and QCD backgrounds by means of a multivariate discriminant. The expected significance for selecting the VBS longitudinal polarized event fraction is 1.4$\sigma$ with 3000~fb$^{-1}$ at the CMS experiment. Comparable results are anticipated for the ATLAS experiment.


\subsection{\textbf{\large Projections with the ATLAS detector}} 
\label{sec:sec:vbs_atlas_proj}

We now summarize the HL-LHC projections at the ATLAS experiment for the $\SSWW$ (Section~\ref{sec:ssWW_atlas}) and
$WZ$ scattering channel (Section~\ref{sec:WZjj_atlas}), 
as well as the sensitivity to anomalous quartic gauge couplings (Section~\ref{sec:aQGC_atlas}).

\subsubsection{$W^\pm W^\pm jj$}\label{sec:ssWW_atlas}
\begin{figure}[t!]
  \centering
  \includegraphics[width=\columnwidth]{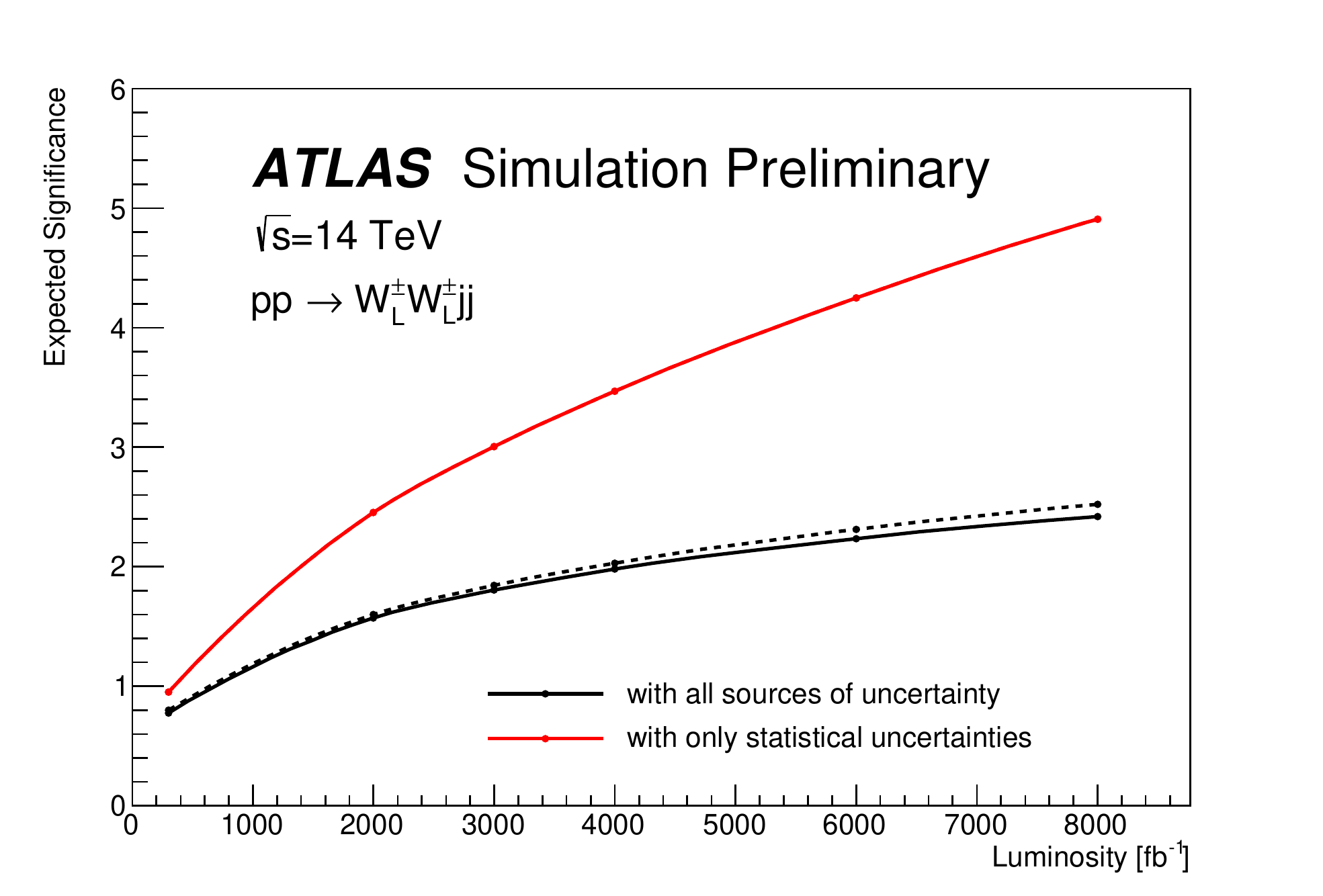}
\caption{Projection of the expected significance of the $W^\pm_\textrm{L} W^\pm_\textrm{L}$ scattering process as function of integrated luminosity, considering all sources of uncertainty (black) or only the statistical uncertainty (red). The dashed line show the expectation for the optimistic scenario~\cite{ATLAS:2017dhm}. 
}
 \label{fig:ssWW_longi}
\end{figure}

\begin{figure*}[t!]
  \centering
  \includegraphics[width=\textwidth]{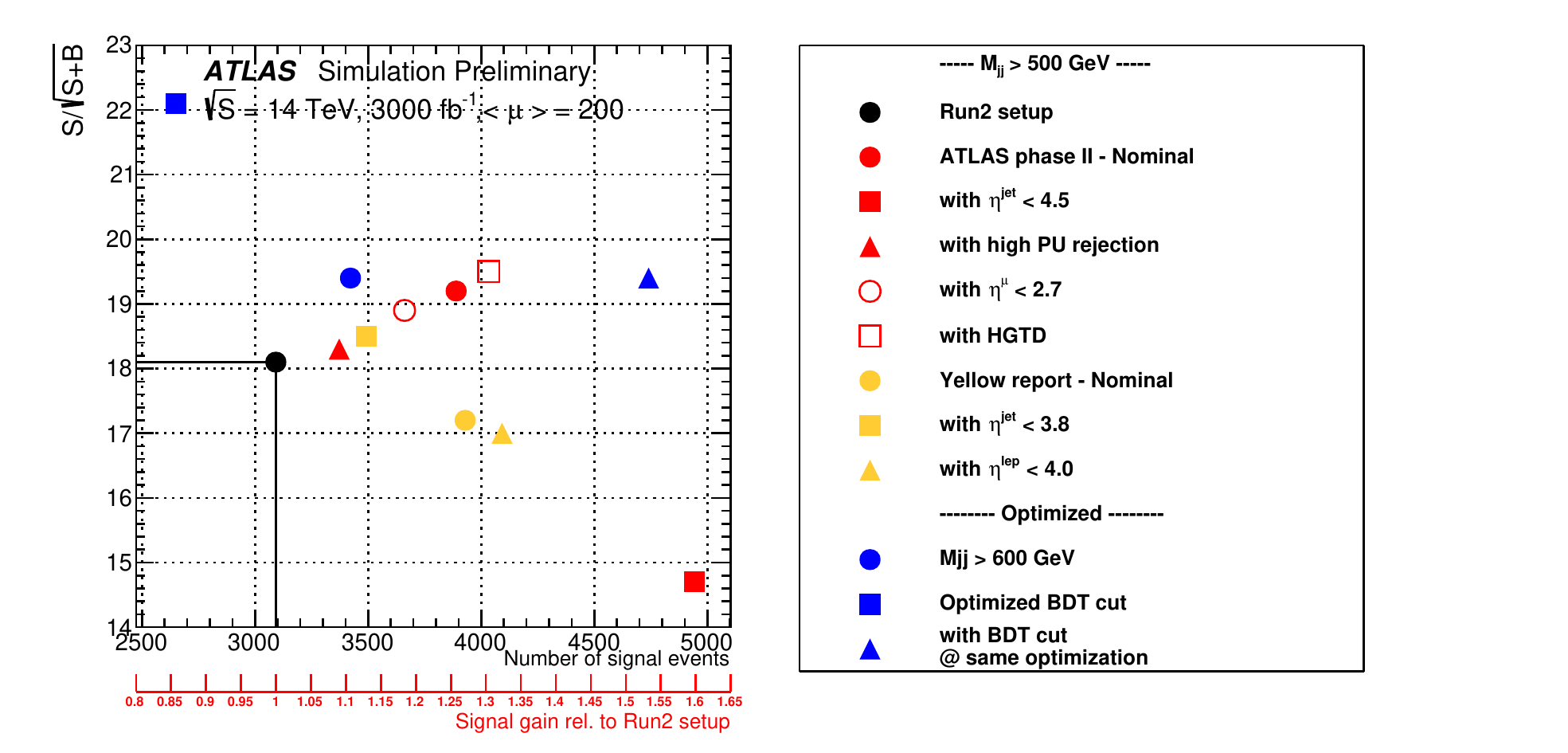}
\caption{Overview over the anticipated sensitivity performance for the $WZ$ process for the HL-LHC phase achieved using the event selections based on the detector performance parameters described in the text. Signal sensitivity as a function of the number of events and relative signal gain with respect to the Run~2 settings. Figure from Ref.~\cite{ATLAS:2018tav}.
}
 \label{fig:WZ_cuts}
\end{figure*}

The most recent HL-LHC projections for $\SSWW$ scattering with the ATLAS detector~\cite{Aad:2008zzm} focus on the comparison of detector coverage and an investigation into the extraction of the longitudinal component. The selection is based on the 8 TeV ATLAS measurement reported in Ref.~\cite{Aad:2014zda}, while the study of Ref.~\cite{ATLAS:2017dhm} uses particle-level simulated samples.

For both studies, the signal as well as the same-sign $WW$ and $WZ$ backgrounds are simulated. Particle-level objects are smeared to achieve a realistic estimate of the event yield, accounting for the detector resolution. Other non-dominant backgrounds are estimated from the 8~TeV measurement. A systematic uncertainty on the background estimate is assumed to be 15\%. Different detector coverage configurations are studied by varying the range over which particle reconstruction and jet-vertex association are possible. The default scenario does not foresee forward tracking for $|\eta|>2.7$. Extending the coverage of jet-vertex association to $|\eta|<3.8$ allows one to lower the $p_{T}$ threshold for jets; this has the largest impact and increases the signal acceptance by up to 10\%. Extending the muon acceptance to $|\eta^\mu|=3.8$ increases the overall signal acceptance by $\sim5\%$, while extending the range for electron reconstruction to $|\eta^e|=3.8$ has only a minor impact. Overall, uncertainties on the signal strength are projected to be in the range of 4.0-4.5\%. 

Reference~\cite{ATLAS:2018uld} presents the extraction of longitudinally  polarized $W^\pm_{L} W^\pm_{L}$ scattering. The event selection is slightly optimized for this purpose by raising the momentum thresholds. This also serves to reject backgrounds from QCD production of $WW$. Experimental systematic uncertainties for the study are taken from the 13 TeV observation of same-sign $WW$ scattering events~\cite{Aaboud:2019nmv}. Alternatively, an optimistic scenario reducing the uncertainties on data-driven backgrounds by 33-50\% is also considered. The total uncertainty on the same-sign $WW$ scattering cross section is projected to be 6\%. A likelihood fit is used to extract longitudinal scattering signal from the distribution of the difference in the azimuth angle between the two jets in two regions of dijet invariant mass: (i) $520<m_{jj}<1100$ GeV and (ii) $m_{jj}>1100$ GeV. The significance is expected to be $1.8~\sigma$ as shown in Fig.~\ref{fig:ssWW_longi}. The expected precision of the measurement of the longitudinal $W^\pm_{L} W^\pm_{L}$ scattering cross section is 47\%. Considerable improvements are achievable using multivariate techniques. However, these have not yet been explored.

\subsubsection{$WZjj$}\label{sec:WZjj_atlas}

The HL-LHC prospects for observing $WZ$ scattering with leptonic decays of the $W$ and $Z$ bosons are investigated in Ref.~\cite{ATLAS:2018tav}. All major detector effects are included using  parametrization. The selection is based on the selection developed for the observation of $WZ$ scattering with the ATLAS detector at $\sqrt{s}=13$~TeV~\cite{Aaboud:2018ddq}.

The ATLAS detector coverages studied are described in detail in Ref.~\cite{ATLAS:2018tav}. Most notable are the Run~2 benchmark scenario (accepting only central leptons, $|\eta|<2.5$, and requiring forward jets, $2.5<|\eta|<4.5$ with $p_{\mathrm{T}}>75$~GeV) and a nominal ATLAS Phase-2 scenario with extended acceptances for leptons and jets. A variation of this selection includes using the high granularity timing detector (HGTD)~\cite{CERN-LHCC-2020-007},  which improves the electron and jet selection efficiencies in the forward region ($2.4 < |\eta| < 3.8$). For all event selections, the signal region is defined using the $m_{jj}>500$~GeV selection criterion. Alternatively, an optimized selection based on a BDT classifier trained using 25 input variables sensitive to the VBS topology is used. Figure~\ref{fig:WZ_cuts} compares these selections using the number of selected signal events and the expected signal significance estimated as $S/\sqrt{S+B}$. Extending the jet acceptance is only possible if a jet-vertex association can be applied or if the jet $p_{\mathrm{T}}$ threshold is increased, which lowers the number of signal events. Otherwise a prohibitive amount of background is selected. The maximal gain in signal events compared to the Run~2 setup is 60~\% using either the BDT or extending the jet acceptance up to $|\eta| = 4.5$ without increasing momentum thresholds. An alternative BDT-based selection achieves the largest signal significance, however with a 15~\% loss of signal events with respect to the Run~2 setup. The investigation projects that the uncertainties will be dominated by systematics related to jets and the $WZ$ background. It is suggested that these could be better constrained using additional data-driven control regions. 

\begin{figure}[t!]
  \centering
   \includegraphics[width=\columnwidth]{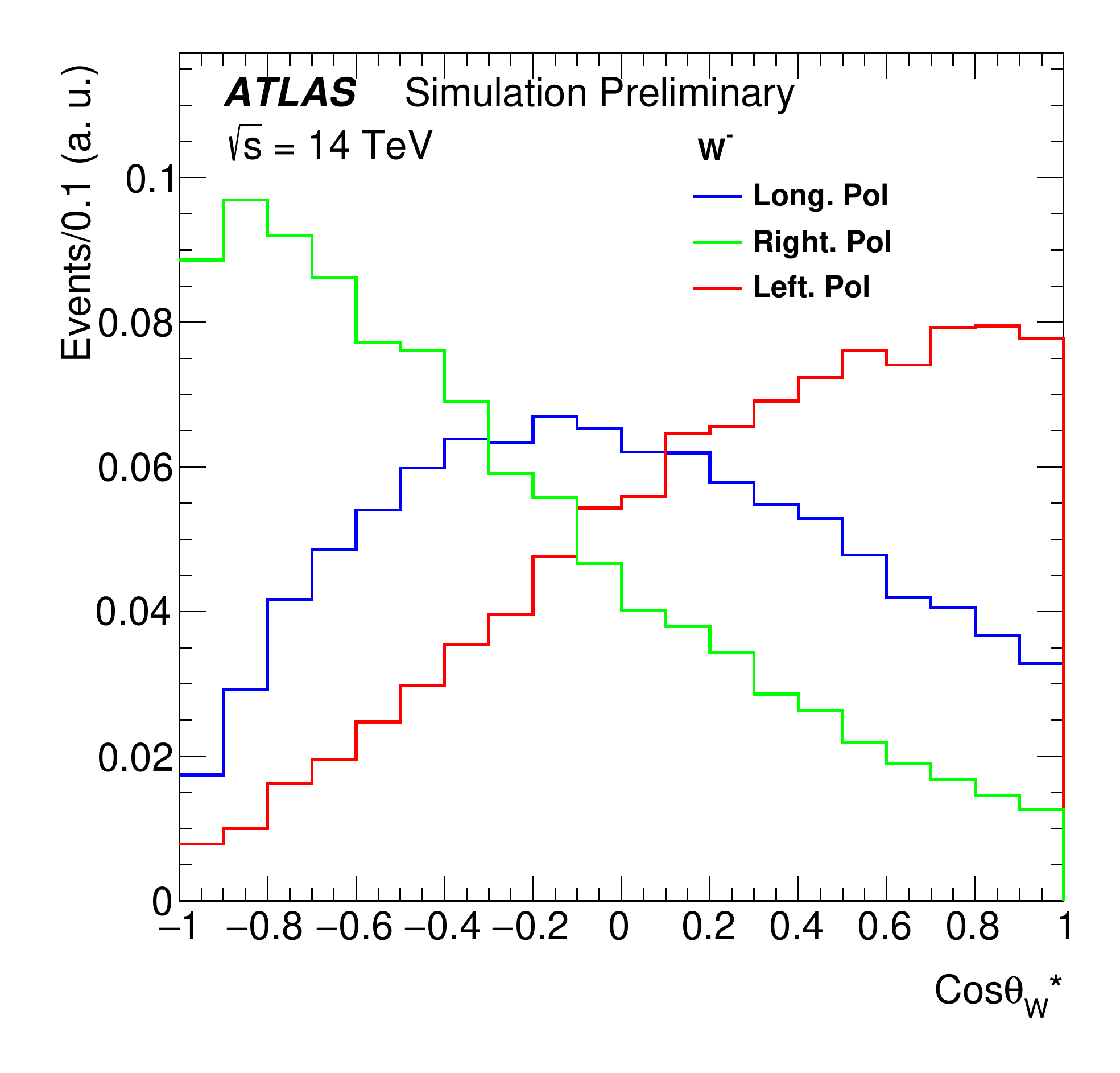}
\caption{Normalized $W^-$ polarization state templates at the reconstruction level and after the final event selection, including requiring $m_{jj} > 600$~GeV. Figure from Ref.~\cite{ATLAS:2018tav}.
}
 \label{fig:WZ_cuts_polar}
\end{figure}

The fraction (F$_\lambda$) of the individual vector boson polarization state $\lambda$ can be extracted using template fits of F$_{L}$ and (F$_{-}$-F$_{+}$) to the $\cos\theta^*$ distribution. The angle $\theta^*$ is defined as the decay angle of the lepton (or anti-lepton for $W^+$) with respect to the boson's direction in the $WZ$ rest-frame. Figure~\ref{fig:WZ_cuts_polar} shows the normalized distributions of the different $W$ polarization states as a function of $\cos\theta^*$ at the reconstruction level. Four different event distributions are fitted depending on the boson that is probed and on the charge of the $W$ boson present in the event. For the systematic uncertainty associated to the total background normalization, three possible scenarios are assumed: 20\%, 5\% and 2.5\%. Different selections are tested, namely the nominal selection requiring $m_{jj}>500, 600, 1100$~GeV, as well as two different BDT cuts. In addition, the effect of quark-gluon tagging on the leading and sub-leading jets to further reduce the backgrounds is studied. Finally, the luminosity is doubled to emulate the combination of the ATLAS and CMS data samples. The expected significance for the  polarization signal is 0.5-3.5 $\sigma$ depending on the selection, with the most sensitive distribution being that of the $Z$ boson in $W^+Z$ events.

The signal significance for the $W_L Z_L$ polarization state is estimated in a simultaneous fit of $\cos\theta^*$ and the scalar sum of the leptons' transverse momenta, without considering the background processes. The estimated significance is less than one standard deviation. The use of the full double differential distribution of $\cos\theta^*(W)$ versus $\cos\theta^*(Z)$ or other more sophisticated methods could however improve these results. 

\subsubsection{Anomalous Quartic Gauge Couplings}\label{sec:aQGC_atlas}

Projections for limits on anomalous quartic gauge couplings and dimension-8 EFT operators~\cite{Eboli:2006wa} have been reported in Ref.~\cite{TheATLAScollaboration:2013rrd}, which uses one-dimensional distributions for both $WZ$ and $WW$ scattering processes in the limit setting. It is instructive to compare the projected limits with those extracted using the LHC Run~2 data to answer the obvious question as to how realistic the HL-LHC projections in general are. The projected limits for an integrated luminosity of 300~fb$^{-1}$ in 14 TeV collision data are about a factor 2-3 worse than what was achieved with Run~2 data~\cite{Sirunyan:2020gyx}. This indicates that the projections are rather conservative.

\subsection{Summary}
VBS production of EW boson pairs are rare processes that allow precision measurements of the EW sector. VBS processes are typically statistically limited at Run~2 and will benefit from the HL-LHC operation conditions (14~TeV, 3000~fb$^{-1}$), resulting in better constraints of known processes ($W^{\pm} W^{\pm}jj, WZjj$) and measurements of those that have not been observed yet ($W^+ W^-jj, ZZjj$). Moreover, the HL-LHC will allow the ATLAS and CMS experiments study of final states other than the fully leptonic ones, such as semi-leptonic or even fully hadronic final states. These modes guarantee higher statistics. In the HL-LHC scenario, the limiting factor to precision measurements is expected to be the systematic uncertainties: a work toward the reduction of systematic, theoretical and experimental, would be of primary interest for the LHC community.

\section[SMEFT in VBS at the HL-LHC]{\large SMEFT in VBS at the HL-LHC}\label{sec:hllhc_smeft}

The Standard Model Effective Field Theory (SMEFT) has become a standard tool used in indirect and semi-model-independent searches for BSM physics at the LHC. A lot of effort, both on the theory and experimental sides, goes into finding the optimal methodological guidelines for applying SMEFT in VBS processes in particular. One vital issue is how to correctly set limits on Wilson coefficients without violating the validity of EFT.  Another lasting question is whether to focus on dimension-6 or dimension-8 SMEFT operators in data analysis.  Some of the most recent progress concerning these two important issues are discussed. This progress involves new developments in setting limits on SMEFT dimension-8 operators based on the LHC Run 2 data collected by the ATLAS and CMS detectors, and new studies of potential effects of SMEFT dimension-6 operators based on current experimental bounds from global fits to data from other processes.
As an example of this progress, we focus in this section on the use of the so-called ``clipping'' technique by CMS during Run 2 and the prospects for probing dimension-6 operators with same-sign $WW$ scattering at the HL-LHC. (For examples of studying SMEFT operators at future colliders, see Section~\ref{sec:smeft_muonCo}.)

\subsection{Full clipping technique in the analysis of Run 2 data}\label{sec:clip}

Recently, the CMS Collaboration published an analysis of the same-sign $WW$ and $WZ$ scattering processes, based on data collected during the LHC Run~2, where limits on dimension-8 operators were derived for the first time using a partial ``clipping'' technique~\cite{Sirunyan:2020gyx}.  For every value of an individual Wilson coefficient, simulated BSM distributions were ``clipped'' at the unitarity limit. Beyond this limit, the high mass tails were replaced with SM tails to compute the EFT prediction. The impact of considering only the unitarity condition on the results was shown to be significant, with the actual limits being relaxed typically by a factor of $4-5$ compared to limits calculated without unitarity considerations.

\begin{figure}[t!]
  \centering
   \includegraphics[width=\columnwidth]{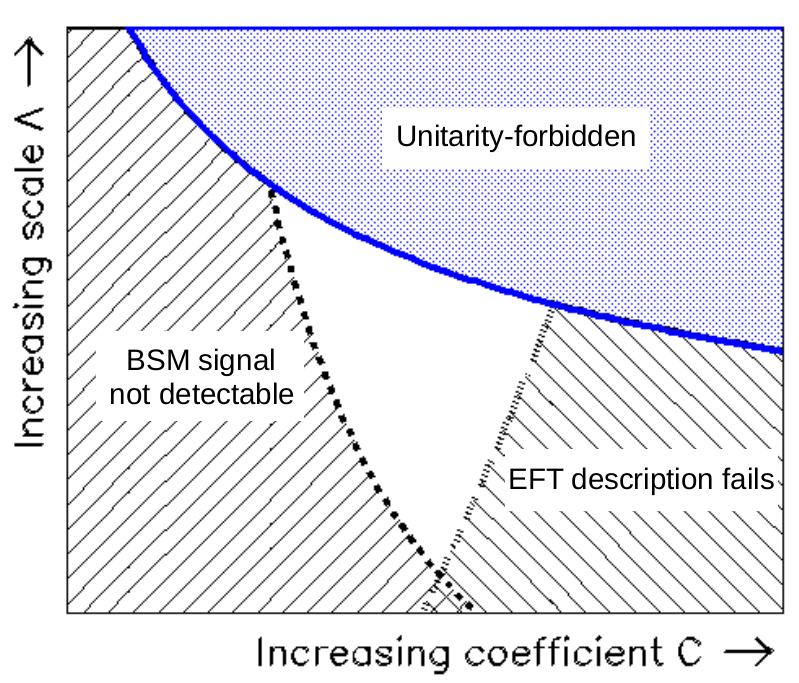}
\caption{For a Wilson coefficient $C$ and EFT cutoff scale $\Lambda$, a sketch of the regions in $\Lambda-C$ space that are forbidden by unitarity (large $\Lambda$ and large $C$), are inconsistent with the assumptions of EFTs (small $\Lambda$ and large $C$), and are inaccessible to experiments (small $C$ and/or large $\Lambda$). Adapted from Ref.~\cite{Kalinowski:2018oxd}.
}
 \label{fig:eftValidity}
\end{figure}

In reality, the scale of new physics can be lower than the unitarity limit.  Physically interpretable limits on individual Wilson coefficients $C$ exist only as a function of the EFT $\Lambda$ cutoff parameter, the latter ranging between the lowest $WW$ mass accessible in the experiment and the appropriate unitarity bound.  Limits then take the form of 2-dimensional exclusion curves, with one axis being the value of $C$ and the other axis being the value of $\Lambda$.
A sketch of such an exclusion curve is shown in Fig.~\ref{fig:eftValidity}.
Also shown in this cartoon are the regions in $\Lambda-C$ space that are forbidden by unitarity (large $\Lambda$ and large $C$), are inconsistent with the assumptions of EFTs (small $\Lambda$ and large $C$), and are inaccessible to experiments (small $C$ and/or large $\Lambda$).

Observed (and expected) experimental limits can be compared to theoretical limits that follow solely from the unitarity condition.  Naturally, experimental limits are meaningful only if they are stronger than theoretical limits.  An encouraging observation is that theoretical exclusion curves, based on the unitarity condition, give a flatter dependence of  $C(\Lambda)$ than experimental exclusion curves based on signal significance.  From this it can be anticipated that the two curves will cross at some value of $\Lambda$ and hence experimental limits for most dimension-8 operators can indeed be meaningful as long as $\Lambda$ is not too high.  Given that unitarity limits for the values of Wilson coefficients obtained in Ref.~\cite{Sirunyan:2020gyx} are typically
around 1.5 TeV, one can expect this critical value of $\Lambda$ be roughly around 2 TeV (this takes into account that limits on  $C$ with a constant $\Lambda$ are more permissive than with a running $\Lambda (f)$).  In the HL-LHC perspective, this range will still widen. 

The analysis also included a study of the most sensitive kinematic variables to each individual operator, assuming the signal process $pp\to W^\pm W^\pm jj \to \ell_i^\pm\ell_j^\pm j j \nu\overline{\nu}$. As observed in several earlier studies, the cluster mass variable
\begin{align}
& M_{o1} =  \nonumber\\
 & \sqrt{\left(
\vert\vec{p}_T^{~\ell_1}\vert
+
\vert\vec{p}_T^{~\ell_2}\vert
+
\vert\vec{p}_T^{~\rm miss}\vert
\right)^2 
- 
\left(
\vec{p}_T^{~\ell_1}
+
\vec{p}_T^{~\ell_2}
+
\vec{p}_T^{~\rm miss}
\right)^2}    
\end{align}
is the most sensitive variable for most dimension-8 operators.
Here, $\vec{p}_T^{~\ell_k}$ for $k=1,2$ are the transverse momentum vectors of the two same-sign charged leptons and $\vec{p}_T^{~\rm miss}$ is the transverse momentum imbalance of the event.
However, for $S$ operators and sufficiently low $\Lambda$ it is outperformed by the kinematic ratio
\begin{equation}
 R_{pT} = \frac{\vert \vec{p}_T^{~\ell_1}\vert \vert \vec{p}_T^{~\ell_2}\vert }{\vert \vec{p}_T^{~j_1}\vert \vert \vec{p}_T^{~j_2}\vert},
\end{equation}
where $\vec{p}_T^{~j_k}$ are the transverse momentum vectors of the two VBS-tagged jets.
For further details, see Ref.~\cite{Kalinowski:2018oxd}.

\subsection{Dimension-6 operators in the same-sign $WW$ process}\label{sec:dim6}

The potential impact of SMEFT dimension-6 operators within their  current experimental bounds on the same-sign $WW$ VBS scattering process at the LHC was recently studied in Refs.~\cite{Gomez-Ambrosio:2018pnl,Araz:2020zyh,Dedes:2020xmo,Ellis:2020unq,Ethier:2021ydt}. Outside of SMEFT, recent studies include Refs.~\cite{Delgado:2017cls,Delgado:2018nnh,Delgado:2019ucx}. In particular, in the generator-level study of Ref.~\cite{Dedes:2020xmo}, potential BSM signals from all the individual operators at dimension 6 in the Warsaw basis \cite{Grzadkowski:2010es} was checked by studying their impact on the distributions of different kinematic variables.  For each operator, the variable with the highest sensitivity was chosen. Signals were normalized to the nominal integrated luminosity for the HL-LHC. The study explicitly included four VBS operators that affect the reaction
$W^\pm W^\pm \to W^\pm W^\pm$: ${\cal O}_W$, ${\cal O}_{\phi\Box}$, ${\cal O}_{\phi D}$ and ${\cal O}_{\phi W}$, neglecting ${\cal O}_{\phi WB}$ which vanishes in the cross section formula to leading order in $s$, and neglecting CP violating operators. In addition, 26 background operators, i.e., those that do not affect the $WW$ scattering reaction -but may affect the full process at the $pp$ level- were explicitly considered.

Current experimental limits on dimension-6 operators other than 4-fermion operators were taken from recent global fits to existing data~\cite{Ellis:2018gqa,Dawson:2020oco} that include all the most recent LHC Run 2 data.  For 4-fermion operators, older bounds~\cite{Domenech:2012ai} were used and found sufficient to claim them negligible.  The following background operators were also found to be negligible within the existing limits:
\begin{eqnarray}
& \calO_{u\phi}, ~\calO_{d\phi}, ~\calO_{uG}, ~\calO^{(1)}_{\phi q}, ~\calO_{\phi u}, ~\calO_{\phi d}, ~\calO_{uu},
\nonumber\\
& 
~\calO_{dd}, ~\calO^{(1)}_{ud}, ~\calO^{(8)}_{ud}, ~\calO^{(1)}_{qq}, ~\calO^{(1)}_{qu}, ~\calO^{(8)}_{qu}, ~\calO^{(1)}_{qd}, ~\calO^{(8)}_{qd}.
\end{eqnarray}
For $\calO_{uG}$, limits are deduced assuming minimal flavor violation.
The effects of background operators for which experimental limits are not available in literature, namely
\begin{equation}
    \calO^{(1)}_{quqd}, ~ \calO^{(8)}_{quqd}, ~ \calO_{uW}, ~\calO_{uB}, ~ \calO_{dG}, ~ \calO_{dW}, ~ \calO_{dB},  ~ \calO_{\phi ud}, ~\calO^{(3)}_{qq},
\end{equation}
were studied up to the strong coupling limit and found either negligible or leading to a distinct event kinematics that are easily distinguishable from that of VBS operators. One background operator, ${\cal O}^{(3)}_{\phi q}$, was found to potentially produce an effect up to $\sim$5$\sigma$ in significance at the HL-LHC if the present experimental bounds are assumed.  It too, however, produces a distinct kinematic signature in which jet $p_T$ is the most affected variable.

The effects of proper VBS dimension-6 operators were studied using their global bounds and applying them to each operator individually.  This procedure is justifiable by the fact that correlations between the four operators of interest range between negligible and mild \cite{Ellis:2018gqa}. In the $WW$ scattering process these operators typically affect different helicity combinations and therefore do not significantly interfere.  Large possible effects at the HL-LHC, by far exceeding the 5$\sigma$ threshold, were found for ${\cal O}_{\phi W}$, ${\cal O}_{\phi\Box}$, and ${\cal O}_W$.  The results show that dimension-6 operators cannot be safely ignored in VBS analyses.
Indeed, if the Wilson coefficient for ${\cal O}_W$ is large, the corresponding experimental signature allows the possibility of improving experimental bounds on this operator by using data already collected during Run~2.


\section[Neutrino BSM with VBS signatures]{\large Neutrino BSM with VBS signatures}\label{sec:nuVBS}

Among the several experimental and theoretical motivations for new physics is the case made by \textit{tiny}  neutrino masses and their large mixing angles~\cite{Ahmad:2002jz,Ashie:2005ik}. To reconcile the SM's gauge structure and the renormalizability with oscillation data, and hence nonzero neutrino masses, there must exist new particles that couple to the SM's lepton and Higgs sectors~\cite{Ma:1998dn}. Such scenarios, which are known collectively as Seesaw models, come in many forms. They range from postulating new gauge symmetries with vastly expanded Higgs sectors to minimal extensions with only a few new fermions or scalars that carry SM gauge charges. Whether neutrino masses are generated in these models at tree-level or at loop-level, the particles responsible for generating neutrino masses can couple to SM states, even appreciably in some scenarios. If these particles are also kinematically accessible, then it may be possible to produce them at collider experiments through a variety of mechanisms, including VBS, thereby establishing a probe of physics beyond the SM~\cite{Cai:2017jrq,Cai:2017mow}.

\begin{figure*}[t!]
    \centering
    \includegraphics[width=.94\columnwidth]{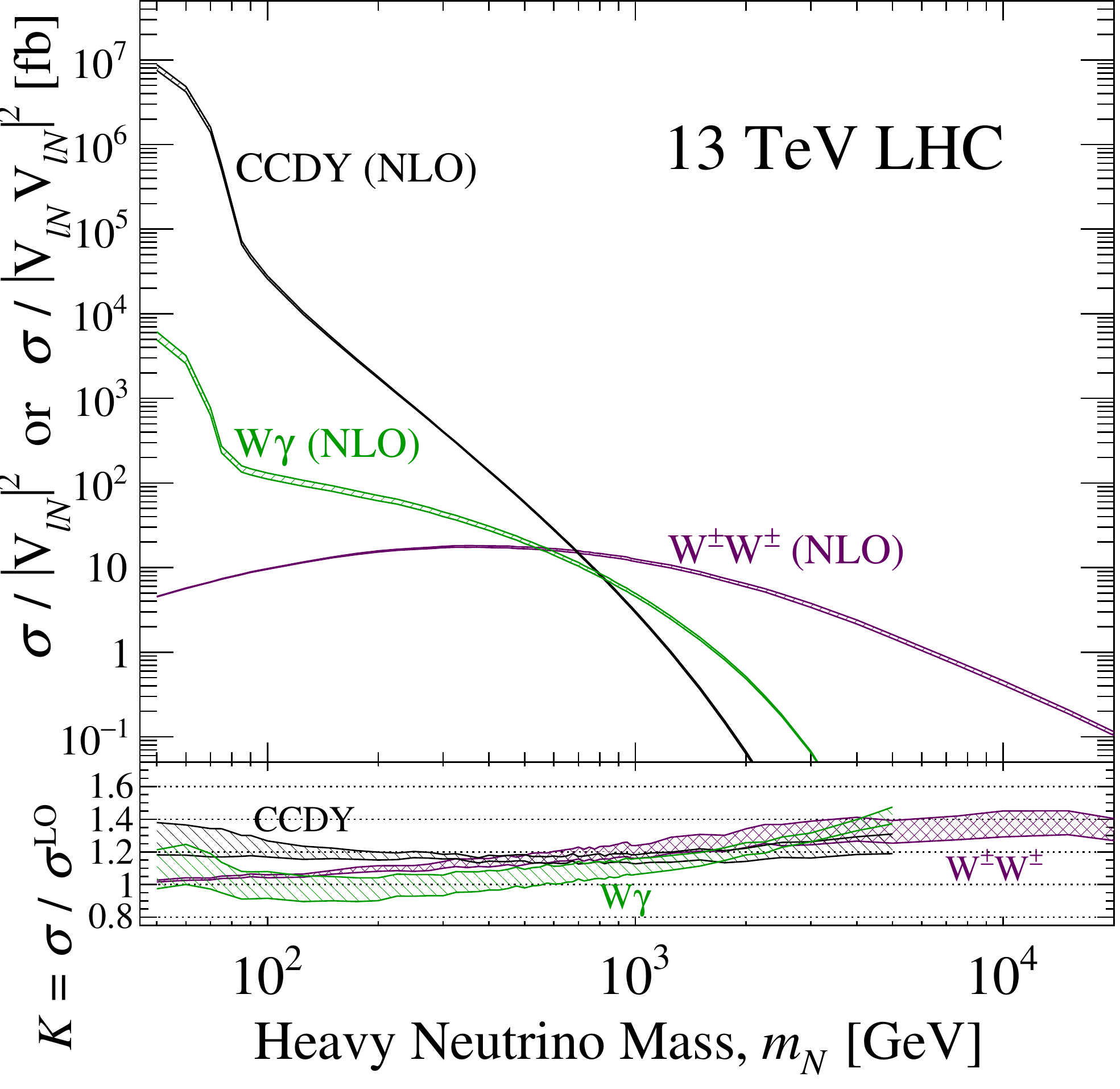}   
    \includegraphics[width=\columnwidth]{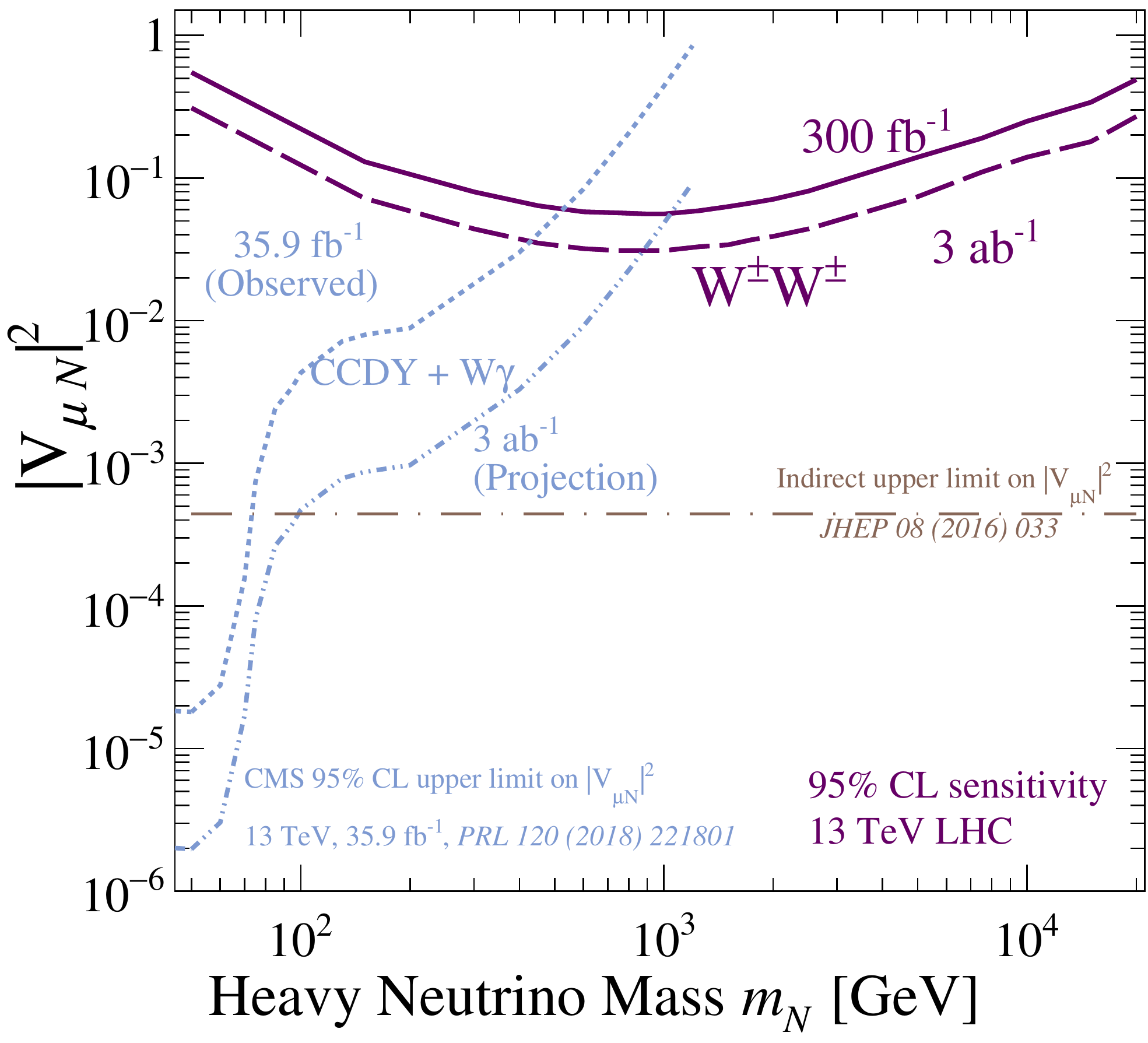}  
        \caption{
        Left: Cross section predictions at NLO in QCD at the $\sqrt{s}=13$ TeV LHC for heavy Majorana neutrinos $N$ production via the charged current Drell-Yan (CCDY), $W\gamma$ fusion, and same-sign $W^\pm W^\pm$ fusion mechanisms, assuming a benchmark active-sterile mixing of $\vert V_{\ell N}\vert =1$. Right: Sensitivity projections for active-sterile mixing of $\vert V_{\ell N}\vert^2$ at the LHC for the $W^\pm W^\pm$ channel in comparison to experimental limits~\cite{Sirunyan:2018mtv,Sirunyan:2018xiv}.
        Figures adapted from Ref.~\cite{Fuks:2020att}. }
    \label{fig:nuVBS_typeI_ww}
\end{figure*}

In recent years, LHC searches for particles hypothesized by neutrino mass models have undergone a transformation. More specifically, the adoption of EW VBS production mechanisms, which were unavailable at predecessor colliders like LEP and the Tevatron, has significantly increased the sensitivity to TeV-scale states at the LHC. This includes, for example, the $W\gamma$ fusion channel in searches for heavy Dirac and Majorana neutrinos $N$ by the CMS experiment~\cite{Sirunyan:2018mtv,Sirunyan:2018xiv} and $WZ$ fusion in searches for charged Higgs $H^{\pm}$ by the ATLAS experiment~\cite{Aad:2015nfa}. The providence of these experimental searches stem from equally transformative developments in the theory community, developments that are the result of systematic investigations into the utility of EW VBS in tests of Seesaw models. The creation of dedicated Monte Carlo tools for studying EW VBS in Seesaw models has also played an important role. For recent reviews of such phenomenological developments, see Refs.~\cite{delAguila:2008cj,Atre:2009rg,Deppisch:2015qwa,Cai:2017jrq,Cai:2017mow,Pascoli:2018heg}.

In the following, a brief summary of these theoretical developments is provided. Also given is an outlook for discovering new particles at HL-LHC and future proton colliders with VBS. In Section~\ref{sec:nuVBS_typeI}, focus is placed on the role of $W\gamma$ and same-sign $WW$ scattering in searches for heavy Dirac and Majorana neutrinos. In Section~\ref{sec:nuVBS_typeII}, singly and doubly charged Higgs bosons from $\gamma\gamma$ and same-sign $WW$ scattering are considered. Finally, in Section~\ref{sec:nuVBS_dim5}, sensitivity of same-sign $WW$ scattering to the sole dimension-five operator in the SMEFT is presented.

\subsection{Heavy neutrinos from $W\gamma$ and $W^\pm W^\pm$ scattering}\label{sec:nuVBS_typeI}

\begin{figure*}[!t]
    \centering    
    \includegraphics[width=\columnwidth]{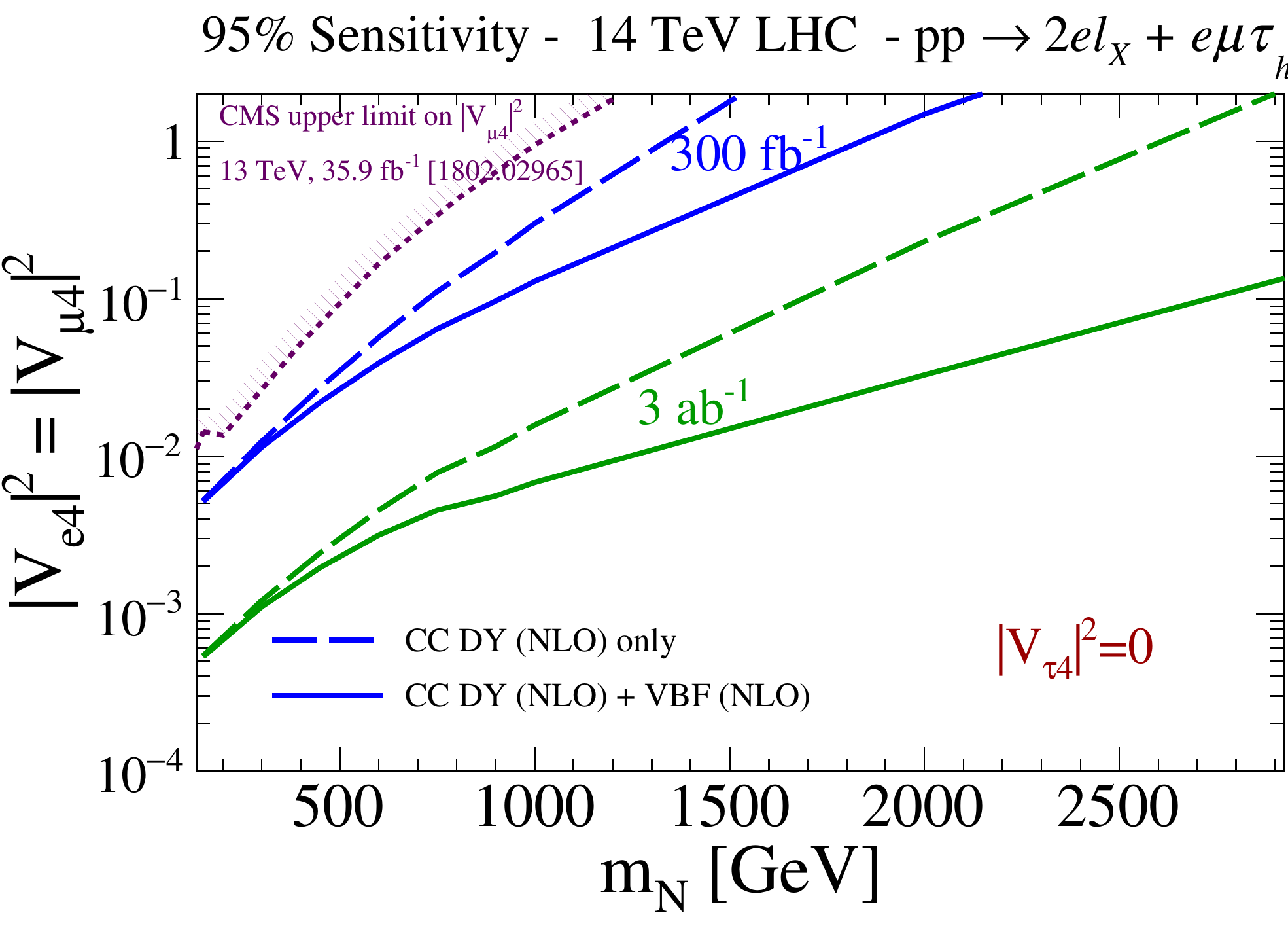}    
    \includegraphics[width=\columnwidth]{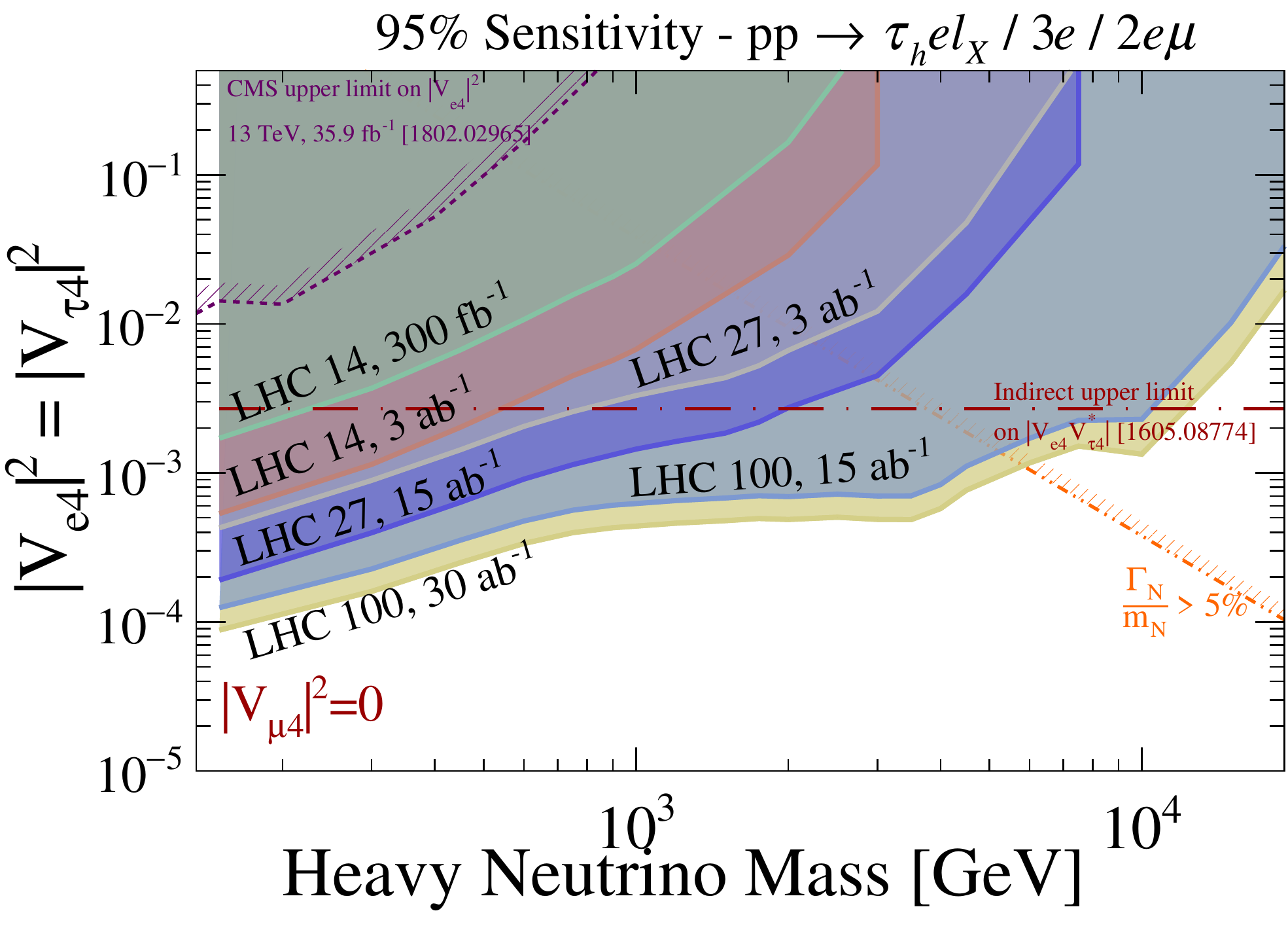}   
    \caption{Left: Sensitivity projections for active-sterile mixing of $\vert V_{\ell N}\vert^2$ at the LHC when using only the Drell-Yan channel (dashed) and when augmented by the $W\gamma$ fusion channel (solid)~\cite{Pascoli:2018heg}. Also shown for reference are LHC limits with $\mathcal{L}\approx36$ fb$^{-1}$ of data~\cite{Sirunyan:2018mtv}.
    Right: Sensitivity projections for active-sterile mixing of $\vert V_{\ell N}\vert^2$ at the LHC and potential successor experiments using the Drell-Yan and $W\gamma$ channels~\cite{Pascoli:2018heg}.}
    \label{fig:nuVBS_typeI_wa}
\end{figure*}

In the SM neutrinos are massless~\cite{Weinberg:1967tq}. This follows from the absence of right-handed neutrinos $\nu_R$, and hence the lack of Yukawa couplings between the SM Higgs field and left-handed lepton doublets. As fermions without any gauge charges, $\nu_R$ are allowed to be their own antiparticles, i.e., be Majorana fermions, and therefore possess right-handed Majorana masses $\mu_R$. Therefore, despite being a natural explanation for nonzero neutrino masses,  hypothesizing the existence of at least two $\nu_R$ is nuanced~\cite{Pilaftsis:1991ug,Kersten:2007vk,Moffat:2017feq}.  For example: Majorana masses can take values well above and below the EW scale while still reproducing oscillation data. Likewise, the possible dynamical origin of $\mu_R$ complicates the low-energy picture because the number of symmetries of the SM with $\nu_R$ increases in the absence of masses, \eg, the emergence of a left-right global symmetry. Whether the $\mu_R$ are much larger, comparable, or smaller to the EW scale also leads to distinct phenomenologies~\cite{Pilaftsis:1991ug,Kersten:2007vk,Moffat:2017feq} that are testable, to a degree~\cite{Cai:2017mow,Pascoli:2018heg}, with VBS.

In searches for Dirac and Majorana neutrinos $N$ with masses above the EW scale, VBS has been indispensable in extending the sensitivity of LHC experiments~\cite{Sirunyan:2018mtv,Sirunyan:2018xiv}. In Fig.~\ref{fig:nuVBS_typeI_ww}~(left), for example, one sees that the NLO cross section (normalized for an active-sterile mixing $\vert V_{\ell N}\vert$ of unity) for the  charged-current Drell-Yan (CCDY) process $q\overline{q'}\to W^{\pm *} \to \ell^\pm N$ is complemented by the $W^\pm\gamma \to \ell^\pm N$ and same-sign $W^\pm W^\pm \to \ell^\pm_i \ell^\pm_j$ scattering processes for $N$ masses around $m_N\sim1$ TeV~\cite{Fuks:2020att}. Fully differential computations up to NLO in QCD with parton shower (PS)-matching are available using the \textsc{HeavyN} Universal \textsc{FeynRules} Object (UFO) model libraries for Majorana~\cite{Degrande:2016aje} and Dirac~\cite{Pascoli:2018heg} $N$. Predictions up to NLO+PS with dimension-six operators are also available if using the \textsc{HeavyN\_vSMEFTdim6} UFO \cite{Cirigliano:2021peb}.

\begin{figure*}[t!]
    \centering
    \includegraphics[width=\columnwidth]{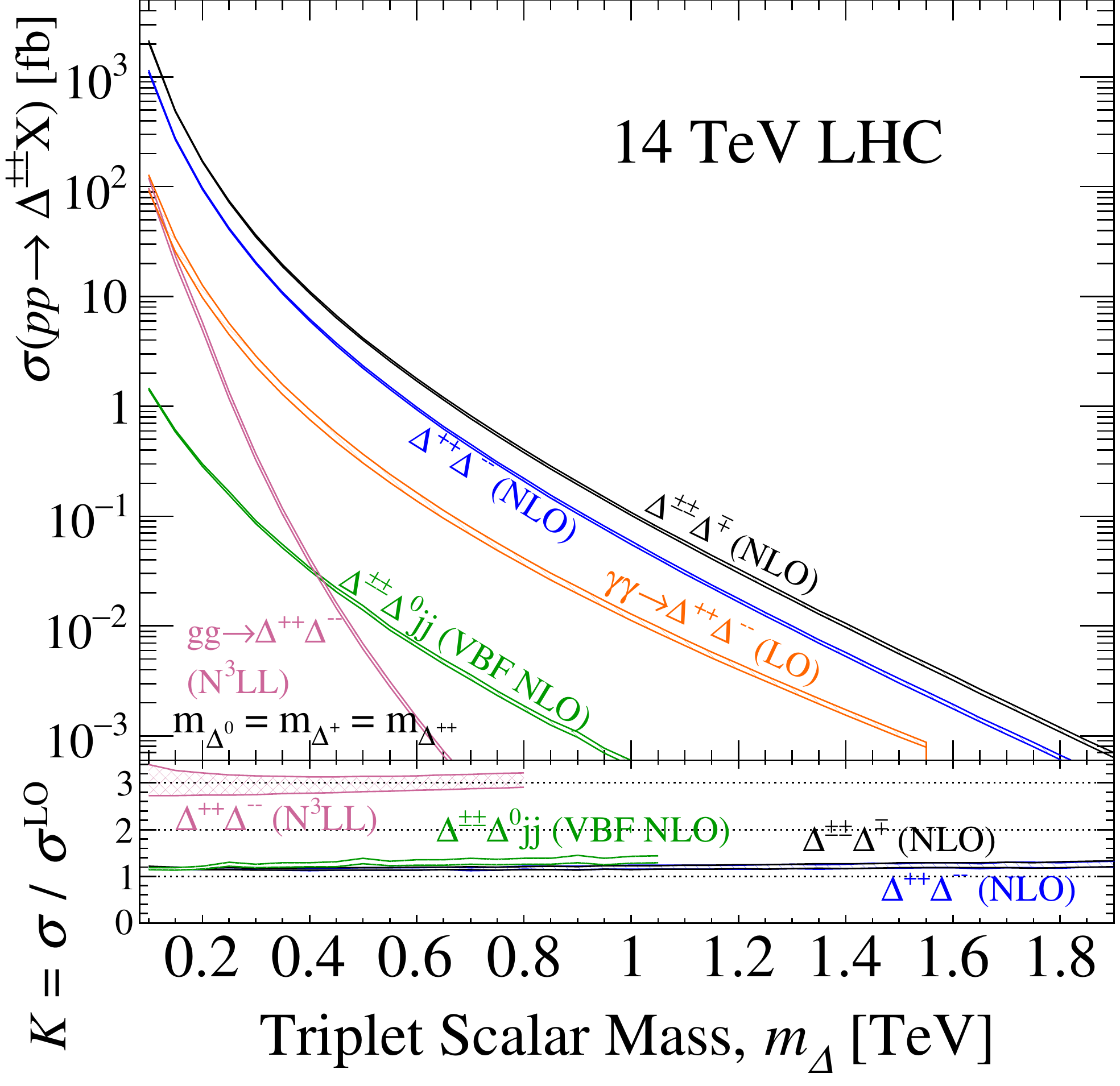}    
    \includegraphics[width=\columnwidth]{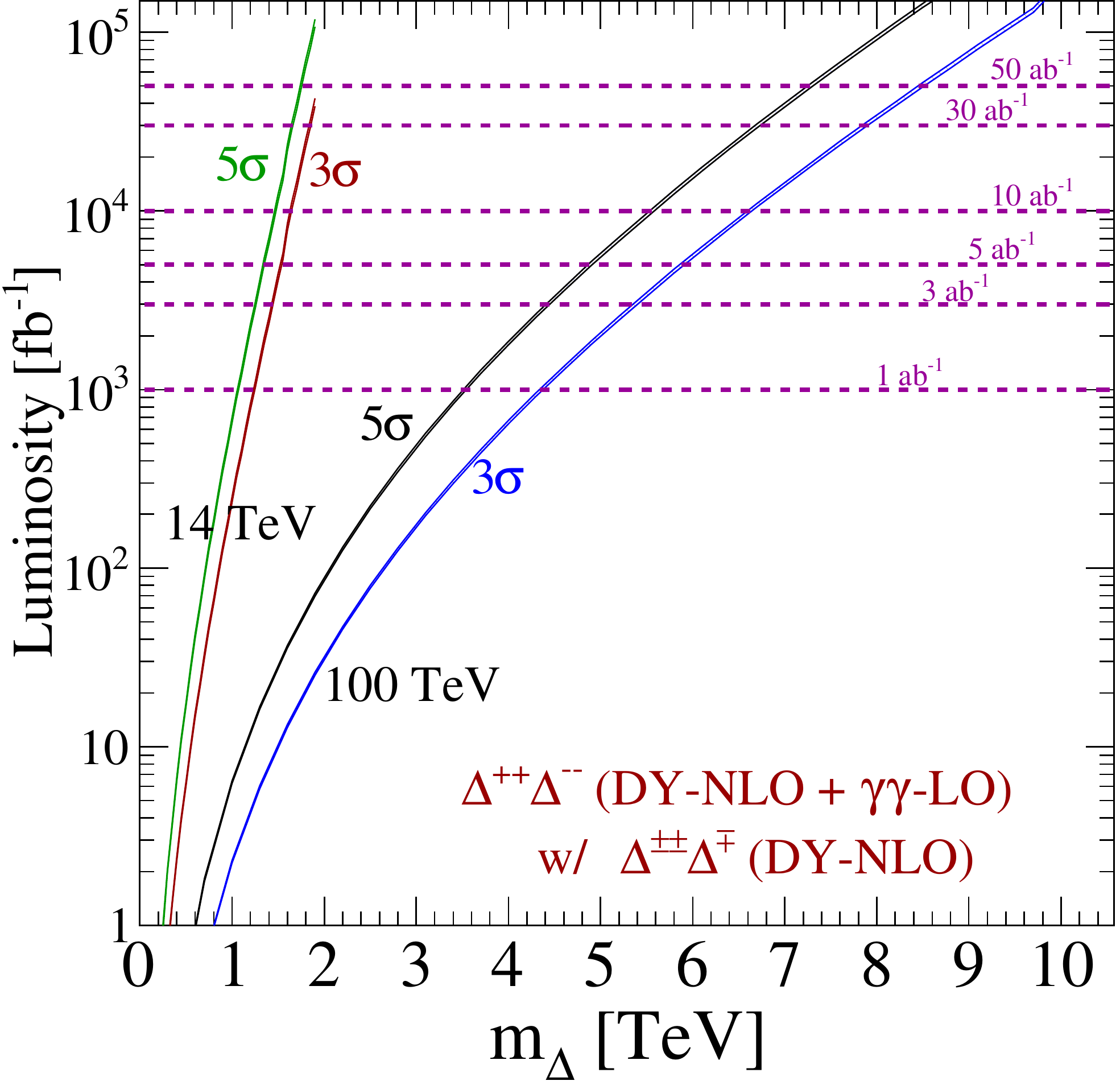}   
    \caption{Left: High-precision cross sections predictions at the $\sqrt{s}=14$ TeV LHC for doubly charged Higgs  $\Delta^{\pm\pm}$ production via various mechanisms in the Type II Seesaw. Right: Sensitivity projections to $\Delta^{\pm\pm}$ for the LHC and a hypothetical collider at $\sqrt{s}=100$ TeV~\cite{Fuks:2019clu}.}
    \label{fig:nuVBS_typeII}
\end{figure*}

The same-sign $W^\pm W^\pm$ process is special for several reasons, including: 
(a) its relationship to the neutrinoless $\beta\beta$ decay process~\cite{Dicus:1991fk,Datta:1993nm}; 
(b) its accessibility to $\ell=\mu$ and $\tau$ lepton flavors, which are not accessible in nuclear decay experiments;
and 
(c) its sensitivity to violations of $s$-wave perturbative unitarity~\cite{Fuks:2020att}.
As shown in Fig.~\ref{fig:nuVBS_typeI_ww}(R), due to the non-resonant nature of $W^\pm W^\pm$ process, LHC experiments are sensitive to $N$ masses around $m_N\sim10$ TeV for $\vert V_{\ell N}\vert^2\sim0.1$, when employing basic VBS selection cuts~\cite{Fuks:2020att}.

Also of special interest is the $W\gamma$ fusion channel. The channel, which features contributions from elastic~\cite{Dev:2013wba}, inelastic~\cite{Alva:2014gxa,Degrande:2016aje}, and deeply inelastic~\cite{Datta:1993nm} initial-state photons, can be added coherently to the CCDY process whenever searches are inclusive to forward jet activity~\cite{Alva:2014gxa}. For TeV-scale Dirac and Majorana neutrinos, Fig.~\ref{fig:nuVBS_typeI_wa}~(left) shows, for example, that including the $W\gamma$ channel can improve sensitivity to $\vert V_{\ell N}\vert^2$ by at least an order of magnitude for TeV-scale $N$~\cite{Pascoli:2018rsg}. Such improvements hold for signal categories with final-state $\tau$ leptons~\cite{Andres:2017daw,Pascoli:2018rsg}. With newer analysis strategies, active-sterile mixing as low as $\vert V_{\ell N}\vert^2\sim10^{-3}$ can be probed at the HL-LHC~\cite{Pascoli:2018rsg}. Even smaller values of active-sterile mixing are also available at prospective successors of the LHC as shown in Fig.~\ref{fig:nuVBS_typeI_wa}~(right). Both VBS channels are also sensitive to dimension-six operators involving heavy neutrinos~\cite{Cirigliano:2021peb}.

\subsection{New Higgs bosons from $\gamma\gamma$ and $W^\pm W^\pm$ scattering}\label{sec:nuVBS_typeII}

In the Type II Seesaw model~\cite{Magg:1980ut,Schechter:1980gr,Cheng:1980qt,Mohapatra:1980yp,Lazarides:1980nt}, light neutrino masses are understood to originate through the spontaneous breaking of lepton number symmetry by a new scalar $\hat{\Delta}$ when it acquires a small vacuum expectation value (vev) $v_{\Delta}=\sqrt{2}\langle\hat{\Delta}\rangle\ll v_{\rm EW}=\sqrt{2}\langle\Phi\rangle\approx 246$ GeV. Here, $\hat{\Delta}$ denotes a gauge eigenstate, whereas $\Delta$ (no hat) denotes a mass eigenstate. $\Phi$ is the SM Higgs field and $v_{\rm EM}$ is its vev. The new field is formally a complex triplet under the SM's SU$(2)_L$ gauge group and carries a hypercharge of $Y=+2$. This means that after EW symmetry breaking one generates the following field content in addition to the SM particles: a doubly charged Higgs boson, $\Delta^{\pm\pm}$; a singly charged Higgs boson, $\Delta^{\pm}$; an electrically neutral, CP even(odd) Higgs boson, $\Delta^0~(\xi^0)$.

EW symmetry breaking proceeds in the usual way for the SM sector. The exception are neutrinos: they obtain left-handed Majorana masses that are proportional to $v_{\Delta}$ and their Yukawa coupling to $\hat{\Delta}$. As a consequence, the decay rates of the charged and neutral $\Delta$ to leptons are proportional to light neutrino masses and the measured values of the PMNS matrix. If the $\Delta$ are realized in nature, then collider experiments can potentially probe the $\nu_\tau$ mixing sector, which is poorly constrained by oscillation experiments~\cite{Parke:2015goa,Ellis:2020ehi}. For a concise summary of this symmetry breaking pattern and phenomenology, see Ref.~\cite{Fuks:2019clu} and references therein. It is important to stress that this model is an example of generating of a neutrino mass model that does not contain right-handed neutrinos.

At the LHC, the production of singly and double charged $\Delta$ can proceed through a variety of mechanisms, including the Drell-Yan ($q\overline{q'}$ annihilation) channels
\begin{equation}
 q\overline{q'} \to \Delta^{+(+)}\Delta^{-(-)}, \Delta^{\pm\pm}\Delta^{\mp}, \Delta^\pm \Delta^0, \Delta^\pm \xi^0, 
\end{equation}
the gluon-gluon fusion channel
\begin{equation}
 g g \to \Delta^{++}\Delta^{--}, \Delta^{+}\Delta^{-}, 
\end{equation}
the photon-photon scattering channel
\begin{equation}
 \gamma\gamma \to \Delta^{++}\Delta^{--}, \Delta^{+}\Delta^{-}, 
\end{equation}
as well as the $WZ/\gamma$ and same-sign $WW$ scattering channels
\begin{align}
 W^\pm Z, W^\pm\gamma &\to \Delta^{\pm\pm}\Delta^{\mp}, \Delta^{\pm} \Delta^0, \Delta^{\pm} \xi^0, \Delta^{\pm}, \\
 W^\pm W^\pm &\to \Delta^{\pm\pm}, \Delta^{\pm\pm}\Delta^0.
\end{align}

Many other channels, particularly via the VBS mechanism, are also possible in the Type II Seesaw. However, a fully comprehensive comparison of all VBS production channels does not yet exist. What does exist is a high-precision comparison of $\Delta^{\pm\pm}$ production mechanisms~\cite{Fuks:2019clu}, and is shown for the $\sqrt{s}=14$ TeV LHC in Fig.~\ref{fig:nuVBS_typeII}(L). From the figure we see that many channels are now known at NLO in QCD or with higher precision. While the total Drell-Yan cross section has been known at NLO in QCD for some time~\cite{Muhlleitner:2003me}, fully differential predictions at NLO and with parton shower matching have only recently been made available using the \textsc{TypeIISeesaw} UFO~\cite{Fuks:2019clu}. 

As with the SM Higgs, the VBS channels possess smaller cross sections than the Drell-Yan modes at the LHC. At the same time, the VBS channels possess more handles to distinguish them from background processes. In particular, the $\gamma\gamma$ channel, whose PDF uncertainties are now under definitive control~\cite{Fuks:2019clu}, can appear in ultra peripheral proton collisions and give rise to exceptionally clean signatures. An outlook for the LHC and a potential successor at $\sqrt{s}=100$ TeV are summarized in Fig.~\ref{fig:nuVBS_typeII}(R). For $\mathcal{L}=3-5$ ab$^{-1}$, the LHC is sensitive to pair and associated production of $\Delta^{\pm\pm}$ with masses up to $m_{\Delta}\sim1.2-1.5~(1.1-1.4)$ TeV at $3~(5)~\sigma$ as reported in Ref.~\cite{Fuks:2019clu}.

\subsection{The Weinberg operator in $W^\pm W^\pm$ scattering}\label{sec:nuVBS_dim5}

\begin{figure}[t!]
    \centering
    \includegraphics[width=\columnwidth]{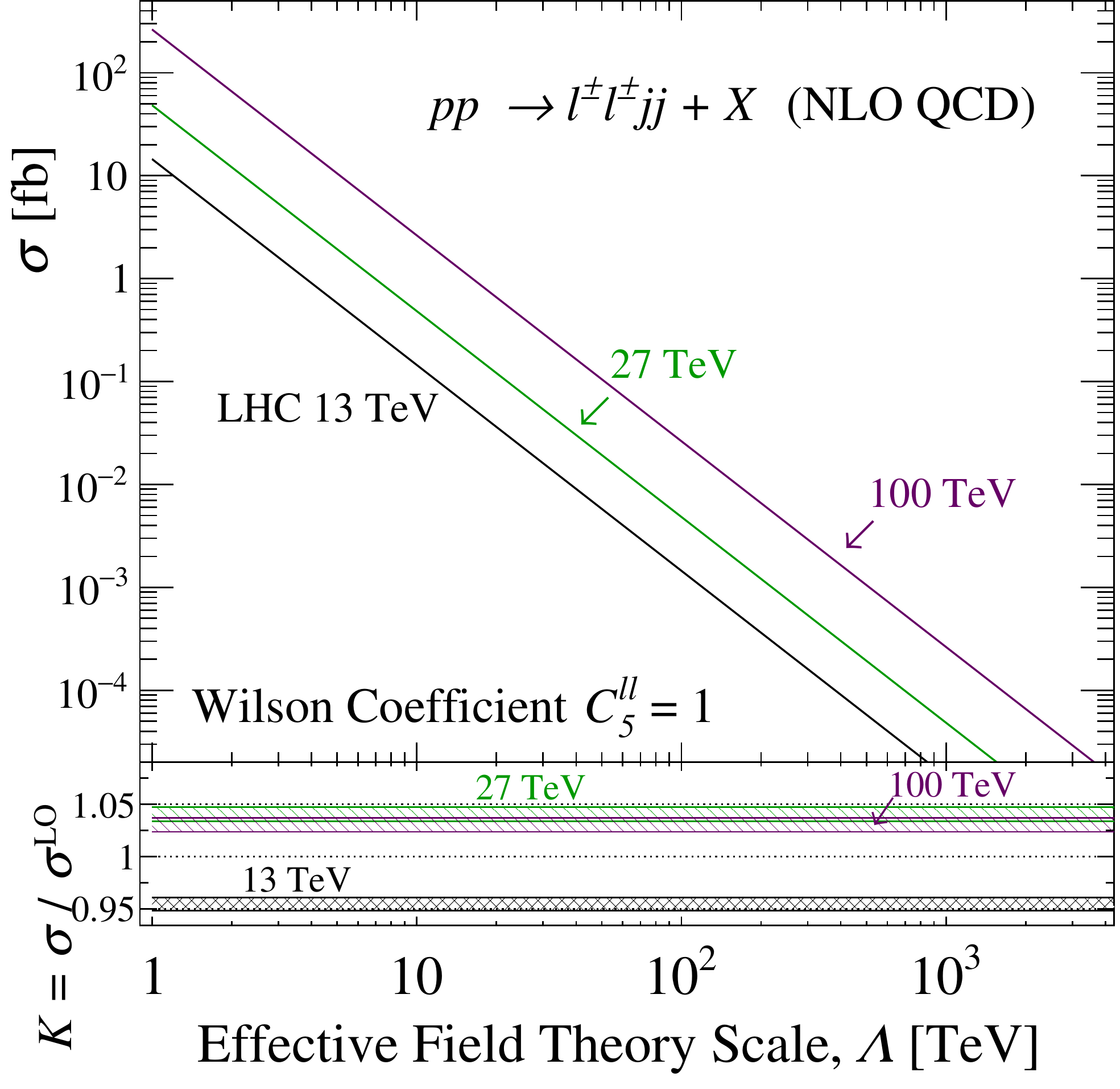}
    \caption{Cross section predictions at NLO in QCD for the $W^\pm W^\pm \to \ell^\pm_i \ell^\pm_j $ process when mediated by the Weinberg operator for representative $pp$ collider energies~\cite{Fuks:2020zbm}.}
    \label{fig:nuVBS_dim5_xsec}
\end{figure}

If those particles that are responsible for generating neutrino masses are too heavy to be directly probed at the LHC, then their impact on SM interactions can still be described through the effective field theory formalism. In particular, the SMEFT framework classifies and arranges the (composite) field operators that remain after decoupling (integrating out) ultra heavy particles into a set of operators that are manifestly invariant under the SM gauge symmetries. For reviews of this framework, see Refs.~\cite{Grzadkowski:2010es,Brivio:2017vri,Brivio:2017btx}. While the power counting of SMEFT extends the SM only by operators that start at dimension $d\geq5$, of notable interest is the solitary operator at dimension $d=5$, which is given by
\begin{equation}
 \mathcal{L}_5 = \ \frac{C_5^{\ell_i\ell_j}}{\Lambda} \big[\Phi\!\cdot\! \overline{L}^c_{\ell_j}\big]
    \big[L_{\ell_j}\!\!\cdot\!\Phi\big].
\end{equation}
Here $C_{5}^{\ell_i\ell_j}$ is the Wilson coefficient for lepton flavors $\ell_i,\ell_j\in\{e,\mu,\tau\}$, $\Lambda$ is the EFT cutoff scale, $\Phi$ is the SM Higgs field, and $L_\ell$ is the SM left-handed lepton doublet of flavor $\ell$.

The so-called Weinberg operator~\cite{Weinberg:1979sa} is capable of generating Majorana neutrino masses after EW symmetry breaking, and therefore serves as a way to parameterize a key aspect of neutrino mass models in experiments. However, other operators at $d=6,7$ or higher may also been needed to fully parameterize other aspects of the theory at low energies. This holds so long as neutrinos are Majorana fermions and that their masses are not generated instead at a higher dimension, which can be the case for some loop-level neutrino mass models~\cite{Bonnet:2012kz,Cai:2017jrq}.

While the ratio $\vert C^{ee}/\Lambda\vert$ is strongly constrained by nuclear neutrino-less $\beta\beta$ decay experiments~\cite{Agostini:2020xta}, the Wilson coefficients for other lepton flavors are actually poorly constrained~\cite{Atre:2005eb,Fuks:2020zbm}. It is also possible that $C^{ee}$ is immeasurably small due to exact symmetries or accidental cancellations, resulting in the so-called ``funnel region'' for normal ordering of neutrino masses. 

In such circumstances, the Weinberg operator can be probed at the LHC through the same-sign $WW$ scattering channel, which is given for $\ell_i=e,\mu,\tau,$ but $\ell_j=\mu,\tau,$ by
\begin{equation}
 W^\pm W^\pm \to  \ell_i^\pm \ell_j^\pm.
\end{equation}
Recent developments allow the Weinberg operator to be efficiently modeled at collider experiments by importing the \textsc{SMWeinberg} UFO into state-of-the-art Monte Carlo event generators~\cite{Fuks:2020zbm}. As shown in Fig.~\ref{fig:nuVBS_dim5_xsec}, the $ W^\pm W^\pm \to  \ell_i^\pm \ell_j^\pm$ cross section at NLO in QCD for the $\sqrt{s}=13$ TeV LHC can reach appreciable values, even if $\Lambda/C^{\ell_i\ell_j}\sim\mathcal{O}(10)$ TeV, and more so at higher collider energies. With a relatively simplified analysis, scale and ratio combinations up to 
\begin{equation}
\vert \Lambda/C^{\mu\mu}\vert \sim8.3~(11)~{\rm TeV} 
\end{equation}
can be probed with $\mathcal{L}=300~(3000)$ fb$^{-1}$ at the LHC~\cite{Fuks:2020zbm}.

\section[Event identification with machine learning]{\large Event identification with machine learning}
\label{sec:MLVBS}

At the HL-LHC, more collisions per bunch crossing and higher granularity data will pose major challenges for triggering and event reconstruction. It will be important to maximize the VBS signal acceptance already at the hardware trigger level in order to ensure sensitivity to the highest-background channels (see Section~\ref{sec:cms_upgrade}). In addition, the large increase in integrated luminosity could allow for several challenging measurements, like the polarization fractions in VBS events or all-hadronic final states. However, disentangling such signatures from the overwhelming background will require improved signal-enhancing algorithms. Given the complex event topology of VBS events and the presence of very similar background processes occurring at significantly higher rates, it is natural to consider Machine Learning (ML) algorithms as a tool for VBS event reconstruction.

\subsection{Triggering of VBS events}
With an average number of collisions per bunch crossing (also known as ``pileup'') increasing from 50 to 200, one of the most crucial aspects for VBS at the HL-LHC will be to maintain or improve the signal acceptance (detector upgrades for HL-LHC are discussed in Section~\ref{sec:cms_upgrade}). A natural candidate for doing so is better triggering on the forward quark jets. To this end, three ingredients will be important in HL-LHC: better pileup mitigation, better jet resolution, and better quark/gluon jet separation.

\subsubsection{Quark versus gluon jet discrimination}
\begin{figure}[!t]
    \centering
    \includegraphics[width=\columnwidth]{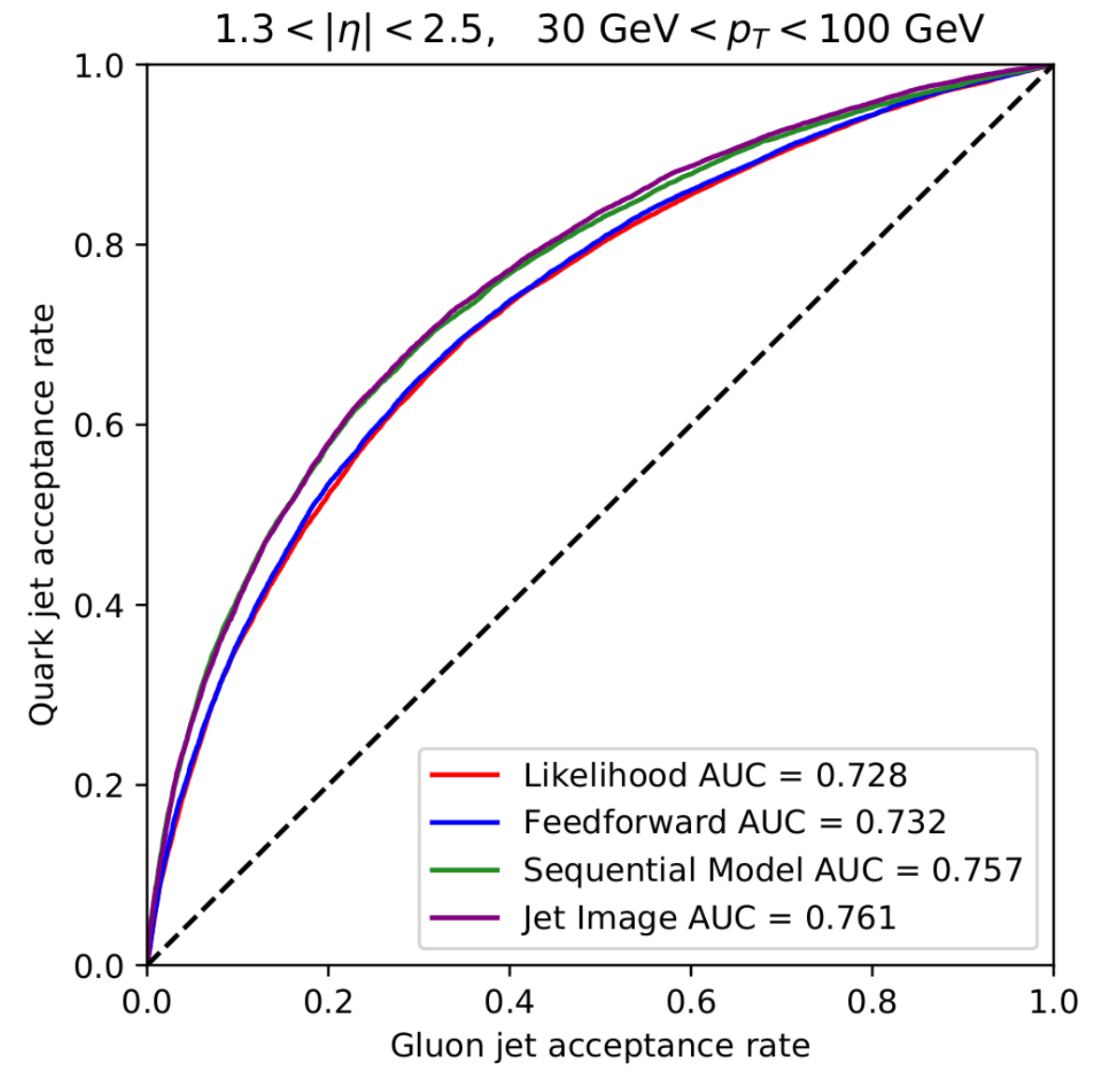}
    \caption{ROC curves showing the quark-jet acceptance rate as a function of the gluon-jet acceptance rate for each quark/gluon discriminator in the detector’s endcap region of $1.3<|\eta |<2.5$ and $30<p_T<100$~GeV, from Ref.~\cite{qgtag}. The red line corresponds to the benchmark likelihood-based discriminator. The blue, green and purple lines correspond to three neural network models.}
    \label{fig:qg}
\end{figure}

In CMS, the new CMS Endcap High-Granularity Calorimeter (HGCAL), which covers a pseudorapidity of $1.5<\vert\eta\vert<3.0$, offers unprecedented transverse and longitudinal segmentation, allowing for accurate measurements of the shower development and narrowness of  VBS jets~\cite{CERN-LHCC-2017-023}.
With this improved forward jet resolution, it is feasible to cleanly trigger  on and reconstruct  narrow VBS jets, thereby serving as a basis for dedicated Level 1 (L1) hardware triggers. The energy resolution is also significantly improved, meaning any potential selections placed on the invariant mass of forward jet pairs are more efficient. In addition to the excellent energy resolution provided by HGCAL, the inclusion of tracking up to $ \eta = 2.4$ and Particle Flow (PF)~\cite{Sirunyan:2017ulk} in the CMS L1 hardware trigger can allow for excellent quark versus gluon discrimination, as variables like charge multiplicity can be computed in real-time. It is therefore feasible, and highly desirable, to design algorithms capable of discriminating between three classes of jets at the hardware triggering level: high-$p_T$ quark-seeded jets, high-$p_T$ gluon-seeded jets, and low-$p_T$ pileup jets. Simple deep neural networks (DNN) for quark $(q)$ and gluon $(g)$ discrimination already exist, as discussed at length in Ref.~\cite{qgtag} and as illustrated in Fig.~\ref{fig:qg}.  Here, simple ML models are compared to a likelihood-based $q/g$ discriminator, the current default in CMS. ML algorithms using the jet constituents as inputs (referred to as ``sequential'' and ``Jet Image'') are notably better in telling quark and gluon jets apart than the likelihood-based equivalent and would be suitable candidates for deployment in the L1 trigger. 

    Due to competing demands, algorithms running in the L1 trigger hardware  must be extremely fast and compact. The \textsc{hls4ml}~\cite{Duarte:2018ite,DiGuglielmo:2020eqx,Coelho:2020zfu,Aarrestad:2021zos} library has been designed to enable the conversion of ML models into ultra low latency, low-resource FPGA firmware, making it feasible to deploy algorithms like those shown in Fig.~\ref{fig:qg} into the trigger. One should caution, however, that this perspective is optimistic. There are many challenges to be able to employ an ML model with competitive performance and low FPGA resources.

\begin{figure*}[!t]
    \centering
    \includegraphics[width=.25\textwidth]{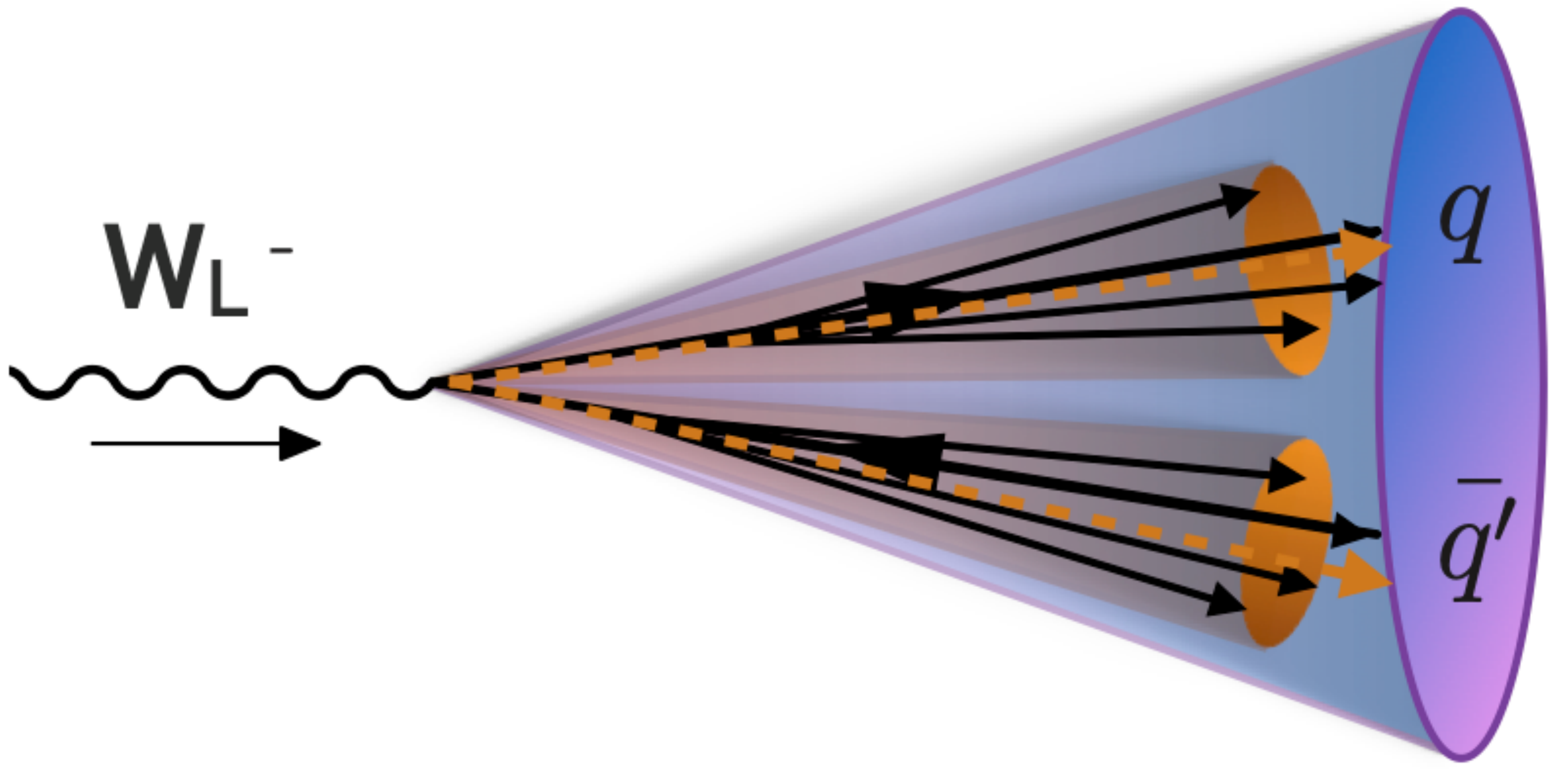}
    \hspace{3cm}
    \includegraphics[width=.25\textwidth]{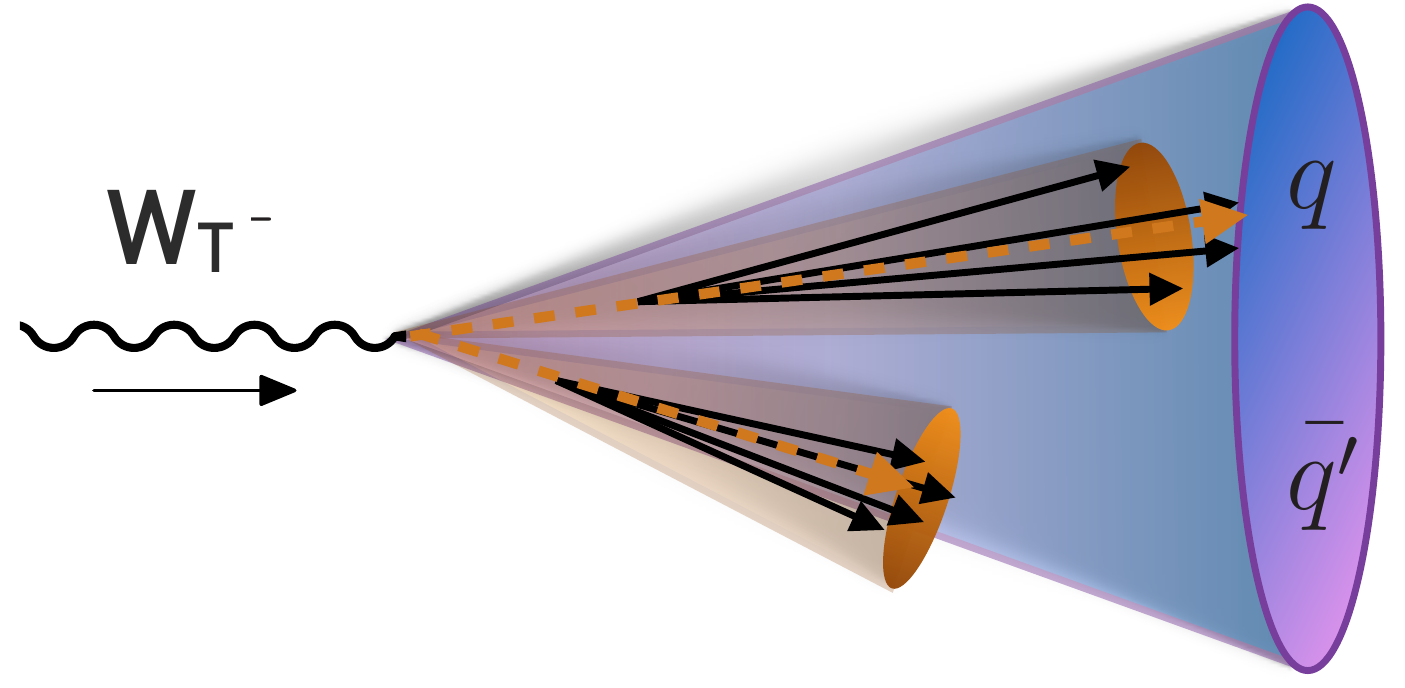}\\
    \includegraphics[width=.95\textwidth]{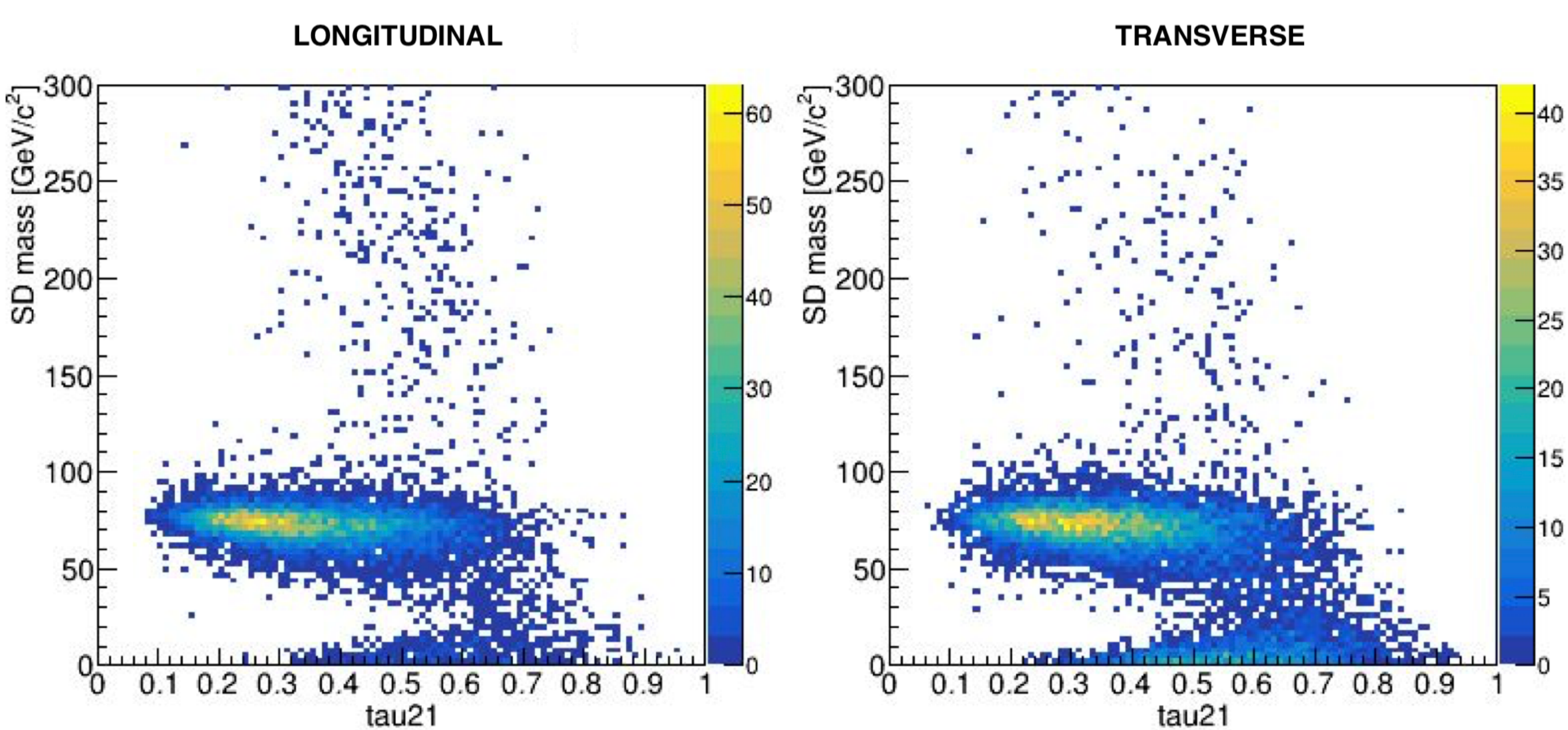}
    \caption{
    Top: Illustration of the decays of high-$p_T$ $W_L$ (Left) and $W_T$ (Right) bosons in the lab frame showing that overlapping decay products are preferentially symmetric (for $W_L$) or asymmetric (for $W_T$) due to spin correlation. 
    Lower: Scatter plot of the soft drop (SD) mass variable as a function of the ``$N$-subjettines'' variable $\tau_{21}$ for $W_L$ (Left) and $W_T$ (Right). Transversely polarized vector bosons (right) more frequently get reconstructed with a soft drop mass of zero and a 1-prong like substructure (i.e. larger $\tau_{21}$ values) due to its decay products being emitted anti-parallel to the $W$ momentum and falling out of the jet cone during the grooming process~\cite{Khachatryan:2014vla,DeBoer:2650187}.
    }
    \label{fig:wtwl}
\end{figure*}

\subsubsection{Jet substructure}
With the PF algorithm available at L1, it is interesting to ask whether there is any gain in identifying and triggering on hadronically decaying vector bosons. This could be done by targeting energetic vector bosons with $p_T$ just above 200 GeV, whose decay products are merged into single large-radius jets, so-called {\em boosted jets}. Several DNNs dedicated to the identification of boosted jets already exist, and some have also been optimized to function in the highly demanding environment of the L1 trigger~\cite{Duarte:2018ite}. However, more detailed studies are required to identify any possible gain with respect to the trigger algorithms currently in place that trigger on high-$p_T$ jets. The additional gain in a dedicated substructure path might be insignificant if most events already pass the standard L1 threshold. The combination of new, high-resolution inputs available in the hardware trigger, the \textsc{hls4ml} workflow, and highly performant networks can lead to significantly higher VBS signal acceptance algorithms in HL-LHC, especially focusing on jet-identification algorithms based on PF candidates.

\subsection{Vector boson polarization}
\label{sec:MLvbspolarization}

With exclusion limits being pushed to higher and higher resonance masses, searches for physics beyond the SM at the LHC are beginning to reach a phase space where hypothetical new particles can no longer be produced directly, though many exceptions to this also exist. In this era, doing precision measurements that target BSM physics is a natural next step. The study of 2$\rightarrow$2 scattering of EW bosons is a direct probe of the quartic gauge coupling, a way to probe the Higgs coupling without the Higgs boson~\cite{Henning:2018kys}, and a way to indirectly look for BSM signatures~\cite{Azatov:2016sqh,Franceschini:2017xkh,Panico:2017frx}.

Of especially high interest is a study of the polarization fraction in VBS events, as these are sensitive to BSM enhancements~\cite{Franceschini:2017xkh}. At scattering energies that are large compared to weak boson masses, VBS can probe BSM interactions that mainly couple to longitudinally polarized vector bosons. With 90\% of SM VBS events being of the $W_T W_T$-type (for $M_{VV} > 250$ GeV), SM $W_T W_T$ becomes an irreducible background when attempting to look for enhancement effects in the longitudinal channel. It is therefore desirable to discriminate transversely polarized vector bosons from longitudinally polarized vector bosons. Reconstructing the vector boson polarization is difficult in leptonic final states due to the missing four-vector of final-state neutrinos. One solution would be to use the all-hadronic final state, where all the final state particles are visible. In addition, the large branching ratio of $W\rightarrow q \bar{q}$  is beneficial when targeting the relatively rare $W_L W_L$ channel. The polarization could then be accessed through both the forward VBS quark-jets, as well as through the quark decay products of the hadronically decaying vector bosons. As the BSM contribution grows with energy, the vector bosons might have a large Lorentz boost resulting in their decay products being contained in a single, large-radius jet. These could be identified as vector-boson jets using dedicated jet substructure techniques. The final state would therefore consist of two central, large-radius jets with masses compatible with the $W$ mass, and two forward quark jets. Extracting the polarization fraction from this is a two-stage problem: First, the EW VBS process must be distinguished from QCD diboson processes. Secondly, one would need to discriminate longitudinally polarized vector bosons from transversely polarized vector bosons (as well as, possibly, the jet charge). Both of these tasks could be solved highly accurately with DNNs, allowing for two powerful tests of the SM: a cross section measurement in a $W_L W_L$ enriched region looking for deviations from the SM prediction, as well as a full measurement of the VBS helicity fractions.

\subsubsection{Jet substructure and polarization}

To identify hadronically decaying vector bosons with $p_T > 200$ GeV, jet substructure variables are usually used. These include methods for improving the jet mass resolution by removing soft and wide angle radiation, called grooming~\cite{Butterworth:2008iy,Khachatryan:2014vla,Sirunyan:2017ezt}, and methods for computing the probability of a jet consisting of two or more sub-jets~\cite{Thaler:2010tr}. For reviews of such techniques, including with the use of ML, see Refs.~\cite{Larkoski:2017jix,Asquith:2018igt,Marzani:2019hun}.

Spin correlation in the decays of boosted objects is important in real life measurements. For example: when a transversely polarized $W$ boson decays in its rest frame, its decay products are back-to-back and preferentially aligned parallel or anti-parallel to the $W$'s spin axis. When boosting the momentum of $W$ and its decay products (partons in the case of hadronic decays) to the lab frame, the result is an asymmetric $p_T$ distribution among the constituents. For highly boosted $W$ bosons, final-state partons may be overlapping. For longitudinally polarized $W$ bosons, decay products are preferentially aligned orthogonally to the $W$'s spin axis, and leads to a softer asymmetry. This is illustrated in the lab frame in Fig.~\ref{fig:wtwl} (top) for a longitudinal $W$ (Left) and transverse $W$ (Right). As jet grooming removes softer, wider angle radiation, it will often completely remove the softer of the two quark sub-jets in the hadronic decays of highly boosted $W$ bosons. 

Beyond jet grooming, spin-correlation also impacts 
$N$-subjettiness $\tau_N$, which measures the numbers of principal axes (or ``prongs'') in a jet~\cite{Thaler:2010tr}. More specifically, the
$N$-subjettiness ratio $\tau_{MN}=\tau_M/\tau_N$ is a measure of whether a jet is better described as an object with $M$ or $N$ prongs (sub-jets). As $\tau_{21}$ is a measure of compatibility for having one vs two axes within a jet, it will not be able to identify the 2-prong structure when overlapping partons are present but incidentally ``groomed away.'' This configuration is illustrated in Fig.~\ref{fig:wtwl} (top). The quantitative effect of this is shown in Fig.~\ref{fig:wtwl} (bottom), where we show scatter plots of the soft drop mass variable~\cite{Larkoski:2014wba}, i.e., the invariant mass after the algorithm has been applied, as a function of $\tau_{21}$ for longitudinal $W$ (Left) and transverse $W$ (Right). A large fraction of transversely polarized $W$'s are reconstructed as having a soft-drop mass close to zero and compatibility with a 1-prong hypothesis. This makes jet substructure algorithms less efficient in detecting transversely polarized, hadronically decaying vector bosons~\cite{DeBoer:2650187,Khachatryan:2014vla}.

\begin{figure}[!t]
    \centering
    \includegraphics[width=\columnwidth]{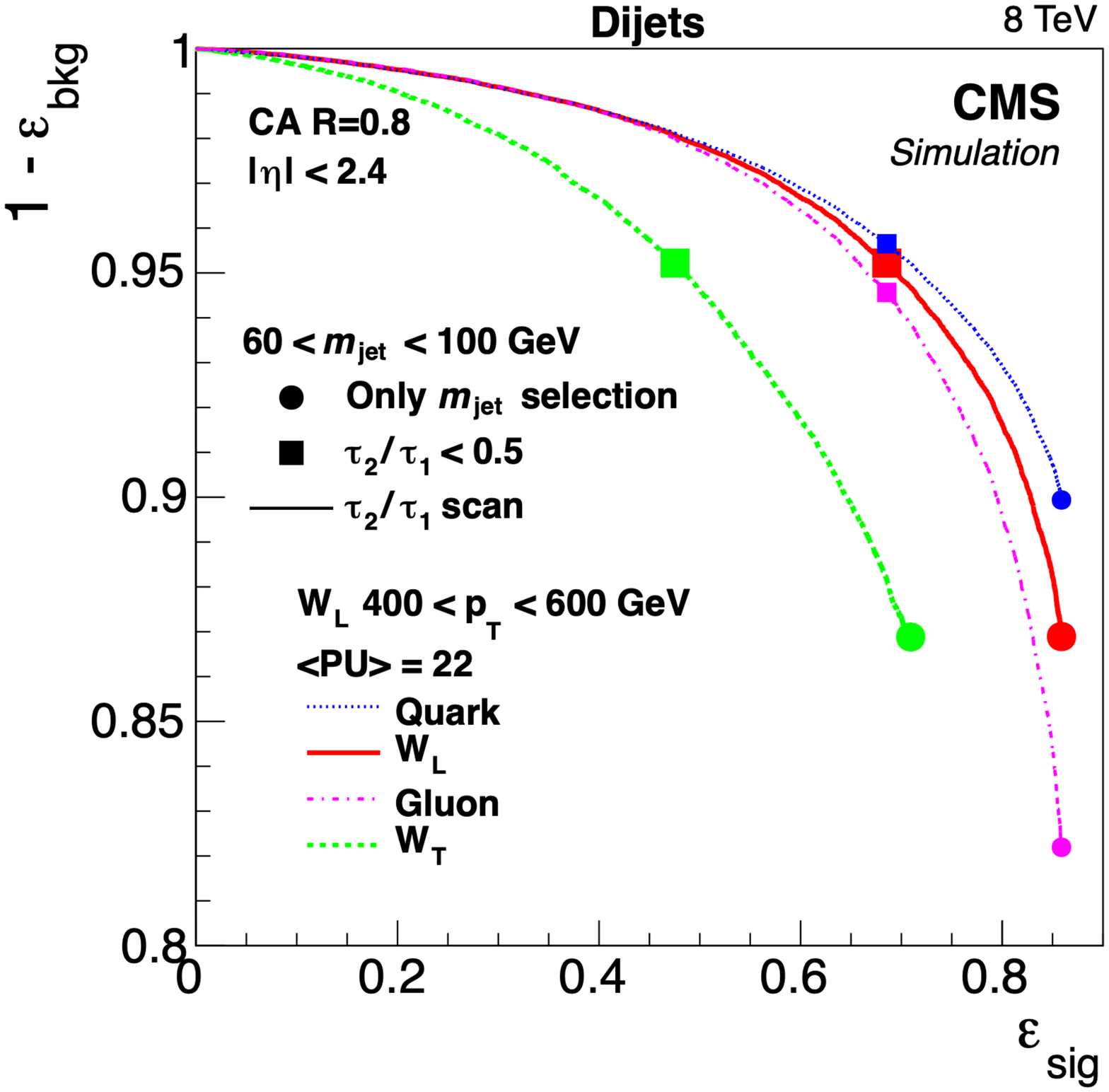}
    \caption{Efficiency for correctly identifying longitudinal (red) and transverse (dotted green) vector bosons versus 1-efficiency for tagging quark (dotted blue) or gluon (dotted pink) jets~\cite{Khachatryan:2014vla}.}
    \label{fig:rocwtag}
\end{figure}

The resulting identification efficiency for transversely polarized vector bosons is significantly reduced, as shown in Fig.~\ref{fig:rocwtag}. If it is desirable to not only enhance the $W_L W_L$ channel, but to measure all polarization components, for instance in a simultaneous fit to extract the polarization density matrix, it is crucial to not only develop algorithms for discriminating between vector boson polarizations, but also substructure algorithms which are equally efficient for both $W_{T}$ and $W_{L}$. Both of these issues can be addressed using particle-based DNNs, for instance permutation-invariant algorithms like graph neural networks~\cite{Moreno:2019bmu,Qu:2019gqs}. Advanced graph neural networks for discriminating between transversely and longitudinally polarized vector bosons is currently being studied. In the proof-of-principle study of Ref.~\cite{DeBoer:2650187}, it has been demonstrated that relatively simple DNNs can obtain a factor of two enhancement in signal sensitivity for longitudinally polarized vector bosons.

Importantly, and as pointed out above, in order to access polarization in the all-hadronic final state, not only a polarization identification algorithm is needed. A ``polarization un-biased'' jet substructure algorithm will additionally be required in order to recover inefficiencies observed with current vector boson tagging algorithm for transversely polarized vector bosons.

In addition, such single-object polarization identification algorithms might not reach the required sensitivity, and it could be necessary and beneficial to include the full event information. This includes correlations between the two vector boson jets and correlations between the forward VBS jets. An example of this can be found in Ref.~\cite{Grossi:2020orx}.

The most challenging part of any such data-driven algorithm will be how to calibrate it using standard candles, developing appropriate systematic uncertainties and corrections, as well as accounting for potential energy-dependence of the algorithm.  This is highly dependent on the final selected algorithm and we leave this discussion for future studies. With these algorithms at hand, several interesting measurements can potentially be considered at HL-LHC. For example: extracting the components of the polarization density matrix using the all hadronic final state, performing differential measurements of the polarization fraction versus energy, and looking for BSM enhancements in high-energy $W_L W_L$ events.

\subsection{Summary}
ML will be a critical part of VBS searches in HL-LHC on multiple fronts. This includes, but is not limited to, algorithms for improving the VBS signal acceptance in the hardware trigger, for reconstructing the complex VBS topology, for discriminating between VBS and background processes, and for discriminating between vector boson polarizations. Exploring the all-hadronic final state might be feasible using advanced DNNs to reconstruct vector boson polarization and charge, and for developing jet substructure taggers with higher efficiency for transversely polarized vector bosons than current algorithms. The development of such ML solutions opens the door for several exciting measurements at HL-LHC, like measurements of the polarization fraction versus energy or a simultaneous fit of the components of the polarization density matrix. Extremely high signal acceptance and low background rates is needed, making highly accurate deep neural networks excellent candidates for the task.


\section[Anomaly detection with machine learning]{\large Anomaly detection with machine learning}
\label{sec:ML}

VBS represents a sensitive probe of both EW symmetry breaking and new physics. If the couplings of the Higgs boson to vector bosons deviate from the SM prediction, the cross sections of VBS processes will increase with center-of-mass energy up to the scale of new physics. In addition, many BSM models predict an extended Higgs sector. The contribution from new resonances can also increase the VBS cross section in certain regions of phase space.
This section is dedicated to a concrete example of the use of ML for the observation of such effects.

Among the many VBS processes, the same-sign scattering, golden process $W^\pm W^\pm jj\to\ell^\pm \ell^\pm \nu\overline{\nu}jj$
has already been observed at the LHC
(see Section~\ref{sec:vbsresults}). This is similar to the case of the trilepton channel, $W^\pm Z jj \to \ell^\pm \nu \ell^+ \ell^-  j j$~\cite{Aaboud:2018ddq,Sirunyan:2020gyx}.  While the production cross sections of the $Z(\to \ell^+ \ell^-) Z(\to \nu \overline{\nu}) j j$ and $Z(\to \ell^+_i \ell^-_i) Z(\to \ell^+_k \ell^-_k) j j$ channels~\cite{Aad:2020zbq,Sirunyan:2020alo} are small, they too have been observed due to their small backgrounds. There is also recent evidence for the $Z \gamma j j \to \ell^+\ell^- \gamma j j $ channel~\cite{Aad:2019wpb}, which benefits from its larger production rate. Even though the opposite-sign $W^+ W^- jj \to \ell^+\ell^-\nu\overline{\nu} jj$ process has the largest production rate, it has not been observed yet, due to the huge $t\bar{t}$ background. 

\begin{figure}[!t]
\includegraphics[width=\columnwidth]{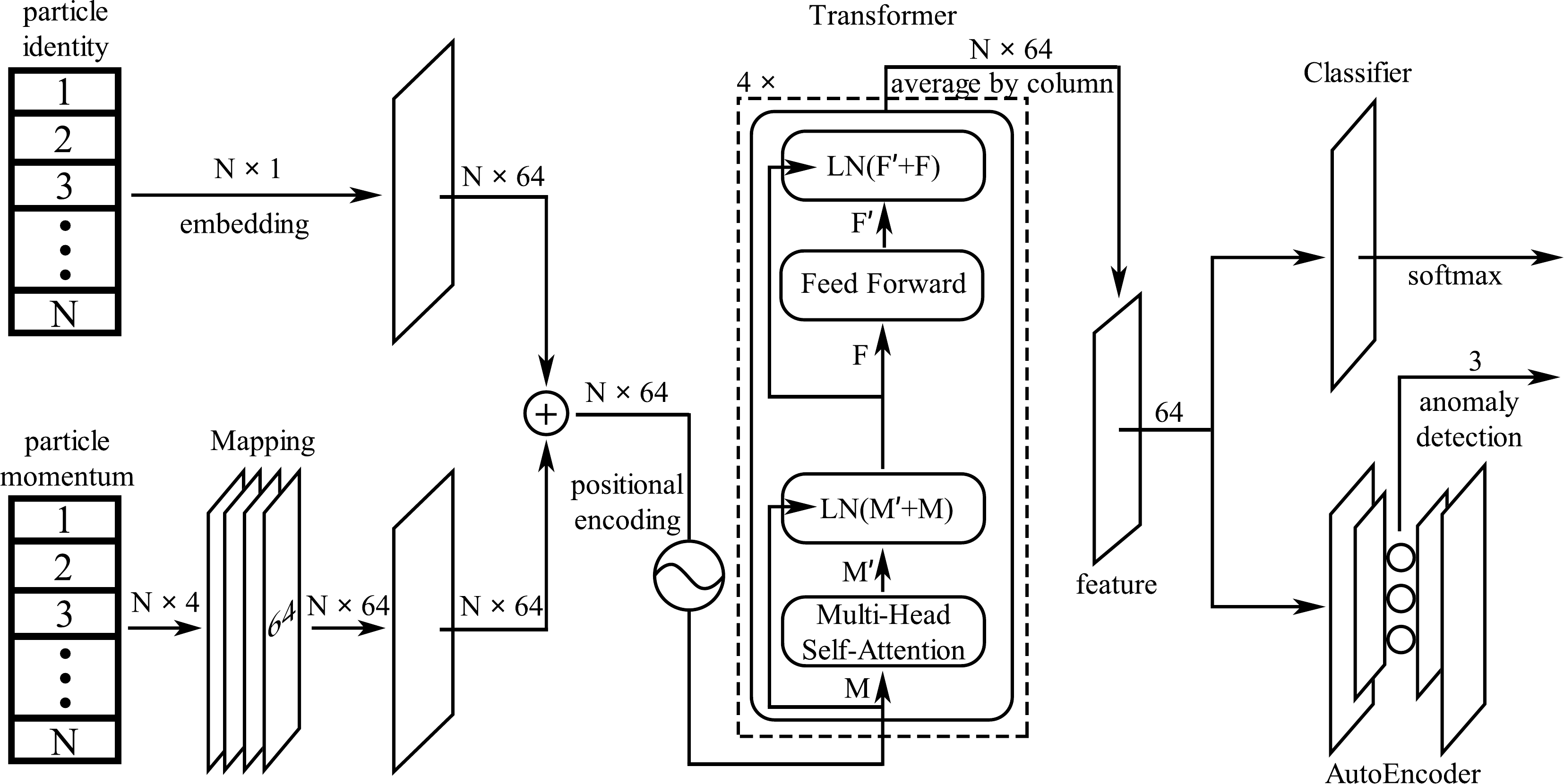}
\caption{Architecture of neural network~\cite{Li:2020fna}.
\label{fig:network}}
\end{figure}

Understanding the polarization of the gauge bosons is an important step after the measurements of the VBS processes. There are studies aiming to determine the polarization of gauge bosons in the $W^+ W^-$ channel~\cite{Han:2009em,Ballestrero:2017bxn,BuarqueFranzosi:2019boy}, in the fully leptonic $W^\pm W^\pm$ channel~\cite{Ballestrero:2020qgv}, in the fully leptonic $WZ/ZZ$ channels~\cite{Ballestrero:2019qoy}, in SM Higgs decay modes~\cite{Maina:2020rgd} and in generic processes with boosted hadronically decaying $W$ boson~\cite{De:2020iwq}. Various kinematic observables have been proposed in these works to discriminate the longitudinal and transverse polarizations of gauge bosons. Several recent studies have shown that deep neural network with input of final states momenta can be used for regression of the lepton angle in the gauge boson rest frame~\cite{Searcy:2015apa,Grossi:2020orx} and classification of events with different polarizations~\cite{Lee:2018xtt,Lee:2019nhm}.

Reference~\cite{Li:2020fna} in particular focuses on the fully leptonic and semi-leptonic channels of the $W^+ W^-$+jets process. There, the authors propose a neural network based on the Transformer architecture~\cite{vaswani2017attention} to learn the features of the VBS process, including its polarization. The SMEFT and 2HDM are considered as examples demonstrating that this method is able to test a wide class of BSM physics which contribute to VBS. 

\subsection{Analysis procedure}\label{sec:vbsML_procedure}

\begin{figure*}[!t]
\includegraphics[width=\columnwidth]{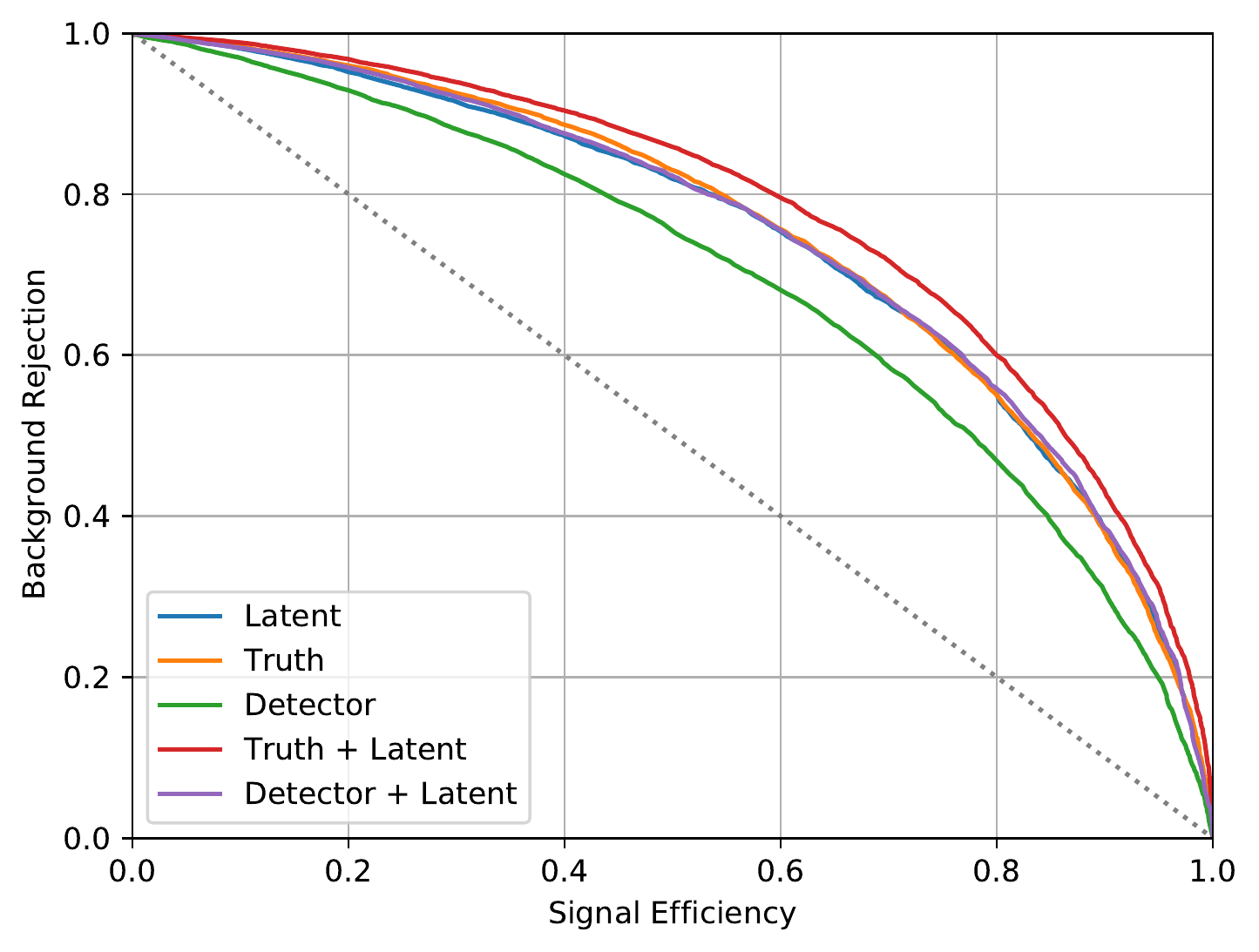}
\includegraphics[width=\columnwidth]{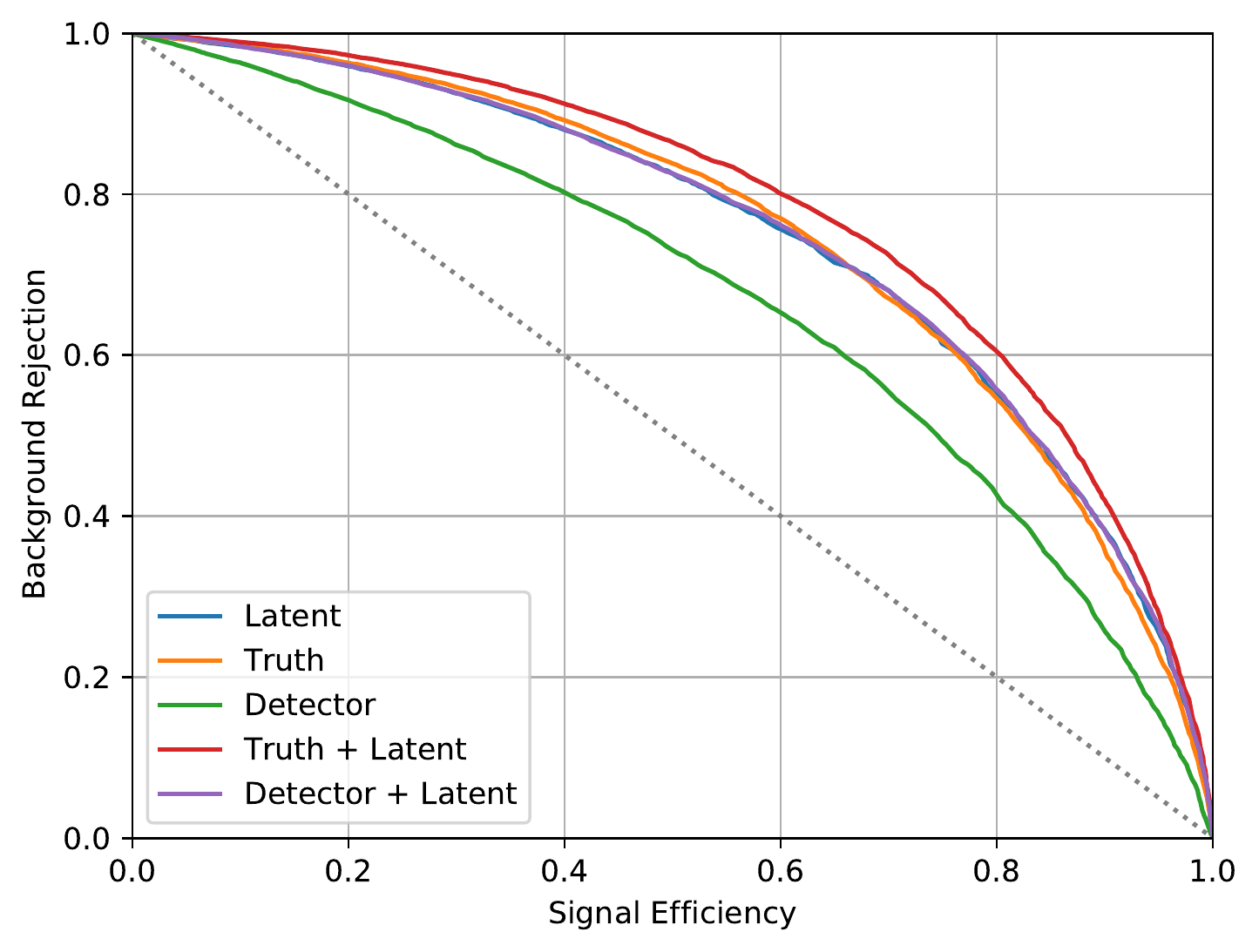}
\caption{Left: Comparison of the discriminating power of methods with different input variables in the dileptonic channel. Right: Same, but for the semi-leptonic channel. Note the variables used for plotting the ROC is different from that in the dileptonic channel. Adapted from Ref.~\cite{Li:2020fna}} 
\label{fig:rocll}
\end{figure*}

The signal and background events in Ref.~\cite{Li:2020fna} are generated with the \textsc{MadGraph5\_aMC@NLO}~\cite{Stelzer:1994ta,Alwall:2014hca} framework, in which the \textsc{Madspin} is used for the decays of heavy SM particles. The interference between the EW amplitude at $\alpha_{\text{EW}}^4$ and mixed EW-QCD amplitude at $\alpha_s^2 \alpha^2_{\text{EW}}$ production is ignored. The partonic center-of-mass frame is taken as the reference frame for defining the vector boson polarization in this work, \ie, the rest frame defined by the two initial parton in the $q q'\to W^+ W^- j j$ process. Focus is placed on the dileptonic and semi-leptonic channels of EW $W^+ W^- jj$ production. The dominant backgrounds are from QCD production of $t\bar{t}+X$ process, single top production, mixed EW-QCD production of $WW/WZ$, and the EW production of $WZ$ pairs.
At this stage of event generation, heavy resonances are not decayed and the transverse momenta of final-state QCD partons must be greater than 20 GeV. The corresponding fiducial cross sections $(\sigma^{\rm fid})$ at $\sqrt{s}=13$~TeV are listed in the second column of Table~\ref{tab:xsec}.

\begin{table}[t!]
\begin{center}
\resizebox{\columnwidth}{!}{
\begin{tabular}{c c c c } 
\hline\hline
  & $\sigma^{\text{fid}}$ [pb] & $\sigma^{\ell \ell}$ [fb] & $\sigma^{\ell j}$ [fb] \\ \hline
  $t t_\ell$                          &     210.3 & 139.8 & 3007.6 \\ 
  $t W_\ell$/$t_\ell W$     &     15.9  &  11.6  &  224.6 \\ 
  $W_\ell W j j^{\rm QCD}$ &    4.68  &  14.7 &  340.5 \\ 
  $W_\ell Z j j^{\rm QCD}$  &    2.20  &   4.49 & 165.7\\ 
    $W_\ell Z j j^{\rm EW}$  &  0.487 &  3.68  & 22.2\\ 
    $W_\ell  Wj j^{\rm EW}$ &   0.738 &  4.36  & 37.3 \\ \hline\hline
\end{tabular}
}
\caption{\label{tab:xsec} Production cross sections of signal and background processes before ($\sigma^{\rm fid}$) and after pre-selections cuts for the dilepton $(\sigma^{\ell\ell})$ and semi-lepton $(\sigma^{\ell j})$ categories, for proton-proton collisions at $\sqrt{s}=13$~TeV. In the column $\sigma^{\rm fid}$, fiducial cuts are applied. Adapted from Ref.~\cite{Li:2020fna}.}
\end{center}
\end{table}

\begin{figure*}[!t]
\centering
\includegraphics[width=\columnwidth]{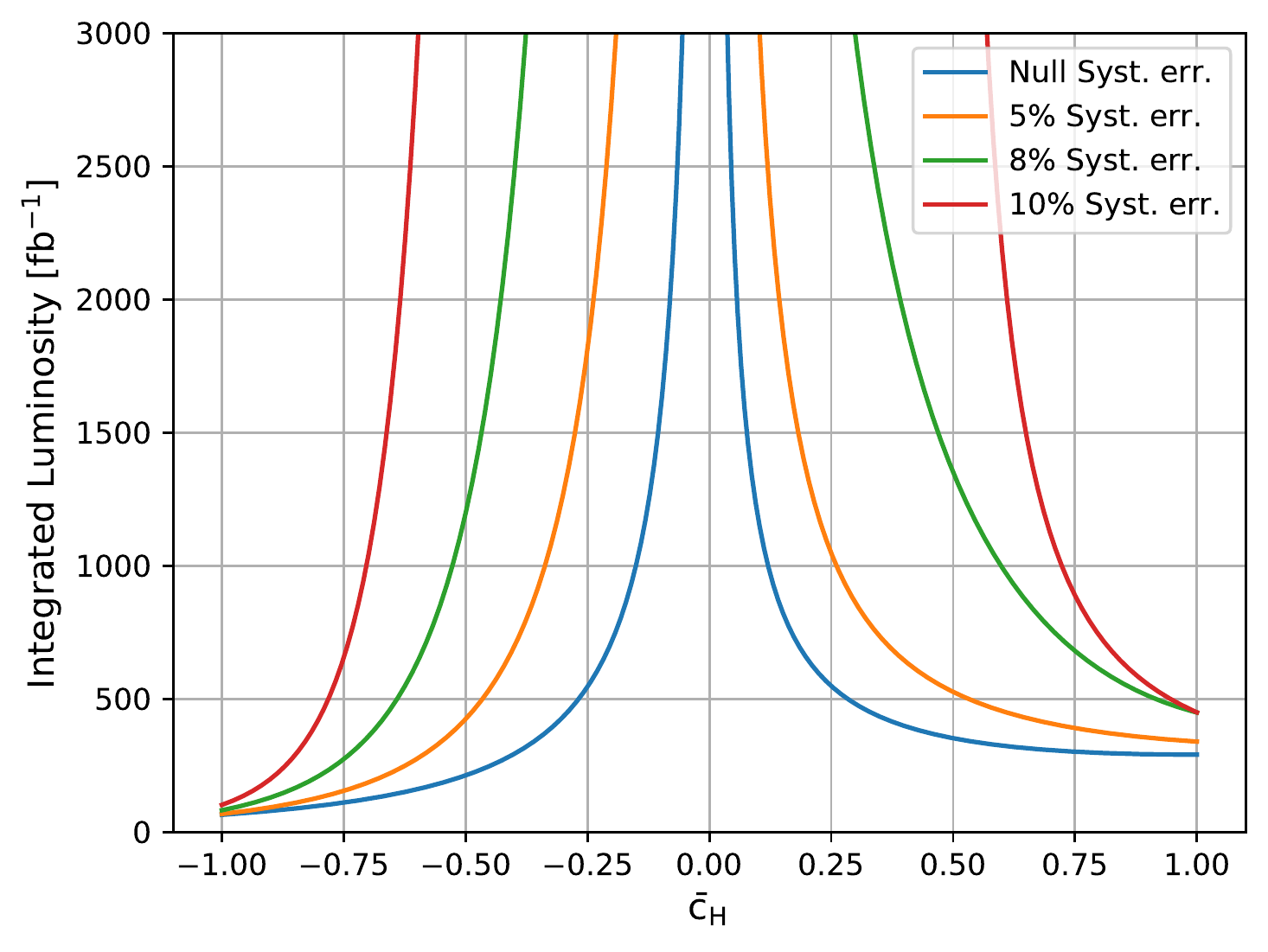}
\includegraphics[width=\columnwidth]{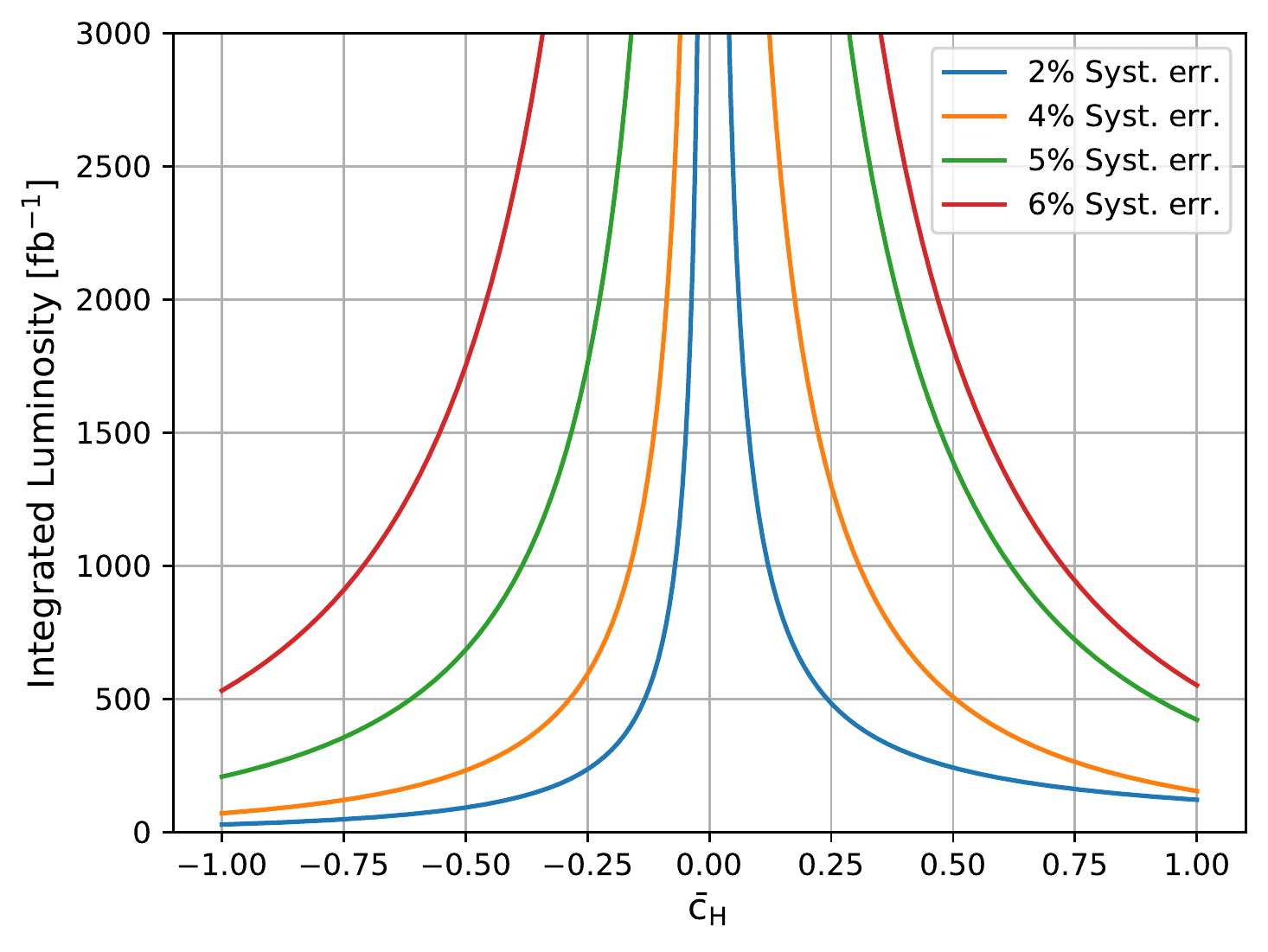}
\caption{Integrated luminosity required to probe the signal (with different $\bar{c}_H$) at 95\% C.L. in the dileptonic channel (Left) and semi-leptonic channel (Right). Several different systematic uncertainties are considered. From Ref.~\cite{Li:2020fna}.}
\label{fig:chpvalue}
\end{figure*}

The events are divided into two classes with the following pre-selection~\cite{Alessandro:2018khj}:
\begin{itemize}
\item \textbf{Dilepton:}  exactly two opposite sign leptons with $p_T(\ell) >20~\text{GeV}, ~ |\eta(l)|<2.5$; at least two jets with $ p_T(j) >20~\text{GeV}, ~ |\eta(j)|<4.5$; the two jets with leading $p_T$ should give large invariant mass ($m_{jj}>500$ GeV) and have large pseudorapidity separation ($|\Delta \eta|_{jj}>3.6$); no $b$-tagged jet in the final state.
\item \textbf{Semi-Lepton:} exactly one charged lepton with $p_T(\ell) >20~\text{GeV}, ~ |\eta(l)|<2.5$; at least four jets with  $ p_T(j) >20~\text{GeV}, ~ |\eta(j)|<4.5$; the pair of jets with the largest invariant mass ($m_{jj}>500$ GeV) that also satisfies $|\Delta \eta_{jj}|>3.6$ is taken as the forward-backward jet pair; (4) among the remaining jets, the jet pair with invariant mass closest to the $W$ boson mass is regarded as the jet pair from $W$ decay.
\end{itemize}
The cross sections for signal and backgrounds after the dilepton and semi-lepton selections are provided in the third and fourth columns of the Table~\ref{tab:xsec}, respectively.

The pre-selected events are fed into the network for learning the features. The input for the network of dileptonic channel consists of the momenta of two leptons, the forward and backward jets, the sum of all detected particles, and the sum of jets that are not assigned as forward-backward jets. The input for the network of semi-leptonic channel consists of the momenta of the lepton, the forward and backward jets, the two jets from $W$ decay, the sum of all detected particles, and the sum of remaining jets.

\subsection{The network}\label{sec:network}

In this section, we discuss that the network in Ref.~\cite{Li:2020fna} is capable of discriminating different polarization modes of the EW $W^\pm W^\mp jj$ production with the low-level inputs.
We go on to discuss the application of this network to SMEFT and the 2HDM as potential LHC searches.

\subsubsection{The polarization of the $W^\pm W^\mp jj$}
\label{sec:polarization}

The neural network proposed in Ref.~\cite{vaswani2017attention} and  illustrated in Fig.~\ref{fig:network}, can be used to  effectively extract the internal connections of features. Reference~\cite{Li:2020fna} trained the network with labeled events with EW production of $W^+_L W^-_L jj$, $W^+_L W^-_T jj$, $W^+_T W^-_L jj$, and $W^+_T W^-_T jj$, respectively. Here $W_L$ ($W_T$) represents longitudinally (transversely) polarized $W$ boson. To assess the discriminating power of the network, the authors performed a comparative study on methods with different input variables. A three dimensional ``latent" space was found to recover the original features of the leptonic and semileptonic channels with a loss of $\mathcal{O}(10^{-4}).$
In machine learning jargon, ``latent'' variables are the quantities that the training network uses  to learn (or parameterize/fit) the training data. For a fixed training data set and learning cycles (epochs), the ability for the training network to reproduce the training data is measured by ``loss.'' Besides the three latent features, two classes of variables are defined: experimentally accessible variables and truth-level variables that can only be obtained in Monte Carlo simulations.

The Gradient Boosting Decision Tree (GBDT) method is adopted to calculate the receiver operating characteristic (ROC) curves with inputs of the variables in a class either with or without including the latent variables. The ROC curves for the dileptonic channel are shown in Fig.~\ref{fig:rocll}(left), where one considers the events of the $W^+_L W^-_L jj$ as the signal and events of other polarization modes as background. One can find that using latent features alone (blue line) already outperforms the GBDT with all detector-level variables (green line). 
For detector variables in the dileptonic channel, a signal efficiency of about 0.3~(0.8) results in a background rejection rate of about 0.9~(0.5); for the semi-leptonic channel, the rejection rates are slightly lower. 

The GBDT that combines the latent variables with the detector-level variables (purple line) does not have better discriminating power than the method with solely latent variables (blue line). The GBDT with truth-level variables (orange line) has slightly improved discriminating power than the method with latent variables (blue line).
It is also interesting to observe that the discriminating power can be improved further by combining truth-level variables and latent variables (red line).
While employing truth-level observables is not applicable in real-life experiments, such an exercise can help estimate and quantify the performance and uncertainties of networks.
Similar ROC curves for the methods with different inputs for the semileptonic channel are presented in Fig.~\ref{fig:rocll}(right). Conclusions similar to the dileptonic channel are drawn.

\begin{table}[!t]
\begin{center}
\resizebox{\columnwidth}{!}{
\begin{tabular}{c c c c c c c } 
\hline\hline
$\bar{c}_H$ &  -1.0 & -0.5 & 0 & 0.5 & 1.0 \\ \hline 
$\sigma^0_{m_{jj}>500}$ [fb] & 440.6 & 421.8 & 419.7 & 426.7 & 436.2  \\ 
$\sigma_{ll}$ [fb] & 4.82  & 4.44 & 4.36 & 4.48 & 4.62  \\ 
$\sigma_{lj}$ [fb]  & 40.2 & 37.7 & 37.3 & 37.9 & 39.3 \\ \hline 
$\sigma^{LL}_{m_{jj}>500}$ [fb] & 46.29 & 29.68 & 25.84 & 28.79 & 34.01  \\ 
$\sigma^{LL}_{ll}$ [fb] & 0.754 &  0.397 &  0.314 &  0.356 &  0.462 \\ 
$\sigma^{LL}_{lj}$  [fb] & 5.28 & 3.04 &  2.40 &  2.79 &  3.50  \\ 
\hline\hline
\end{tabular}
}
\caption{\label{tab:ch_xsec}   $\sigma^0_{m_{jj}>500}$ and $\sigma^{LL}_{m_{jj}>500}$ are the production cross sections (requiring the invariant mass of forward backward jets to be greater than 500 GeV at parton level) for the total and longitudinal polarized EW $W^+ W^- jj$ productions. $\sigma^{(LL)}_{ll/lj}$ correspond to the cross sections of the dileptonic channel ($ll$) and the semi-leptonic channel ($lj$) after preselection cuts. From Ref.~\cite{Li:2020fna}. }
\end{center}
\end{table}

\subsubsection{Application to EFT and 2HDM}
\label{sec:models}

The SMEFT contains a complete set of independent, gauge-invariant operators made up by the SM fields. Ref.~\cite{Li:2020fna} considered the following operator~\cite{Giudice:2007fh,Contino:2013kra}
\begin{align}
\mathcal{O}_H = \frac{\bar{c}_H}{2 v^2} \partial^\mu [\Phi^\dagger \Phi]  \partial_\mu [\Phi^\dagger \Phi] \Rightarrow  \frac{\bar{c}_H}{2} \partial^\mu h \partial_\mu h
\end{align}
since it is less constrained by the EW precision data. The $\Phi$ field is Higgs doublet and $h$ denotes the Higgs boson field with the vacuum expectation value $v\approx246.2$ GeV. The $\mathcal{O}_H$ operator contributes to the Higgs boson kinetic term, and an appropriate field redefinition is required to bring back the kinetic term to its canonical form:
\begin{align}
h \to h ~ [1 - \frac{1}{2} c_H].
\end{align}
This leads to the following changes to the Higgs couplings
\begin{align}
\mathcal{L}_{H}  \supset & \frac{g m_W}{c^2_W} [1- \frac{1}{2} \bar{c}_H]  Z_\mu Z^\mu h  + g m_W [1-\frac{1}{2}\bar{c}_H ] W^\dagger_\mu W^\mu h  \nonumber \\
 & + [ \frac{y_f}{\sqrt{2}} [1-\frac{1}{2}\bar{c}_H] \bar{f} P_R f h +{\rm H.c.}] \label{eq:hcouplings}
\end{align}
The updated global fit to the EFT coefficients constrains $\bar{c}_H \lesssim 0.4$ (marginalizing over all other operators)~\cite{Dawson:2020oco}. Future lepton colliders, such as the ILC, will constrain the $\bar{c}_H$ to the 1\% level~\cite{Jung:2020uzh}.

Here, its effects on the EW $W^+ W^- jj$ production at the LHC are summarized. The production cross section of the EW $W^+ W^- j j$ process (with different choices of $\bar{c}_H$) before and after the pre-selection are given in Table~\ref{tab:ch_xsec}. The $\bar{c}_H=0$ case corresponds to the SM. One can find the fraction of the longitudinal $W$ production increases with $|\bar{c}_H|$ as cancellations among VBS amplitudes become less exact. Pre-selection cuts can raise the fraction of the longitudinal $W^+_L W^-_L j j$, especially for the dileptonic channel. After the pre-selection, the production rate of the semi-leptonic channel is an order of magnitude large than that of the dileptonic channel.

To measure the consistency of the SM and EFT with non-zero $\bar{c}_H$, a binned log-likelihood test in the latent space is performed.
That is to say, the three latent variables determined from each training event (which collectively populate a three-dimensional space) are compared to the three latent variables determined from each signal event (which collectively populate a second three-dimensional space).
Each collection of points makes up a shape and the two shapes are compared using a binned log-likelihood test.

The null hypothesis is the SM backgrounds plus SM EW $W^+W^- j j$ and the test hypothesis is the SM backgrounds plus EFT EW $W^+W^- j j$ with a non-zero $\bar{c}_H$. The required integrated luminosity to achieve 95\% {C.L.} probing for different $\bar{c}_H$ are presented in Fig.~\ref{fig:chpvalue}. It can be seen that the semi-leptonic channel outperforms the dileptonic channel if the systematic uncertainty can be controlled below $\sim$ 5\%. Due to higher backgrounds in the semi-leptonic channel, the sensitivity drops quickly when the systematic uncertainty is larger than 5\%. With systematic uncertainty around 5\%, the method will be able to constrain the $\bar{c}_H$ to [-0.2,0.1] at the HL-LHC.

\begin{table}[!t]
\begin{center}
\resizebox{\columnwidth}{!}{
\begin{tabular}{c c c c c c c} 
\hline\hline
$(m_{h_2}, \sin(\beta - \alpha))$ &  (300,0.7) & (300,0.9) & (700, 0.7) & (700,0.9) \\  \hline
$\sigma^0_{m_{jj}>500}$ [fb] & 636.2 & 492.5 & 461.9 & 428.5  \\ 
$\sigma_{ll}$ [fb] & 8.362 & 5.853 &  5.527 & 4.842 \\ 
$\sigma_{lj}$ [fb]  & 64.07 & 46.52 & 43.70 &  39.33 \\ \hline 
$\sigma^{LL}_{m_{jj}>500}$ [fb] & 170.75 & 79.81 & 71.58 & 42.65 \\ 
$\sigma^{LL}_{ll}$ [fb] & 2.91 &  1.27 &  1.30 &  0.676 \\ 
$\sigma^{LL}_{lj}$ [fb] & 20.78  & 9.35 &  9.50 &  5.06 \\ 
\hline\hline
\end{tabular}
}
\caption{\label{2hdm_xsec}  Similar as Table~\ref{tab:ch_xsec}, but for the 2HDM model. The corresponding parameters are given in the first row. From Ref.~\cite{Li:2020fna}.}
\end{center}
\end{table}

The 2HDM~\cite{Branco:2011iw,Aoki:2009ha} is one of the simplest extension to the Higgs sector of the SM. The Type II variant was considered in Ref.~\cite{Li:2020fna}. There, the authors show that their method is sensitive to changes of the polarization and kinematic properties of the EW $W^+ W^- j j$ production in the 2HDM. Comparing the latent features of the $W^+ W^- j j$ process in the 2HDM with those from measurement, constraints on the parameters of the 2HDM can be obtained.

There are six parameters in the Type II 2HDM: the mass of scalars ($m_{H_1}, m_{H_2}$, $m_A$ and $m_{H^\pm}$), the mixing angle between two CP-even scalars $\alpha$, and the ratio between two vevs $\tan \beta$. The parameter $m_{H_1}$ has been measured to be close to 125 GeV. The $m_A$ and $m_{H^\pm}$ are not relevant in $W^+ W^- j j$ production. Their mass is set to 3 TeV to forbid the decays of $H_2$ into those states. In Table~\ref{2hdm_xsec}, we summarize the production cross sections of the EW $W^+ W^- jj$ process for a few points in the 2HDM for illustration.

\begin{figure*}[!t]
\centering
\includegraphics[width=\columnwidth]{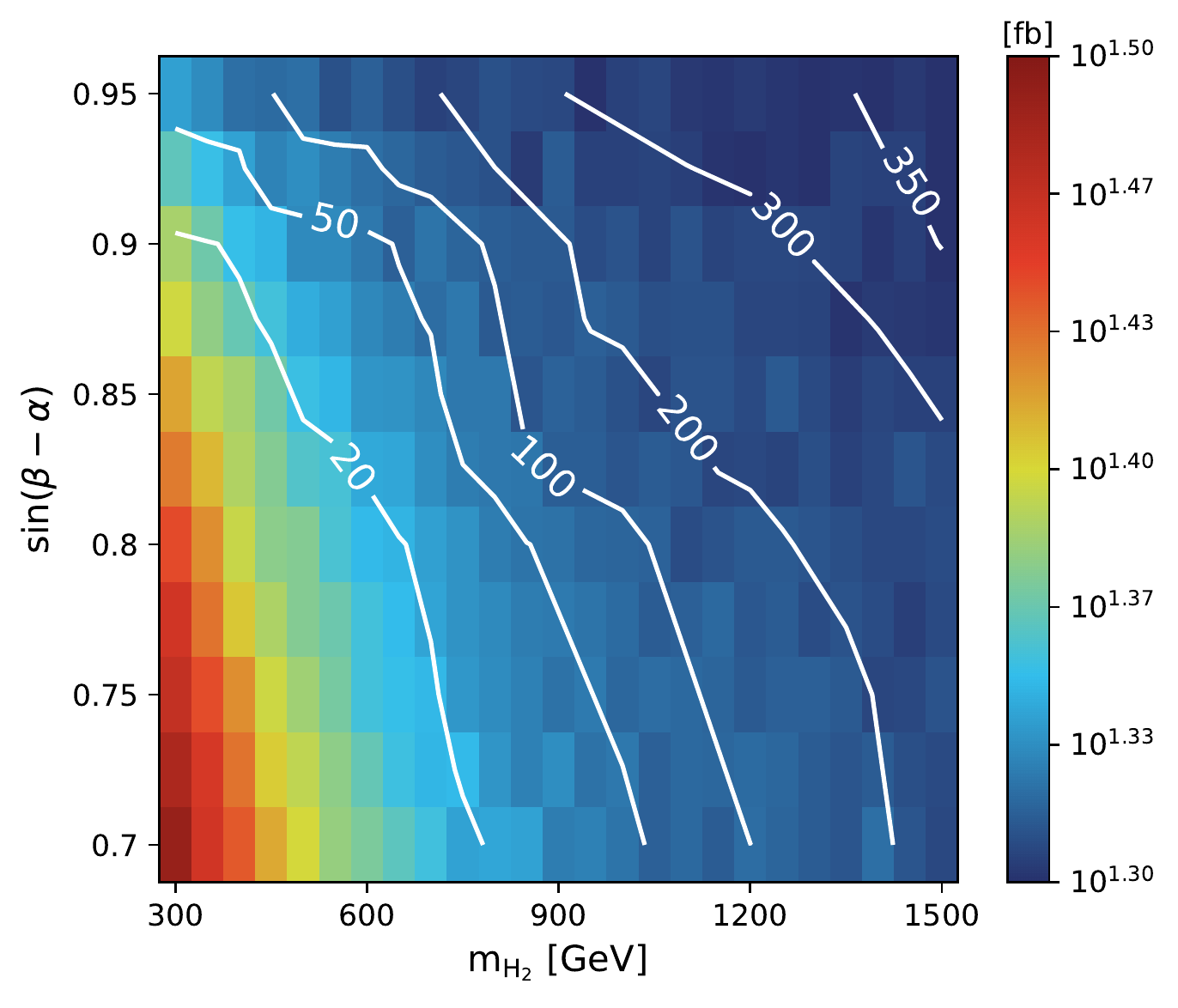}
\includegraphics[width=\columnwidth]{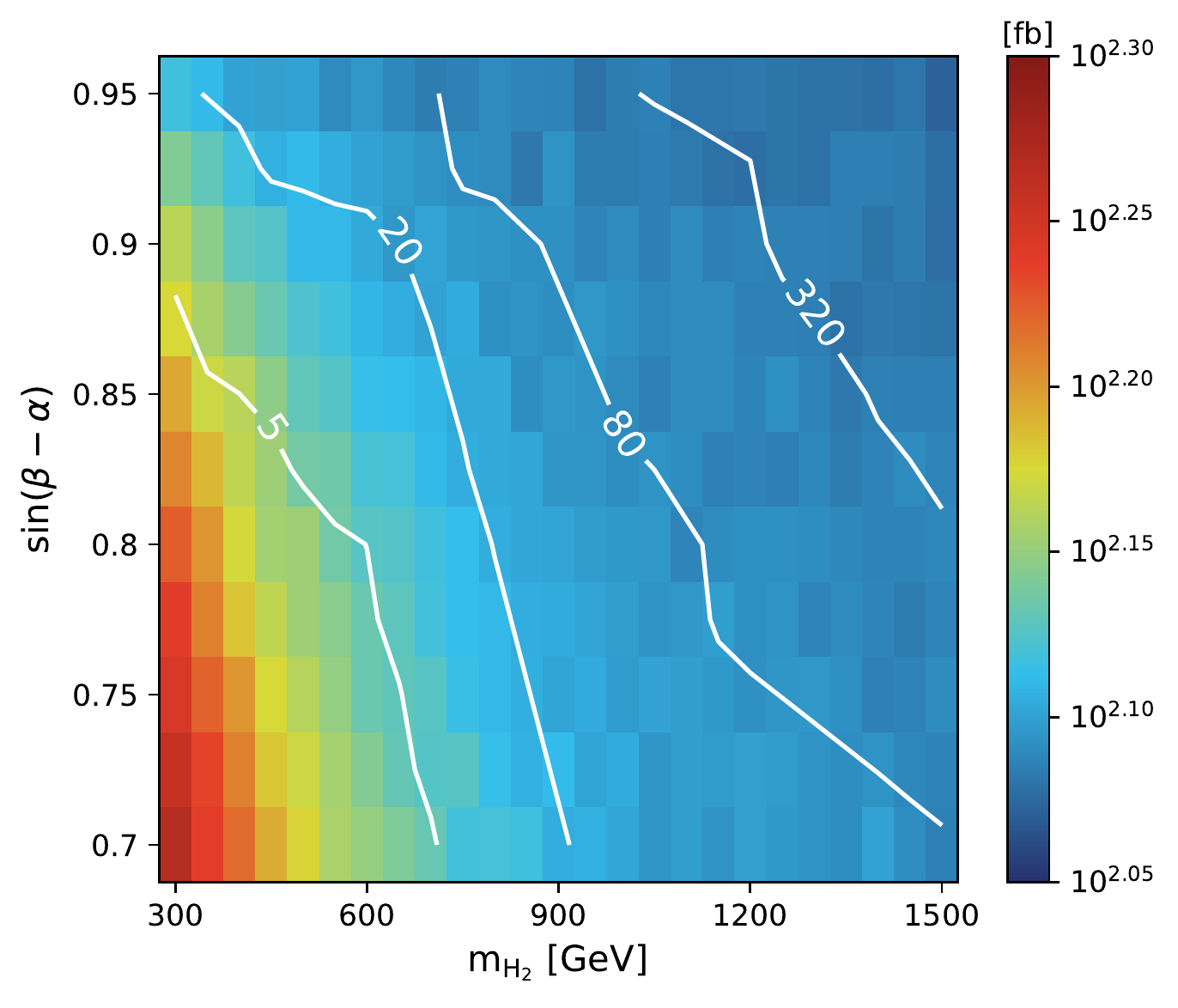}
\caption{\label{fig:2hdmpvalue} Contours corresponding to the required integrated luminosity to probe the signal (with different $\sin(\beta - \alpha)$ and $m_{H_2}$) at 95\% C.L. Color grades correspond to the fiducial cross sections (requiring $m_{jj} >500$ GeV at the parton level) times the branching rates. The systematic uncertainties are set to 5\% for both the dileptonic channel (Left) and semi-leptonic channel (Right). Adapted from Ref.~\cite{Li:2020fna}.}
\end{figure*}

Due to the fact that the cancellation between the amplitudes with and without Higgs exchange are delayed to the scale of $m_{H_2}$ and the heavy scalar dominantly decays into longitudinally polarized vector boson, the fraction of $W^+_L W^-_L jj$ is considerably larger than that of the SM. For relatively light $H_2$ and small $\sin(\beta-\alpha)$, which means the contribution of $H_2$ is significant, the fraction of $W^+_L W^-_L jj$ can reach $\sim$ 30\% before applying  pre-selection cuts, while the number is 6\% in the SM.
Pre-selection can increase the fraction even further. This feature renders the above network very sensitive to the signals in the 2HDM.

The required integrated luminosity for achieving 95\% C.L. sensitivity to a particular point in the $m_{H_2}$-$\sin(\beta-\alpha)$ plane is shown in Fig.~\ref{fig:2hdmpvalue}, for the dileptonic and semi-leptonic channels, respectively.
The method of Ref.~\cite{Li:2020fna} proposes to probe both the resonant feature and the modification to Higgs couplings simultaneously.
This is unlike traditional searches for heavy Higgs resonances~\cite{Aaboud:2018bun,Sirunyan:2019pqw}, which quickly lose sensitivity  at large $m_{H_2}$ due to the suppressed production rate. The parameter space with $H_2$ as heavy as 1.5 TeV can be probed with relatively low integrated luminosity provided the $\sin(\beta-\alpha)$ is not too close to one. The production cross sections of both channels before applying pre-selection cuts are indicated by the color grades in the figure. One can find the sensitivity of the method is roughly determined by the cross section, even though a slightly better sensitivity can be achieved in the small $\sin(\beta - \alpha)$ region, \eg, by comparing to the point ($m_{H_2}=300~\text{GeV}, \sin (\beta -\alpha)=0.9$). Lower integrated luminosity is required to probe the point ($m_{H_2}=550~\text{GeV}, \sin (\beta -\alpha)=0.7$), even though their production cross sections are similar. The improvement on the sensitivity is attributed to the fact that points with smaller $\sin (\beta -\alpha)=0.7$ contains larger fractions of longitudinally polarized $W$ bosons.

\subsection{Summary}
We report that  Ref.~\cite{Li:2020fna} has constructed a neural network consisting of a classification network and an autoencoder. With the input of low level information (4-momenta and the identities of particles in this case), the network is capable of reducing the dimensionality of the feature space for $W^+W^-jj$ production, without losing much discriminating power. This means discriminating the EW production of $W^+W^-jj$ from other processes, as well as discriminating different polarization modes of the EW $W^+W^-jj$. Reference \cite{Li:2020fna} finds the feature space of both dileptonic and semi-leptonic channels can be compacted into three dimensions. Performing the binned log-likelihood test on the distributions of latent features, one can draw the conclusion whether the data is consistent with the SM prediction. One finds that those latent features are sensitive to various  new physics that can possibly contribute to the VBS.

\section[Detector \& performance upgrades for the HL-LHC]{\large Detector \& performance upgrades for the HL-LHC}
\label{sec:cms_upgrade}

	\begin{figure}[!t]
	    \centering
	    \includegraphics[width=\columnwidth]{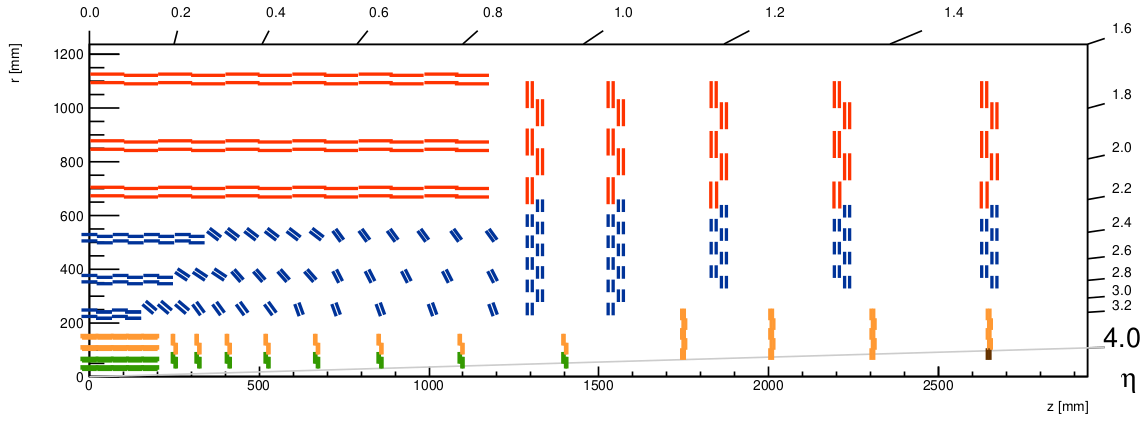}
	    \caption{Layout of one quarter longitudinal section of the CMS Tracker for the HL-LHC. The upgraded tracking detector will extend the coverage for charged particles up to $|\eta|=4$. In green (yellow) the silicon pixel detectors of the Inner Tracker with two (four) readout chips per module. In blue (red) the PS (2S) silicon detector modules of the Outer Tracker~\cite{CMSCollaboration:2015zni}.}
	    \label{fig:phase2_tracker}
	\end{figure}

	\begin{figure*}[!t]
	    \centering
	    \includegraphics[width=\textwidth]{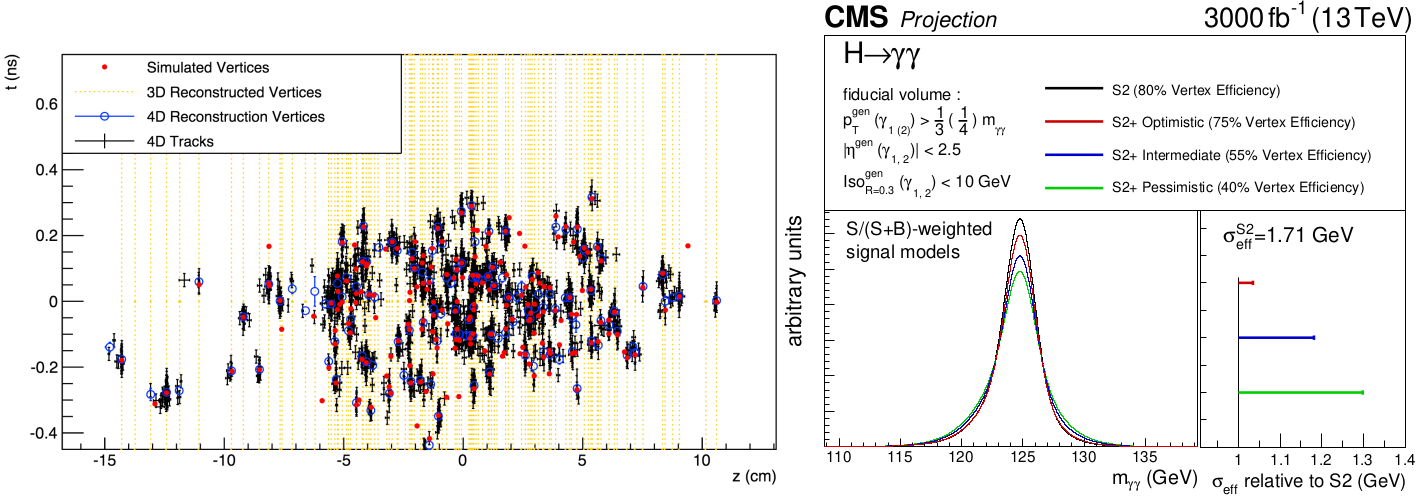}
	    \caption{Left: Proton-proton interaction vertices: time versus longitudinal coordinate.
	    Simulated and reconstructed vertices in a 200 pileup event assuming a MIP timing detector covering the barrel and the endcaps. Right: H $\longrightarrow \gamma\gamma$ mass resolution under different vertex reconstruction efficiency scenarios. Plots from Refs.~\cite{CERN-LHCC-2017-027,Butler:2019rpu}.}
	    \label{fig:mtd_timing}
	\end{figure*}

	After Run 3, the LHC itself will be upgraded in order to deliver a much larger data set for physics to the LHC detector experiments. The HL-LHC upgrade \cite{Apollinari:2015bam} will see the LHC operating at 5.0-7.5 times the nominal instantaneous luminosity, reaching values up to 5 $\times$ 10$^{34}$cm$^{-2}$s$^{-1}$ (7.5 $\times$ 10$^{34}$cm$^{-2}$s$^{-1}$ in the ultimate luminosity scenario). A total delivered integrated luminosity of 3000~fb$^{-1}$ (4000~fb$^{-1}$) is expected by the end of the HL-LHC operation. 

	The luminosity increase is experimentally challenging  for both the accelerator and the detectors. With an increase in instantaneous luminosity with respect to Run 3, the ATLAS and CMS detectors need to be upgraded in order to cope with the increased radiation levels and event multiplicity. While the current average number of pileup collisions, i.e., the number of collisions per bunch crossing, is 30-40, in the HL-LHC phase an average pileup of 140-200 collisions per bunch crossing is expected. Such pileup levels would be unsustainable for the current Phase-1 detectors and go beyond their reconstruction capabilities. Therefore, the ATLAS and CMS detectors are planned to be upgraded after Run~3 for the Phase-2 upgrade.
	 The detector upgrades will help maintain or improve particle identification and event reconstruction in the high-pileup environment, including up to large pseudorapidities. Such upgrades will certainly benefit the reconstruction of VBS processes, which as discussed above is characterized by objects with large $p_T$ and/or large rapidities. 
	
	In what follows, we summarize the planned detector upgrades and anticipated detector performance for the HL-LHC program. As an example, the case of the Phase-2 upgrades of the CMS detector is discussed~\cite{CMSCollaboration:2015zni} . A comparable upgrade program and outlook for detector and physics performance are anticipated for ATLAS~\cite{ATLASCollaboration:2012ilu}.

	\subsection{Current CMS tracker in HL-LHC environment}
	The current Phase-1 tracker is designed to sustain an instantaneous luminosity of 1$\times$10$^{34}$cm$^{-2}$s$^{-1}$. Simulations show that the tracker is expected to start degrading at 500 fb$^{-1}$~\cite{CMSCollaboration:2015zni}, with spatial resolution decreasing due to decreased charged sharing between neighboring pixels. The HL-LHC radiation levels would cause the increase of the depletion voltage as well as the leakage current, compromising the functioning of many tracking modules.

	\subsection{Phase-2 tracker upgrade}
	The Phase-2 tracker is designed to withstand $>$3000~fb$^{-1}$ and increased radiation levels. In order to cope with the increased L1 trigger rate of 750 kHz, the tracking information needs to be provided to the L1 trigger. The new Tracker is composed of the Inner and Outer trackers.

	The Outer Tracker (OT) design is driven by the new trigger demands. The modules are composed of two single-sided, closely-spaced silicon strip sensors read out by a common set of front-end ASICs that correlate the signals in the two sensors providing a local transverse momentum $(p_T)$ measurement. ``Hit'' pairs are selected to be compatible with particles above the chosen $p_T$ threshold (2~GeV). The OT modules are placed in a tilted geometry in order to keep the tracking efficiency high also for large $\eta$ tracks.

	The Inner Tracker (IT) is composed of pixel sensors with a narrower pitch than Phase-1 tracker, with a pixel size of 25 $\times$ 100 $\mu$m$^2$ or 50 $\times$ 50 $\mu$m$^2$, and are expected to exhibit the required radiation tolerance and two-track separation capability. The new electronics is highly segmented with 1$\times$2 and 2$\times$2 readout chips. The addition of tracking modules in the forward region extends the tracking coverage up to $|\eta|=4$, as shown in Fig.~\ref{fig:phase2_tracker} \cite{Paoletti:2723307}.

	\subsection{Phase-2 tracker performance}
	Simulations of the Phase-2 tracker~\cite{CERN-LHCC-2017-009} show that a high tracking efficiency is maintained over the full $\eta$ coverage for  10 GeV muons and $t\bar{t}$ events ($p_T >$0.9~GeV), with efficiencies ranging around 95-100\% in the first case, and 85-95\% in the latter. The level of mis-identification (``fake'') rates is expected to be smaller than 4\% for $t\bar{t}$ events, even in the ultimate pileup conditions (200 simultaneous collisions). The estimated $p_T$ and impact parameter ($d_0$) resolutions are also expected to improve with respect to the Phase-1 tracker, and $b$-tagging can be extended up to $|\eta|$=3.5. The new tracker will feature a substantial reduction in material budget in terms of radiation lengths, thus reducing the inactive material within the tracking volume.

	\subsection{MTD: A new precise timing detector}

A precise timing detector will bring a completely new capability to CMS: the ability to measure precisely the production time of minimum ionizing particles (MIP) to disentangle the approximately 200 nearly simultaneous pileup interactions that will occur in each bunch crossing. The MIP Timing Detector (MTD) is a new detector, placed outside of the tracker and inside the calorimeter, planned for the CMS experiment during the HL-LHC era.  The MTD will aim at providing a time resolution of about 30-50~ps over the period of operation at the HL-LHC. It will also provide new capabilities for charged hadron identification and to the search for long-lived particles.

The timing upgrade of the CMS detector will mitigate the high pileup HL-LHC environment and improve the performance to a level comparable to the Phase-1 CMS detector, exploiting the additional information provided by the precision timing of both tracks and energy deposits in the calorimeters. The  event  display  in  Fig.~\ref{fig:mtd_timing} (left) shows the  power  of  space-time  4-dimension event reconstruction in 200 pileup collisions. The time information is essential to distinguish vertices spread along the beam axis. Figure \ref{fig:mtd_timing} (right) shows the Higgs boson diphoton mass peak for different vertex reconstruction efficiencies. An efficiency of 80\% was achieved in Phase-1, while only 40\% is expected for Phase-2 if no timing information is provided; the other two cases include different degrees of precision timing. Based on different technologies, the MTD is divided into a barrel (LYSO scintillator coupled to silicon photomultipliers, $|\eta|<1.5$) and two endcap sections (silicon detectors, Low Gain Avalanche Diodes (LGADs)) covering up to $|\eta|=3.0$. Details can be found in Refs.~\cite{CERN-LHCC-2017-027,Butler:2019rpu}.
	
	\subsection{The calorimeters: ECAL and HCAL}
\begin{figure}[!t]
	    \centering
	    \includegraphics[width=\columnwidth]{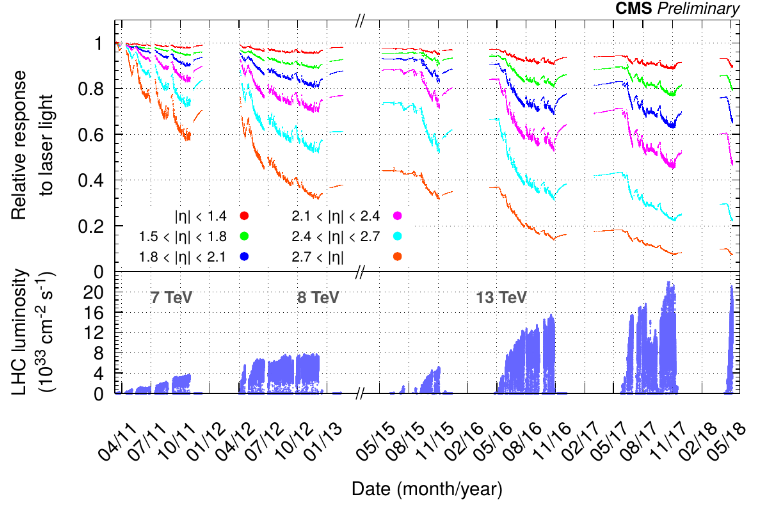}
	    \caption{Upper panel: relative ECAL response as a function of time at different $\eta$ regions as monitored by laser calibrations since 2011. Lower panel: corresponding LHC luminosity. From Ref.~\cite{CMS-DP-2017-023}.}
	    \label{fig:ecal_response}
	\end{figure}

	The electromagnetic and hadronic calorimeters (ECAL and HCAL, respectively) responses need to be calibrated throughout data taking. In fact, the scintillators are degraded by radiation and their response progressively decreases. Figure~\ref{fig:ecal_response} shows the ECAL crystal response to laser light during Run~1 and Run~2. The crystals in the forward region of the detector are the most affected by radiation, with the response reduced to less than 20\% its nominal value at the end of Run~2. The studies suggest that the barrel ECAL and HCAL can survive up to a luminosity of 3000~fb$^{-1}$, while the endcaps need to be replaced before the HL-LHC. Details can be found in Ref.~\cite{CERN-LHCC-2017-011}.

	\subsection{Phase-2 ECAL upgrade}
	The current crystal geometry of ECAL is maintained throughout Phase-2, while the front-end and readout electronics are replaced to cope with trigger requirements. Single-crystal information is made available to L1 triggers, instead of the previous standard of 5$\times$5 crystal towers. Faster readout electronics allow discrimination between an electromagnetic (EM) shower signal and anomalous spikes in Avalanche Photodiodes (APDs), this latter originated by the dark current. A discrimination on the different signal shape is employed to retain EM shower signals only.

	Projections show that the transparency loss of crystals at 3000 fb$^{-1}$ will be limited to 50\% of the end of Phase-1 value \cite{CERN-LHCC-2017-011}, therefore the current lead tungstate (PbWO$_4$) crystals may be kept until the end of the HL-LHC phase. The increased radiation level causes the increase of dark current in the APDs. However, the dark current can be kept under control by further cooling down the detector to 9$^\circ$C.
	
	The new electronics will allow CMS ECAL detector to reach a precise time resolution ($\sim$30~ps) of the signal in the APDs, which is essential to distinguish the interaction vertices in an extreme pileup scenario. Beam tests with electrons show that a resolution below 2\% is achievable, down to 1\% for a cluster energy larger than 50~GeV~\cite{Marinelli:2019fxa}.

	\subsection{High-Granularity Calorimeter (HGCAL)}
	The existing forward calorimeters will be replaced by a new High-Granularity Calorimeter (HGCAL)~\cite{CERN-LHCC-2017-023}, providing a unique fine grain in view of a multi-dimensional shower reconstruction up to a pseudorapidity $|\eta|=3.0$.

\begin{figure}[!t]
	    \centering
	    \includegraphics[width=0.8\columnwidth]{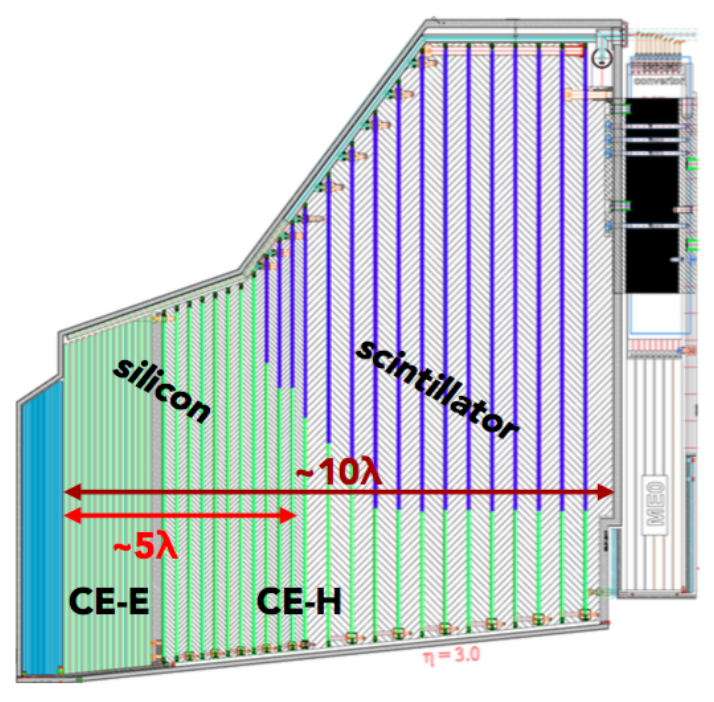}
	    \caption{Schematic drawing of the CMS HGCAL design~\cite{CERN-LHCC-2017-023}.}
	    \label{fig:hgcal}
	\end{figure}

	The HGCAL consists of a sampling calorimeter with silicon and scintillators as active material, including both electromagnetic (EE) and hadronic (FH+BH) sections. A schematic view is shown in Fig.~\ref{fig:hgcal}. The active elements are 320~$\mu$m-thick silicon sensors. The detector is transversely segmented into hexagon cells of about 1~cm$^2$ surface, for a total of over 6 million channels. Plastic scintillator tiles are used in the outermost regions of FH and BH, alternated with stainless steel absorbers. The thickness of the EE part amounts to about 25$X_0$ and about 1$\lambda$. The thickness of the hadronic part corresponds to about 3.5$\lambda$ and 5.7$\lambda$ for the FH and BH, respectively, for a total of about 9$\lambda$ for the 24 layers. The whole system needs to be kept at -30$^\circ$C in order to function.

	\subsection{HGCAL performance}
\begin{figure}[!t]
        \centering
        \includegraphics[width=\columnwidth]{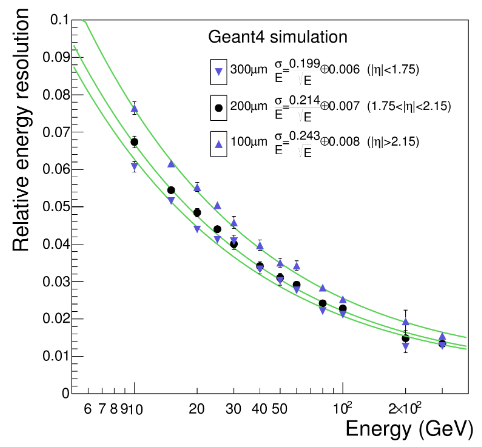}
        \caption{Energy resolution as a function of energy from a standalone simulation of incident electrons on HGCAL, for different thickness of silicon sensors~\cite{CMSCollaboration:2015zni}.}
        \label{fig:hgcal_resolution}
    \end{figure}

	The expected performance of HGCAL were first evaluated with a standalone GEANT4 simulation. Figure \ref{fig:hgcal_resolution} shows the electron energy resolution as a function of the electron energy for various active thickness of the silicon sensors. The stochastic term ranges from 20 to 24\% but the constant term is targeted to be low (1\%) \cite{Ochando:2017fie}. The time resolution of HGCAL is expected to be 30~ps for clusters with $p_T>$5 GeV. The timing performance combined with the high granularity of this detector allow to use it as an imaging calorimeter.

    \subsection{Muon detector upgrade}
	The muon detector electronics will be replaced to cope with increased radiation and meet trigger/readout latency requirements. Performance in the forward region will be enhanced by adding a resistive plate chamber and gas electron multiplier (GEM) chambers in the forward region to improve efficiency, fake rejection, and resolution in the region 1.6 $<\eta<$ 2.4. A new GEM chamber (ME0) will finally extend the coverage up to $|\eta|$=2.8, adding a muon trigger signal in the very forward region \cite{CERN-LHCC-2017-012}.
	
	\subsection{Summary}
	In the HL-LHC era, a new tracking detector will increase the overall acceptance of CMS up to $|\eta|$=4 and, for the first time, tracking information will be provided to the L1 trigger.
	CMS will include a new electromagnetic and hadronic endcap calorimeter with high granularity. The Phase-1 ECAL crystals are maintained during the HL-LHC phase while the front-end electronics are replaced, allowing for precise timing of signals. The muon detector coverage is extended to match the extended tracker. More broadly, simulations and beam tests suggest that both ATLAS and CMS will at least maintain the current performance in the high-pileup environment, or even improve in certain aspects, \eg, object reconstruction, particle identification, and background rejection.

\part[VBS at future colliders]{VBS at future colliders}\label{part:future}


As the HL-LHC program will conclude in the late 2030s or early 2040s, the community has turned to discussing opportunities for a successor to the LHC~\cite{Strategy:2019vxc,EuropeanStrategyGroup:2020pow}.
Presently, several colliders are being discussed. This includes: circular and linear $e^+e^-$ colliders with center-of-mass energies at or below a couple TeV, $\mu^+\mu^-$ colliders with energies at a few TeV, more energetic proton colliders with energies reaching up to 100s of TeV, as well as others. 

The international particle physics community has developed a
plan~\cite{Strategy:2019vxc,EuropeanStrategyGroup:2020pow} and the
top priority is the construction of an $e^+e^-$ collider to study the Higgs and other less-known parameters of the SM. 
For these facilities, the required technology is mature.
Other collider scenarios are possible, and are certainly great leaps forward in terms of knowledge and the energy frontier, but are farther away due to technological obstacles. Interestingly, at the electron, muon, and proton machines presently under discussion, the VBS process manifests differently. In all cases, VBS remains a powerful tool to exploring the EW sector and beyond.

Part~\ref{part:future} continues by first discussing VBS at electron colliders (Section~\ref{sec:future_vbs_ee}), and then EW parton distribution functions (Section~\ref{sec:muon_ewPDF}) and EW parton showers (Section~\ref{sec:ewVincia}), both of which are intimately related to VBS at ultra high-energy colliders.
It continues with VBS studies of SMEFT (Section~\ref{sec:smeft_muonCo}) and BSM scenarios (Section~\ref{sec:bsm_muonCo}) at multi-TeV muon colliders, and finally, VBS at a 100 TeV proton collider  (Section~\ref{sec:future_vbs_pp}) is discussed. We note that discussions here are only representative of a rich body of literature.


\section[VBS at $e^+e^-$ colliders]{\large VBS at $e^+e^-$ colliders}\label{sec:future_vbs_ee}

As described above, the 2020 Update for the European Strategy of Particle Physics has designated an $e^-e^+$ Higgs factory as one of the highest priorities of the high energy physics community, particularly as a staging platform to an even higher energy hadron collider ~\cite{Strategy:2019vxc,EuropeanStrategyGroup:2020pow}. There are four proposals for such a next-generation machine: two circular variants, the CEPC~\cite{CEPCStudyGroup:2018ghi} and the FCC-ee~\cite{Abada:2019zxq}, and two linear variants, the ILC~\cite{Baer:2013cma,Behnke:2013lya} and CLIC~\cite{Linssen:2012hp,Aicheler:2012bya}. While the circular options can cover measurements at the $Z$ pole as well as the $W^+W^-$, $ZH$, and $t\overline{t}$ thresholds with ultra-high integrated luminosity, only the linear machines can be extended into energy regions which are relevant for VBF and scattering measurements. Therefore, we focus here on the last two options. A fifth proposal, a many-TeV $\mu^+\mu^-$ collider, is discussed in Sections \ref{sec:muon_ewPDF}, \ref{sec:smeft_muonCo}, and \ref{sec:bsm_muonCo}, .

Roughly speaking, VBS becomes interesting at lepton collider energies of $\sqrt{s}=500$ GeV and beyond. The benefit of VBS studies at an electron-positron collider is the well-defined, quasi-partonic initial-state emission of EW bosons, where the dominating theoretical uncertainties are under better control. There are well-defined, well-separated and clean events that allow for a triggerless operation, full coverage of final states, and high-precision measurements. Spin, isospin, and CP quantum numbers of SM particles can be resolved with high precision due to excellent particle identification using particle-flow techniques. Due to the low hadronic environment, electron-positron colliders naturally have discriminating power for light quark flavors in jets, \ie, for charm and potentially even for strange quarks.

The ILC project is in a mature state, entering the so-called ``prelab'' phase in 2022~\cite{1866338}, prepared itself by the international development team, whose main deliverable is the engineering design report (EDR) of the machine. The machine is based on superconducting radiofrequency (RF) cavities with a design gradient of 31.5 MV/m, hosted in a 20.5~km long tunnel in the Kitakami region of the Iwate prefecture in northern Japan. The tunnel is extendable to 30 km and beyond, such that the machine can be upgraded based on the superconducting RF technology within the given infrastructure to 1 TeV or even a little higher. The energy range is tunable, which allows measurements at different energy stages. Such features are interesting, for example, for constraining or measuring EFT parameters at different energies. On the other hand, CLIC is based on normal conducting RF cavities using a drive beam concept which allows to reach 100 MV/m. There are three different energy stages foreseen, 380 GeV, 1.5 TeV, and 3 TeV.

Another important asset of linear $e^+e^-$ colliders is the possibility of polarized beams. These help enhance event yields for specific signal processes. The ILC and CLIC both foresee 80\% electron polarization. While CLIC considers unpolarized positron beams, the ILC baseline design uses rotating targets and undulators to generate 30\%  positron polarization. This  rate is expected to deteriorate to 20\% at the 1 TeV stage. The staged running foresees large data samples for ILC and CLIC at the highest energies: $\mathcal{L}=1.5~(5)$ ab${}^{-1}$ for $\sqrt{s}=1.5~(3)$ TeV  for CLIC, and $\mathcal{L}=4~(8)$ ab${}^{-1}$ for $\sqrt{s}=500~(1000)$ GeV for the ILC.

In comparison to proton colliders, the main topology of VBS processes remains unchanged at $e^+e^-$ colliders. At high energies, the VBS signature is characterized by high-$p_T$ activity that is accompanied by two forward partons. These forward partons are either neutrinos when VBS involves $W^\pm$ scattering, or charged leptons when involving $Z/\gamma$ scattering. In order to accept forward charged leptons or to  veto them, a very good, low-angle coverage by the ILC and CLIC detectors is crucial. This is achieved by using the luminosity and beam calorimeters (LumiCal and BeamCal), which have coverage down to 38-110 mrad (LumiCal) and 15-38 mrad (BeamCal), respectively. Hence, they allow one to go down to rapidities as low as $\vert\eta\vert \approx 4.9$ in the forward direction~\cite{Idzik:2015oja}. Another asset of electron-positron colliders is the possibility to perform VBS measurements with fully hadronic final states of the EW vector bosons $W^\pm, Z$. This is achieved using the particle flow algorithm as it was shown for the CLIC detector~\cite{Marshall:2012ry}. This allows a hadronic $W/Z$ discrimination with an efficiency of 88\%, as the particle flow algorithm almost completely removes the photon-induced hadron background for energy ranges between 100 and 1200 GeV. The efficiency is slightly reduced to 71-79\% when considering the photon-induced background in the high-energy regime of CLIC. 

\subsection{Anatomy of $W^+W^-$ scattering at $e^+e^-$ colliders}

\begin{figure}[t!]
    \centering
    \includegraphics[width=\columnwidth]{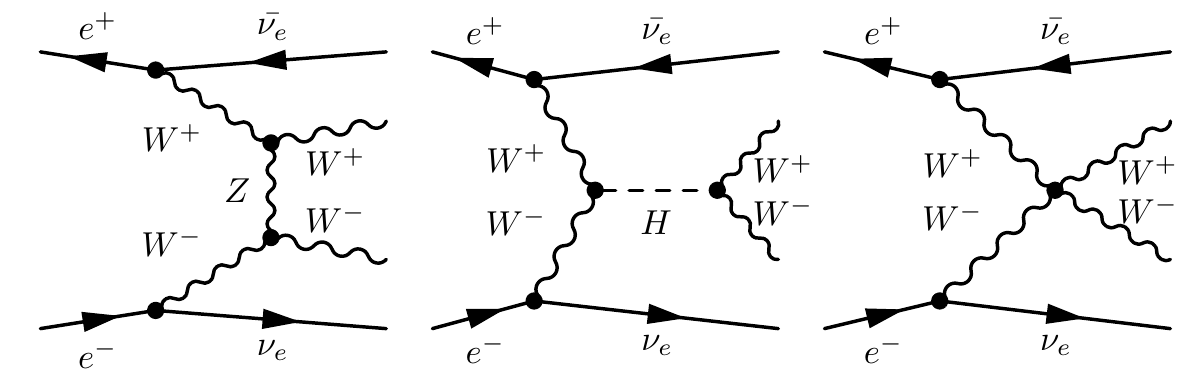}\\
    \includegraphics[width=\columnwidth]{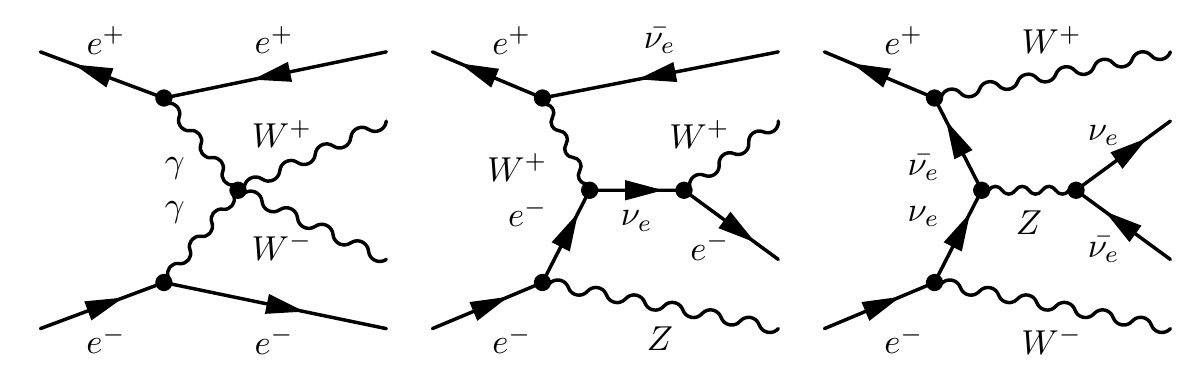}\\
    \includegraphics[width=\columnwidth]{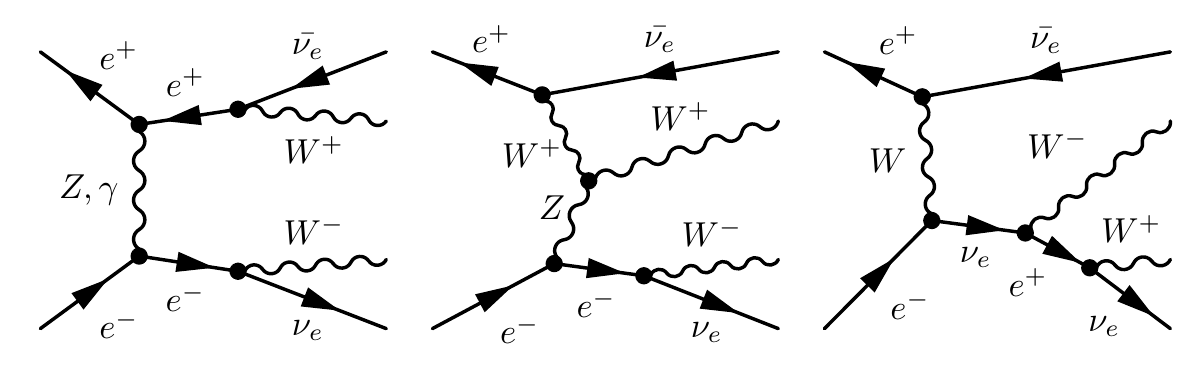}    
    \caption{Topologies for $W^+W^-$ scattering processes at $e^+e^-$ colliders: signal diagrams (top row), reducible background diagrams (middle row), and irreducible diagrams (bottom row), respectively.}
    \label{fig:ee_vbs_diags}
\end{figure}

While at hadron colliders the EW diboson production with additional QCD radiation constitutes the largest background contamination for VBS processes, at $e^+e^-$ colliders the dominant background is from EW radiation and other resonant EW production processes. Figure~\ref{fig:ee_vbs_diags} shows examples of Feynman diagrams for $W^+W^-\to W^+W^-$ scattering and its backgrounds at $e^+e^-$ colliders: In the top row there are the signal diagrams for $W^+W^-\to W^+W^-$ containing triple and quartic gauge vertices and Higgs exchange. The middle row shows the reducible background from photon fusion as well as from triple boson production and multi-peripheral diagrams. The bottom row finally shows the irreducible background diagrams, mostly from EW radiation.
Here, the term ``reducible'' backgrounds refers to backgrounds with characteristic final states, topologies, or kinematics that can facilitate their removal with selection cuts and other analysis-level requirements. ``Irreducible'' backgrounds are those that survive such selection cuts and requirements on final-state particles. Note that diagrams with initial- and final-state EW radiation can be enhanced by large EW logarithms; however, this regimes just begins  in the multi-TeV range. Similar classes and types of diagrams to those in Fig.~\ref{fig:ee_vbs_diags} are present for other VBS processes in high-energy lepton collisions.

\begin{table}[!t]
\begin{center}
\resizebox{.9\columnwidth}{!}{
  \begin{tabular}{l c c c}
  \hline\hline\noalign{\smallskip}
  Process $(e^+e^-\to X)$ & $\sigma_{1.4~\rm TeV}$ & $\sigma_{3~\rm TeV}$ & Mis-ID Factor\\
  \hline\noalign{\smallskip}
  \multicolumn{4}{c}{Signal and background for $VV\to W^+W^-$}\\
  \hline
  $W^+W^-\nu\bar{\nu}$  & 0.119 & 0.790 & 1 \\
  $W^+W^-e^+ e^-$       & $<1$ ab & $<1$ ab & 1\\
  $W^\pm Z e^\mp \nu$   & 0.269 & 1.200 & 0.136\\
  $ZZe^+ e^-$           & $<1$ ab & $<1$ ab & 0.019\\
  $W^+ W^- (Z \to \nu\bar{\nu})$ & 0.039 & 0.610 & 1 \\
  \hline\noalign{\smallskip}
    \multicolumn{4}{c}{Signal and background for $VV\to ZZ$}\\
    \hline
  $ZZ\nu\bar{\nu}$  & 0.084 & 0.790 & 1\\
  $ZZe^+ e^-$       & $<1$ ab & $<1$ ab & 1\\
  $W^\pm Z e^\mp \nu$       & 0.288 & 1.590 & 0.136\\
  $W^+W^-e^+ e^-$           & $<1$ ab & $<1$ ab & 0.019\\
  $ZZ (Z \to \nu\bar{\nu})$ & $<1$ ab & $<1$ ab & 1\\
  \hline\hline\noalign{\smallskip}
  \end{tabular}
  }
  \caption{SM cross sections in fb ($\pm 1 \%$ error) of signal and background processes for $VV\to W^+W^-$ (upper) and $VV\to ZZ$ (lower) scattering,  
  with selection cuts for the fully hadronic final state in $e^+e^-$ collisions at $\sqrt{s}=1.4\;$TeV and $3\;$TeV. Both  particle  beams  are  unpolarized. Detection efficiencies  and  branching ratios  are  not  included. All cross sections have to be multiplied by the factors in the fourth column to take the misidentification of vector bosons into account. For further details, see Ref.~\cite{Fleper:2016frz}.
  }    \label{t:totcsw} 
   \end{center}
\end{table}

There is quite some literature on the study of VBS at electron (-positron) colliders, which started with the first interests in linear colliders in the late 1980s~\cite{Gunion:1987ta,TofighiNiaki:1988bz,Barger:1995cn}. There were also quite some publications triggered by the TESLA preparations~\cite{Han:1997ht,Boos:1997gw,Boos:1999kj,Abe:2001swa}. Among those were the ones first pointing out the interplay and gauge dependence between VBS and triple boson diagrams. At this time  the EW corrections to (on-shell) VBS had been calculated~\cite{Denner:1996ug,Denner:1997kq}. Later on, the leading logarithmic corrections for the off-shell process were calculated~\cite{Accomando:2006hq}. Also off-shell effects for Higgs processes in VBS have been studied~\cite{Liebler:2015aka}. Many studies were devoted to the question whether a strongly coupled EW sector could be discovered in VBS at an $e^+e^-$ collider~\cite{Dominici:1997zh,Kilian:2000pg,Chierici:2001ar,Rosati:2002wqa,Beyer:2006hx}. Within these, some studies were also made for same-sign $e^-e^-$ colliders in order to access same-sign $WW$ scattering.

In the following, we will describe the technical details how to extract the VBS signal at a lepton collider from the EW background and the subtleties in how to look for deviations from the SM. The first, rather complete assessment of all backgrounds to VBS and triple boson production has been made in~\cite{Beyer:2006hx}, for 1 TeV, assuming 40\% positron polarization. The main processes are $e^+e^- \to \nu\bar{\nu} q\bar{q}q\bar{q}$, which contains the signal subprocesses $W^+W^- \to W^+W^-, ZZ$ as well as the triple-boson production $V\to VVV$. The process $e^+e^- \to \nu e q\bar{q}q\bar{q}$ consists of the signal subprocess $WZ\to WZ$, and the process $e^+e^- \to e^+e^- q\bar{q}q\bar{q}$ contains the subprocess $ZZ\to W^+W^-, ZZ$. Cross sections range between a few fb up to 130~fb for the $\nu e jjjj$ final state. The last three processes are also populated by radiative Bhabha events. Six-(and eight-) fermion final states are contaminated by top pair production and $ttH$, which peaks at around 800 GeV center-of-mass energy. The largest individual background is from four-jet processes by diboson production (and additional QCD radiation) which reaches up to almost 4 pb. Further backgrounds come from single-$W$ production, \ie, $e^+e^- \to e\nu q\bar{q}$, and radiative $Z$ production, $e^+e^- \to e^+e^- Z$. Another substantial background comes from di- and multi-jets which amounts to 1.6~pb.

These older studies have been revisited in Ref.~\cite{Fleper:2016frz}. Table~\ref{t:totcsw}, using the simulation framework of~\cite{Kilian:2007gr,Brass:2018xbv}, shows the total cross sections for two different high-energy stages of CLIC at $\sqrt{s}=1.4$ and 3 TeV for on-shell vector boson production. 
In this table SM cross sections in fb ($\pm 1 \%$ error) of signal and background processes for $VV\to W^+W^-$ (upper) and $VV\to ZZ$ (lower) scattering,   with selection cuts for the fully hadronic final states. 
Selection cuts include: 
requiring that reconstructed $W/Z$ bosons are central through polar angle and $p_T$ cuts; 
requiring that the $(WW)$- or $(ZZ)$-system carry a large invariant mass and large transverse momentum;
and
requiring final-state leptons be central or carry a large invariant mass.
For more specifics, see Ref.~\cite{Fleper:2016frz}. The last column of Table~\ref{t:totcsw} is the correction factor due to the mis-identification efficiency for hadronic decays of vector bosons. It takes into account a probability for mis-identification between $W$ and $Z$ bosons, and has to be taken into account experimentally. It is apparent that the signal cross sections rise significantly from 1.4 to 3 TeV. The contamination from $WZ$ scattering is particularly noteworthy. Also note that the background from triboson production is irreducible as VBS and triboson production are in the same class of EW gauge-invariant processes. 

\begin{figure}[!t]
 \centering
  \includegraphics[width=\columnwidth]{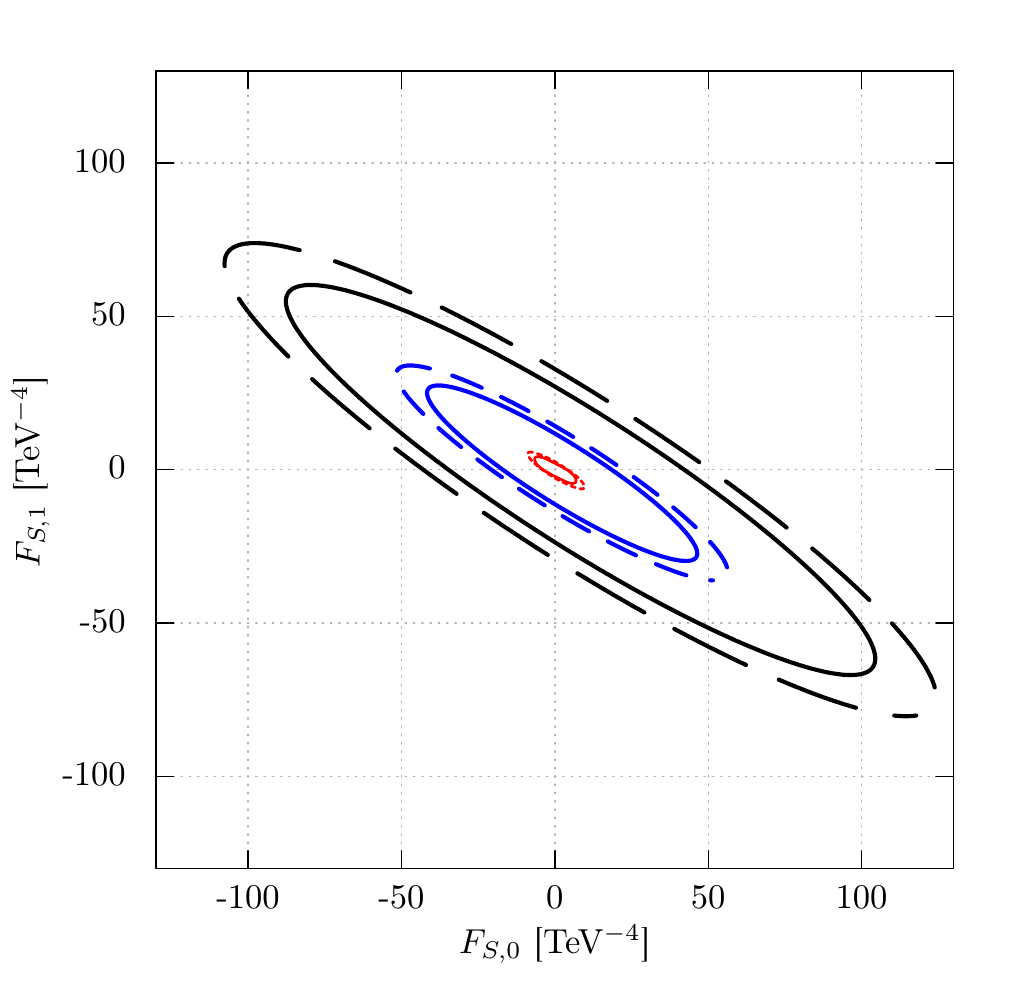}
  \caption{The 90\;\% exclusion sensitivities for polarized (solid) and    unpolarized (dashed) particle beams at energies of    $\sqrt{s}=1\,\text{(black)},\ 1.4\,\text{(blue)},\ 3$\,TeV (red) combined, assuming integrated luminosities    of $5\;\text{ab}^{-1}$ at $\sqrt{s}=1$ TeV, $\ 1.5\;\text{ab}^{-1}$ at $\sqrt{s}=1.4$ TeV, and    $2\;\text{ab}^{-1}$ at $\sqrt{s}=4$ TeV. Adapted from Ref.~\cite{Fleper:2016frz}.}
  \label{fig:lc_reach}
\end{figure}

One possibility to enhance the VBS component is by a selection cut on the fiducial phase space. The following cuts can be used to reduce the different backgrounds, which are adapted in their numerical values to the different energy stages of ILC and CLIC: (1) Backgrounds from $Z\to\nu\bar\nu$, $W^+W^-$ diboson production, and the QCD 4-jet continuum can be significantly reduced by an invariant mass cut on the neutrino/invisible system, $M_{inv}(\nu\bar\nu)$; (2) Backgrounds from $t$-channel, multi-peripheral sub-processes in the production can be reduced by cuts on the transverse momentum of single EW bosons, $p_{\perp,W/Z}$, and simultaneously on the (beam) angle of the EW boson, $\cos\theta(W/Z)$; (3) Photon-induced backgrounds from the effective photon approximation (EPA)/ISR-like  setups can be suppressed by transverse momentum cuts on the diboson systems, $p_\perp(WW,ZZ)$, respectively; furthermore, very effective suppression is reached by vetoing visible forward electrons $\theta(e) > 15$~mrad; (4) Finally, massive EW radiation can be eliminated selecting certain (high-energy) windows of the diboson-system invariant masses, $M_{inv}(WW,ZZ)$, respectively. For details, see Ref.~\cite{Fleper:2016frz}.

\subsection{Constraining New Physics in VBS at $e^+e^-$
  colliders}\label{sec:cons_np_ee}

\begin{figure}[!t]
      \includegraphics[width=\columnwidth]{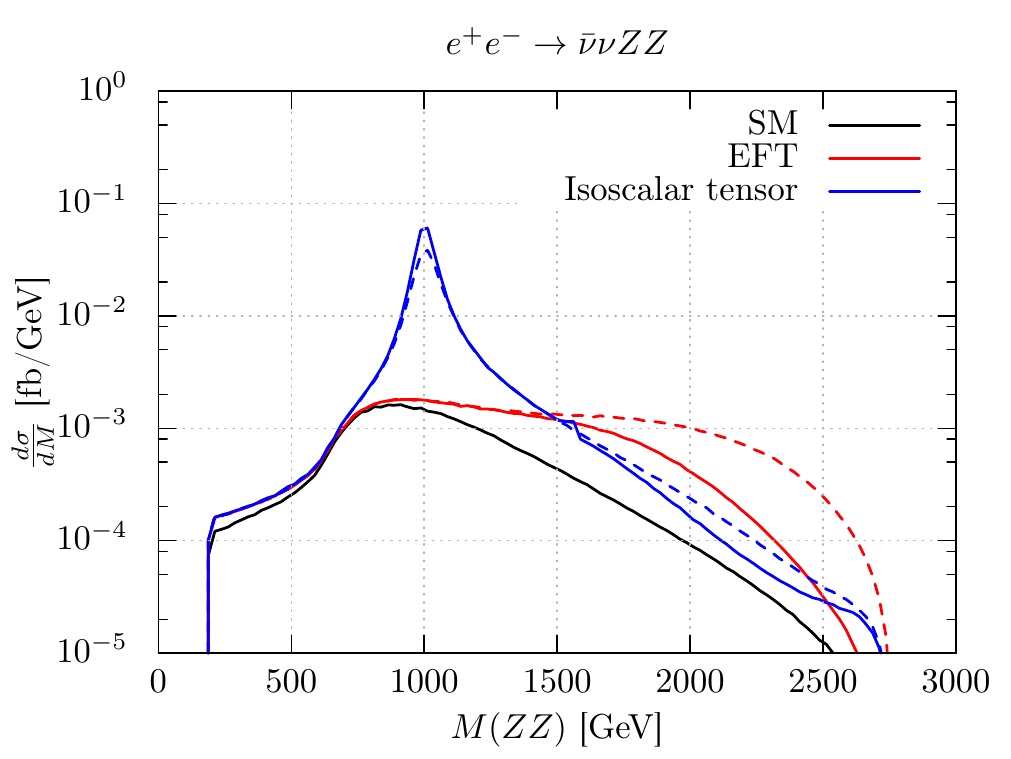}
   \caption{Differential cross sections depending on the invariant mass of the $ZZ$ system in $e^+e^-\to ZZ\nu\overline{\nu}$ at $\sqrt{s}=3$\,TeV.
     The solid lines show the signal process
     $\bar{\nu}\nu ZZ$ with SM values
     $F_{S,0}=F_{S,1}=0$.
     The red lines show matched EFT results with $F_{S,0} = 150.8\;\text{TeV}\,^{-4}$ and $F_{S,1}=-50.3\;\text{TeV}\,^{-4}$.
     The blue lines show an isoscalar tensor resonance $f$ with $m_f = 1$ TeV, $F_f = 17.4\;\text{TeV}\,^{-1}$, and $\Gamma_f=100$ GeV.
    (Dashed lines: naive EFT results, solid lines: unitarized results). 
     Adapted from Ref.~\cite{Fleper:2016frz}.}
   \label{fig:isoscalar_tensor_res}
\end{figure}

VBS is one of the tools to look for deviations from the SM in the EW sector. In a modern framework after the Higgs discovery, deviations are described in terms of higher-dimensional operators. To derive constraints on these operators or finding a significant deviation of experimental data from the SM by non-zero Wilson coefficients is the main task of these measurements.

There are several stages of describing new physics beyond the SM, as it was discussed in detail for VBS in~\cite{Gallinaro:2020cte}:  (1) in terms of an effective field theory expansion,  (2) by a simplified model with generic new resonances, and (3) specific BSM models. In~\cite{Fleper:2016frz} the sensitivity of ILC-1.0, CLIC-1.5 and CLIC-3.0 have been derived for dimension-eight operators in the SMEFT framework, shown in Fig.~\ref{fig:lc_reach}.  These results take into account a signal model that always stays within the unitarity bounds of the amplitudes~\cite{Alboteanu:2008my,Kilian:2014zja}. There have been also studies on simplified models comprising resonances that couple dominantly to the EW diboson systems, such that their production via Drell-Yan processes are too faint~\cite{Fleper:2016frz} (resonances coupling to transverse modes had been introduced in~\cite{Brass:2018hfw}, for LHC). This framework can be mapped in a leading-order power expansion to the standard SMEFT operators. An example is shown for an isoscalar spin-2 resonance in Fig.~\ref{fig:isoscalar_tensor_res}. For more details, see Ref.~\cite{Fleper:2016frz}. 

In summary, these studies show that the clean environment of an electron-positron collider at highest energies is a fantastic opportunity for searches for new physics in the EW sector. In general, the reach is much enhanced by being able to go to the highest-available energies -- there are proposals for plasma-wakefield driven accelerators that could reach tens of TeV -- while polarization of at least the electron beam is desirable in order to reduce backgrounds to provide discriminating analysis power for the effects of new physics. This applies both to polarization measurements as well as to the resolution of CP quantum numbers.

\section[EW parton distribution functions]{\large EW parton distribution functions}\label{sec:muon_ewPDF}

\begin{figure*}[!t]
\includegraphics[width=\columnwidth]{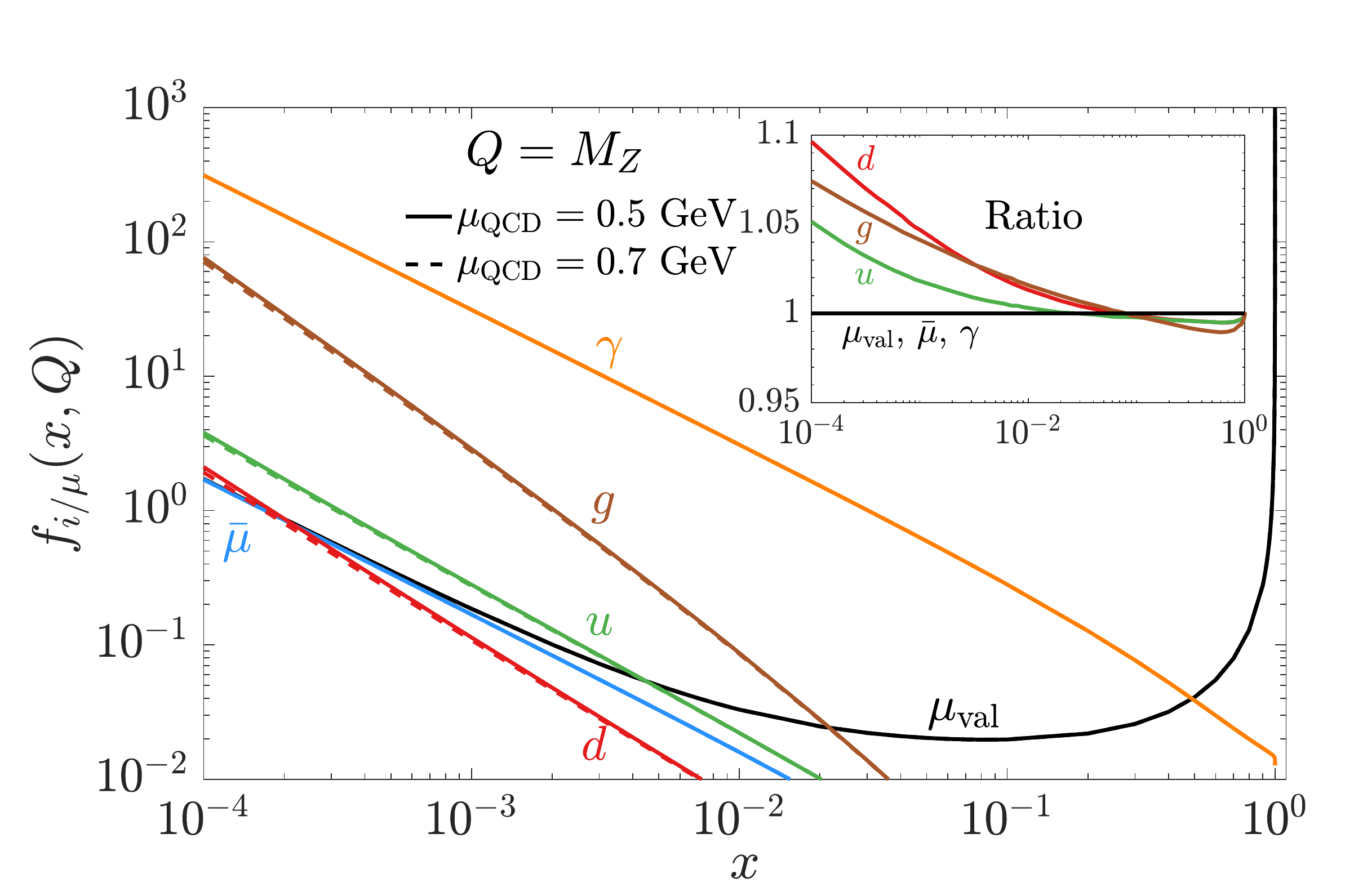}
\includegraphics[width=\columnwidth]{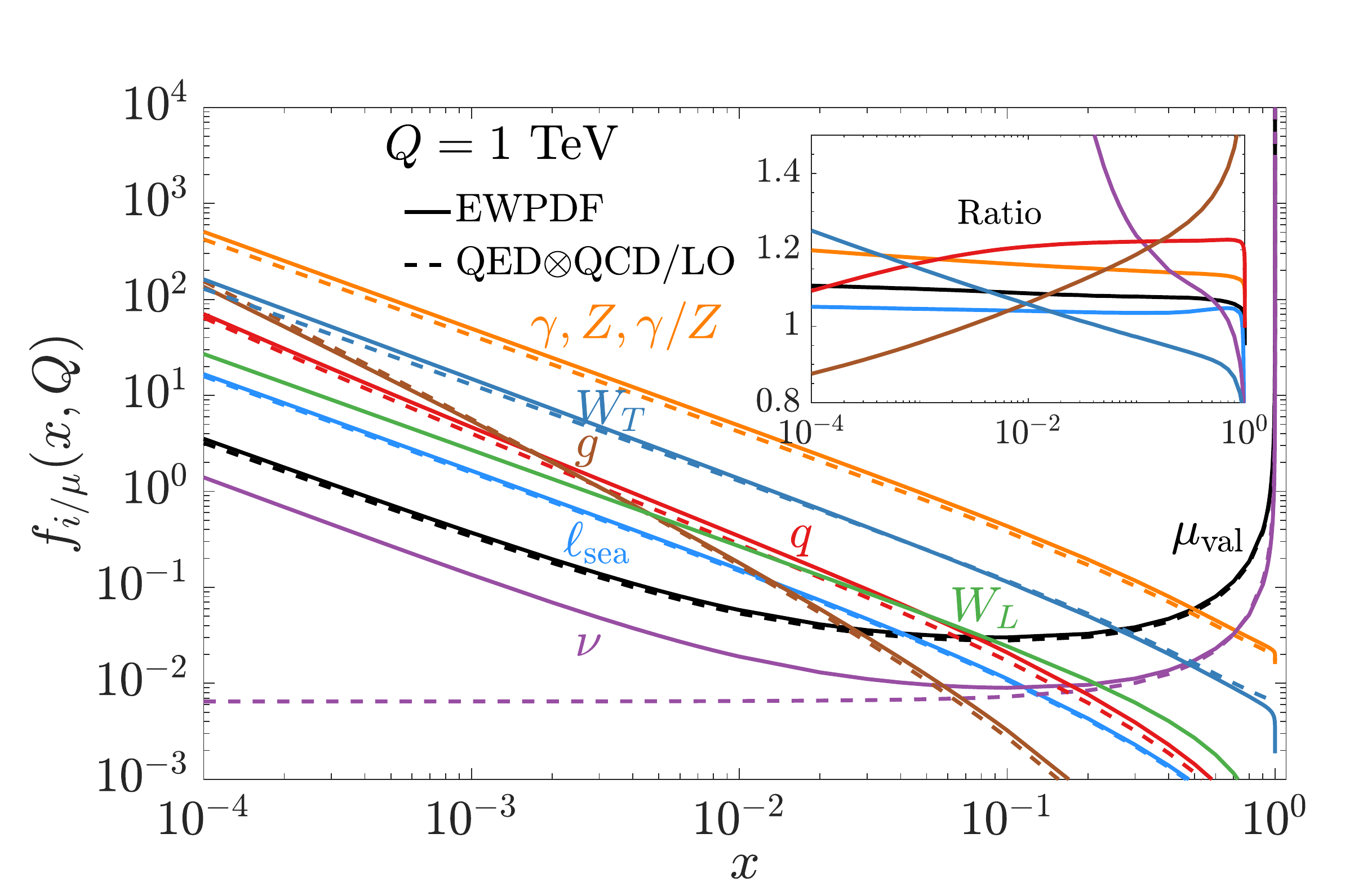}
\caption{PDFs of a muon at $Q=M_Z$ (Left) and 1 TeV (Right). Left: The solid (dashed) lines correspond to the PDFs with matching scale $\muQCD=0.5~(0.7)$ GeV. Right: The solid lines indicate the complete EW PDFs, while the dashed lines are the pure QED$\otimes$QCD evolution for $\gamma,\lsea,q,g$ and the LO splittings for $\nu,W$, respectively. Inset: the ratios of the solid to dashed lines with the same colors.}
\label{fig:PDFs}
\end{figure*}

The discovery of the Higgs boson at the LHC~\cite{Aad:2012tfa,Chatrchyan:2012ufa} symbolizes a great triumph of the SM, and understanding the properties of the Higgs remains the community's top priority.
The next targets on the particle physics road map~\cite{Strategy:2019vxc,EuropeanStrategyGroup:2020pow}
require reaching a new energy frontier. This can be achieved by building new colliders with partonic center-of-mass energies that are well beyond $\mathcal{O}(1)$ TeV. 

There are several pathways to reach such energies. For example: recently, breakthroughs in the cooling technology for a muon beam provides a potential pathway to construct a $\mm$ collider with a collision energy up to $\sqrt{s}\sim\calO(10)$ TeV~\cite{Delahaye:2019omf}. A future 100 TeV proton-proton collider, such as the FCC-hh promoted by CERN~\cite{Benedikt:2018csr} or the SppC promoted by IHEP~\cite{Tang:2015qga}, and which can readily reach $\calO(10)$ TeV partonic collisions, is another appealing option to push the energy frontier up to an unprecedented level.  

Regardless of the specific collider, at collision energies well above the EW scale $\muEW\approx M_Z$, all  SM particles are essentially massless and the SM gauge symmetry SU(2)$\otimes$U(1)$_{Y}$ is approximately restored \cite{Chen:2016wkt}. In analogy to quark/gluon splitting in hadron collisions, the collinear splitting mechanism involving the EW bosons becomes a dominant phenomenon in this kinematic regime. For initial-state radiation, the EW parton distribution function (PDF) formalism should be adopted as a proper description, which resums potentially large collinear logarithms~\cite{Bauer:2017isx,Bauer:2018arx,Han:2020uid,Han:2021kes}. Similarly, final-state EW radiation and associated logarithms should be resummed by  fragmentation functions~\cite{Bauer:2018xag}, or equivalently by the Sudakov form factor implemented in the parton showering \cite{Chen:2016wkt} (see also Section~\ref{sec:ewVincia} for related details).
From this perspective, one can think of VBS as being initiated by a pair of initial-state  EW bosons instead of a pair of initial-state leptons or quarks (and intermediate EW bosons).

In this section, we present results of recent work~\cite{Han:2020uid,Han:2021kes} on EW PDFs and their corresponding application to high-energy lepton colliders. In Section~\ref{sec:PDFs}, we lay out the EW PDF formalism up to double-log accuracy, and present some numerical results.  Application to representative SM processes at a multi-TeV muon collider is discussed in Section~\ref{sec:XSec}. We summarize and conclude in Section~\ref{sec:concEWPDF}. This formalism is equally applicable for an $e^+e^-$ collider, with an enhancement by a factor of $\log(m_\mu^2/m_e^2)$. It is also straightforward to extend this formalism to high-energy hadron colliders, which is left for a future work.

\subsection{The electroweak parton distribution functions}
\label{sec:PDFs}
Parton distribution functions $f_i(x,Q)$ evolve according to the well-known DGLAP equations \cite{Dokshitzer:1977sg,Gribov:1972ri,Lipatov:1974qm,Altarelli:1977zs}
\begin{equation}
\frac{\dd f_{i}}{\dd\log Q^2}=\sum_{I}\frac{\alpha_I}{2\pi}\sum_{j}P_{i,j}^I\otimes f_j,
\end{equation}
where $I$ runs over the SM gauge and Yukawa interactions, and the indices $i,j$ run over all particle species. (In principle, the evolution for particles of different chiralities are different. However, this notation is suppressed here for clarity.) In the above expression, $Q$ is the evolution scale, $\alpha_I(\mu_r=Q)$ is the renormalized coupling of gauge interaction $I$, and $P_{i,j}^I$ is the $j\to i$ splitting function under $I$. The operator $\otimes$ denotes the usual convolution over momentum fractions $x$. The full SM spectrum is characterized by two scales, $\LambdaQCD\sim200$ MeV and $\LambdaEW\sim250$ GeV. To validate the QCD perturbativity, we adopt a matching scale $\muQCD$ to separate the pure QED region at low energy from the QED$\otimes$QCD region at an intermediate scale. Below $\muQCD$, only photon and light charged fermions are active through the electromagnetic interactions. Above $\muQCD$, the QCD interaction enters, and the gluon becomes an active parton content as well. In this approach, the non-perturbative effect is parameterized with this threshold variable $\muQCD$, which in principle should be determined through data or lattice simulation. In practice, we take $\muQCD=0.5$ GeV, inspired by the critical scale in Ref.~\cite{Drees:1994eu}. Its variation quantifies the corresponding QCD threshold uncertainty.  

\begin{figure*}[!t]
\includegraphics[width=\columnwidth]{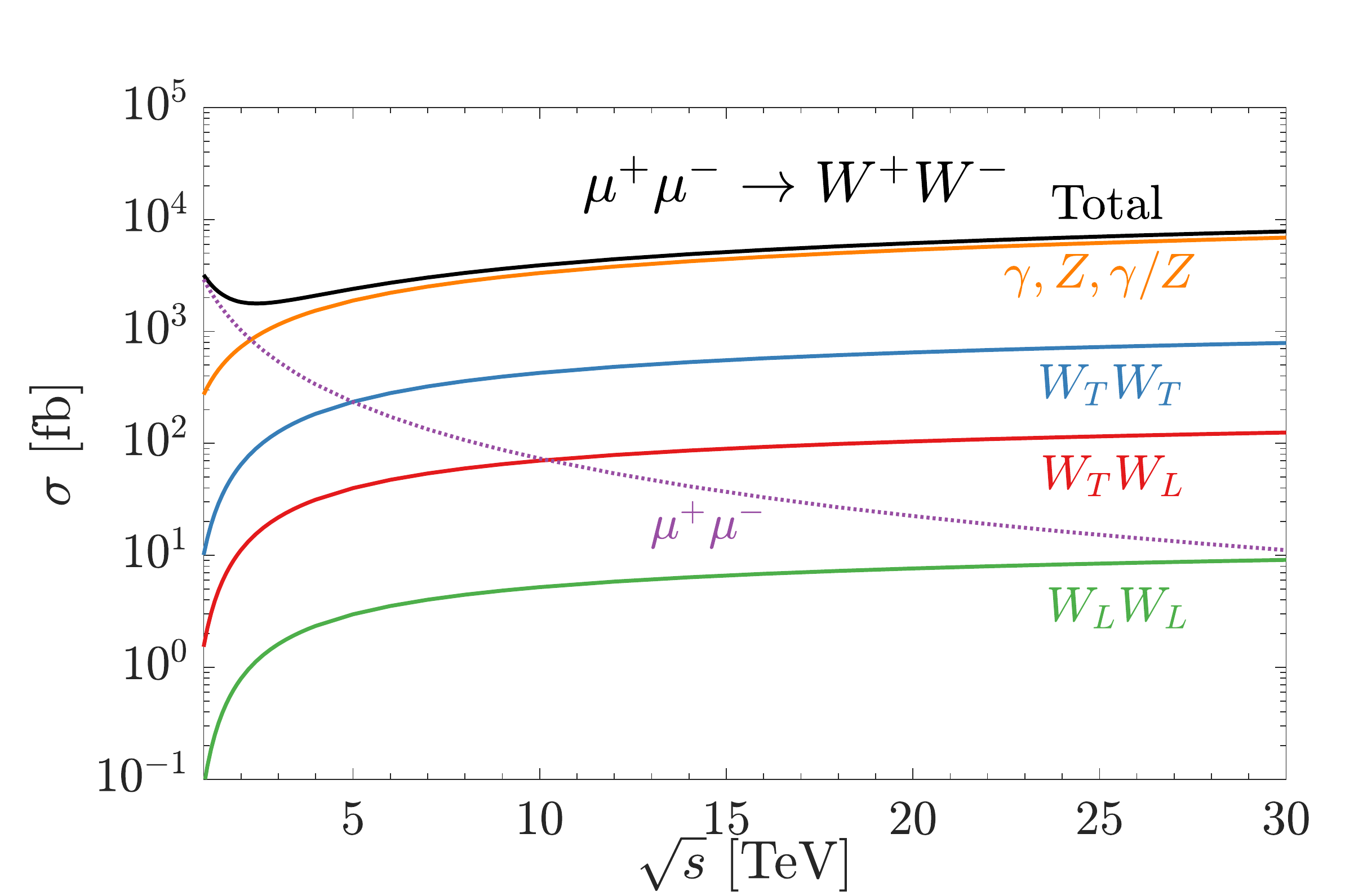}
\includegraphics[width=\columnwidth]{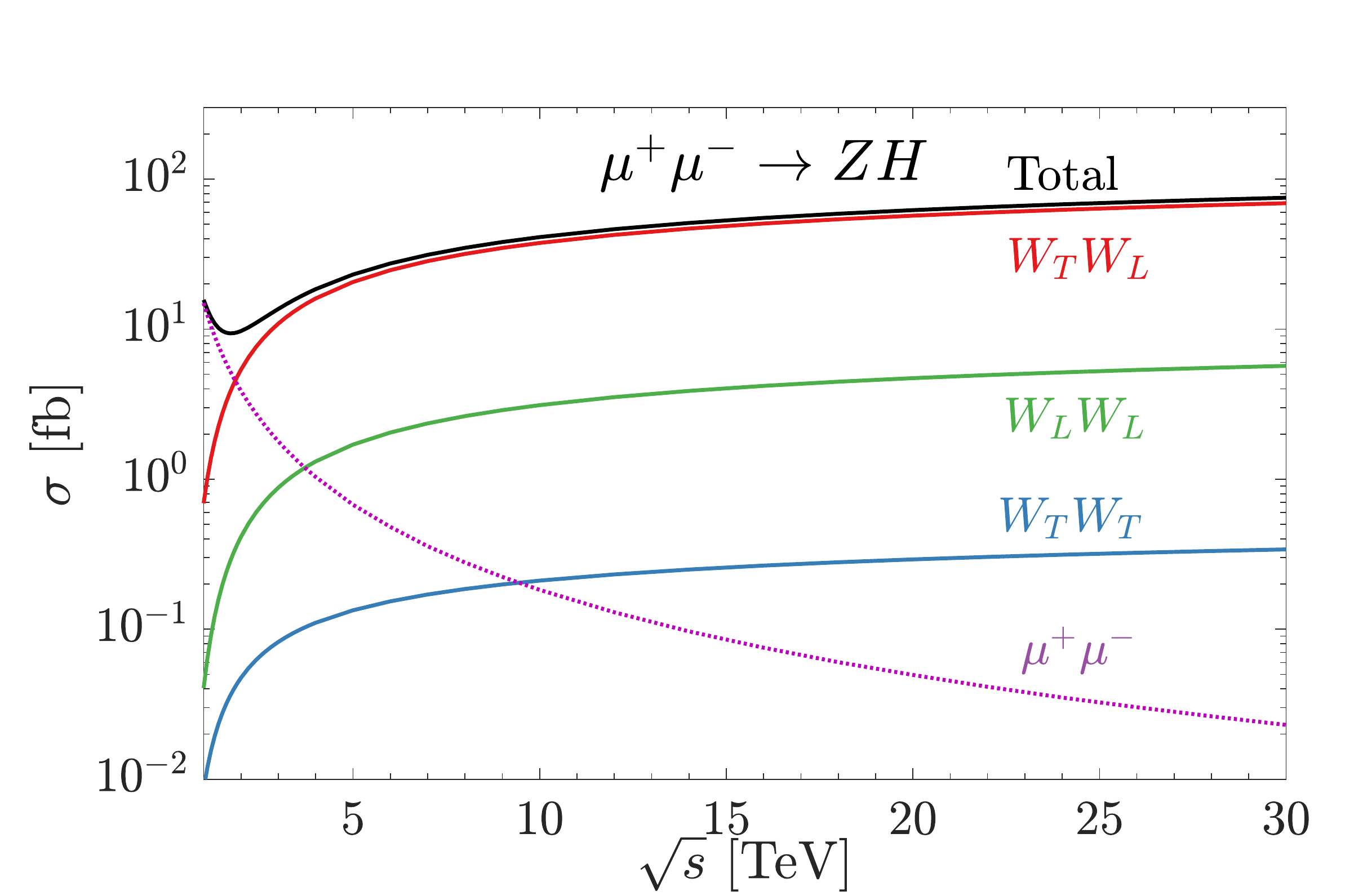}
\caption{The semi-inclusive cross sections of the $W^+W^-$ (Left) and $ZH$ (Right) production and the corresponding decomposition channels at high-energy muon colliders with collision energy from 1 to 30 TeV.}
\label{fig:WWZH}
\end{figure*}

Below the EW scale $Q^2 \ll \muEW^2$, the effects of the massive EW gauge bosons are suppressed by $g^2 Q^2/M_Z^2$ and the corresponding partons are inactive. Above the EW scale, all the EW states are activated and evolve according to the unbroken SM gauge group. It is more convenient to work in the gauge basis with the EW partons $B,W^{\pm,3}$. At the EW scale $\muEW$, a matching condition bridges the PDFs in the QED$\otimes$QCD to EW regions with a general relation
\begin{equation}\label{eq:rotation}
\begin{pmatrix}
f_B\\ f_{W^3}\\ f_{BW^3}
\end{pmatrix}=\begin{pmatrix}
c_W^2 & s_W^2  & -c_W s_W\\
s_W^2 & c_W^2  & c_W s_W\\
2c_W s_W & -2c_W s_W & c_W^2-s_W^2 
\end{pmatrix}
\begin{pmatrix}
f_\gamma \\ f_Z\\ f_{\gamma Z}
\end{pmatrix},
\end{equation}
where $s_W=\sin\theta_W$ is the sine of the weak mixing angle and 
\begin{equation}
f_{\gamma}(x,M_Z)\neq0, ~f_Z(x,M_Z)=0, ~f_{\gamma Z}(x,M_Z)=0.
\end{equation}
The mixed PDF $f_{\gamma Z}$ (or $f_{BW^3}$) represents the coherently mixed state, resulted from the interference between the $\gamma,Z$ (or $B,W^3$)~\cite{Chen:2016wkt,Bauer:2017isx,Ciafaloni:2000df}.

As the QCD/EW partonic formalism is set up, the PDFs of a leptonic beam are fully calculable in a perturbative framework. For a momentum fraction $x$ and factorization scale $\mu_f$, the initial conditions start from the lepton mass $m_\ell$ at the leading order as
\begin{equation}
f_{\ell/\ell}(x,\mu_f = m_{\ell})=\delta(1-x),
\end{equation}
while all other partons are zero at the initialization scale $Q_0^2=m_{\ell}^2$.
We solve the DGLAP equations and obtain PDFs for a high-energy muon beam, with two typical scales $Q=M_Z$ and 1 TeV shown in Fig.~\ref{fig:PDFs}. The PDFs at some other scales and the corresponding parton luminosities for muon colliders at various energies can be found in Refs.~\cite{Han:2020uid,Han:2021kes}. Many important features for the light-flavor PDFs are discussed in Ref.~\cite{Han:2021kes} and the EW PDFs in Refs.~\cite{Han:2020uid}. Here, we want to emphasize the threshold uncertainty estimated by varying the matching scale as $\muQCD=0.7$ GeV \cite{Drees:1994eu}. The overall size is a few percent for quark and gluon PDFs. Compared to up-type PDFs, the relative variation of the down-type PDFs is larger, because the corresponding absolute PDFs are smaller resulted from the smaller electromagnetic charges. For an electron beam, the $\muQCD$ uncertainty is expected to be larger due to the larger logarithm $\log(\muQCD^2/m_\ell^2)$ \cite{Han:2021kes}. The impact on the lepton and photon PDFs is negligible, as the QCD interactions only enter as a higher-order effect.

In Fig. \ref{fig:PDFs}(right), the complete EW PDFs at $Q=1$ TeV are shown as solid curves. The sea-flavor fermions can be expressed in terms of valence-flavor (val) fermions and are given by the summations
\bea
f_{\lsea} &=&f_{\bar{\ell}_{\textrm{val}}}+\sum_{i\neq\lval}^{N_{\ell}} (f_{\ell_i}+f_{\bar{\ell}_i}),\\
f_{\nu}&=&\sum_{i}^{N_{\ell}} (f_{\nu_i}+f_{\bar{\nu}_i} ),\\  
f_{q}&=&\sum_{i}^{N_u}(f_{u_i}+f_{\bar{u}_i})+\sum_{i}^{N_d}(f_{d_i}+f_{\bar{d}_i}) .
\label{eq:sea}
\eea
Here $N_u=3$ as the top quark becomes active as well above $\muEW$. For the neutral current components, we convert the PDFs back to the mass basis and sum over $\gamma,Z$ and the mixing $\gamma Z$. Besides the EW PDFs, we also show the ones obtained through the pure QED$\otimes$QCD evolution for $\lsea,q,g$ and $\gamma,Z,\gamma Z$ and the leading order splittings for neutrino and $W_T$ PDFs.
The difference between the solid and dashed curves quantifies the (higher-order) EW corrections. At $Q=1$ TeV, the overall size of EW corrections is about $10\%\sim20\%$, depending on the specific components. If we look at a specific flavor, such as the down-type quarks, the EW corrections can be much larger, due to the relatively large SU(2)$_\textrm{L}$ gauge coupling compared with the corresponding electromagnetic one. The EW corrections to the neutral current PDFs are as large as 20\%, mostly due to the new components, $Z$ and $\gamma Z$. The LO splittings for $W_T$ and $\nu$ come from $\mu\to W\nu$, which correspond to the Effective $W/Z$ Approximation \cite{Kane:1984bb,Dawson:1984gx,Chanowitz:1985hj}. At a small $x$, the resummed neutrino PDF $f_{\nu}$ deviates from the LO $1/(1-x)$ behavior due to the higher-order returns in the splitting $W\to\ell\nu$. For the $f_{W_T}$ PDF, the higher-order corrections are positive at small $x$ and negative at large $x$. In Fig. \ref{fig:PDFs}, we also demonstrate the $W_L$ PDF, which does not evolve with scale $Q$ at the leading order, as a remnant of the EW symmetry breaking.

\begin{figure*}[!t]
\includegraphics[width=\columnwidth]{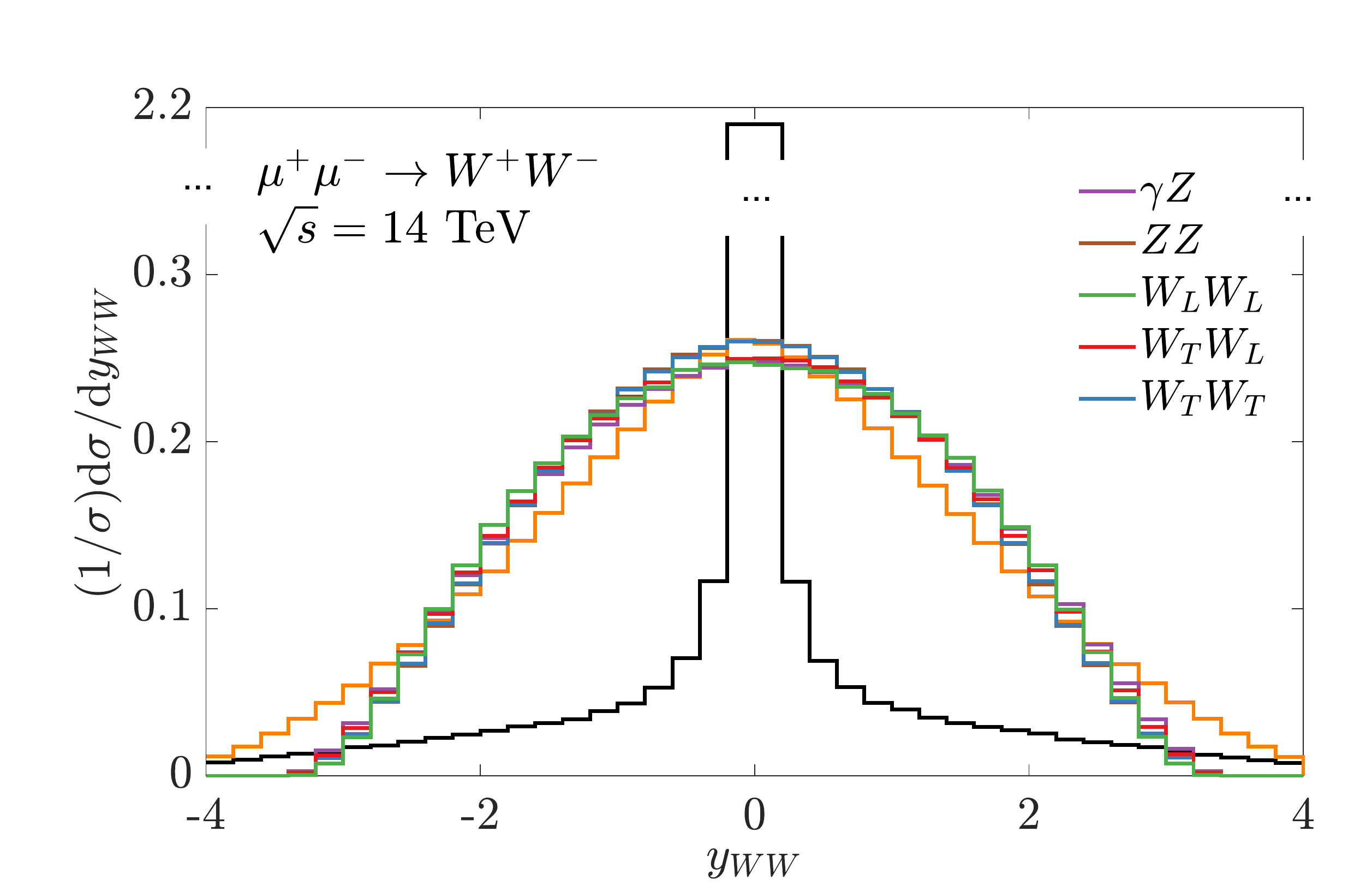}
\includegraphics[width=\columnwidth]{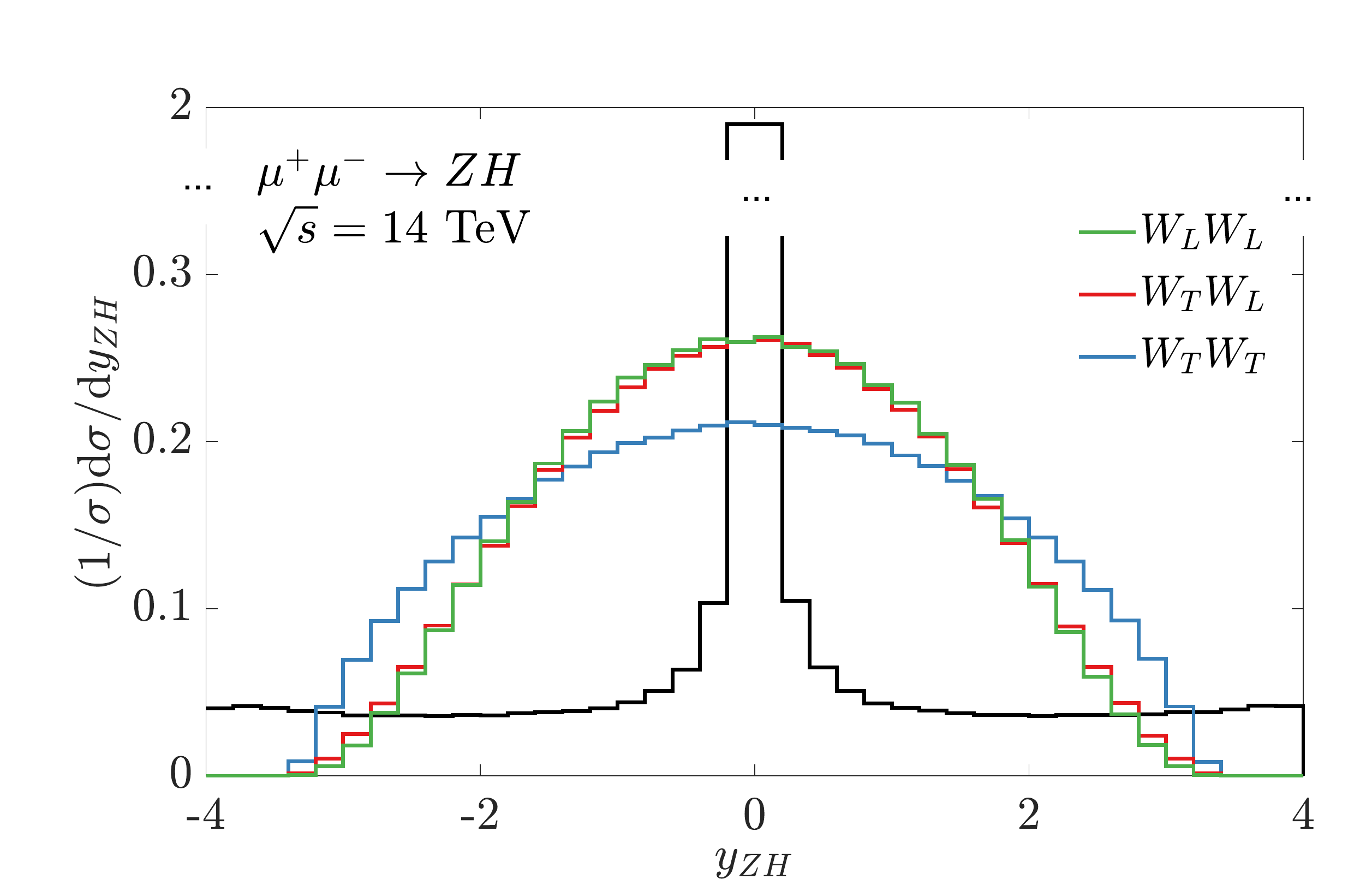}
\caption{Normalized di-boson rapidity distributions for  $W^+W^-$ (Left) and $ZH$ (Right) production at a muon collider at $\sqrt{s}=14$ TeV. Annihilation ($\mu\mu$, black) and diboson fusion (colored) processes are compared.}
\label{fig:rap}
\end{figure*}

\subsection{The standard candle cross sections at a $\mu^+\mu^-$ collider }
\label{sec:XSec}
In this section, we consider $W^+W^-$ and $ZH$ production at a multi-TeV $\mu^+\mu^-$ collider as standard candles to illuminate the effects of EW PDFs. This is done by comparing the cross sections for $\mu^+\mu^-$ annihilation and $VV$ scattering to the analogous final state. The impact on the other SM processes such as $t\bar{t}, HH$, and $t\bar{t}H$ production at high invariant mass can be found in Refs. \cite{Han:2020uid} and di-jet production at low invariant mass in Ref. \cite{Han:2021kes}. In Fig. \ref{fig:WWZH}, we show the semi-inclusive cross sections and the corresponding decomposition versus the collider center-of-mass energy $\sqrt{s}$ from 1 to 30 TeV. Tight cutoffs $|\cos\theta|<0.99$ and $\sqrt{\hat{s}}>500$ GeV,
where $\sqrt{\hat{s}}$ is the invariant mass of the diboson system, have been imposed for the $W/Z$ initiated processes in the center-of-mass frame to assure the validity of EW PDF formalism. 
Compared with the annihilation processes, VBF takes over at higher energies around $\sqrt{s}\approx2$ TeV for both $W^+W^-$ and $ZH$ production. At $\sqrt{s}=30$ TeV, the VBF cross sections are 3 orders of magnitude larger than  direct annihilation. 

When examining the decomposition, we see the $W^+W^-$ production is mainly contributed from the $\gamma\gamma$ and $W_TW_T$. This is a general feature for EW gauge boson productions, due to the self-coupling of the non-Abelian gauge group SU(2)$_{\rm L}$. The neutral current channels $\gamma\gamma,ZZ,\gamma Z$  dominate over the $W_TW_T$ one due to the relatively large PDFs, which is explicitly shown in Fig. \ref{fig:PDFs} at $Q=1$ TeV. $W_T$ gives a larger contribution than $W_L$ due to the large logarithm enhancement in the $W_T$ PDF, while $W_L$ does not evolve with scales. In comparison, we see  $ZH$ production mainly comes from the $W_TW_L$ component, which contributes over 90\% of the total cross section. It can be understood in terms of the Goldstone equivalence theorem \cite{Chen:2016wkt,Cuomo:2019siu}. The Higgs is largely mediated by the longitudinal gauge $W_L$, due to the large scalar self-coupling. Compared with the $W_TW_T$ component, the $W_LW_L$ one is about one order of magnitude larger, as a result of the large interaction to the longitudinal $Z$ boson.

As we know, the kinematic distribution of the direct annihilation behaves very differently from the fusion processes.  The key features are the invariant mass and rapidity of the final-state particle system. In an annihilation process, the final-state particle system carries the full collider energy, while the invariant mass of a fusion system starts from low energy around the threshold and drops very fast. The corresponding distributions are already explicitly shown in Ref.~\cite{Han:2020uid} for $t\bar{t}$ production and in Ref.~\cite{Han:2021kes} for di-jet production. The general features remain the same for $W^+W^-$ and $ZH$ production. The normalized rapidity distributions of the di-boson systems of $W^+W^-$ and $ZH$ production from various components are shown in Fig.~\ref{fig:rap} for a possible $\mu^+\mu^-$ collider at $\sqrt{s}=14$ TeV. As expected, the annihilation ($\mu\mu$, black histogram in the plot) sharply peaks around $y_{ij}\sim0$, with long tails at larger rapidities due to initial state radiation (ISR). In contrast, the fusion processes show a much larger spread in rapidity. Comparing different fusions to $W^+W^-$ production, we see the $\gamma\gamma$-initiated processes (orange histogram) show larger tails while the EW gauge boson-initiated ones exhibit smaller rapidity values. For the $ZH$ production (Fig.~\ref{fig:rap}, right), the transversal $W_TW_T$ spreads out more than the longitudinal components. 

\begin{figure*}[!t]
\includegraphics[width=\columnwidth]{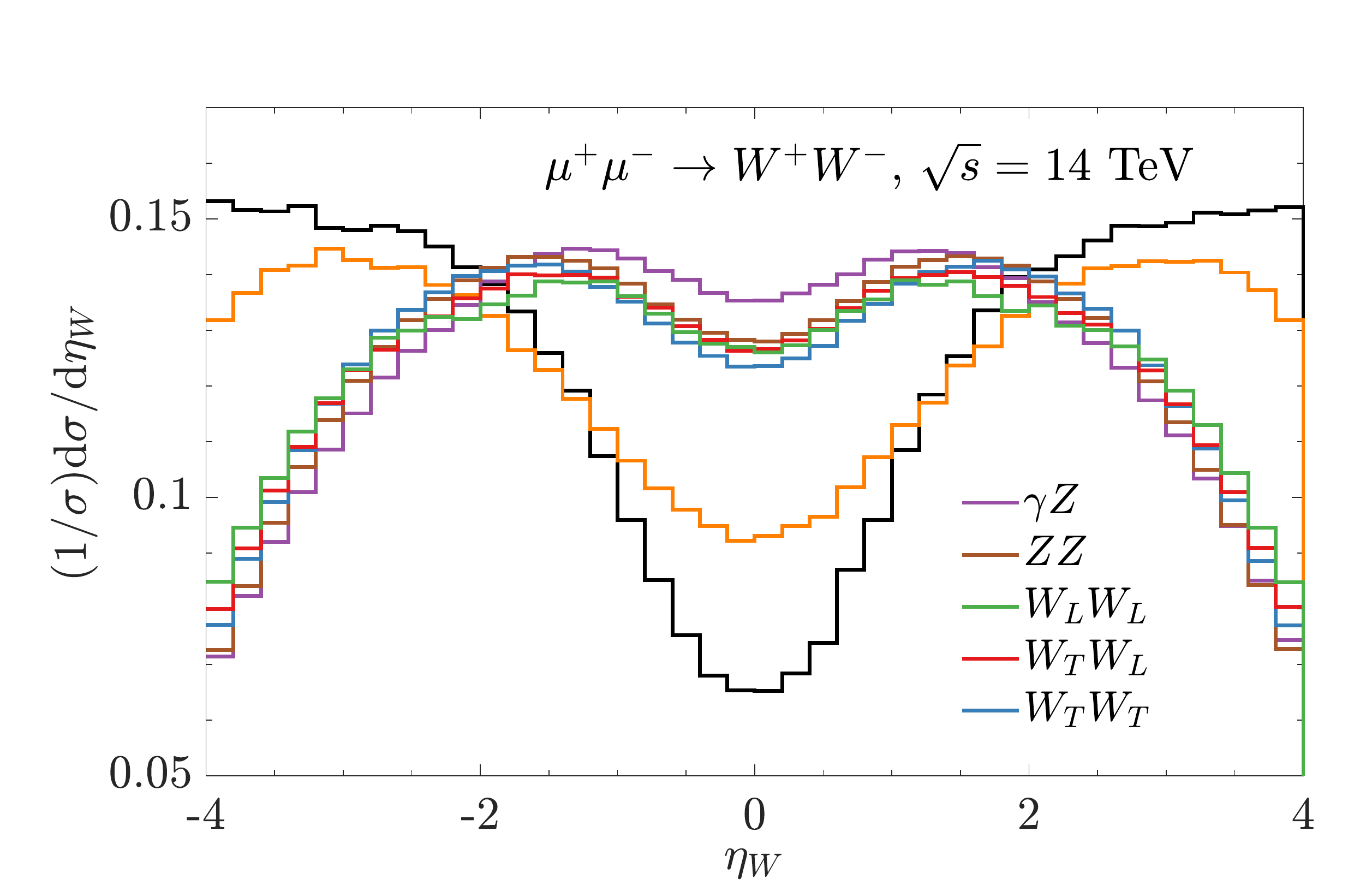}
\includegraphics[width=\columnwidth]{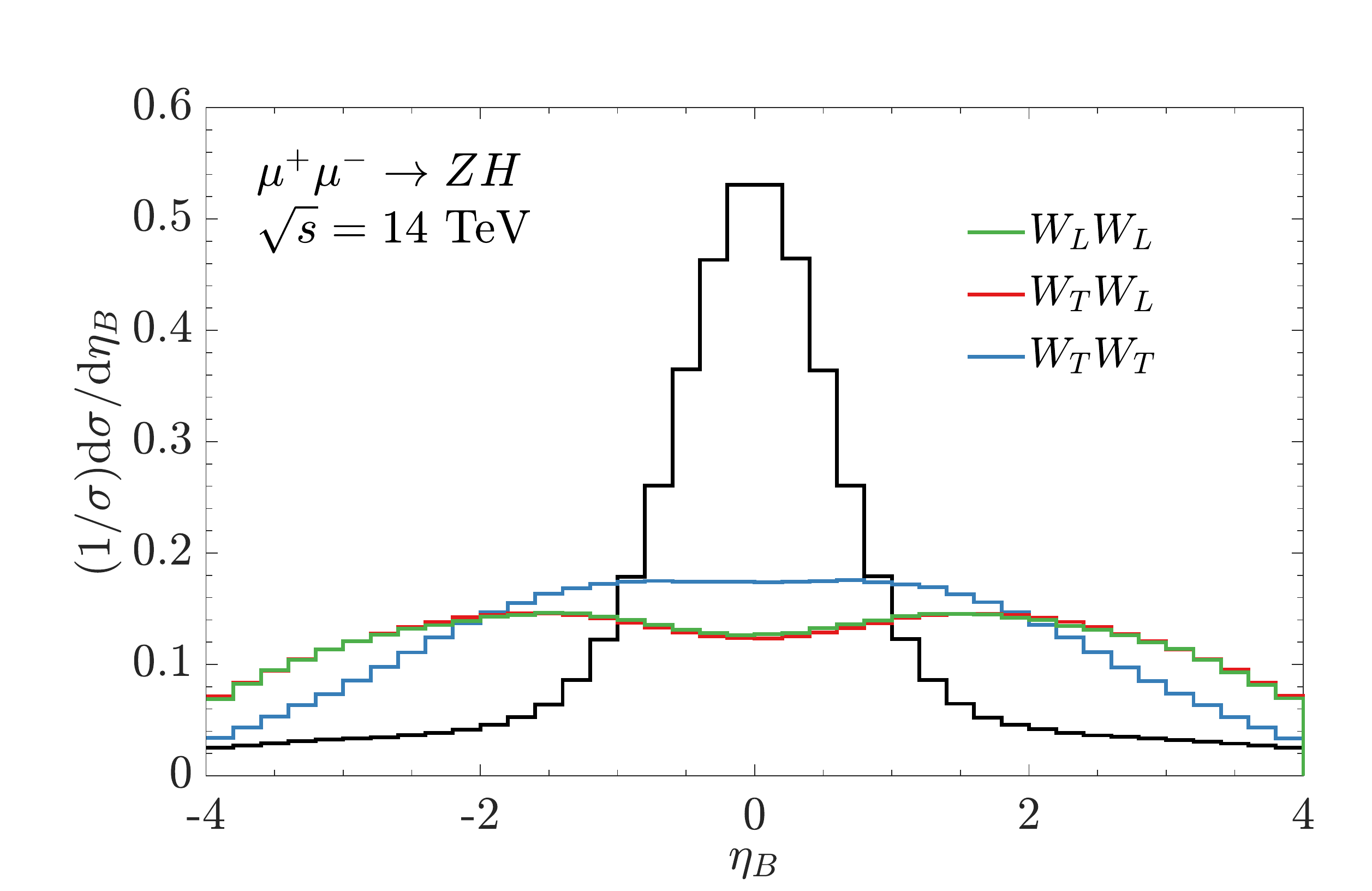}
\caption{In  $W^+W^-$ (Left) and $ZH$ (Right) production for a muon collider at $\sqrt{s}=14$ TeV, the normalized pseudo-rapidity distributions of each boson ($\eta_W$ on the left, and $\eta_B$ on the right). Annihilation ($\mu\mu$, black histograms) and diboson fusion (colored histograms) processes are compared.}
\label{fig:eta}
\end{figure*}

Finally, the normalized pseudo-rapidity ($\eta$) distributions of the final-state bosons in the $W^+W^-$ and $ZH$ production are shown in Fig. \ref{fig:eta}. We note that both bosons are included in the $\eta$ distributions. For the $W^+W^-$ production, we see a large rate of the $W$ bosons are produced in the forward region, especially for the annihilation and $\gamma\gamma$ fusion processes. In direct annihilation, the $W^+W^-$ can be produced through $t$-channel neutrino exchange, which gives the collinear $1/(1-\cos^2\theta)$ behavior of the final $W$ bosons\footnote{The singularity is regulated by the $W$ boson mass.}. A similar $t$-channel exists for the $\gamma\gamma$ fusion process through the $W$ boson exchange. This collinear feature is largely regulated in the EW gauge boson-initiated processes by the mass effect. In the $ZH$ production, we see the $s$-channel $(1-\cos^2\theta)$ behavior for the annihilation process. In the fusion ones, the $W_T$-initiated process gives much more central final states while the $W_L$-initiated one is relatively forward.

\subsection{Summary}
\label{sec:concEWPDF}

In high-energy collisions at future colliders, splitting phenomena involving weak bosons begin to dominate because of the logarithmic enhancement in $t,u$-channel exchanges of weak bosons. The EW PDF formalism should be adopted, which factorizes the scattering processes into the hard partonic cross sections and the collinear PDFs. The PDFs evolve according to the SM DGLAP equations, which effectively resums  potentially large logarithms. The EW PDFs for high-energy leptonic beams are completed in the recent works Refs.~\cite{Han:2020uid,Han:2021kes}. In this report, some follow-up and supplementary results are presented. 

In Section~\ref{sec:PDFs}, we present the EW PDFs at the EW matching scale $Q=M_Z$ and one higher scale $Q=1$ TeV, and explore the QCD threshold uncertainty by varying the matching scale within $\muQCD=0.5\sim0.7$ GeV~\cite{Drees:1994eu}. The impacts on the quark and gluon PDFs of a muon beam at $Q=M_Z$ are less than 10\%, while no impact on the lepton and photon PDFs are observed. This uncertainty becomes smaller when energy evolves to higher scales. Above the EW scale at $Q=1$ TeV, we explicitly demonstrate the EW corrections to light-flavor PDFs are about $10\%\sim20\%$. The higher-order correction to the Effective $W/Z$ Approximation is positive (negative) at small (large) momentum fraction $x$. The corrections to the neutral current PDFs are mainly due to the new contribution from the $Z$ and $\gamma Z$ components.

In Section~\ref{sec:XSec}, we take $W^+W^-$ and $ZH$ production at a high-energy muon collider as standard candles to illuminate the effects of EW PDFs. We see the VBFs take over the annihilation processes around the center-of-mass collision energy $\sqrt{s}\approx 2$ TeV. The EW PDF formalism gives the advantage to examine the decomposition from various sub-channels. We find that the transverse gauge bosons initiated processes dominate the $W^+W^-$ production, while the $ZH$ mainly comes from $W_LW_T$ due to the Goldstone equivalence theorem \cite{Chen:2016wkt,Cuomo:2019siu}. Some important kinematic distributions are explicitly presented. 

In this report, we mainly explored the EW PDFs for a high-energy muon beam and the corresponding effects at muon colliders. It can be easily extended to an electron collider, which is examined in the recent work Ref.~\cite{Han:2021kes}.
The EW PDF formalism is equally applicable to the proton beam, with quarks and gluons as the radiation source, which is expected to make an impact on the future 100 TeV proton-proton colliders~\cite{Benedikt:2018csr,Tang:2015qga}.


\section[EW parton showers]{\large EW parton showers}
\label{sec:ewVincia}

Monte Carlo event generators have become an indispensable component of the particle physics toolkit. In most cases, they are the only valid option for the translation of theoretical models to predictions for data measured at collider experiments. At the LHC Run 2, analyses are already frequently limited by theoretical uncertainties, originating from either incomplete modeling of the underlying physics or from insufficient statistics due to the inefficiency of current codes.  With the upcoming luminosity upgrade of the LHC and future colliders in mind, the improvement of the current state of the art in terms of both accuracy and efficiency is thus critically important.

Fortunately, recent years have seen a number of developments on both fronts. In the area of computational efficiency, major issues are being tackled. An example is the improvement in the treatment of negative event weights, which often occur in the context of matching fixed-order calculations to parton showers and lead to lower statistical power of an event sample. They may be tackled by improvement of the matching procedure \cite{Frederix:2020trv,Danziger:2021xvr}, re-sampling techniques \cite{Andersen:2020sjs,Nachman:2020fff}, or they may be used as training data for generative machine learning models which can then be used to sample unweighted events \cite{Verheyen:2020bjw,Butter:2019eyo}. Those same generative models offer an exciting new avenue for the sampling of multi-leg, fixed-order calculations \cite{Gao:2020vdv,Gao:2020zvv,Bothmann:2020ywa,Verheyen:2020bjw}, which are currently a major computational bottleneck for analyses that require matrix element merging up to high multiplicities. Complementary perspectives include avoiding negative event weights altogether~\cite{Nason:2004rx}.

\begin{figure}[!t]
    \centering
    \includegraphics[width=\columnwidth]{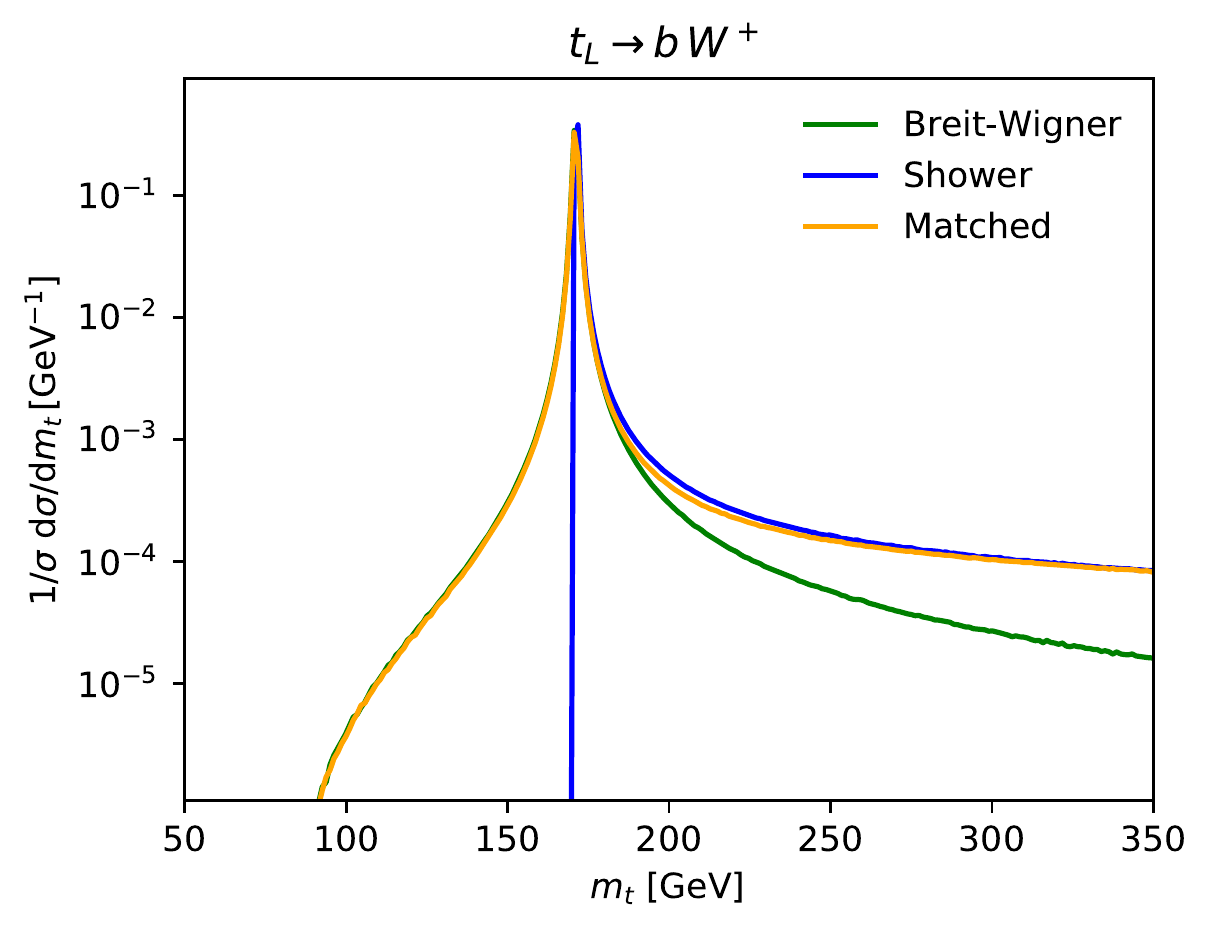}
    \caption{Mass spectra of the decay of a 1 TeV left-handed top quark. The matched spectrum (orange) is the numerical combination of the Breit-Wigner distribution (green) for small virtualities, and the shower distribution (blue) for large virtualities.}
    \label{fig:resonance_spectrum}
\end{figure}

In the area of theoretical accuracy, some of the remaining challenges are of a technical nature. For instance, NNLO calculations are becoming increasingly available, but their matching to parton showers and the merging of higher-multiplicity LO or NLO samples is a very complex task, see for example Refs.~\cite{Monni:2019whf,Lombardi:2020wju}. Even if it is solved in a general sense, the resulting event generators may prove to be computationally prohibitive. On the other hand, the challenges in the field of parton shower development are of a more theoretical nature, and much progress has been made there in recent years. The requirements for formal NLL accuracy were set out in Refs.~\cite{Dasgupta:2020fwr,Nagy:2020rmk,Forshaw:2020wrq}, and progress on the inclusion of the higher-order branching kernels is being made~\cite{Li:2016yez,Hoche:2017hno}. Much work has also been done on the inclusion of sub-leading color effects~\cite{Hamilton:2020rcu,Nagy:2015hwa,Platzer:2018pmd,Forshaw:2019ver,Isaacson:2018zdi}, although they are not yet part of standard event generator codes. 

\begin{figure*}[!t]
    \centering
    \includegraphics[width=\columnwidth]{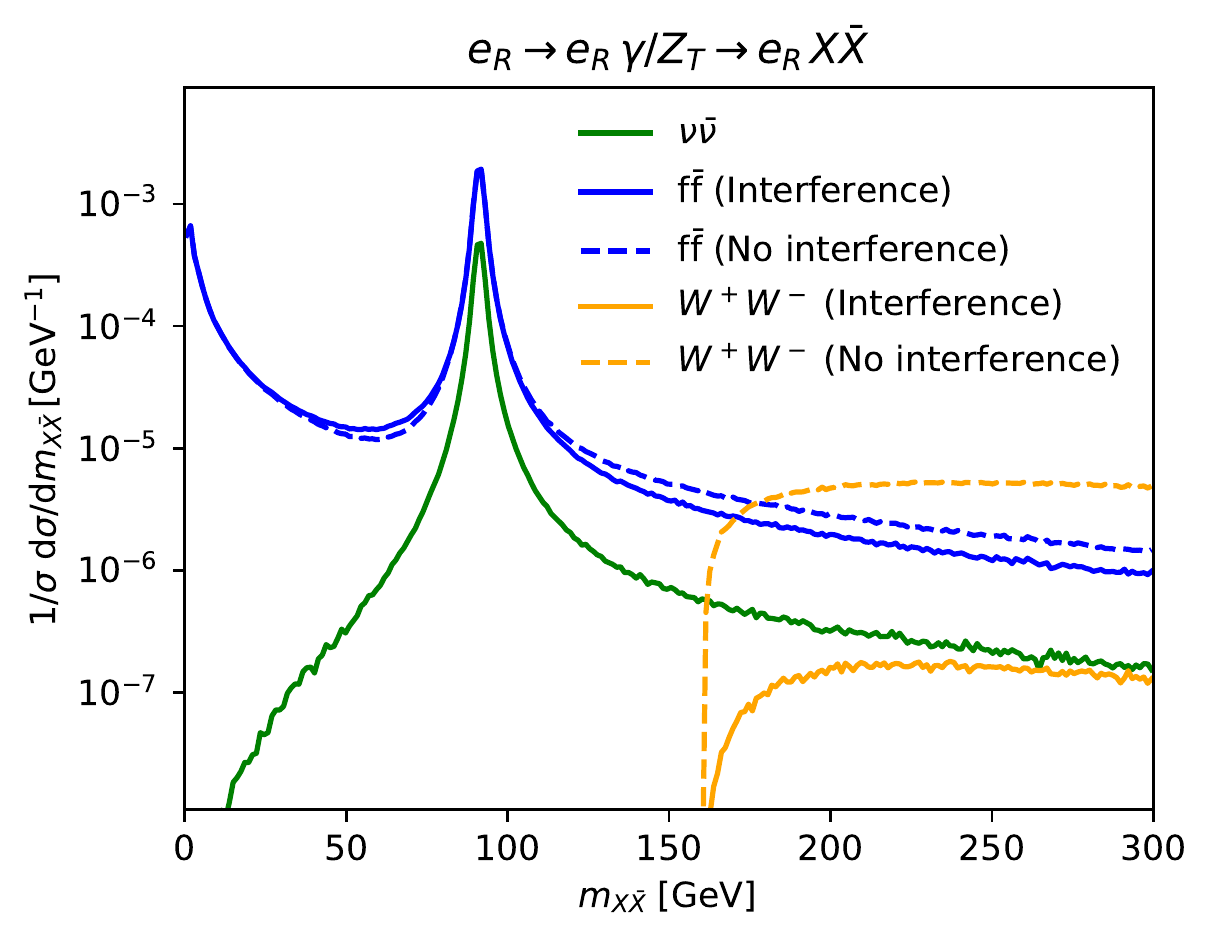}
    \includegraphics[width=\columnwidth]{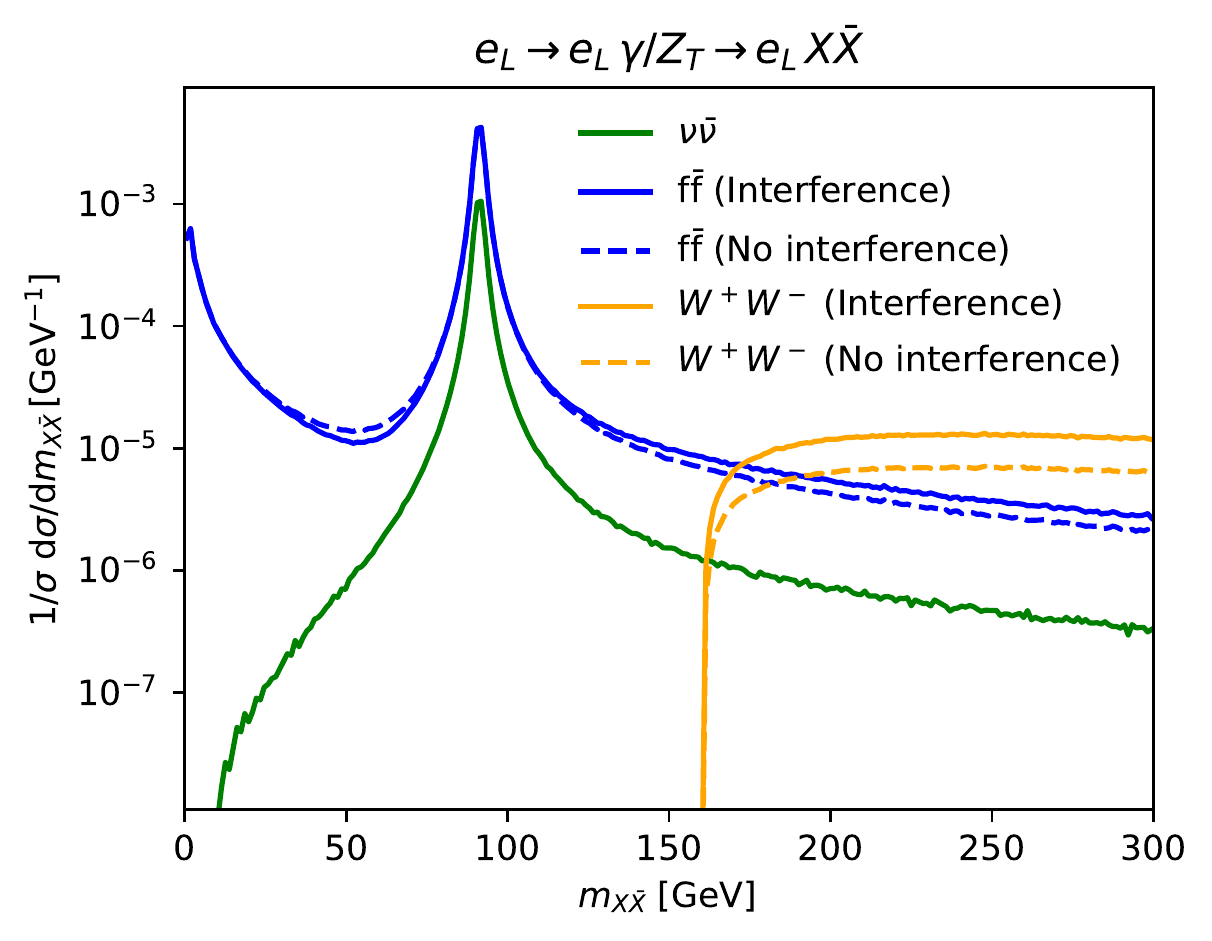}
    \caption{Invariant mass spectra resulting from including neutral boson interference in the showering of a 10 TeV left-handed electron (Left) and a right-handed electron (Right).}
    \label{fig:interference_spectra}
\end{figure*}

Finally, progress towards including EW effects in parton showers is currently being made~\cite{Christiansen:2014kba,Krauss:2014yaa,Chen:2016wkt,Kleiss:2020rcg}. In the EW sector, the large logarithms resummed by the parton shower are regulated by the EW scale, but at sufficiently large energies still lead to sizable contributions. Negative virtual EW corrections have been computed and resummed for many processes and observables (see for instance Ref.~\cite{Bothmann:2020sxm} and  references therein), and their incorporation into a parton shower offers a process-independent way of including such corrections systematically.

Similarly, at large enough energies, EW vector bosons, top quarks, and even Higgs bosons start to appear as parts of jets and contribute to fragmentation functions~\cite{Bauer:2016kkv,Bauer:2018xag,Bauer:2020jay} and parton density functions~\cite{Dawson:1984gx,Kane:1984bb,Bauer:2017isx,Bauer:2018arx,Han:2020uid,Han:2021kes}.
It is the connection to these last two points that make EW parton showers particularly relevant to VBS at higher energies.
As described in Section~\ref{sec:muon_ewPDF}, it becomes meaningful to discuss the $W/Z$ content of high-energy leptons and protons. This means that the VBS process can be modeled, with some uncertainty, as the scattering of two initial-state EW bosons that propagate along the beam axis and carry no transverse momentum.
In real life, initial-state $W/Z$ bosons carry some $p_T$ since they are generated perturbatively, \ie, through $f_i\to f_j V$ splitting where $f_i$ and $f_j$ are leptons or quarks.
This absence of $p_T$ is a source of uncertainty but is precisely described by backwards evolution / fragmentation in EW parton showers~\cite{Ruiz:2021tdt}. (This is analogous to matching $gg\to t\overline{t}$ to $qg\to t\overline{t}q$ and  $qq\to t\overline{t}qq'$ via backwards evolution in QCD.) Therefore, using EW boson PDFs to describe VBS at high energies requires an EW parton shower to correctly describe particle kinematics. Eventially, all of these effects must be included in the general-purpose event generators.  A particular case study is the implementation of EW corrections into the Vincia parton shower~\cite{Brooks:2020upa}, which highlights some of the unique features that appear in the EW sector at high energies.

\begin{figure*}[!t]
    \centering
    \includegraphics[width=\textwidth]{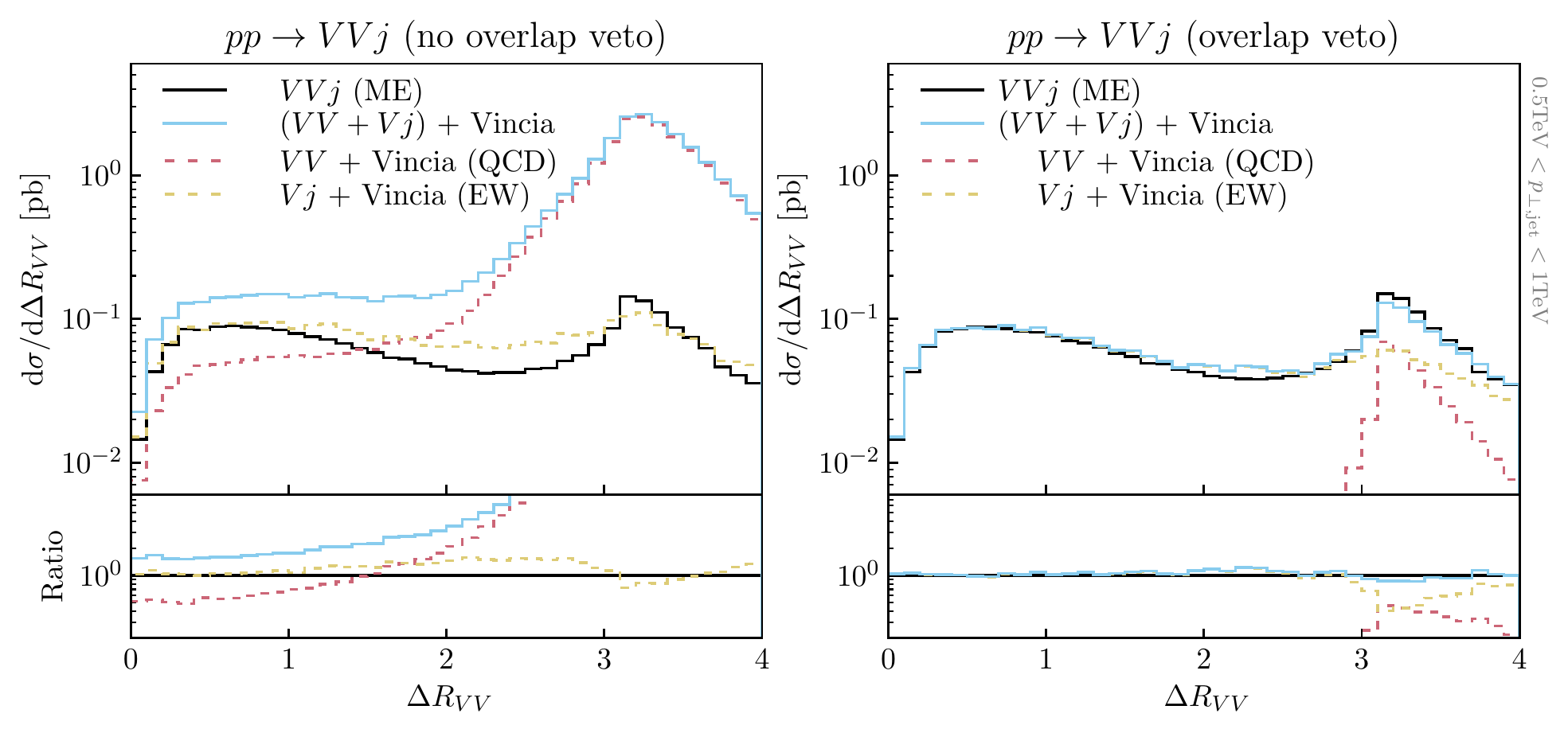}
    \caption{Upper: As a function of angular separation between the vector bosons $\Delta R_{VV}$, the LO matrix element (ME) for $pp \rightarrow VVj$ (solid black) and the \textsc{Vincia} prediction (solid blue), which consists of the sum of $pp \rightarrow VV$ + Vincia QCD (dashed red) and $pp \rightarrow Vj$ + Vincia EW (dashed yellow), evaluated at 14 TeV when the overlap veto is disabled (left) or enabled (right). Lower: Ratio with respect to LO ME.
    All simulations require jet $p_T$ in the range $p_T\in(0.5~{\rm TeV},1~{\rm TeV})$.
    }
    \label{fig:EWoverlapVeto}
\end{figure*}

\subsection{Vincia: Electroweak showers in simulation}\label{sec:secName}

Parton showers are a critical component of Monte Carlo event generators, serving as the link between high-scale, fixed-order calculations and low-scale, non-perturbative physics. They offer a process-independent and fully differentiable resummation framework, incorporating the large logarithms associated with soft and collinear branchings of the partons produced in the hard scattering~\cite{Contopanagos:1996nh}. The Vincia shower is one of many showers currently available, which all make slightly different choices in their modeling of the same physics. Vincia is based on the antenna subtraction formalism \cite{Kosower:1997zr,Kosower:2003bh}, of which implementation details may be found in Ref.~\cite{Brooks:2020upa}. One of Vincia's unique features is that it allows for the evolution of states of definite helicity~\cite{Larkoski:2013yi,Fischer:2017htu}. Due to the chiral nature of the EW sector, this type of propagation of spin information is crucial for the correct modeling of EW corrections.  In the following, a qualitative outline is given on the modeling a number of features that are unique to the EW sector. Technicalities of the implementation are detailed in Refs.~\cite{Kleiss:2020rcg}.

\subsection{Branching kernels}
Branching kernels are one of the core components of a parton shower. They contain the dynamics of the soft and collinear branchings it generates. The calculation of EW branching kernels, which are antenna functions in case of Vincia, mostly proceeds analogously to their QCD counterpart. The EW sector contains a rich physical landscape, leading to a vast number of collinear $1 \rightarrow 2$ branchings involving both spin-1 and spin-0 states. These include triple vector boson branchings, like $W^{\pm*} \rightarrow W^{\pm} Z$; branchings that involve a Higgs, like $Z^* \rightarrow Z h$; and resonance-type branchings, like $t \rightarrow b W^+$ or  $Z^* \rightarrow t \bar{t}$. Particular care needs to be taken in the calculation of branching rates that involve longitudinal polarizations, where the scalar component can lead to spurious, unitarity-violating terms. These may be removed by performing the calculation in a suitable gauge and using Goldstone-gauge boson equivalence~\cite{Chen:2016wkt,Chen:2017ekt,Chen:2019dkx}, or by using the spinor-helicity formalism and isolating the terms directly \cite{Kleiss:2020rcg}. The results, computed for all possible spin states, reveal many features that are a direct consequence of EW symmetry breaking. For example, one observes the usual mass-suppression in the branching $Q_L \rightarrow Q_R Z_T$ in correspondence with the QCD branching $Q_L \rightarrow Q_R g$, but additional terms  proportional to $m_z^2$ also appear. On the other hand, the branching $Q_L \rightarrow Q_R Z_L$ is not mass-suppressed, but it instead behaves like a scalar branching in correspondence with Goldstone boson equivalence.

\subsection{Resonance matching}
The EW shower includes branchings like $Z \rightarrow f \bar{f}$, which would normally be associated with a resonance decay. When these branchings occur at virtualities $Q^2 \gg Q^2_{EW}$, the resonance mass is a small correction and the EW shower offers an accurate description of the underlying physics. However, closer to the EW scale, it is well-known that the branching spectrum should follow a mass-dependent Breit-Wigner distribution.  Interestingly, the Breit-Wigner distribution involves a different kind of resummation through the Dyson summation of the width. It is theoretically unclear how these two resummations should be matched, so the Vincia implementation makes use of a numerical matching procedure that complements \textsc{Pythia}~\cite{Sjostrand:2014zea}'s usual treatment of resonances. That is, any resonances produced by the EW shower are assigned a mass according to a Breit-Wigner distribution. The resonance-type branchings of EW shower are then suppressed close to the EW scale, and the resonance is decayed when the shower passes the off-shellness scale of the resonance. An example of a top quark resonance decay spectrum is shown in Fig.~ \ref{fig:resonance_spectrum}.

\subsection{Neutral boson interference}
The EW sector contains multiple neutral bosons that may interfere with each other. For instance, the EW shower paradigm models the process $e_{L/R} \rightarrow e_{L/R} \, Z_T/\gamma \rightarrow e_{L/R} \, W^+ W^-$ as an incoherent sum of two separate contributions corresponding with intermediate $Z_T$ or $\gamma$. However, in this process the interference between the neutral boson can lead to $\mathcal{O}(1)$ effects. To see this, one may consider that the $Z_T/\gamma \rightarrow W^+ W^-$ coupling is purely of SU(2) nature. An $e_R$ has small SU(2) content, so the coherent $W^+ W^-$-contribution should be suppressed,  while the $e_L$ has large SU(2) content, leading to an enhancement.  For reasons of computational efficiency, the Vincia implementation incorporates this effect at leading order only by applying an event weight that corrects the incoherent sum of branching kernels to a coherent one.  The result of the application of this event weight is shown in Fig.~\ref{fig:interference_spectra}. Note that the neutrino rate is not modified because it does not couple to the intermediate photon. The $W^+ W^-$ rates are affected due to the SU(2) nature of the $\gamma/Z \, W^+ W^-$ coupling.

\subsection{Double counting of hard processes}
When QCD and EW branchings are both enabled, a technical double counting problem may appear. To see this, consider for example the process $pp \rightarrow V V j$. For any point in phase space, this final state may be reached either by starting from $pp \rightarrow VV$ and adding an initial-state QCD emission. The same configuration can also be obtain by starting from $pp \rightarrow Vj$ and adding an EW emission. Including both types of emissions leads to a double counting issue that needs to be addressed explicitly.

To that end, \textsc{Vincia} generalizes a procedure used in Ref.~\cite{Christiansen:2014kba}, wherein a veto based on a $k_T$-like measure is applied to ensure that every phase space point is populated by the path that describes the physics the most accurately. A validation of this procedure is shown in Fig.~\ref{fig:EWoverlapVeto}, where \textsc{Vincia} is compared to the LO matrix element of $pp \rightarrow VVj$ as a function of the angular separation between the vector bosons $\Delta R_{VV}$ and with the requirement $0.5 \text{ TeV} < p_{\perp,\text{jet}} < 1 \text{ TeV}$. The \textsc{Vincia} prediction consists of the sum of $pp \rightarrow VV$ with a QCD emission and $pp \rightarrow Vj$ with an EW emission. For small angular separations, the vector bosons are collinear and the EW shower should perform best, while at large angular separation the vector bosons are back-to-back and the QCD shower should be preferred. In absence of the overlap veto, the \textsc{Vincia} prediction clearly overshoots the direct matrix element significantly, while closely matching it when it is enabled.

\subsection{Summary}
The incorporation of EW corrections in parton showers is one of the many promising recent developments in the development and improvement of Monte Carlo event generators. The EW sector offers a rich physics landscape that leads to a large number of inspiring challenges.
In this section,  features of EW corrections in parton showers, and particularly in Vincia, have been highlighted. While the most significant contributions are included in this implementation, many outstanding problems such as soft coherence, spin-interference effects and Bloch-Nordsieck violation, have yet to be solved. \textsc{Vincia}'s EW shower has been part of \textsc{Pythia} since its 8.304 release.


\section[SMEFT with VBS at $\mu^+\mu^-$ colliders]{\large SMEFT with VBS at $\mu^+\mu^-$ colliders}\label{sec:smeft_muonCo}

\begin{figure*}[t!]
\centering
\includegraphics[width=\columnwidth]{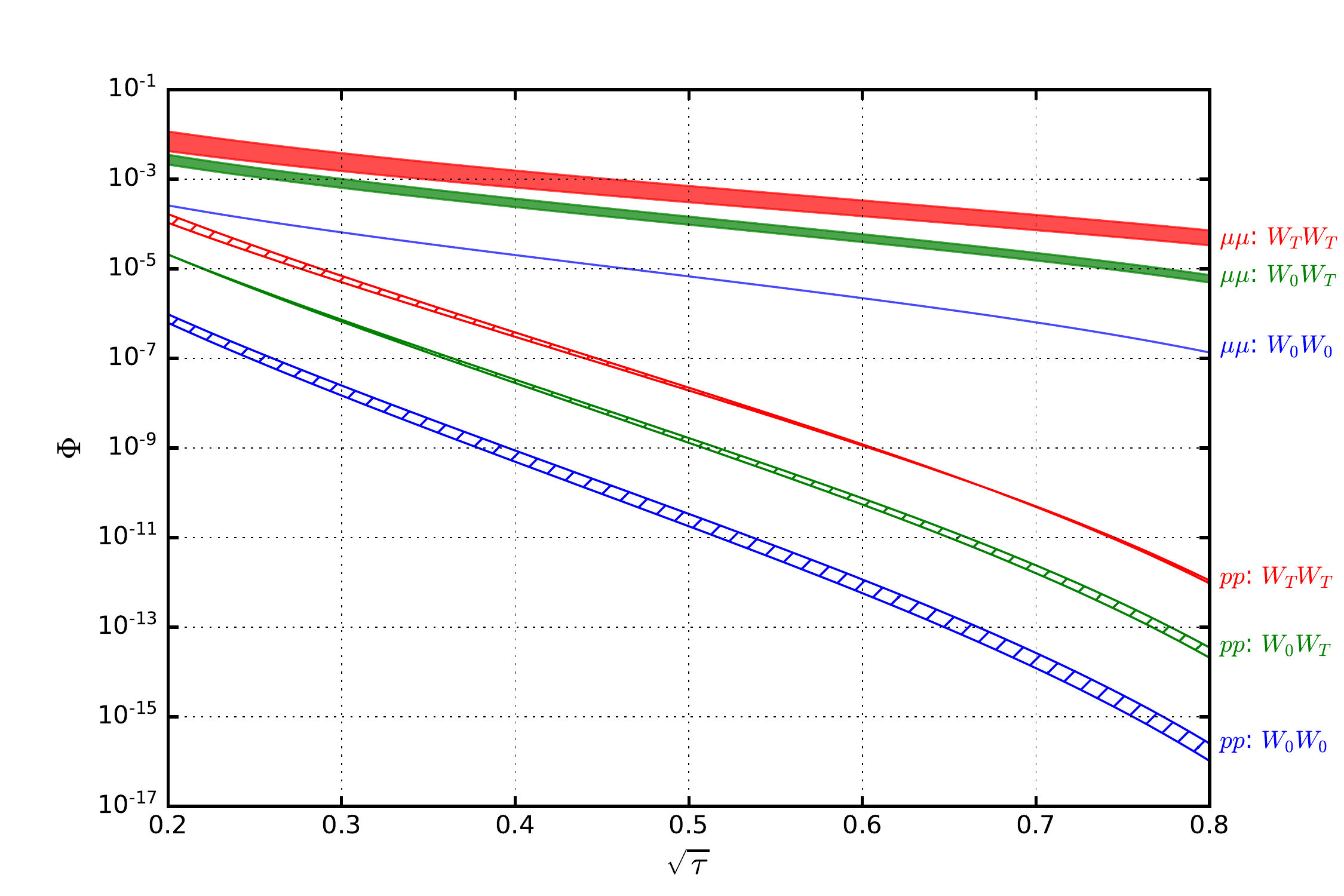}
\includegraphics[width=\columnwidth]{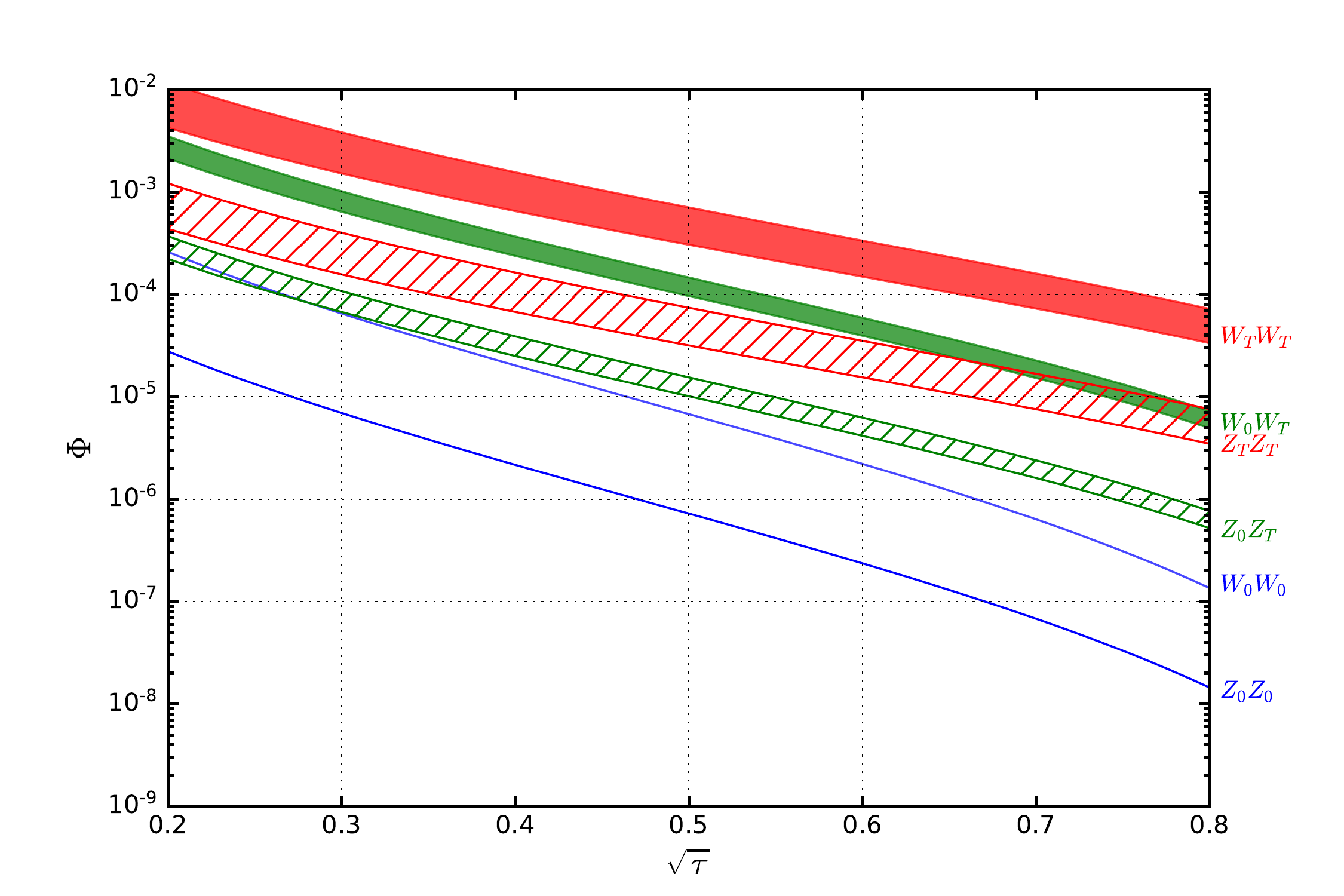}
\caption{\label{fig:EWlum} 
Parton luminosities $\Phi$ as  a  function  of  the scattering  energy fraction $\sqrt{\tau}=M_{VV^\prime}/\sqrt{s}$ (see text).
Comparison between the $WW$ parton luminosities 
at muon and proton colliders (Left), and the difference between $Z$ and $W$ bosons luminosities at a muon collider (Right). The parton luminosities have been decomposed among the various helicity combinations. Bands are given by variations of the factorization scales. Adapted from Ref.~\cite{Costantini:2020stv}.}
\end{figure*}

 As discussed above, the community has started to discuss the possibility of a multi-TeV muon collider~\cite{Delahaye:2019omf,Costantini:2020stv,Lu:2020dkx,Ali:2021xlw,Franceschini:2021aqd}. As their mass is about $200$ times heavier than the electron mass, they emit much less synchroton  radiation, suggesting that muons could be ideal candidates to explore the high-energy frontier. At the same time, being fundamental particles, they do not entail the challenges of proton colliders in terms of QCD background, and could potentially allow for a rather clean environment. The main challenges however all stem from the fact that muons are not stable particles, and their decay complicates both beam production and detector designs~\cite{Palmer:1996gs,Ankenbrandt:1999cta,Delahaye:2019omf}. Despite the serious R\&D challenges, there has been a recent increase of interest with many studies assessing the physics potential of these machines~\cite{Costantini:2020stv,Lu:2020dkx,Ali:2021xlw,Franceschini:2021aqd,Capdevilla:2021fmj,Han:2021udl,Liu:2021jyc,Capdevilla:2021rwo,Buttazzo:2020uzc,Yin:2020afe,Buttazzo:2020eyl,Bandyopadhyay:2020otm,Han:2020uak,Han:2020pif,Han:2020uid,Capdevilla:2020qel,Chiesa:2020awd,Ruiz:2021tdt,Capdevilla:2021kcf}.

In particular it has been observed that at sufficiently high energies, one expects VBS to become the dominant production mode with respect to $s$-channel production. This seems to hold not only for SM processes, but also for not too heavy BSM scenarios~\cite{Costantini:2020stv}. Leaving aside the discovery potential, a muon collider seems to be also an ideal machine to probe new physics indirectly, taking advantage of the high energy scatterings of EW bosons to scout tails of distributions and exploit enhanced energy behavior. In this perspective, the SMEFT offers a flexible, semi- model- independent framework to parameterize  new physics and it is worth exploring the prospects of indirectly finding BSM at muon colliders through VBS.
Such a program would build on and complement the on-going program at the LHC (for additional details on the LHC program, see also Sections \ref{sec:lhc_results_smeft}, \ref{sec:hllhc_smeft}, and \ref{sec:nuVBS_dim5}).

\subsection{EW bosons luminosities}\label{sec:smeft_muonCo_EWbosons} 
\label{sec:EWlumimumu}

Since it can be shown that above $\sim 5$ TeV a muon collider becomes effectively an EW boson collider~\cite{Costantini:2020stv,Ruiz:2021tdt}, it is worth having a comparison of the potential for EW VBS between muon and proton colliders. In order to make this comparison, we employ the language of parton luminosity and make use of the Effective $W/Z$ Approximation~\cite{Dawson:1984gx,Kane:1984bb}, or sometimes also called the Effective Vector Boson Approximation. This allows us to describe the emission of EW bosons on the same footing as QCD partons in protons, i.e., describe the $W/Z$ content of a lepton or proton beam. Under this approximation, splitting functions can be used to describe the likelihood of collinear $f\to V f'$ splitting, where $f$ is any lepton or quark and $V$ is an EW gauge boson.

In particular, for $W/Z$ bosons  of helicity $\lambda$ and longitudinal energy fraction $\xi=E_V/E_f$, the leading order PDFs for a transversely polarized 
$(f_{V_T/f})$ and longitudinally polarized  $(f_{V_0/f})$ $V$ are given by the expressions
\begin{align}
f_{V_T/f}&(\xi, \mu_f) = \frac{C}{16\pi^2} 
\nonumber\\ 
& \quad \times \frac{(g_V^f \mp g_A^f)^2 +(g_V^f \pm g_A^f)^2(1-\xi)^2}{\xi}\log{\left(\frac{\mu_f^2}{M_V^2}\right)}, 
\label{eq:ewa_VT}
\\
f_{V_0/f}&(\xi, \mu_f) = \frac{C}{4\pi^2} (g_V^{f~2} + g_A^{f~2})\left( \frac{1-\xi}{\xi}\right)\,.
\label{eq:ewa_V0}
\end{align}
Here, $\mu_f$ is the collinear factorization scale, $M_V$ is the mass of $V=W/Z$, and $C,~g_V^f,$ and $g_A^f$ represent the appropriate weak gauge couplings of $f$. Explicitly, these are given by
\begin{align}
\text{for}~V=W      : \quad & C=\frac{g^2}{8}, \quad           g_V^f=-g_A^f=1 \, ,
\\
\text{for}~V=Z      : \quad & C=\frac{g^2}{\cos^2{\theta_W}}, \quad g_V^a=\frac{1}{2}\left(T^3_L\right)^f- Q^f\sin^2{\theta_W},
\nonumber\\  & g_A^f = -\frac{1}{2} \left(T^3_L\right)^f ,  
\end{align}
where $Q^f$ is the electric charge of $f$, $(T_L^3)^f$ is its weak isospin charge, $g\approx0.65$ is the weak coupling constant, and $\theta_W$ is the usual weak mixing angle.
Note that the PDFs here are not resummed, unlike those discussed in Section~\ref{sec:muon_ewPDF}.

Given these PDFs, we can define the parton luminosity $(\Phi)$ for helicity-polarized $W^+W^-$ pairs carrying polarizations $(\lambda_1,\lambda_2)$, and a squared invariant mass of $M_{WW}^2=(p_{W^+}+p_{W^-})^2 \equiv \tau s$ in a muon collider of energy $\sqrt{s}$ as
\begin{equation}
 \Phi_{W^+_{\lambda_1}  W^-_{\lambda_2} }(\tau, \mu_f) = \int_{\tau}^1 \frac{d\xi}{\xi} ~ f_{W_{\lambda_1} /\mu}\left(\xi, \mu_f\right) \, f_{W_{\lambda_2} /\mu}\left(\frac{\tau}{\xi}, \mu_f\right)  \, .
\end{equation}
For a proton collider, the analogous expression is given by additional convolutions of $W/Z$ PDFs in quarks with quark PDFs in protons. Analytically, this given by
\begin{align}
&\Phi_{V_\lambda V'_{\lambda'}}(\tau, \mu_f) = \frac{1}{1+\delta_{V_\lambda V'_{\lambda'}}} \int_\tau^1 \frac{d\xi}{\xi}\int_{\tau/\xi}^1 \frac{dz_1}{z_1}\int_{\tau/(\xi z_1)}^1 \frac{dz_2}{z_2} \sum_{q, q'} 
\\
&f_{V_{\lambda}/q}(z_2)f_{V'_{\lambda'}/q'}(z_1)
\left[
f_{q/p}(\xi)f_{q'/p}\left(\frac{\tau}{\xi z_1 z_2}\right) 
+ 
f_{q/p}\left(\frac{\tau}{\xi z_1 z_2}\right)f_{q'/p}(\xi)\right] \, .
\nonumber
\end{align}
Here, $f_{q/p}$ are the usual PDFs for quarks in a proton.

As it can be seen in Fig.~\ref{fig:EWlum}, where we plot the luminosities as function of the hard scattering energy fraction $\tau=M_{WW}^2/s$, at a muon collider we would benefit from much higher parton luminosities. This is particularly striking at high energy, where the proton luminosities are significantly more suppressed. On the right of the figure, we also show a comparison between $W$ and $Z$ luminosities at a muon collider, observing an expected order of magnitude difference in favor of the $WW$ components.

\begin{figure}[t!]
\centering
\includegraphics[width=\columnwidth]{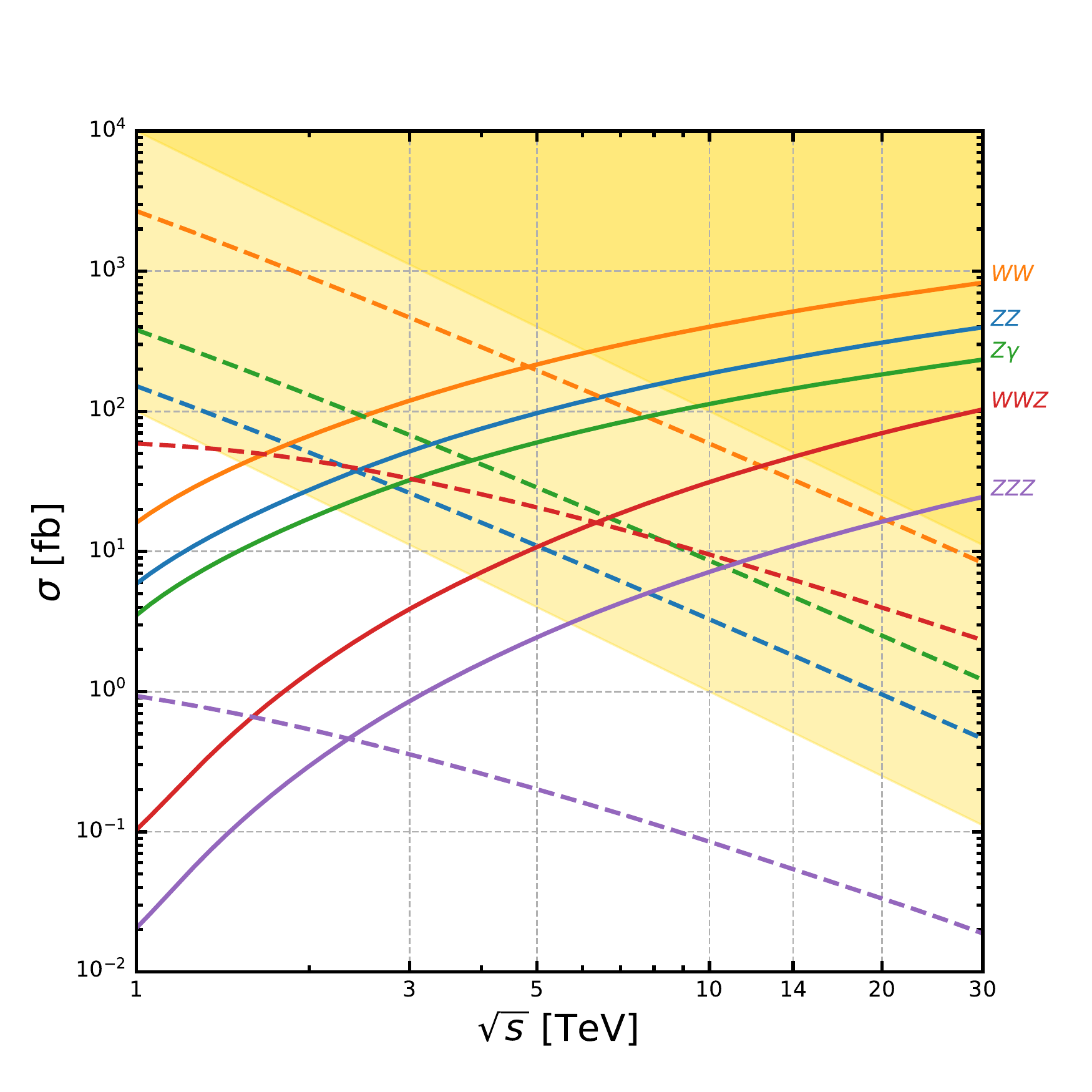}
\caption{\label{fig:VBS}Multi-boson production cross sections at a muon collider as a function of colliding energy. Solid lines
show VBS production while dashed lines show $s$-channel ones. The former benefits from a logarithmic growth with respect to 
the latter that decreases with energy. The light-shaded region corresponds to at least 10k expected events, while the darker shaded region to 1M. Adapted from Ref.~\cite{Costantini:2020stv}.}
\end{figure}

\subsection{VBS prospects}

The aforementioned analysis seems to indicate that a high-energy muon collider is the perfect stage to fully exploit VBS and explore the EW sector. As a matter of fact, VBS provide unique sensitivity at high energies and is a fundamental  part of a successful program of EW precision measurements. At the LHC there are significant challenges since  signal-to-background ratios are not favorable. In particular, it is  diboson production which dominates, but this class of processes probe very different  kinematic regimes and interactions. At a muon collider the situation is reversed: diboson production is significantly smaller, and becomes negligible at high energy (see Fig.~\ref{fig:VBS}). In terms of SMEFT operators at dimension six, the most relevant ones that affect VBS physics are listed in Table~\ref{tab:eftList}.

\begin{figure*}[t!]
\centering
\includegraphics[width=\columnwidth]{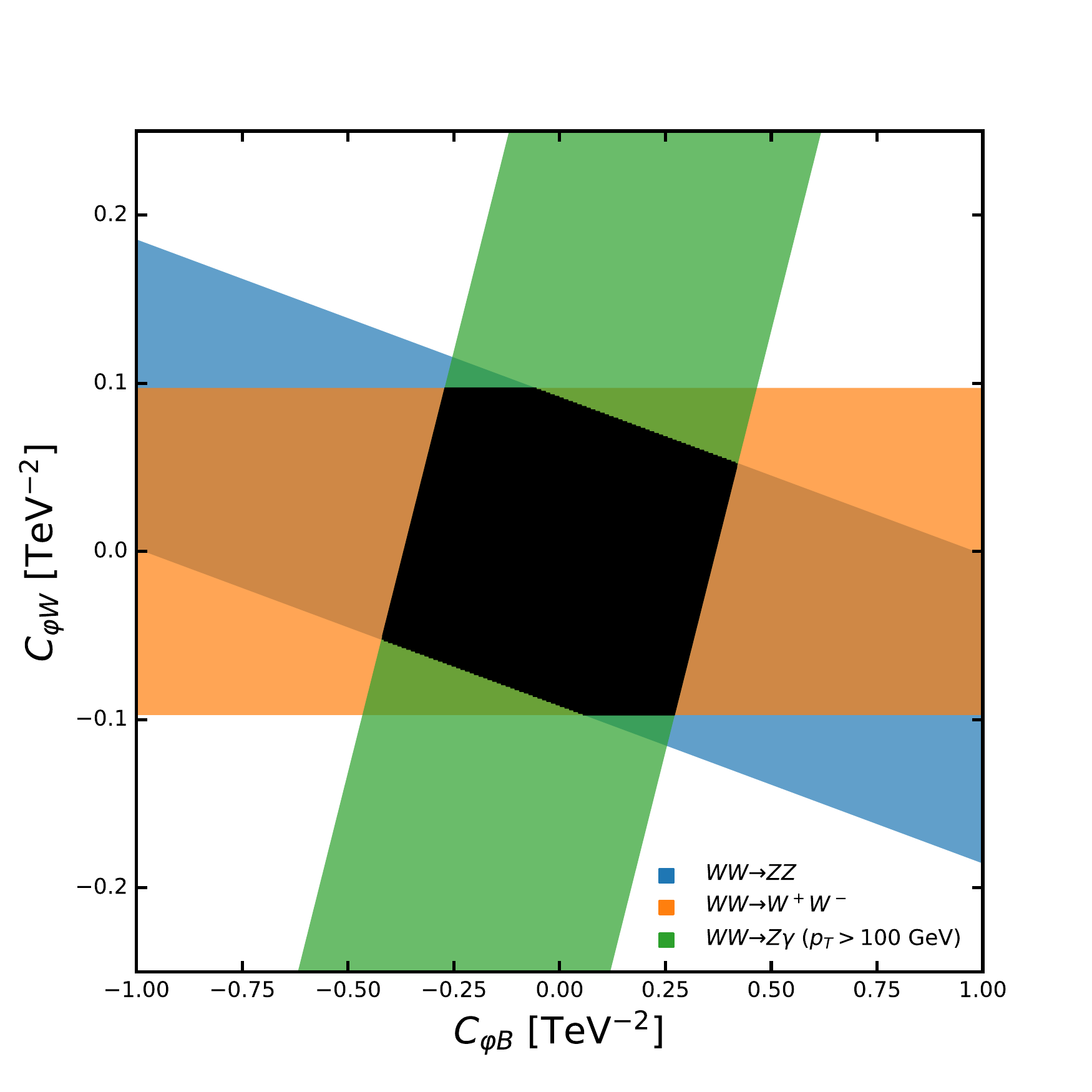}
\includegraphics[width=\columnwidth]{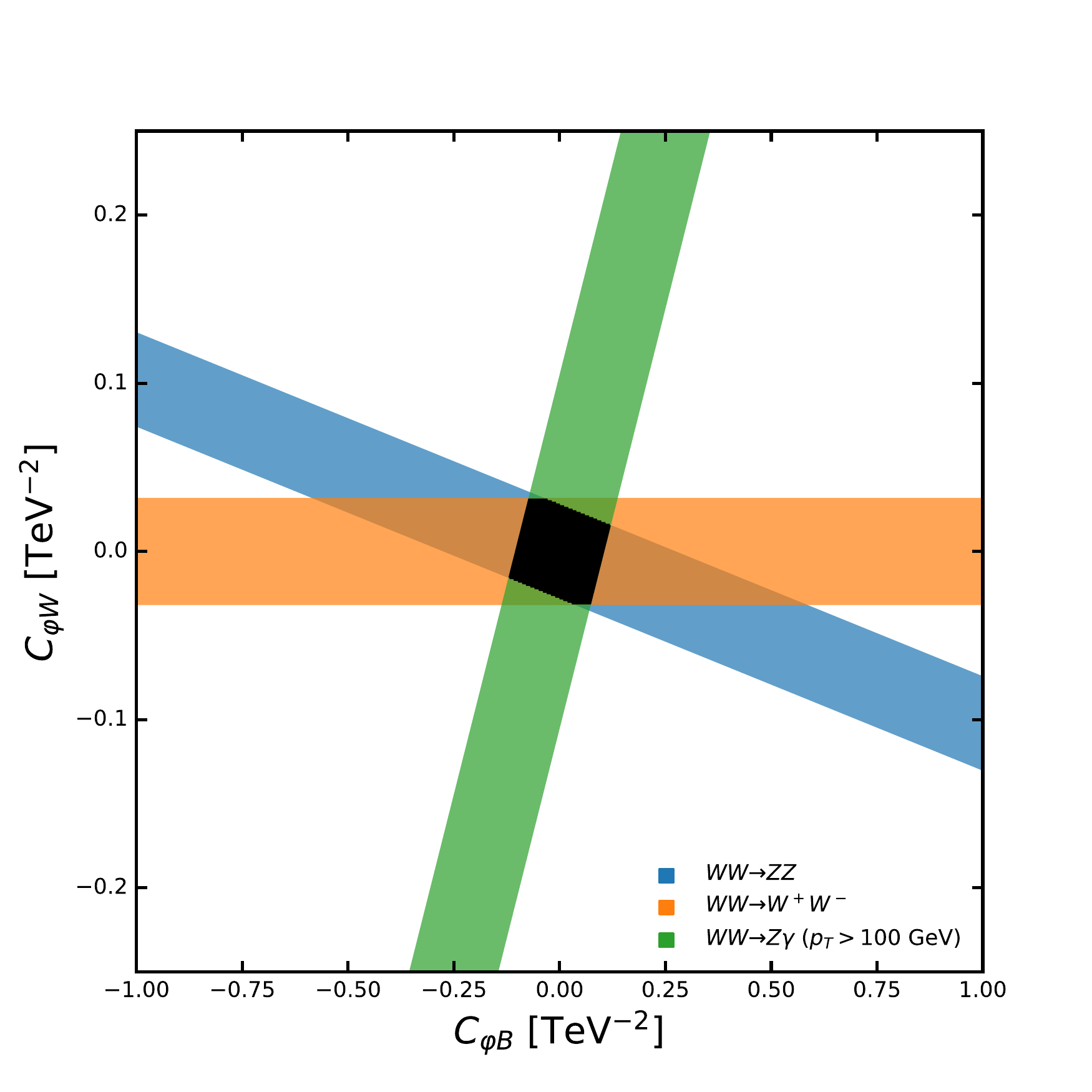}
\caption{
\label{fig:limits}
Projections for the limits imposed on the Wilson coefficients $\mathcal{C}$ of the operators $\Op{\varphi B}$ and $\Op{\varphi W}$ at a future muon collider at 3~TeV (Left) and 14~TeV (Right). Adapted from Ref.~\cite{Costantini:2020stv}.}
\end{figure*}

The listed operators modify in particular the self- interaction of the EW bosons and the coupling to the Higgs as well. They do so  with different Lorentz structures with respect to the SM ones. It is important to remember that the SMEFT should be treated globally, since it is unrealistic that only certain operators are generated by the full UV theory. However, currently no global analysis including VBS exist but simplified analysis can already bring useful information to the table. Recently in Ref.~\cite{Ethier:2021ydt} a first step to include VBS processes in a global fit has been taken. In particular, they perform a joint analysis with diboson data at the LHC and observe that VBS can improve the sensitivity to several dimension-six operators, while opening up new windows on operators like $\Op{\varphi B}$ and $\Op{\varphi W}$, which are left basically unconstrained by
diboson data. We can give a first estimate of the potential to constrain these operators at a high-energy muon collider, by assuming that a cross section measurement compatible to the SM is obtained. The total cross section in the SMEFT at linear level is given by
\begin{equation}
\sigma = \sigma_{SM} + \sum_i \frac{C_i}{\Lambda^2}\sigma_i \, ,
\end{equation}
where $i$ runs through the different operators affecting the process. One can project limits at $95\%$ CL on the Wilson coefficients $C_i$ by a simple signal-to-background estimate
\begin{equation}
\mathcal{S} = \frac{S}{\sqrt{B}} = \frac{|\mathcal{L}\cdot(\sigma - \sigma_{SM})|}{\sqrt{\mathcal{L} \cdot \sigma_{SM}}} \leq 2 \, .
\end{equation}
Here, $\mathcal{S}$ is the statistical significance of the simplified Gaussian estimator, and $S~(B)$ is the number of signal (background) events for an integrated luminosity $\mathcal{L}$. (See Ref. \cite{Delahaye:2019omf} for details on the projected luminosities at a muon collider.) In Fig.~\ref{fig:limits} we show the projections for benchmark collider energies $3$ and $14$ TeV, combining  three different processes.

\begin{table}[!t]
\begin{center}
\resizebox{\columnwidth}{!}{
\begin{tabular}{lll}
\hline\hline
  Operator $\qquad$ & Coefficient $\qquad$ & Definition \\
  \hline
$\Op{\varphi B}$ & $c_{\varphi B}$ & $\left(\pdp\right)B^{\mu\nu}\,B_{\mu\nu}$\\ 
$\Op{\varphi W}$ &$c_{\varphi W}$ & $\left(\pdp\right)W^{\mu\nu}_{\sss I}\,
W_{\mu\nu}^{\sss I}$ \\ 
$\Op{\varphi W B}$ &$c_{\varphi W B}$ & $(\varphi^\dagger \tau_{\sss I}\varphi)\,B^{\mu\nu}W_{\mu\nu}^{\sss I}\,$ \\ 
$\Op{\varphi D}$ & $c_{\varphi D}$ & $(\varphi^\dagger D^\mu\varphi)^\dagger(\varphi^\dagger D_\mu\varphi)$ \\ 
 $\mathcal{O}_{W}$&   $c_{WWW}$ & $\epsilon_{IJK}W_{\mu\nu}^I W^{J,\nu\rho} W^{K,\mu}_\rho$ \\
\hline\hline
\end{tabular}
}
\caption{Dimension-six SMEFT operators most relevant to VBS physics.}\label{tab:eftList}
\end{center}
\end{table}

The importance of combining multiple processes and measurements cannot be possibly understated. A single data point presents  flat directions in the Wilson coefficients parameter space and the combination of them is crucial to break the degeneracies and gain sensitivity. As expected, we see a striking improvement when we consider a multi-TeV machine as compared to a few-TeV one. In particular, this is the result of both a higher cross section (more statistics) and the energy enhanced operator effects, that concur in increasing dramatically the sensitivity.

\subsection{Summary}

Muon colliders are among the most promising machines to probe the EW sector by exploring both the precision and high-energy frontiers. In this section we discussed how muon colliders are effectively EW boson colliders, allowing us to fully exploit the sensitivity to new physics that the scattering of vector bosons can probe. In particular, we showed projections on the bounds that can be imposed on two operators that are still left unconstrained by diboson data at the LHC. This analysis is however very simplistic, including only inclusive cross sections. Since no observable optimization was performed, nor differential distributions used, this is arguably a conservative projection.


\section[BSM with VBS at $\mu^+\mu^-$ colliders]{\large BSM with VBS at $\mu^+\mu^-$ colliders}\label{sec:bsm_muonCo}

 \begin{figure*}[t!] 
\centering
\includegraphics[width=\columnwidth]{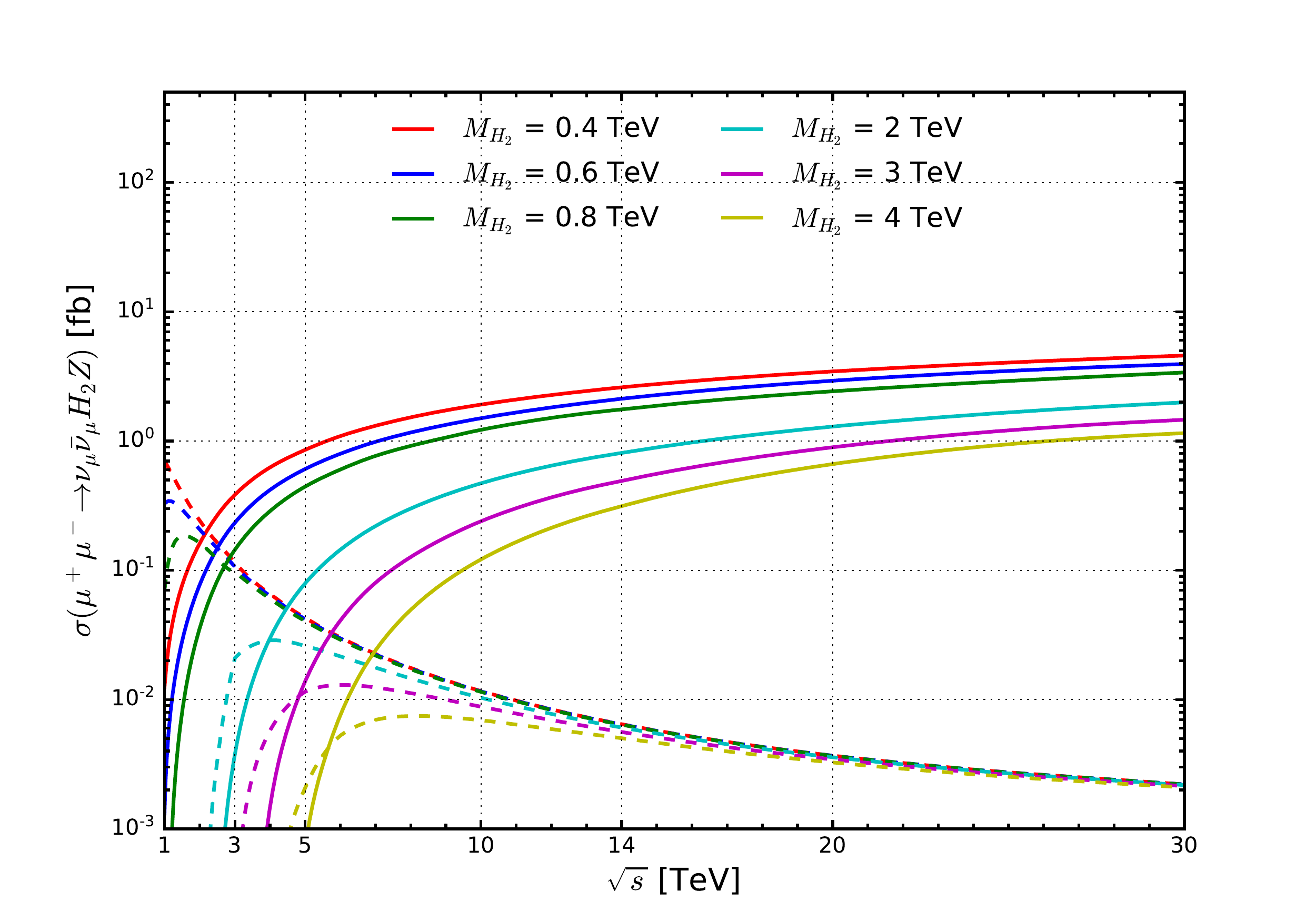}\label{fig:vbfschh2s_2hdm}
\includegraphics[width=\columnwidth]{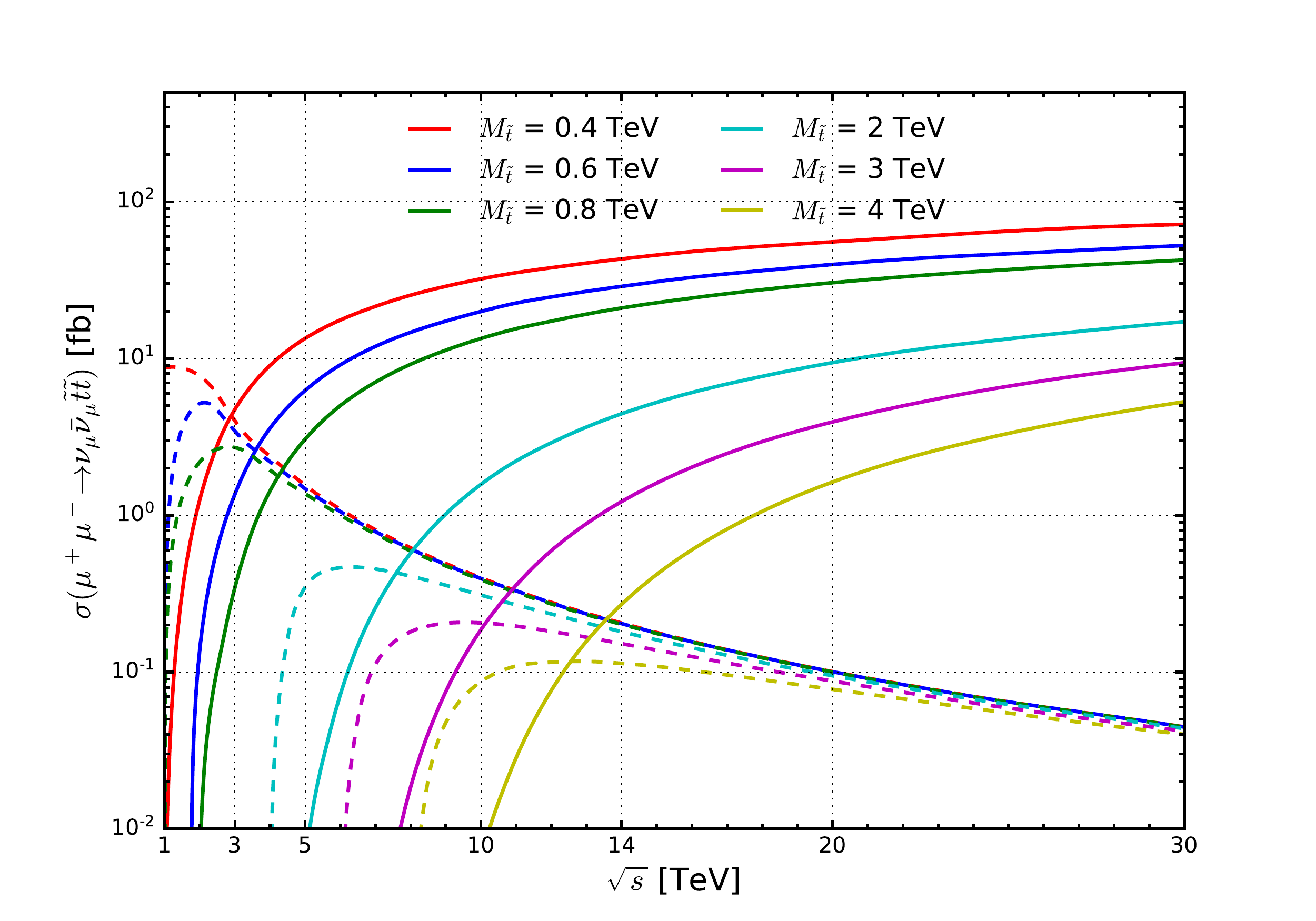}\label{fig:vbfschsttp_stop}
\caption{Cross section [fb] via VBF (solid lines) and  $s$-channel annihilation (dashed lines) for $H_2 Z$ associated production in the 2HDM (left) and $\tilde{t}\overline{\tilde{t}}$ pair production in the MSSM (right), from Ref.~\cite{Costantini:2020stv}.}
\label{fig:bsm_vbf}
\end{figure*}

The starting point of this section is the observation~\cite{Costantini:2020stv,Ruiz:2021tdt} that at sufficiently high energies  the EW VBS/VBF become the dominant production mechanisms at a multi-TeV lepton collider. We anticipate this holding for all SM final states relevant to studying  the EW sector and/or the direct search of (not too heavy) new physics. 

The main interest of this section is the BSM capabilities of a multi-TeV muon collider but we present first some results of relevant production processes via VBF, namely
\begin{align}
\mu^+\mu^-\to X\, {\nu}_{\mu}\overline{\nu}_{\mu}
\end{align}
 and their counterpart from muon annihilation (through the $s$-channel). Some cross sections of SM processes are shown in Table~\ref{tab:neutralSM}. As already stated, the VBF processes shown here and also their partner with muons in the final state ($\mu^+\mu^-\to X\, \mu^+\mu^-$) exhibit a logarithmic growth in the cross section with increasing collider energy, and eventually overcome the $s$-channel production processes~\cite{Costantini:2020stv}.

\subsection{New Physics Potential at a multi-TeV $\mu$ Collider}\label{sec:bsm_muonCo_bsmMu}

\begin{table*}[!t]
\begin{center}
\resizebox{.7\textwidth}{!}{
\begin{tabular}{l cc cc cc cc}
\hline\hline
\multirow{2}{*}{$\sigma$ [fb]} &        \multicolumn{2}{c}{$\sqrt s=$ 1 TeV} &    \multicolumn{2}{c}{$\sqrt s=$ 3 TeV} &    \multicolumn{2}{c}{$\sqrt s=$ 14 TeV} &      \multicolumn{2}{c}{$\sqrt s=$ 30 TeV}\\
		&VBF&s-ch.&VBF&s-ch.&VBF&s-ch.&VBF&s-ch\\
\hline
$t \bar{t}$           	&  4.3$\cdot 10^{-1}$ &1.7$\cdot 10^{2}$ &  5.1$\cdot 10^{0}$ &1.9$\cdot 10^{1}$ &  2.1$\cdot 10^{1}$ &8.8$\cdot 10^{-1}$ &  3.1$\cdot 10^{1}$ &1.9$\cdot 10^{-1}$ \\
$H$                   	&        2.1$\cdot 10^{2}$ &- &        5.0$\cdot 10^{2}$ &- &        9.4$\cdot 10^{2}$ &- &        1.2$\cdot 10^{3}$ &- \\
$H H$                 	&        7.4$\cdot 10^{-2}$ &- &        8.2$\cdot 10^{-1}$ &- &        4.4$\cdot 10^{0}$ &- &        7.4$\cdot 10^{0}$ &- \\
$W W$		&1.6$\cdot 10^{1}$&2.7$\cdot 10^{3}$&1.2$\cdot 10^{2}$&4.7$\cdot 10^{2}$&5.3$\cdot 10^{2}$&3.2$\cdot 10^{1}$&8.5$\cdot 10^{2}$&8.3$\cdot 10^{0}$\\
$Z Z$		&6.4$\cdot 10^{0}$&1.5$\cdot 10^{2}$&5.6$\cdot 10^{1}$&2.6$\cdot 10^{1}$&2.6$\cdot 10^{2}$&1.8$\cdot 10^{0}$&4.2$\cdot 10^{2}$&4.6$\cdot 10^{-1}$\\
\hline\hline
\end{tabular}
}
\caption{$W^+W^-$ fusion and analogous $s$-channel annihilation cross sections $\sigma$ [fb] for various VBF and s-channel processes in the SM as a function of collider energy $\sqrt{s}$ [TeV].}
\label{tab:neutralSM}
\end{center}
\end{table*}

In this section, we present a survey of  BSM models and the potential sensitivity of a $\mpmm$ collider. We note that BSM searches at future lepton colliders have been studied previously in the literature \cite{Buttazzo:2018qqp}. However the energy range considered was $\sqrt{s}\leq 3$ TeV. Conversely, we extend the energy up to $\sqrt{s}=30$ TeV to better exploit the cross section growth with energy due to VBF. Explicitly, we consider the $s$-channel annihilation and VBF processes
\begin{equation}
\mpmm ~\to~ X \quad\text{and}\quad
\mpmm ~\to~ X \ell \ell'.
\end{equation}
Here, $\ell\in\{\mpm,\overset{(-)}{\nu_\mu}\}$ and $X$ is some BSM final state, which may include SM particles. We focus on the complementary nature of the two processes because while $s$-channel annihilation grants accesses to the highest available partonic center-of-mass energies, it comes at the cost of a cross section suppression that scales as $\sigma\sim1/s$ when far above production threshold. On the other hand, in VBF,  the emission of transversely polarized, $t$-channel bosons gives rise to logarithmic factors that grow with the available collider energy (see, \eg, Section~\ref{sec:EWlumimumu} for additional details). Thus,  VBF  probes a continuum of mass scales while avoiding a strict $1/s$-suppression, but at the cost of EW coupling suppression.

As shown here and throughout previous sections, VBF production cross sections $(\sigma^{\rm VBF})$ grow with increasing $\sqrt{s}$, a phenomenon that follows from the propensity for forward emission of transverse gauge bosons at  increasing collider energies. While the precise dependence of $\sigma^{\rm VBF}$ on collider energies of course  depends  on the precise BSM signature, for example on the particles involved, their underlying dynamics, and their kinematics, it nevertheless contrasts with $s$-channel, annihilation processes. These processes feature cross sections $(\sigma^{s-ch.})$ that instead decrease with collider energy as $\sigma^{s-ch.}\sim1/s$, when well above kinematic thresholds. Hence, just as in the SM, we find a commonality in all VBF process here:  assuming fixed model inputs, then for sufficiently high collider energies, VBF cross sections exceed those of analogous, $s$-channel production modes.

Moreover we can roughly estimate the collider energy $\sqrt{s}$ at which $\sigma^{\rm VBF}$ surpasses $\sigma^{s-ch.}$ for a given final-state mass $M_X$.
Essentially, one must solve for when~\cite{Costantini:2020stv}
\begin{equation}
\frac{\sigma^{\rm VBF} }{\sigma^{s-ch.}} \sim 
\mathcal{S} \left(\frac{g_W^2}{4\pi}\right)^2   \left(\frac{s}{M_X^2}\right) \log^2\frac{s}{M_V^2}\log\frac{s}{M_X^2} > 1.
\label{eq:bsm_vbf_scaling}
\end{equation}
Here, $s$ is the total collider energy and the multiplicity factor  $\mathcal{S}$ accounts for the number of transverse polarization configurations contributing to the scattering process. This behavior is observed in the associated production of $H_2 Z$ in a 2HDM depicted in Fig.~\ref{fig:bsm_vbf}.

\begin{figure}[!t]
\centering\includegraphics[width=\columnwidth]{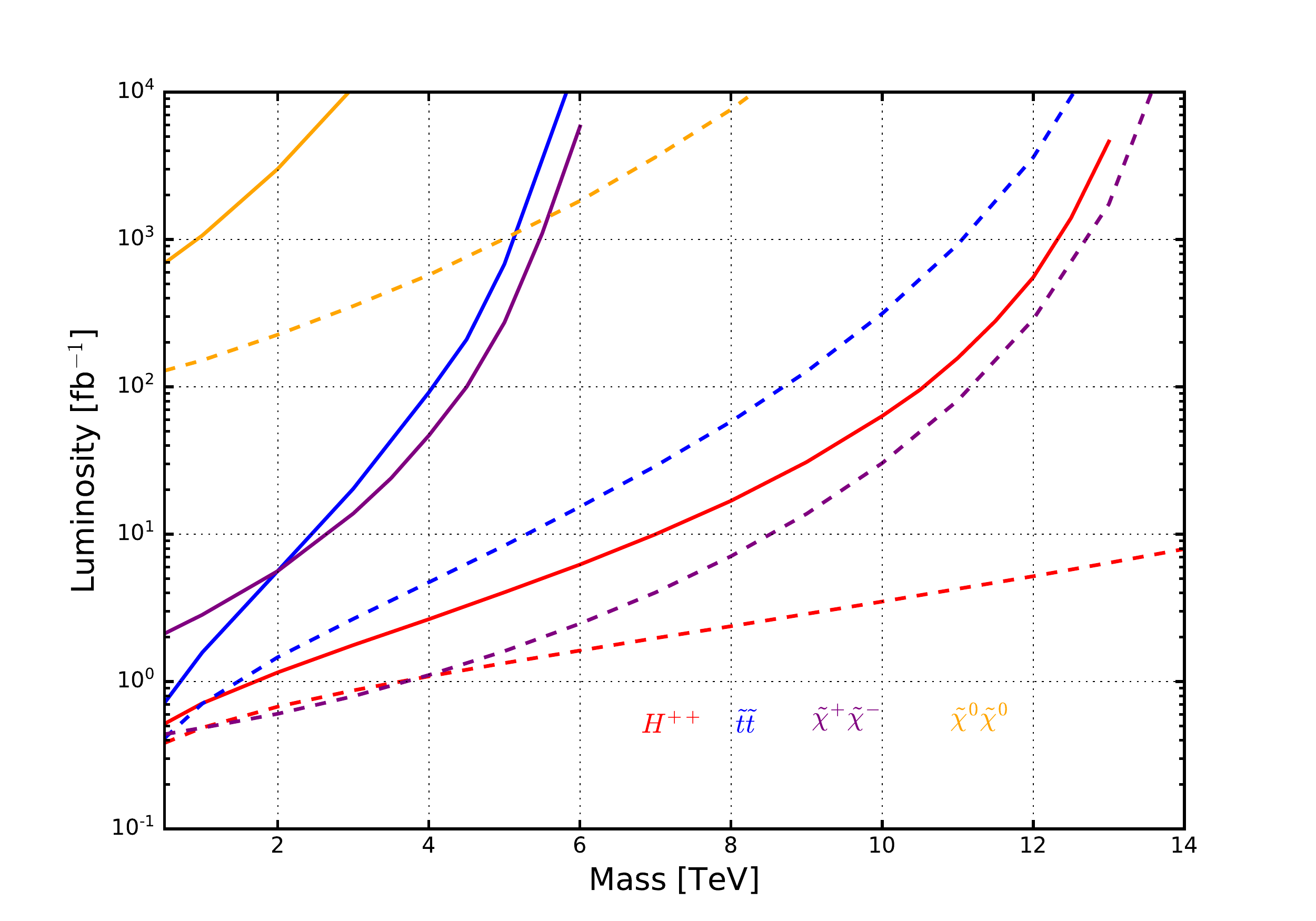}
\caption{Required luminosity [fb] for a  $5\,\sigma$ discovery of  $H^{++}$ (red) in the Georgi-Machacek model; $\tilde{t}\overline{\tilde{t}}$ (blue),  $\tilde{\chi}^+\tilde{\chi}^-$ (purple), and $\tilde{\chi}^0\tilde{\chi}^0$ (yellow)  from in the MSSM, using VBF in $\sqrt{s}=14$ TeV (solid) and $30$ TeV (dashed)  muon collisions. Adapted from Ref.~\cite{Costantini:2020stv}.    
}
\label{fig:sensi}
\end{figure}

Finally, Ref.~\cite{Costantini:2020stv} investigated the sensitivity of EW VBS to a variety of BSM scenarios at multi-TeV muon colliders. In order to give an overview picture of this reach, we present in Fig.~\ref{fig:sensi} the requisite integrated luminosity $\mathcal{L}$ [fb$^{-1}$] for a $5\sigma$ discovery as a function of new particle mass in $\sqrt{s}=14$ TeV (solid) and $30$ TeV (dashed) muon collisions. There, the authors considered specifically  the doubly charged Higgs $H^{++}$ (red) from the Georgi-Machacek model, $\tilde{t}\overline{\tilde{t}}$ (blue), $\tilde{\chi}^+\tilde{\chi}^-$ (purple), and $\tilde{\chi}^0\tilde{\chi}^0$ (yellow) pairs from the MSSM. As dedicated signal and background analyses are beyond the scope of this document, a zero background hypothesis and full signal acceptance was assumed. Likewise, Ref.~\cite{Costantini:2020stv} used as a simple measure of statistical significance $(\mathcal{S})$ the formula, $\mathcal{S}=\sqrt{\mathcal{L}\times\sigma}$. 
While this estimate is optimistic, it nevertheless gives a global picture of the ultimate physics capabilities of the collider.

\subsection{Complex Triplet Extension of the Standard Model}\label{sec:bsm_muonCo_cTSM}
In this Section we discuss the extension of the SM with a complex triplet with $Y=0$, which we label the complex Triplet extension of the SM (cTSM). The gauge and fermion sectors are identical to the SM ones, whereas the scalars of the model before EW symmetry breaking are
\begin{align}
\Phi=\left(
\begin{array}{c}
\phi^+\\
\Phi_0
\end{array}
\right),
\qquad
T=\frac{1}{\sqrt2}\left(
\begin{array}{cc}
t_0&\sqrt2\, t_1^+\\
\sqrt2\, t_2^-&-t_0
\end{array}
\right),
\end{align}
where $\Phi$ is a scalar doublet under SU$(2)_L$ and $T$ is a scalar triplet under SU$(2)_L$.
We describe here the main feature of the model and the interested reader could find more detail in Ref.~\cite{Bandyopadhyay:2020otm}. As already stated, the only difference with the SM relies in the scalar sector. Apart from the SM(-like) Higgs boson $h_D$, there is a second neutral Higgs boson $h_T$, a massive pseudoscalar $a_P$, and two massive charged scalars $h^\pm_{T,P}$. Due to symmetry reasons, the pseudoscalar and one of the charged Higgs bosons are aligned with their gauge eigenstates, even after EW symmetry breaking. The most important consequence is that the massive pseudoscalar is a natural dark matter candidate. In order to study the model one can apply the phenomenological constraint on relevant mass $(m_{h_D})$ and Higgs-mixing $(\mathcal{R}^S_{11})$ ~\cite{Bandyopadhyay:2020otm}:
\begin{align}
\label{scan2}
m_{h_D}=125.18\pm0.16\,\textrm{GeV},\quad \big|\mathcal{R}^S_{11}\big|>.99,
\end{align}
and also make sure that the loop-induced interaction $h_D\,\gamma\,\gamma$, which are modified by the charged scalars $h^\pm_{T,P}$, is consistent with the Higgs in diphoton signal strength \cite{Aaboud:2018xdt, Sirunyan:2018ouh}:
\bea\label{diphosign}
\mu^{\textrm{ATLAS}}_{\gamma\gamma}=0.99^{+0.15}_{-0.14}\;,\quad \mu^{\textrm{CMS}}_{\gamma\gamma}=1.10^{+0.20}_{-0.18}.
\eea

\begin{figure}[t!]
	\centering
	\includegraphics[width=.9\columnwidth]{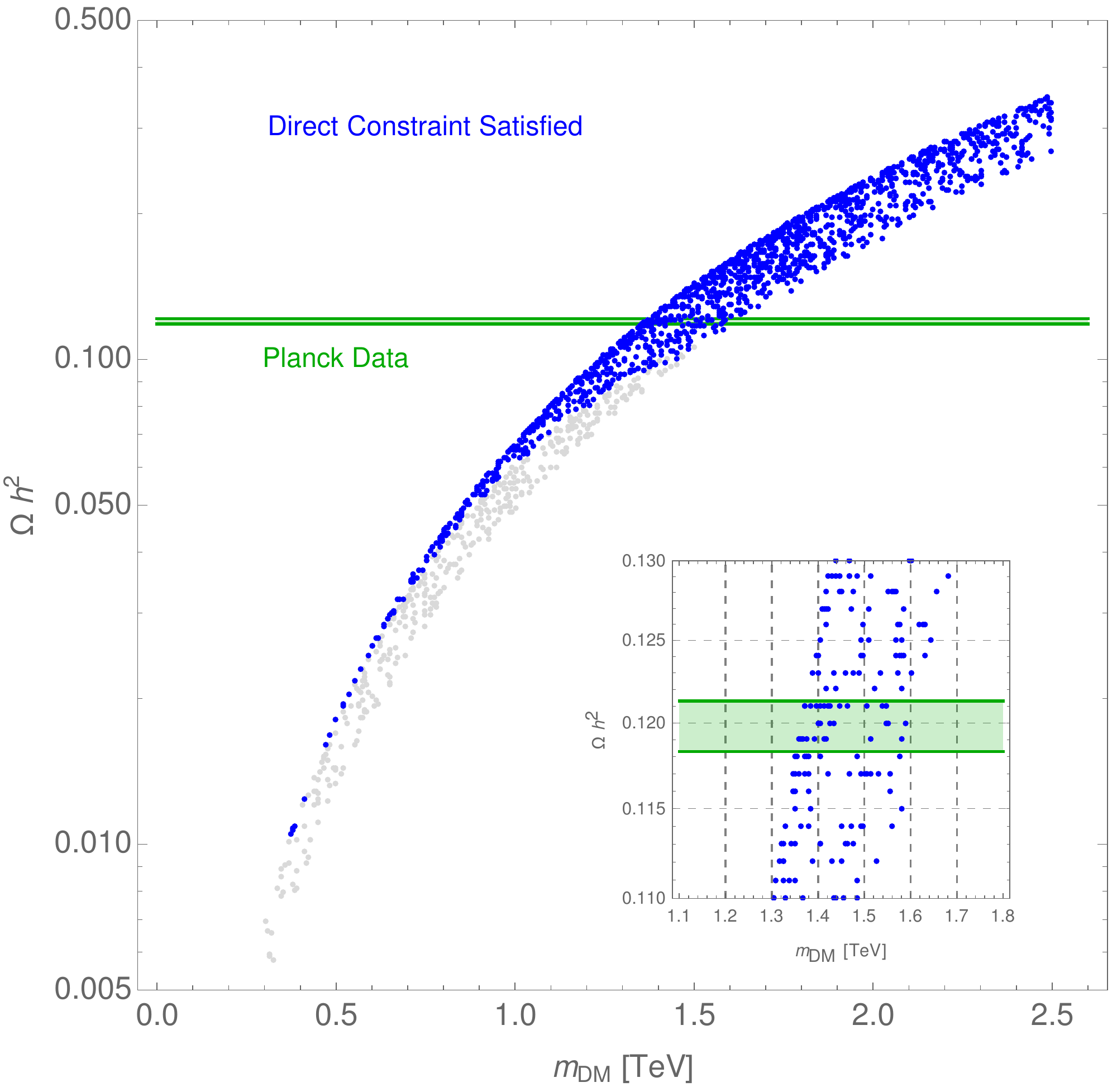}
	\\
	\includegraphics[width=.9\columnwidth]{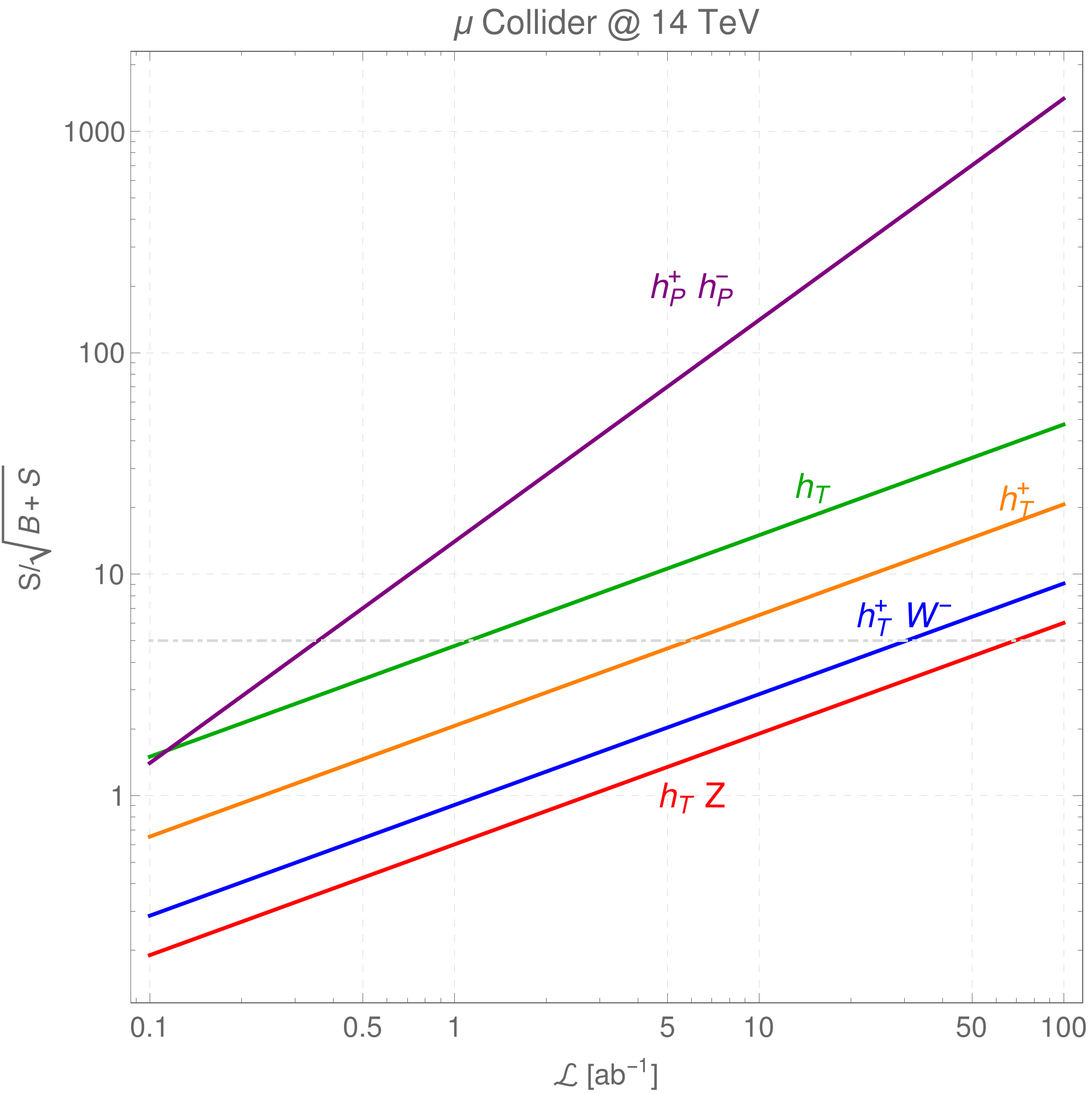}
	\caption{Upper: Relic density as a function of the DM. In blue (gray) the points that are allowed (disallowed) by the direct searches \cite{Aprile:2017iyp,Amole:2017dex,Akerib:2017kat}. The green area represent the Planck results \cite{Ade:2015xua}.
	 Lower: Significance vs luminosity for the $h_T$, $h_T^\pm$, $h_T^+\, W^-$, $h_T\, Z$ and $h_P^+ h_P^-$ production processes through VBF at a 14 TeV muon collider. The production cross sections are multiplied by the branching ratios $Br(h_T\to W^+W^-)$ or $Br(h_T^+\to W^+Z)$, depending on the channel considered. The background considered is the VBF production of $W^+ W^- Z$ in the SM, with $M(W^+ W^-)=m_{h_T}$ or $M(W^+ Z)=m_{h_T^+}$. The process $h_P^+ h_P^-$ is considered background-free because $h_P^\pm$ is a long-lived  state. Adapted from Ref.~\cite{Bandyopadhyay:2020otm}.}\label{fig:relicF}
\end{figure}

Results are presented in Fig.~\ref{fig:relicF}. In particular in the upper plot of Fig.~\ref{fig:relicF} we show $\Omega h^2$ versus $m_{DM}$, computed with \textsc{MadDM v.3.0} \cite{Ambrogi:2018jqj}. The blue points satisfy Eq.~\ref{scan2} together with the constraint from the diphoton signal, and are allowed by the constraint on direct dark matter searches. By looking at the zoomed plots, one can see that the value for the DM mass for which the correct dark matter relic is achieved is 
\bea
m_{DM}\equiv m^{min}_{a_P}\sim1.5\,\textrm{TeV},
\eea

Reference~\cite{Bandyopadhyay:2020otm} has also considered the physics reach for some processes at a muon collider. These results are presented in Fig.~~\ref{fig:relicF} (lower plot), where the significance is plotted as function of the luminosity for the $h_T$, $h_T^\pm$, $h_T^+\, W^-$, $h_T\, Z$ and $h_P^+h_P^-$ production processes through VBF at a 14 TeV muon collider. In the definition of the significance, $\sigma=S/\sqrt{S+B}$, $S$ and $B$ stand for the number of events for the signal and the background respectively,
\begin{align}
S:& \;\sigma(\mu^+\mu^-\,\to h_T \;\nu_\mu\bar\nu_\mu) \times Br(h_T\to W^+W^-) \cdot \mathcal{L},\\
B:& \;\sigma(\mu^+\mu^-\,\to W^+W^- \;\nu_\mu\bar\nu_\mu) \cdot \mathcal{L},
\end{align}
with $M(W^+ W^-)=m_{h_T}\pm5$ GeV.

A similar strategy is applied to the single production of $h_T^\pm$ and the pair-production $h_T Z$ and $h_T^+W^-$. For the charged scalar Higgs $h_T^\pm$, the branching ratio $Br(h_T^+\to W^+Z)$ was considered. This give a conservative estimate on the significance vs luminosity not because of the signal but for the higher cross section (via VBS) of $W^+W^-Z$ compared to $W^+W^-H$ \cite{Costantini:2020stv}. The pair production $h_P^+h_P^-$ has been considered background-free. The pure charged triplet $h_P^\pm$ has a single decay channel, namely $h_P^+\to a_P (W^+)^*$. Whereas the pseudoscalar is undetectable, the process will give rise to displaced off-shell $W$ boons: there is no SM process that have this particular final state.  This also gives rise to displaced leptons/jets plus missing energy in the final-state.

\subsection{Summary}\label{sec:bsm_muonCo_conc}
In this section, we explored a variety of simplified extensions of the SM and have shown how large VBS luminosities can maximize the direct search for new physics. This feature is similar in all the BSM models considered and constitutes a motivation for a multi-TeV muon collider, together of course with the sensitivity to SM and SMEFT that a muon collider can achieve. Also considered was a specific model with a scalar dark matter candidate showing the interplay between cosmological constraint and collider searches at a multi-TeV muon collider.

\section[\textit{HH} production from new particles at 100 TeV]{\large \textit{HH} production from new particles at 100 TeV}\label{sec:future_vbs_pp}

\begin{figure*}[!t]
\begin{center}
\includegraphics[width=\columnwidth]{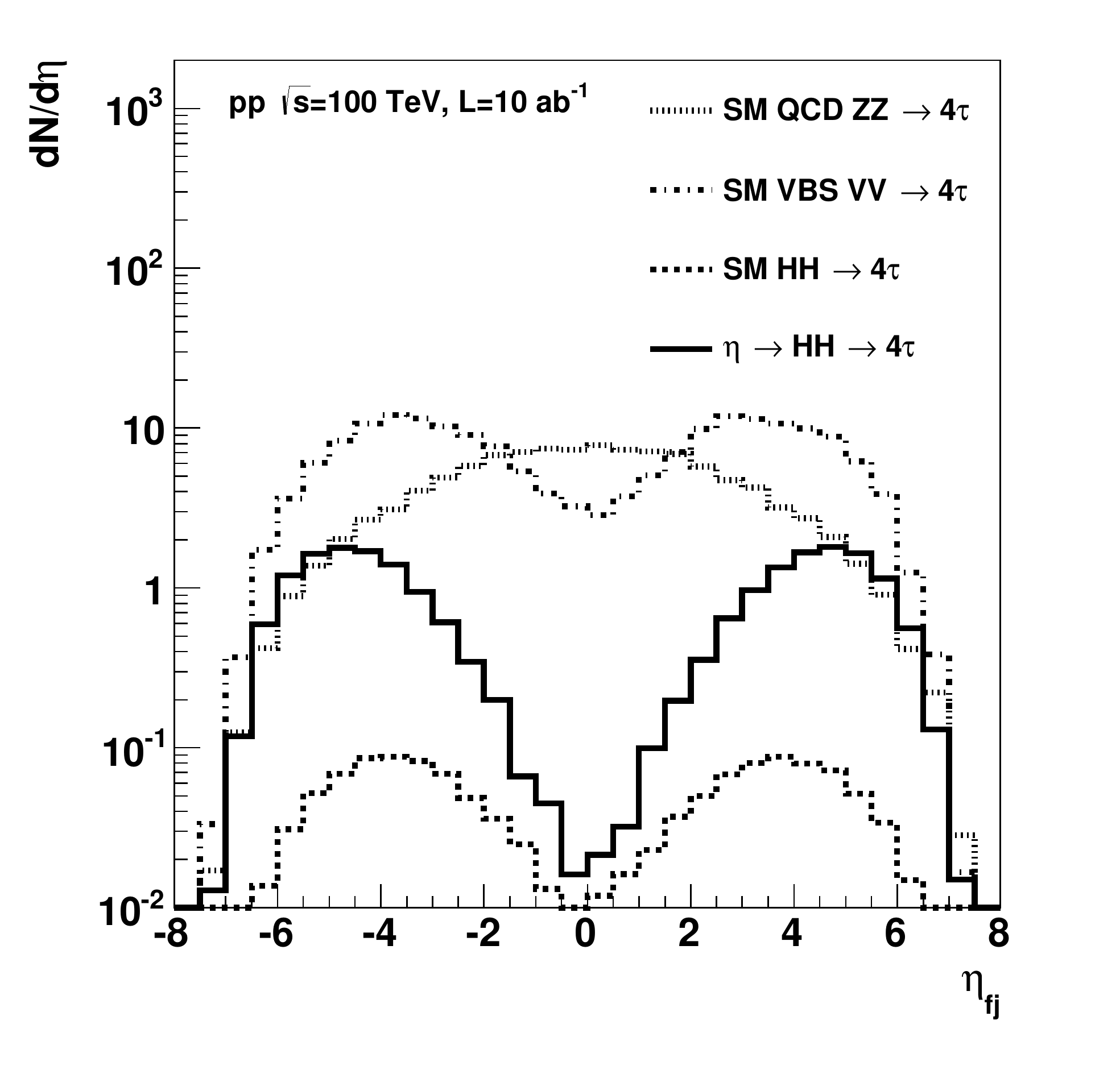}
\includegraphics[width=\columnwidth]{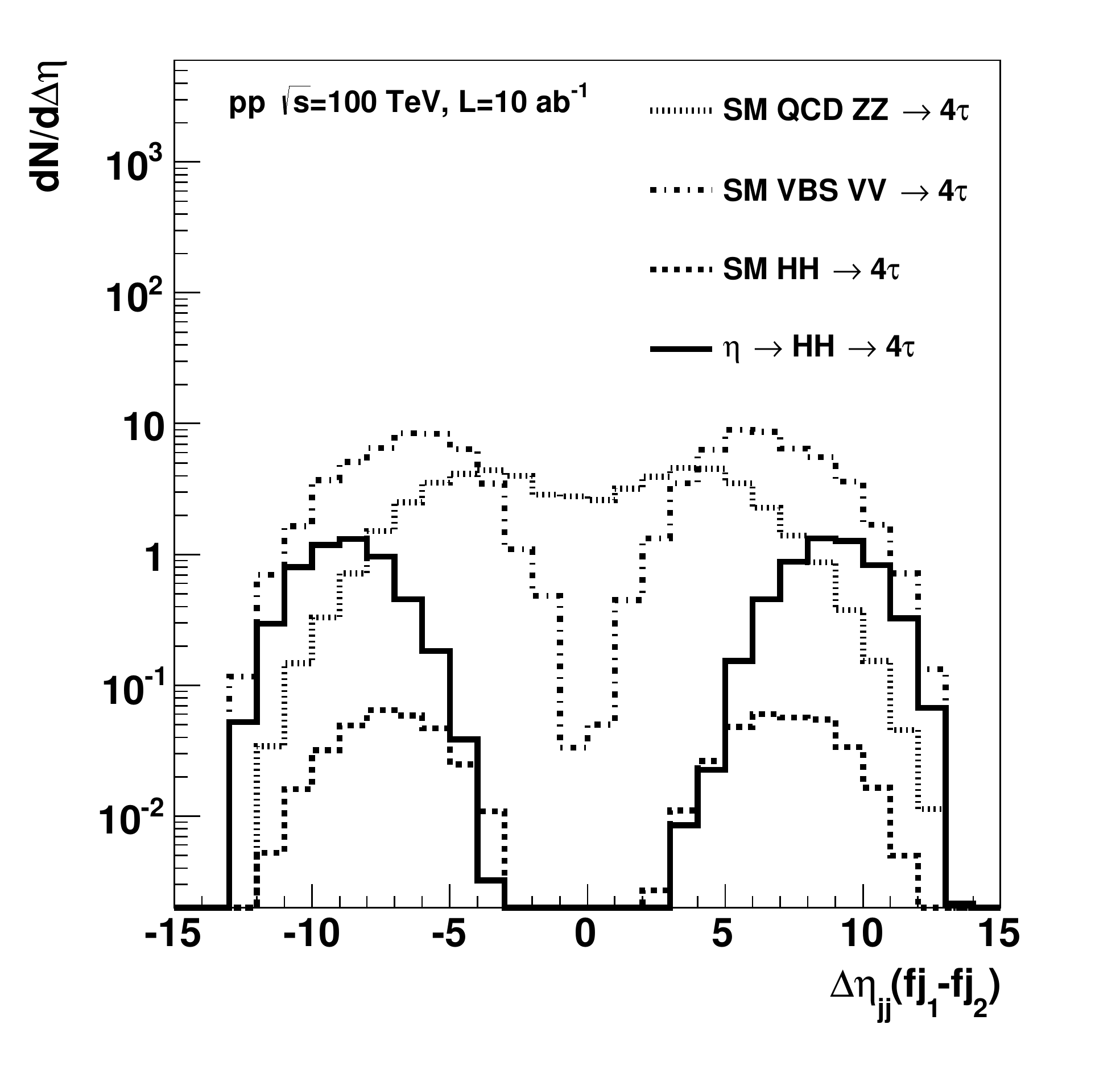}
\caption{Left: The pseudo-rapidity distributions of the forward jets. Right: The  distribution of the difference in pseudo-rapidities of the two forward jets. Figures are reproduced from Ref.~\cite{Kotwal:2015rba}. }
\label{VBSfigs1}
\end{center}
\end{figure*}

\begin{figure*}[!t]
\begin{center}
\includegraphics[width=\columnwidth]{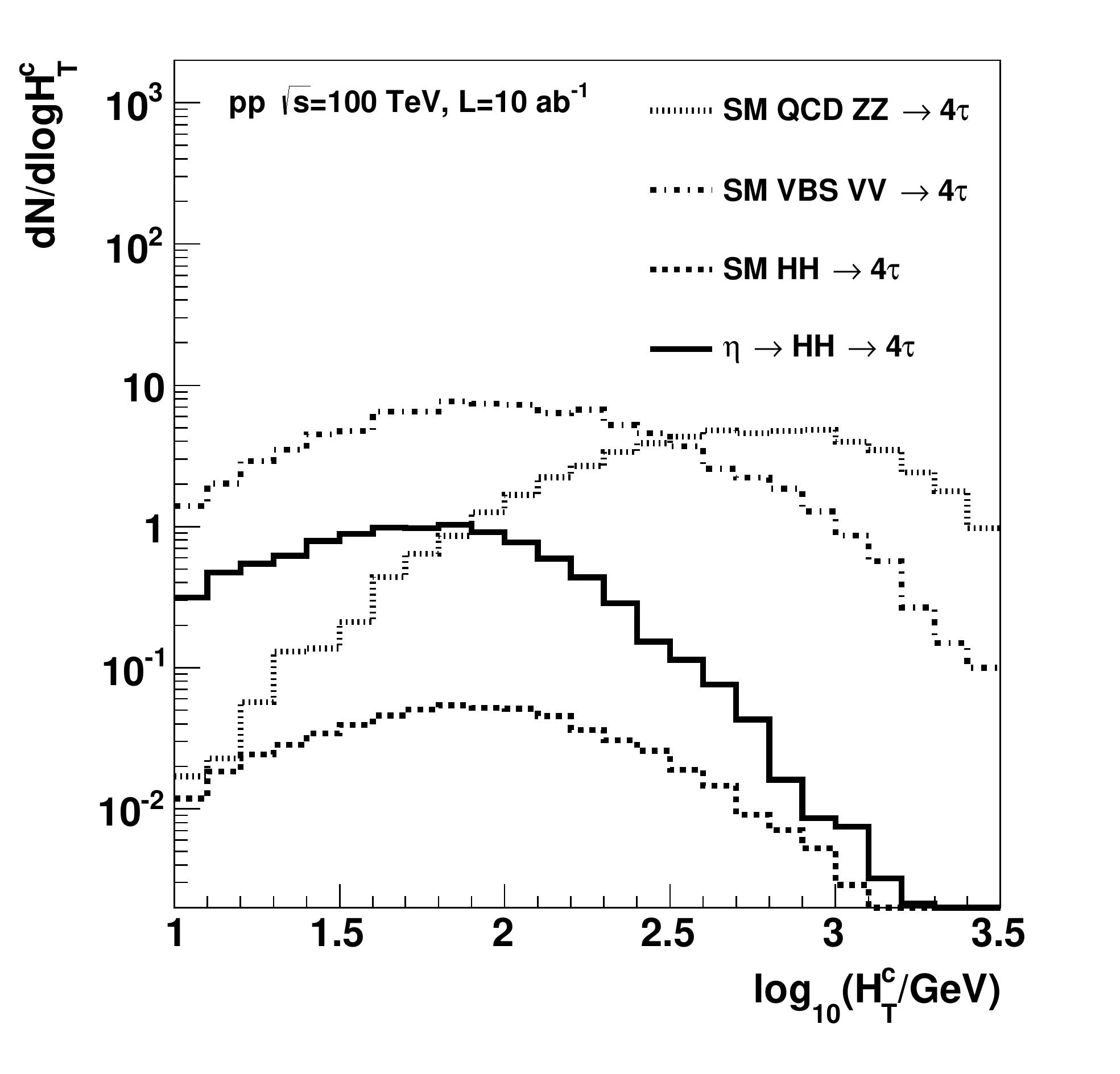}
\includegraphics[width=\columnwidth]{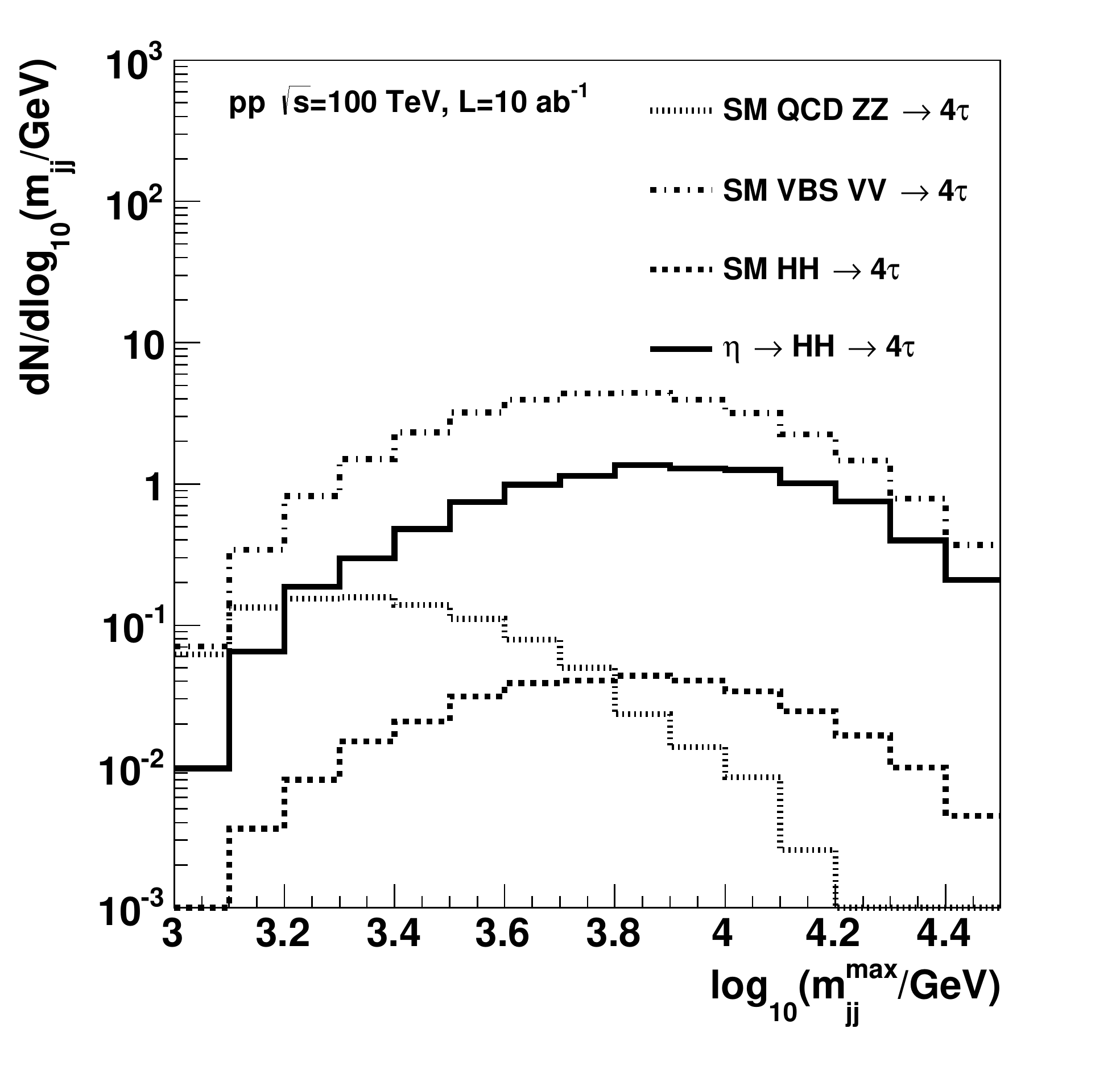}
\caption{Left: The distribution of $H^c_T$, the scalar sum of the transverse momenta of the central jets.
Right: The distribution of dijet mass of the forward jets. Figures are adapted from Ref.~\cite{Kotwal:2015rba}.}
\label{VBSfigs2}
\end{center}
\end{figure*}

In this final section, we turn to the outlook for $HH$ production in the decays of new resonances at a hypothetical 100 TeV $pp$ collider. In general, VBS offers a unique window into new dynamics in the Higgs sector.  In a class of models~\cite{Georgi:1984af, Georgi:1984ef, Kaplan:1983fs,Dugan:1984hq, Agashe:2004rs, Contino:2006qr},  the SM Higgs doublet field, which contains four real fields,  is itself a set of Goldstone modes, and are generated  from the spontaneous breaking of a larger global symmetry at a  new compositeness scale $f_\eta$. The strong dynamics at this scale would produce resonances which couple to the longitudinal polarizations of $W$ and $Z$ bosons and the Higgs boson, leading to typical VBS processes as well as double Higgs production via VBF~\cite{Bishara:2016kjn}.  In phenomenological model of Ref.~\cite{Contino:2011np}, the Lagrangian 
\begin{equation}
\mathcal{L} = \mathcal{L}_{\rm SM} + \frac{1}{2} \partial^\mu \eta \partial_\mu \eta - \frac{1}{2} m_\eta^2 \eta^2 + 
 \frac{a_\eta}{f_\eta}\eta \partial^\mu \pi^a \partial_\mu \pi^a,
\label{etaLagrangian}
\end{equation}
describes the interaction of a new scalar resonance $\eta$ with $\pi^a$, the quartet of  Goldstone modes  in the Higgs doublet. The resemblance of the notation to chiral perturbation theory is intentional. In the high-mass limit, the decay width of the $\eta$ resonance is given by~\cite{Contino:2011np}
\begin{equation}
\Gamma_\eta = \frac{a_\eta^2 m_\eta^3}{8 \pi f_\eta^2}
\label{etaWidth}
\end{equation} 
If the dimensionless coupling $a_\eta$ is set to $a_\eta  = 1 $, the longitudinal VBS process is completely unitarized by the combination of Higgs boson and $\eta$ boson exchange.

\begin{table}[!t]
\resizebox{\columnwidth}{!}{
\begin{tabular}{lcccccc}
\hline\hline
  & \multicolumn{6}{c}{$m_\eta$ (TeV)} \\
{$\cal L$} [ab$^{-1}$] & \multicolumn{2}{c} {$\sqrt{s} = 50$~TeV} & \multicolumn{2}{c} {$\sqrt{s} = 100$~TeV} & \multicolumn{2}{c} {$\sqrt{s} = 200$~TeV} \\
\hline
& $ 20$\% & $ 70$\% & $ 20$\% & $ 70$\% & $ 20$\% & $ 70$\% \\
\hline
1   & 1.26 & 1.89 & 1.75  & 2.81 & 2.27 & 3.85 \\
3   & 1.58 & 2.31 & 2.25  & 3.42 & 2.88 & 4.65 \\
10  & 2.02 & 2.83 & 2.90  & 4.18 & 3.66 & 5.63 \\
30  & 2.49 & 3.36 & 3.56  & 4.94 & 4.44 & 6.60 \\
100 & 3.06 & 3.97 & 4.33  & 5.83 & 5.38 & 7.74 \\
\hline\hline
\end{tabular}
}
\caption{$5 \sigma$ discovery mass reach for the $\eta \to HH \to 4 \tau$ resonance, as a function of the 
 $\sqrt{s}$ of a $pp$  collider. The fractional resonance width $\Gamma_\eta / m_\eta$  is fixed at 20\% and 70\%
 respectively. These results are reproduced from~\cite{Kotwal:2015rba}. }
\label{energyReach}
\end{table}

\subsection{Results}
A sensitivity study for this model was performed in Ref.~\cite{Kotwal:2015rba} with the di-Higgs production channel in VBS, using the {\sc Madgraph5}~\cite{Stelzer:1994ta,Alwall:2014hca}  and {\sc Pythia8}~\cite{Sjostrand:2014zea} generators to simulate $pp$ collisions. The study focused on the $H \to \tau \tau$ decay mode, using a $\tau$-tagging efficiency of 60\%. The  following irreducible background processes were included; (i) $VVjj \to 4\tau jj$ production ($V = Z, \gamma^*$) via purely EW couplings,  (ii) $ZZ jj \to 4 \tau jj$ production  via the presence of the strong coupling in the Feynman amplitudes,   and (iii) $HH jj \to 4 \tau jj$ production via purely EW couplings. Misidentified backgrounds were estimated and expected to be  negligible due the to large $\tau$ multiplicity.
More specifically, a $j\to\tau$ mis-identification rate of $\varepsilon_{j\to\tau}=2\%$ would lead to a suppression factor of  $(\varepsilon_{j\to\tau})^4 \lesssim 2\cdot10^{-5}\%$ for multi-jet processes, even before the application of strong selection cuts. In this analysis, all $\tau$ candidates were required to have $p_T(\tau_i)>100$ GeV for $i=1,\dots,4$, with the leading candidate having $p_T(\tau_1)>300$ GeV,
and a pseudorapidity of $\vert\eta(\tau_i)\vert < 3$.
A multivariate analysis was used to combine the information from various kinematic quantities associated with the resonance decay products and the forward jets produced by the scattered partons. The discovery potential  is shown in Table~\ref{energyReach}. Distributions of interest for the signal and background processes are shown in  Figs.~\ref{VBSfigs1} and~\ref{VBSfigs2}, which are reproduced from~\cite{Kotwal:2015rba}.

\subsection{Summary}
Using longitudinal VBS, a high-energy $pp$ collider with energies beyond the LHC would be able to probe high-mass resonances due to new strong dynamics in the Higgs sector. The discovery reach extends into the multi-TeV mass range for such resonances, depending on the intrinsic width of the resonance and the collider energy and luminosity. At a 100~TeV $pp$ collider with an integrated luminosity of 30 ab$^{-1}$, a scalar resonance of mass between 3.5 and 5~TeV (corresponding to a fractional width between 20\% and 70\%) is   discoverable using the $H \to \tau \tau$ channel alone~\cite{Kotwal:2015rba}.

The trade-off between collider energy and integrated luminosity for a given discovery mass reach favors higher luminosities for a narrow, weakly-coupled resonance and higher energies for a wide, strongly-coupled resonance. For a fractional resonance width of 20\% (70\%), a factor of two in collider energy is equivalent to a factor of 4.3 (8.7) in integrated luminosity~\cite{Kotwal:2015rba}.


\part[Conclusions]{Conclusions}\label{part:final_conclusion}

As outlined in the European Strategy Update for Particle Physics~\cite{Strategy:2019vxc,EuropeanStrategyGroup:2020pow}, many open questions and mysteries exist in high energy physics. These range from the more definitive, with question such as, ``How strongly does the Higgs couple to itself and other elementary particles?'', to the more open-ended, with questions such as, ``How does the EW sector behave in the massless limit?'' In many cases, these mysteries can be understood through the presence of new particles and new interactions. That is to say, through the existence of new physics.

At TeV-scale colliders like the LHC, the HL-LHC, or their proposed successor experiments, these problems and solutions can be studied directly using the vector boson scattering (and fusion) mechanism~\cite{Rauch:2016pai,Green:2016trm,Alessandro:2018khj,Bellan:2019xpr,Baglio:2020bnc,Covarelli:2021gyz}. More specifically, with VBS one can study at the highest attainable energies:  the self-coupling of the Higgs and EW bosons,  the polarization of EW bosons and the respective roles played by longitudinal and transverse degrees of freedom, the existence of new particles and new interactions, boosted topologies and jet substructure, and even the behavior of top quarks and weak bosons in the near-massless limit. Exploring such rich physics, however, involves extreme challenges for detector development, which have since led to breakthroughs in hardware design, trigger design, and applications of machine learning.
Such needs have also led to a revolution in theoretical developments, including: 
advances in EW parton showers and EW boson PDFs,
advances in high-precision computations and simulation tools beyond NLO in QCD with parton shower matching,
advances in Monte Carlo support for studying new physics models and EFTs at colliders,
and novel investigations of VBS at future colliders.

These achievements and others have been made by a sizable component of the high-energy community. A snapshot of ongoing efforts to study VBS at current and future colliders is summarized in this report. The outlook is encouraging for the LHC and HL-LHC, and outright inspiring for future, multi-TeV colliders. With VBS, one hopes and anticipates to at last resolve some of the deepest questions surrounding how nature works.


\section*{Acknowledgments}

The VBSCan (COST Action CA16108) and its companion networks are thanked for organizing the VBS at Snowmass meeting, which inspired this work, and for their financial support.  All the members of the Action and also thanked for inspiring discussions. Talks presented at the VBS at Snowmass meeting are available from the URL:\\  \href{https://indico.cern.ch/e/VSBCanSnowmass}{https://indico.cern.ch/e/VSBCanSnowmass}. 

AC acknowledges support from FRS -- FNRS under project T.04142.18.
AD acknowledges financial support by the German Federal Ministry for Education and Research (BMBF) under contract no.\ 05H18WWCA1 and the German Research Foundation (DFG) under reference numbers DE 623/6-1 and DE 623/6-2. 
AVK acknowledges support from the U.S.\ Department of Energy, Office of High Energy Physics grant no.\ DE -- SC0010007.
DBF acknowledges financial support from the Knut och Alice Wallenberg foundation under the grant KAW 2017.0100 (SHIFT project).
JL and RZ acknowledge the support of National Natural Science Foundation of China (NNSFC) under grant number 11905149.
JRR acknowledges the support by the Deutsche Forschungsgemeinschaft (DFG, German Research Association) under Germamny's Excellence Strategy -- EXC 2121 "Quantum Universe" -- 39083330.
KL acknowledges support by the European Research Council under the European Union’s Horizon 2020 research and innovation program (grant agreement No. 715871, DIMO6fit). Section~\ref{sec:sec:vbs_atlas_proj} is submitted by KL on behalf of the ATLAS Collaboration, Copyright 2020 CERN. 
MG acknowledges the support of the Funda\c{c}\~ao para a Ci\^encia e a Tecnologia, Portugal.
RR acknowledges the support of the Polska Akademia Nauk (grant agreement PAN.BFD.S.BDN. 613. 022. 2021 - PASIFIC 1, POPSICLE). This work has received funding from the European Union's Horizon 2020 research and innovation program under the Sk{\l}odowska-Curie grant agreement No.  847639 and from the Polish Ministry of Education and Science.
RR also acknowledges the support of Narodowe Centrum Nauki under Grant No. 2019/ 34/ E/ ST2/ 00186. 
RV acknowledges support by the Science and Technology Facilities Council (STFC) via grant award ST/ P000274/ 1 and by the European Research Council (ERC) under the European Union's Horizon 2020 research and innovation program (grant agreement No.\ 788223, PanScales).
SD and CS acknowledge support by the state of Baden- W\"urttemberg through bwHPC and the DFG through grants no.\ INST 39/963 -- 1 FUGG and DI~784/3.
This work has received funding from the European Union's Horizon 2020 research and innovation program as part of the Marie Sk{\l}odowska-Curie Innovative Training Network MCnetITN3 (grant agreement no. 722104). 
This work was supported in part by the U.S. Department of Energy under grant No.\ DE -- FG02 -- 95ER40896, U.S. National Science Foundation under Grant No. PHY -- 1820760, and in part by the PITT PACC.
The reproduction of this review or parts thereof is allowed as specified in the CC-BY-4.0 license.

\bibliography{vbsCanSnowmass_main.bib}
\end{document}